# Knitting quantum knots

## Topological phase transitions in Two-Dimensional systems

Santosh Kumar Radha

I have worked on innumerable problems that you would call humble, but which I enjoyed and felt very good about because I sometimes could partially succeed...No problem is too small or too trivial if we can really do something about it.

–Richard P. Feynman

# Abstract

Before 1980, the principle of broken symmetry was the key concept for the classification of states of matter. The discovery of quantum Hall effect by Klaus von Klitzing, started a new paradigm in physics by providing the first example of a quantum phase transition where no spontaneous symmetry was broken. The unique feature of these quantum phases is the fact that even though two ground states share the same symmetry, they still belong to different phases (i.e. cannot be continuously tuned to each other through a smooth deformation of the Hamiltonian). This is a result of the fact that the wave functions that are defined on the n-dimensional Brillouin zone are "knotted" in non-trivial ways, despite having the same symmetry.
In this thesis, we start by describing a symmetry enforced nodal line semi-metal (NLSM) in the 2D flat form of honeycomb Group - $V$ and its non trivial thermo-electric response. We will then proceed to show that, upon buckling, the system undergoes its first phase transition from NLSM to 2 sets of oppositely wound unpinned Dirac cones protected by $\mathscr{C}_2$ symmetry. Further buckling leads to these unpinned Dirac cones annihilating in pairs at two distinct critical angle leading to a second topological phase transition to an insulating state. We then show that this seemingly innocuous insulating state is indeed a weak topological crystalline insulator. Furthermore, upon closer look, this insulating state turns out to be a Higher Order Topological Insulator (HOTI) that is protected by $\mathscr{S}_6$ symmetry. HOTIs are d-spatial dimensional systems featuring topologically protected gap-less states at their $(d-n)$-dimensional boundaries with d>1. In a broader context, we will see that the the topological properties of buckled Group - $V$ stem from the fact that they topologically belong to the class of Obstructed Atomic Limit (OAL) insulators. Combining all these, we will prove that annihilating pairs of Dirac fermions necessitate a topological phase transition from the critical semi-metallic phase to an OAL insulator phase. We also uncover the rich set of phases in the phase diagram in case of annihilating Dirac fermions and study their entanglement properties using entanglement entropy. We will demonstrate that these phases can be distinguished by a $Z_2 \times Z_2$invariant. Finally, based on the non-trivial topology of these systems, we propose the conceptual design of a quantized switch that is protected by topology and a mechanism to create configurable 1D wire channels by breaking the symmetry that is protecting the edge spectrum - the inversion ($\mathscr{I}$).
Last part of the thesis involves the remarkable discovery of a spin polarized 2D electron/hole gas at the surfaces of a well known system - LiCoO2. By mapping the first-principles computational results to a minimal tight-binding model, we will show that these surface states are related to a non-chiral 3D generalization of the quadripartite Su-Schriefer-Heeger (SSH4) model and has a topological origin.

# Thank you,

It was the end of my first year of graduate school, I entered room 104 Rockefeller. The next thing I knew I was crouched in a chair in Walter's office with a notebook on my knee and focused with everything I had on an impromptu lecture he was giving me on a a subject I didn't have the slightest clue on - Density Functional Theory. I had barely introduced myself when he'd started banging out equations on his trusted $A4$ sheets that often lay scattered in his desk. Trying to follow was like learning a new game, with strangely shaped pieces and arbitrary rules. It was a challenge, but I was excited to be talking to a real physicist about his real research. At the end of that meeting, I told him about my inexperience in condensed matter physics (Being an engineer who happened to work in gravitational physics, I barely knew what a band structure was), he smiled and said - *you can pick it up*, and so begin the journey. For encouraging and guiding me in this uncharted territory, ***thank you***.

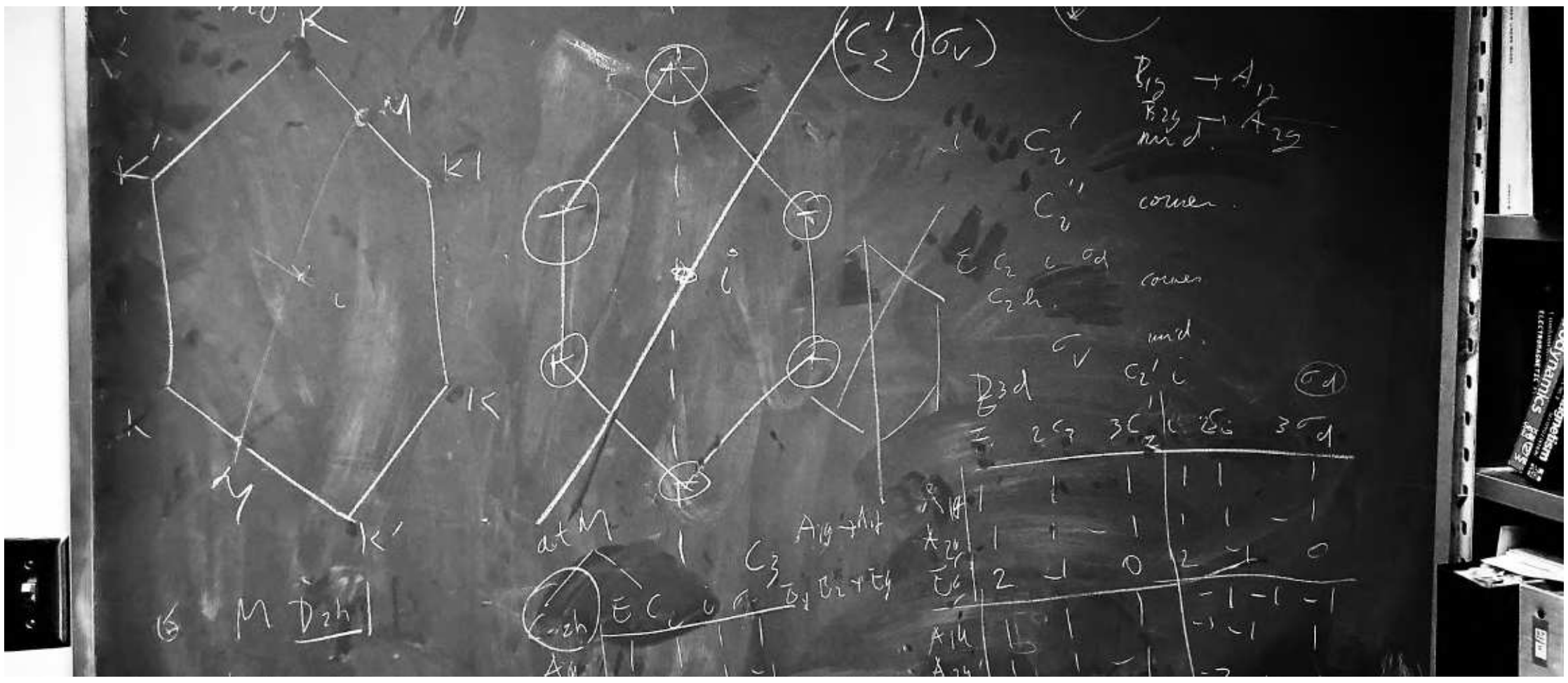

**Figure 1:** Figuring out the symmetries in $D_{3d}$ group at Rock-104

From then, throughout my 4 years here, we would convene in Walter's office and, like our first meeting, I would focus on following his logic and asking questions while he paced back and forth, thinking out loud, and banging out equations on the board. There where probably a lot of times when we had to close the door to not let the excitement of our discussions disturb our next door neighbors. At some point, after three or four hours, he might say something like "Well let's think about it for some time!" that signaled that he was happy enough with the direction he had found to let me forge ahead the path on my own, meaning I would spend the next day or two doing the detailed calculations that we speculated would take us to the next landmark. Sometimes, the path would be clear, but more often than not, there would be obstacle in the way. Either way, I would report back and then we would sink back into another session, only to realize thorough a call from his home, that it is half past 7. To the long hours spent explaining and exploring physics, ***thank you***.

*Thank you, Walter*

# Acknowledgments

This work would not have been possible without the support of my advisor, Walter R.L Lambrecht, who had guided me through this amazing four year journey. I am grateful to all of those with whom I have had the pleasure to work during this and other related projects. I would like to express my deepest appreciation to my committee members Xuan Gao , Harsh Mathur and Alp Sehirlioglu. Starting in gravitational physics, it was the a conversation I had with Prof.Mathur, four years back, that motivated me to explore condensed matter theory. I'm extremely grateful to Prof.Gao for engaging me whenever I randomly pop into his room with ideas/questions. I thank Prof.Alp for the wonderful discussions during our collaborative meetings. I would also like to thank my physics professor, Ravi Shankar, who introduced me to the idea of experiencing "intellectual discomfort".

Nobody has been more important to me in the pursuit of this project than the members of my family. I would like to thank my mother, Radha and grandmother Prema without whose love, support and understanding I could never have completed this doctoral degree and also for giving me the opportunity to choose a career I love the most - physics. I also wish to thank my Uncle (KVS) , who sparked the idea of having a career in *science* back in school. Most importantly, I'm deeply indebted to my grandaunt Vanaja, and granduncle Perry, for making this foreign place a second home to me. Finally, I must express my very profound gratitude to my mother's brother and sisters for providing me with unfailing support and continuous encouragement throughout my life. This accomplishment would not have been possible without them. Thank you.

I would like to thank the funding agencies who supported this work - Department of Energy, Devision of Basic Sciences and U.S. Air Force Office of Scientific Research along with High Performance Computing Resource in the Core Facility for Advanced Research Computing at Case Western Reserve University.

Finally, I would also like to extend my appreciation to the janitors of Rockefeller, who gave me company and kept me engaged during my late night stay in my office.

*This thesis is dedicated to my grandmother, Prema Rangachari.*

# Contents

# List of Figures

# List of Tables

# Introduction 1

The establishment of quantum mechanics in the 1900's yielded a framework towards understanding the physical properties of atomic, molecular, and solid-state systems. Following that, Bloch focused on the behavior of metals and their electrical resistance[1]. In his thesis, he formulated the physics for electrons in a periodic crystal potential which leads to states extended over the crystal that electrons can occupy. These electronic states represent the (un)occupied bands that can be used to explain many physical properties of crystals and formed the basis to understand the behavior of solid-state matter that we use in our daily life.

Until thirty years ago, to fundamental physicists, the boundary of a quantum system was largely just considered an annoying irrelevance that one wanted to get rid of [or make as small as possible] so one could focus on the bulk properties. However, a concept developed by Thouless and Kosterlitz in the early 1970's[2] led to the understanding that the boundary [edge] can actually tell us something fundamental about the bulk and is interesting in its own right. This made a beautiful marriage between Condensed Matter Physics and a truly abstract field of mathematics - Topology. This idea showed that, contrary to usual condensed systems, there are systems with electronic ordering that cannot be destroyed by thermal fluctuations/disorder. The ordering is realized due to the presence of "non-trivial" topology, which cannot be removed (at-least easily) to yield the ultimate ground state since they are topologically distinct.

It all started in Feb 1980, when Klaus von Klitzing was trying to understand the mobility of electrons in MOSFETs at the High Magnetic Field Laboratory in Grenoble[3]. He found that upon applying strong magnetic fields to the samples, the Hall resistance did not follow what Drude's theory predicted. Instead of increasing linearly with magnetic field, it developed a series of plateaus where it remained constant for a range of fields. Moreover, the Hall conductance was quantized in multiples of $\frac{e^2}{h}$, with $e$ the elementary charge and $h$ Planck's constant. Latter, von Klitzing was awarded the Nobel Prize in Physics 1985 for the discovery of the quantum Hall effect. A year later, Robert B. Laughlin came up with a truly remarkable argument to explain the physics of the quantum Hall effect that relied on gauge invariance and static disorder[4]. Immediately, based on Laughlin's paper, Bertrand I. Halperin pointed out the existence of edge states in the mobility gap that arises due to static disorder and showed that these states formed one-dimensional chiral wires that carried current disipationlessly and where responsible for the sharp quantization of the conductance[5]. Little did he know at that time, that his was the first such evidence of a topologically non-trivial state of matter.

Although the "robustness" of this phenomenal Hall conductance was semi-explained by Halperin, there still remained a mystery that this quantization seemed to be oblivious to the shape or geometry of the sample. In 1982, Thouless showed that such a quantization should stem from the bulk material and using periodic potential in a square lattice threaded by a perpendicular magnetic field, he showed a simple model example where the conductivity was quantized in integer multiples of $\frac{e^2}{h}$[6].

It was serendipitous that just a couple of years later, Berry[7] pointed out an (as yet) unrelated idea that the phase of a wave function around a close loop carries non-trivial physics. Barry Simon[8], made the connection between the Berry phase and the formulation of Thouless integer quantization to form the mathematical frame work of topology in condensed matter systems. He showed that the integer found by Thouless and collaborators was the first Chern number[1] and is a topological invariant. Topological invariants are truly useful in topology to be able to decide whether two objects can be continuously connected or not, a result of which they are insensitive to details and provide information about global properties of such objects.

1: We will see this in a basic mathematical constriction later on

In 1988, Haldane[9] made a landmark contribution in which he demonstrated that breaking time-reversal symmetry (TRS) was the only true prerequisite for achieving a nontrivial quantum Hall conductance. He proposed a tight binding model of a honeycomb lattice with complex next-nearest-neighbor hopping, where by breaking time-reversal symmetry, he was able to open up a gap and render the system an insulator. He showed that if such a model is solved on a 1D periodic chain, instead of being solved in the bulk, chiral edge states appear whenever TRS is broken. Soon afterwards, Y. Hatsugai[10] rushed to establish a formal connection between Chern number and edge states, giving us the first glimpse of *"bulk boundary correspondence"*. *Bulk boundary correspondence* is the idea that there is a correspondence between the non-trivial topological nature of the bulk and the physics of the boundary/surface.

After a slew of research, in 2005, Charles Kane and Eugene Mele[11] came up with an idea that would put the field of topological systems at the forefront of modern condensed matter physics. They saw that SOC interaction in graphene, would lead to two time-reversed copies of the Haldane model, one for each spin. Although in this case, where time-reversal symmetry is unbroken, making the Chern number identically zero with no quantized conductivity, each spin shows opposite non-zero Chern number. This implies that one can obtain a quantized *spin Hall conductivity* response. This invariant number is called the $Z_2$ invariant and the systems with non-zero $Z_2$ are called quantum spin Hall insulators. Sadly, it turns out that due to small SOC, an experimental observation of the quantum spin Hall effect in graphene is very hard. Andrei Bernevig, Taylor L. Hughes *et.al.* came up with an interesting solution inspired from Pankratov's 1987 work[12] for experimentally observing the quantum spin Hall effect[13] using HgTe quantum wells which was verified a year latter experimentally by Markus König[14].

**Remark 1.0.1** Even before Haldane, Kane and Mele showed the occurrence of edge states in the graphene 1D nano-ribbon, it was already known that graphene could host edge states depending on the edge termination even without spin-orbit or TRS breaking perturbations[15, 16]. In a very interesting paper, Hatsugai and Ryu[17] showed that the origin of this edge state indeed has again a topological origin, which we will discuss later on. It turns out, although the global topology of the system is trivial in the spin-less case, the edge states steam from the localized non-trivial topology of the system, namely the existence of Dirac cones with opposite winding numbers.

Fast forwarding a few years, there has been a proliferation of new types of topological insulators, such as mirror symmetric insulators[18], non-symmorphic insulators[19], multipole insulators[20], shift insulators[21] and boundary obstructed topological phases [22], to name a few. In a more systematic approach in 2017, Barry Bradlyn *et.al.*[23] established a link between the topology of a material and its crystal symmetry, position and content of the orbitals that can encompass the idea of topological non-triviality to much more general symmetries than TRS. The basis of the approach is elementary band representations (EBRs)[24], an idea introduced by Zak[25] in 1969. They provided rules for all of the ways in which electron bands can connect in the Brillouin zone. They showed that topologically non-trivial insulators can be defined by Definition 1.0.1.

**Definition 1.0.1** *An insulator (or, more generally, a filled group of bands) is topologically non-trivial if it cannot be continued to any atomic limit without either closing a gap or breaking a symmetry.*

They also classified all possible phases one can access, or so they thought. Shortly after around 2018, Ashvin, HC Po and Watanabe[26] found yet another class of insulators which do not posses any atomic limit, but rather posses one when added to a trivial set of bands and called it the *fragile insulators*.

Our main focus on this thesis will be on 3 kinds of topological systems

- Obstructed Atomic Insulators
- Dirac Semimetals
- Higher Order Topological Insulators/ Multipole Insulators

## 1.1 Two-Dimensional Group - V systems

Since the discovery of graphene,[27] the world of 2D atomically thin materials keeps expanding. Of special interest are the elemental 2D materials, such as silicene, germanene, and the recently realized and earlier theoretically predicted antimonene and arsenene.[28–34] Unlike their isovalent analog phosphorene (monolayer black phosphorus)[35] which has a more complex buckled structure with fourfold symmetry, monolayer Sb and As are found to prefer the buckled honeycomb structure, known as $\beta$-Sb, which is also found in silicene and germanene.[36, 37] Interestingly, an almost completely flat honeycomb form was reported to be stabilized epitaxially on a Ag(111) substrate.[38] Thus, flat monolayer Sb and As may be the closest analog to graphene but with the interesting difference that there is one additional valence electron which places the Fermi level in between the usual $p_z$ derived Dirac point at $K$ (as in graphene) and a higher lying $\{p_x, p_y\}$ derived Dirac point.

The electronic structure studies thus far report an indirect band gap for equilibrium buckled $\beta$-Sb but which undergoes a transition to a semimetallic state under tensile in-plane strain.[39, 40] It is related to a transition from a trivial to a non-trivial band gap inversion at Γ. Topological aspects of the band structure of various group-IV and V systems were studied by Huang *et al.* [41] and were also studied in few layer Sb films as function of thickness.[42–44] Flat honeycomb Sb was shown by Hsu *et al.* [45] to be a topological crystalline insulator.

In this thesis, we study this form of 2D group V systems as a function of buckling. In Chapter 4, we show that the Fermi level position near the intersection of two Dirac cones leads to a number of interesting topological features, from a uniquely shaped nodal line to several new Dirac points which are allowed to move as buckling increases and can mutually annihilate in pairs beyond a critical buckling angle. We also show that the nodal line in the flat form has a very unique transport property where the carrier type changes from electron to hole type every 60° as we go around the Fermi surface, an effect that has been named goniopolarity.[46]. This work was published in Phys. Rev. B[47].

Next in Chapter 5, we point out that the completely buckled insulator form that resulted from annihilation of all Dirac cones, is not a trivial insulator, but rather a *weak topological crystalline insulator*.[2] Moreover we also expose the hidden higher-order topology in this buckled form, where one gets protected corner states states in the 0-dimensional islands of the system. This work was posted on arxiv:2003.12656 and is currently under review.

2: Systems where the *detached* edge states are protected from gapping out by crystalline symmetry instead of time reversal symmetry

Finally we exploit this weak topological nature of buckled group - V systems in Chapter 6 to propose two new devices concepts

1. Topological quantum switch
2. Configurable quasi 1D semi-conducting wires

This research was posted on arXiv:2005.06096 and is under review.

After exploring the buckling of Sb systems, in Chapter 7, we show the universal behavior of annihilating Dirac cones, namely that it leads to Obstructed Atomic Limit (OAL) insulators (which will be explained in the next chapter).

## 1.2 (Li/Na)$CoO_2$ and surface states

$LiCoO_2$ is currently the predominant material used as a cathode in rechargeable Li batteries. The utility of $LiCoO_2$ lies in the fact that Li may be reversibly intercalated. Therefore, it is desirable to have a fundamental understanding of this system. Aside from being an important battery material, $LiCoO_2$ has gained a major interest in the fundamental physics community for two main reasons,

1. A mysterious anomalous first-order metal insulator transition in $LiCoO_2$[48]
2. close relation to its superconducting cousin $NaCoO_2$

The first mystery was solved by Chris Marianetti and Gabriel Kotliar[49], where they showed that the highly mobile Li impurities were responsible for the first order correlated transition.

However, the second mystery presides till to-date. It is well known that $NaCoO_2$ is superconducting, but only under certain conditions, namely, that it needs to be hydrated and a specific Na concentration must be present for the superconductivity. Surprisingly, though the chemical and electronic environment is exactly the same for $LiCoO_2$ , there exists no sign of superconductivity in it.

Here, in this thesis, motivated by experimental studies of Pachuta *et.al.*[50][51], where they were able to successfully exfoliate two dimensional forms of $LiCoO_2$ and $NaCoO_2$ , we show the remarkable occurrence of a spin polarized 2D electron/hole gas in this system. By mapping the first-principles computational results to a minimal tight-binding model corresponding to a non-chiral 3D generalization of the quadripartite Su-Schriefer-Heeger (SSH4) model, we show that these surface states have topological origin.

In Chapter 2, we provide some background information on the mathematical concepts of topology useful in the later chapters for the benefit of the reader, unfamiliar with these recent developments and their application in condensed matter physics. The goal here is not to be comprehensive but to be pedagogical.

In Chapter 3, we briefly describe the computational methods used throughout the later chapters. Since these are in large part well known and well described in textbooks and review articles, we provide only a brief factual description of the specific implementations of the methodology used here and refer the reader to the relevant literature.

Chapters 4-6 contains our results on the group-V 2D systems. Chapter 7 contains a generalization of the findingins about Dirac cone merging and chapter 8 presents our results on LiCoO2 surface states and their topological origin.

# Mathematical Introduction 2

## 2.1 Geometry of Quantum systems

### Differential geometry

Differential geometry is a topic in mathematics concerned with the description of manifolds, things like lines, surfaces, and volumes. In particular, Riemannian geometry (geometry of curved space) generalizes the concepts of Euclidean geometry (geometry of flat space). We will illustrate this concept using an example - the 3-Sphere

Let us start by considering a sphere in 3D space ($S_3$) which can be written as

$$\mathbf{r} = (x, y, z) = (r\sin(\theta)\cos(\phi), r\sin(\theta)\sin(\phi), r\cos(\theta)) \tag{2.1}$$

Now to each point $\mathbf{r}$, we will define a tangent plane[1] . Tangent planes are nothing but the usual tangents (the ones we are familiar from circles), but instead of being a line, it's a plane. Intuitively, they are a real vector space that contains the possible directions in which one can tangentially pass through at that $\mathbf{r}$.

1: Formally, Tangent space for a given point $x$ is a set of tangent vectors, where tangent vectors are an equivalence class of curves passing through $x$ while being tangent to each other at $x$.

$$\begin{aligned} \frac{\partial \mathbf{r}}{\partial \theta} &= (r\cos(\theta)\cos(\phi), r\cos(\theta)\sin(\phi), -r\sin(\theta)) \\ \frac{\partial \mathbf{r}}{\partial \phi} &= (-r\sin(\theta)\sin(\phi), r\sin(\theta)\cos(\phi), 0) \end{aligned} \tag{2.2}$$

Equation 2.2 defines a set of points that form a plane that is tangent to the defined sphere Figure 2.1 shows an example of such tangent planes.

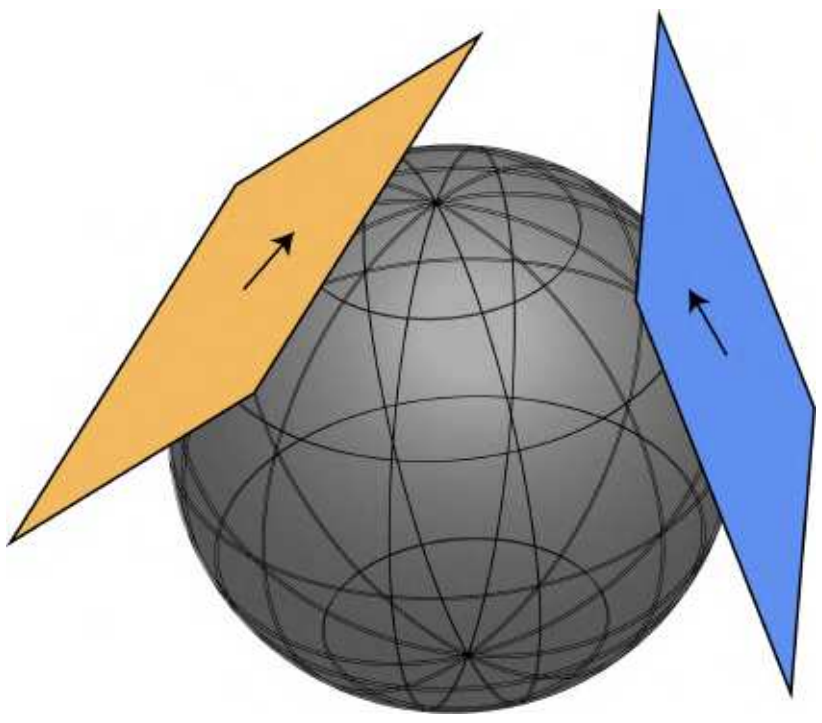

Figure 2.1: Example of tangent planes on 3D sphere

We can now consider a new object which is a set formed by both the sphere and tangent planes $s = \{sphere, Tangents\}$. The construction of this set might seem silly at first, but note that this is needed as we now have tangents that are in different planes which need not be parallel to each other ( as seen in Figure 2.1).[2] The construction we have is

2: This construction is not explicitly needed in Euclidean geometry (flat space), as all tangent planes are parallel,for example of a two-dimensional plane, all tangent vectors lie inside that same plane, hence base space and the tangent space can be thought of as same.

1. The points in manifold that live in base space
2. These points are accompanied by vectors that live in the tangent space that is attached to the manifold at the point where the vector has its base.

The manifold can in general have any dimension and shape, but the tangent spaces are always Euclidean spaces with the same number of dimensions as the manifold.

In flat space, it is obviously possible to move from one point in the manifold to another along a vector. But, since the vectors no longer lie within the manifold itself in Riemannian (curved) geometry, it may seem like this possibility is lost once more general spaces are considered. However, it should be noted that vectors can be considered to lie within the manifold as long as they are infinitesimal. [3] . It is therefore still possible to move from one point in the manifold to another following tangent vectors, as long as we take a series of infinitesimal steps along the surface. Each step moves us to a nearby point, at which a new tangent vector in the tangent space attached to that point leads us on to the next point. The guide to do this is given by the object called "metric tensor" which adds one more structure to the previous set, telling us how to move along this base space and tangent space. It turns out that the metric contains enough information to uniquely relate any vector in the tangent space at one point, to the correct tangent vector at a point infinitesimally close[4] . The last mathematical ingredient needed for the next discussion is the idea of "Gaussian curvature". It turns out that using the notion of metric tensor (which adds structure to our base space and tangent space), one can also construct an entity called curvature[5] .

3: The reason is that the component perpendicular to the manifold goes to zero as the square of the infinitesimal length, while the parallel component only goes to zero linearly (a beautiful effect of limits in calculus).

4: In-fact, there exists a construct called a connection, which can be computed from the metric, and which facilitates parallel transport. When given any tangent vector and a direction in which to move, the connection tells us which tangent vector corresponds to it in the tangent space at the nearby point

5: The fact that this is called curvature is not coincidence and indeed reduces to the intuitive notion of curvature for 1D curves

Finally, we are now ready to get to the beautiful fact that we are just going to state and not prove - For two dimensional surfaces embedded in 3D space (like sphere), the curvature ($E$) when integrated over the entire surface turns out to be $4\pi$. This quantity remains unchanged under any continuous deformation of the sphere's shape.[52]

$$\int E dS = 4\pi \tag{2.3}$$

## Topology

Topology is a field of mathematics concerned with the classification of objects into classes, such that objects are considered equivalent if they can be continuously deformed into each other. This leads us to the ubiquitous example of the fact that topology cannot differentiate between a coffee cup and a doughnut. But a more relevant idea that is needed for this thesis is that the topological classes will not only depend on the objects, but also on the spaces they are embedded in.

This is shown in Figure 2.2. In the top part, the right torus (doughnut) belongs to a different equivalence class from the left one, because it is threaded by a string, which prevents it from being continuously deformed into the other one. In the bottom part, the right sphere cannot be smoothly deformed to the left sphere because of a obstruction inside of it. Clearly when put in the same base space[6] , these two objects on the left and

6: Base space are nothing but the space the object lives in. For ex. A 3-Sphere, lives in a 3 Dimensional Euclidean space, while special relativity has space-time as its base space

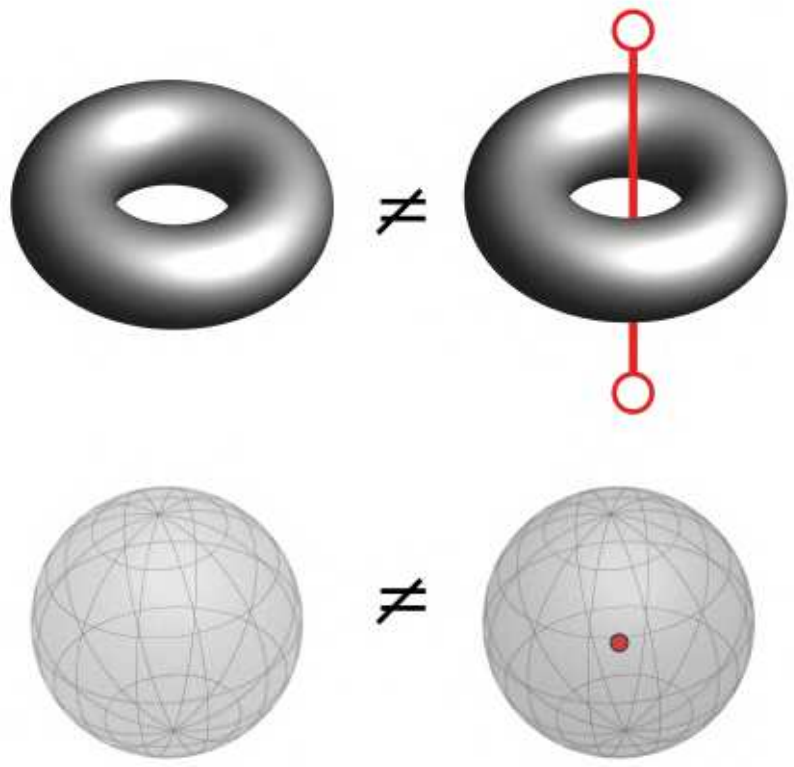

**Figure 2.2:** Topologically different base space. Left and right objects cannot smoothly be deformed to each other (top) because of threaded string, (bottom) because of obstruction inside the sphere

right are topologically equivalent and it is indeed just the base space that differentiates them. We can in this case think of the *embedding space* as consisting of ordinary three-dimensional space minus the points along the string (or) point inside the sphere. The number of holes is in this case still a topological invariant, as it stays the same under continuous deformations of the objects, and two objects that belong to the same class still necessarily have the same invariant. This shows that in the presence of a non-trivial base space, it is no longer guaranteed that two objects with the same topological invariant belong to the same equivalence class.

**Remark 2.1.1** A general feature of topological invariants is that, they are often necessary but not sufficient indicators of two object's topological equivalence. For further discussion on this topic, a beautiful reference would be Ref.[53]

[53]: Nakahara (2003), *Geometry, topology and physics*

Having seen the above examples of topological invariant (or rather lack of equivalence), we consider another example. In the previous section on differential geometry, we mentioned that the integral of the Gaussian curvature over the sphere remained invariant under continuous deformations of the sphere. It turns out that for any surface without borders, the integral is related to the number of holes $h$ through the Euler characteristic[52]

$$\chi = \frac{1}{2\pi}\int EdS = (2 - 2h) \tag{2.4}$$

This expression is useful, because it allows us to calculate the topological invariant, without having to rely on our ability to identify the number of holes by inspection. Thus, for a more generic system, it is common for a topological invariant to be calculated as an integral of some curvature over some manifold.[7] This is part of a more general framework, where a topological index, which is a special type of topological invariant, is related to an integral over a characteristic class[8] . In this case the index is the Euler characteristic, while the characteristic class is the Gaussian curvature. Very similar to this, what will be of interest to us is another topological index, the first Chern number and its relation to the first Chern class. The first Chern class turns out to be directly related to a curvature on what is known as a complex fiber bundle and integrating this over the manifold we arrive at the first Chern number.

7: A manifold is mathematically defined as a topological space that locally resembles Euclidean space near each point. More precisely, each point of an $n$-dimensional manifold has a neighborhood that is homeomorphic to the Euclidean space of dimension $n$.

8: characteristic class is intern derived from a curvature

## Fiber bundles

A fiber bundle is a generalization of the differential manifold. In particular, we saw that, to any given point in a manifold, a tangent space is attached. Now fiber bundles are nothing but a manifold together with the set of all tangent spaces with the manifold as base space and the tangent space as fiber. However, in contrast to the differential manifolds, where the fiber(tangent space) always is a Euclidean tangent space of the same dimension as the base manifold, fiber bundles are allowed to have any type of space as fiber. This thus elevates the structure of manifold and other things we saw previously to a more general framework. In the last section, we saw that tangent spaces contain vectors that, among other things, can point us in directions within the manifold base space, which lead us to the idea of "connections". Connections are nothing but objects that tell us how the various tangent spaces at points in the manifolds are *connected*. In the language of fiber bundle, since we use the notion of Riemannian distances[9] , connections tell us how nearby "fibers" are *connected*. When it comes to fiber bundles, we are often not interested in the base manifold (except for some interesting cases), or distances on that manifold, instead we are often interested in the behavior of the fibers themselves. Thus, often we start with a connection defined on the manifold, rather than a metric. This connection tells us how to "parallel transport" values in the fiber to nearby fiber

9: This is again because fibers can be any object. One can even attach shapes like square or circle of different radii to each point in space and they would form a fiber bundle!

The idea of fiber bundle is quite abstract. A simple example of the structure of fiber bundle would be a structure like (real-space,density) where our real space forms a 3D manifold, to which at each point a value of density is attached. Here the density forms the fibers. Since our fiber is a trivial object $\in \mathbb{R}$, there is no useful "topological" structure we can encode in this object. But even in case of simple objects like these, reformulating in terms of fiber bundles become beneficial. To illustrate this, we use the help of a brilliantly written book[54] to explore the use of fiber bundle structure in a example other than physics.

[54]: Schwichtenberg (2019), *Physics from Finance: A gentle introduction to gauge theories, fundamental interactions and fiber bundles*

Consider a one-dimensional manifold that represents time, say years. At each point of this one-dimensional manifold, we attach a one-dimensional real fiber, which can be used to indicate the median salary that particular time (year). With some thought, we can see that it is not insightful to directly compare values from different fibers, because the income one year cannot be sensibly compared to an income another year, if we do not also know the inflation rate in between those two years. To compare two incomes, we have to transport either of them from one year's fiber, to the other year's fiber, along lines that correctly adjust for inflation. The correct connection, is the one which is determined by parallel transporting between the fibers. This is shown in Figure 2.3, where we compare two fibers, one at time $t_0$ and another at time $t_1$, for which to make a meaningful comparison, one would want to parallel transport the fiber at $t_0$ to $t_1$ and then use what ever distance measure (like percentage change) one would like. More over, things get interesting if we want to do calculus in this space, say we want to calculate $\frac{dI}{dt}$ where $I$ is the income. We need to take into account the connection/parallel transport (or

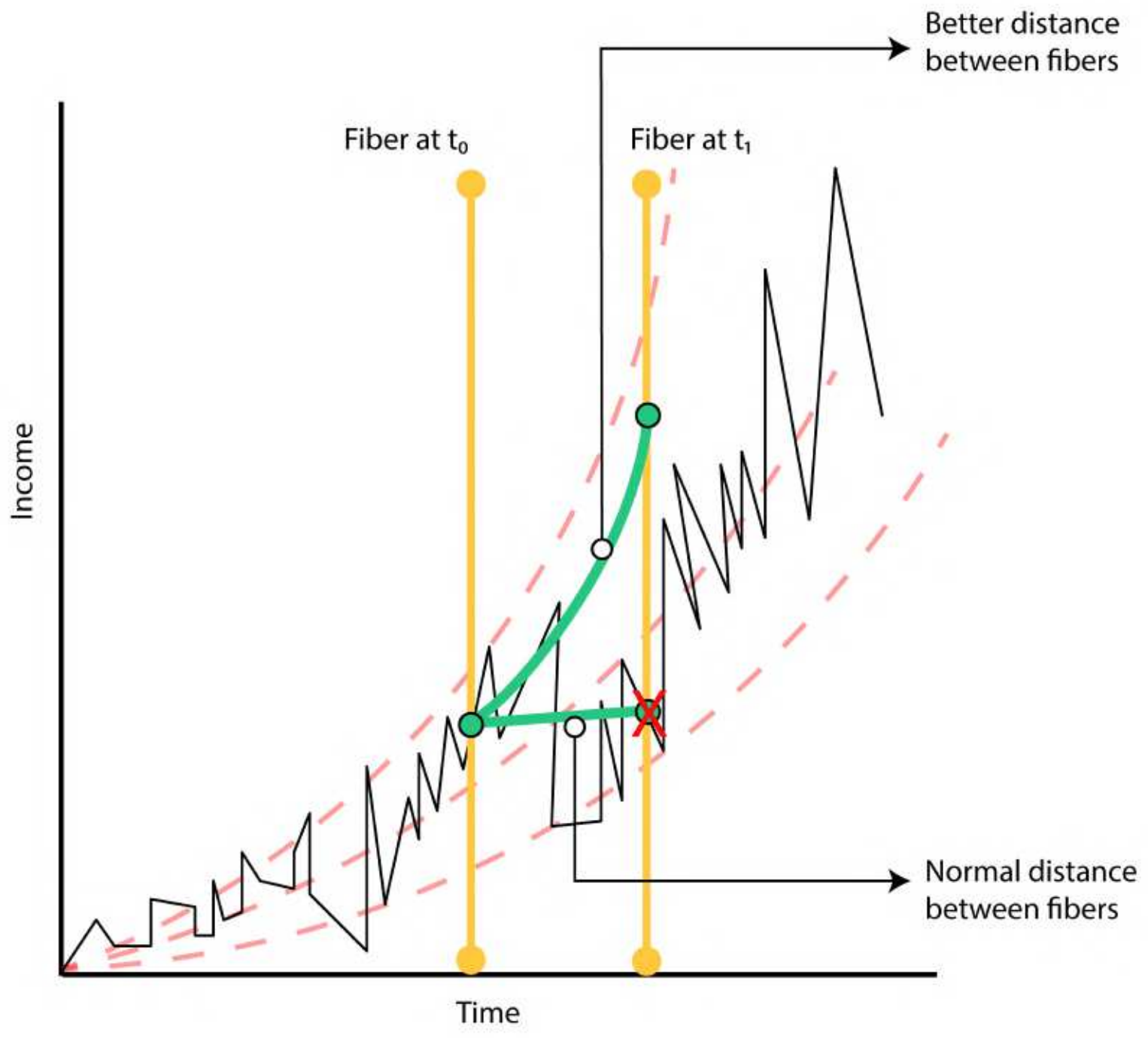

**Figure 2.3:** Black stochastic curve shows the income while the red curves show the inflation adjust parallel transported curves. Comparing Income at two different time point (two fibers) at time $t_0$ and $t_1$ the distance measure with circle and $X$ is the usual metric distance while the distance between the circles (accounting for inflation) is more relevant

equivalently inflation) and thus we need to define a new "covariant derivative" given by

$$D_t = \partial_t - a \tag{2.5}$$

where we have assumed an inflation given by $I_0 e^{at}$ (which gives us an inflation rate of $a$). This gives us a covarient derivative of 0 for constant inflation adjusted income. !

Let us take an example from quantum mechanics, the fiber is the complex line on which the wave function takes values. In QM we are used to the idea that the global phase is irrelevant and that we are free to set this phase to whatever we want, so let us say we have,

$$\Psi(\mathbf{k}, \mathbf{x}) = ce^{i\mathbf{k}\cdot\mathbf{x}} \tag{2.6}$$

where $c$ can be any complex number with $|c|^2 = 1$, in a more general gauge field theory, we can even have $c \longrightarrow c^{i\alpha(x)}$. Just like in the example of income, our derivative for the wave function now changes (to account for the "inflation" or gauge freedom) by

$$\partial_\mu \longrightarrow \partial_\mu + i\partial_\mu \alpha(\mathbf{x}) \tag{2.7}$$

This thus gives us our generalized definition of derivative in this gauge invariant space and tells us that the "connection" connecting the nearby fibers (wave functions) in the bundle ($\mathbf{x}$ space) is $iA_\mu(\mathbf{x}) = i\partial_\mu \alpha(\mathbf{x})$.

## Geometry of quantum space

We are now ready to dive into constructing geometry inside the world of quantum mechanics. In the previous section, where we introduced the idea of having connection for a QM wave function, one should note that the base space of the wave function (or fibers) is nothing but a trivial 3-dimensional space and the connection we are talking about, are for the fibers. Thus all the ideas, that we developed in the previous sections about curvature, has nothing to do with the trivial base space, but instead the wave functions. We adopt the generalized point of view that for an object to be curved, it involves the failure of a value in the fiber to come back to itself when it is parallel transported around a closed loop.

For the simple wave function we defined above with a connection $iA_\mu(\mathbf{x}) = i\partial_\mu \alpha(\mathbf{x})$, we see that the entire fiber bundle is flat[10] . But for a more general connections, parallel transport can result in arbitrary changes in the phase as a value is carried around a closed loop.

Let us now focus on a more relevant set of fiber bundles: a Hermitian matrix defined on a $n-$dimensional torus[11] which is nothing but the Brillouin zone (BZ). Here, our fibers are the $n$-dimensional complex state vectors that live on the BZ. Let $H(k_x, k_y)$ be a Hermitian matrix, for which we have a non-degenerate eigenspectrum[12] . As $H$ is Hermitian, we can diagonalize it and along with the assumption that $H$ is non-degenerate, we have an ordered set of $n$ eigenvectors $\left|\Psi^{(\lambda)}\left(k_x, k_y\right)\right\rangle$ at each point $(k_x, k_y)$ where $\lambda$ enumerates the eigenstates in increasing order of eigenvalues. Our goal would be now to relate this seemingly unrelated set of $\left|\Psi^{(\lambda)}\left(k_x, k_y\right)\right\rangle$ by defining parallel transport whereby an eigenvector in one fiber is transported into the eigenvector in the nearby fiber which has the same index $\lambda$, but at a point $\mathbf{k} + d\mathbf{k}$.

As discussed before, we want to construct a connection ($A$), a quantity that makes the covariant derivative zero through the relation (or essentially compensates for the inflation, from the inflation example), giving us[13]

$$\left(\partial_\mu + iA_\mu^{(\lambda)}\right)\left|\Psi^{(\lambda)}\right\rangle = 0 \tag{2.8}$$

(We have implicitly assumed the $\mathbf{k}$ dependence of $\left|\Psi^{(\lambda)}\right\rangle$) Now, multiplying this with $\left\langle\Psi^{(\lambda)}\right|$, we get

$$A_\mu^{(\lambda)} = \frac{i\left\langle\Psi^{(\lambda)}|\partial_\mu\Psi^{(\rho)}\right\rangle}{\left\langle\Psi^{(\lambda)}|\Psi^{(\lambda)}\right\rangle} \tag{2.9}$$

If we have a normalized set of wave functions, we get the result that

$$A_\mu^{(\lambda)} = -\operatorname{Im}\left(\left\langle\Psi^{(\lambda)}|\partial_\mu\Psi^{(\lambda)}\right\rangle\right) \tag{2.10}$$

This connection is called the Berry connection[55]. With this definition of connection, one can import the ideas of curvature back from the general theory and thus define Berry curvature,

10: This can be seen by seeing that the value of covarient derivative at each point $x_t$ along path from $x_0$ to $x_t$ is given by $e^{i(\alpha(\mathbf{x_0})-\alpha(\mathbf{x_t}))}$ (assuming at $x_0$ wave function has a value 1), thus at $x_t = x_0$ for a closed loop, the value remains 1 (a constant)

11: a Hermitian matrix representing the $\mathbf{k}$-dependent Hamiltonian for the periodic parts of the Bloch functions in a periodic system. This H(kx,ky) is defined for a base space which is the torus defined by -$\pi$/a <$k_x$<$\pi$/a, -$\pi$/b <$k_y$<$\pi$/b, in other words, the Brillouin zone. We will choose $a = b = 1a$ lattice constants.

12: We will later relax the idea of non-degeneracy

13: It should be noted that a general connection $A$ has 3 indices, but since we want the parallel transport to occur only between the same states labeled $\lambda$, $A^\rho_{\mu\lambda}$ is going to be diagonal in $\lambda$ and $\rho$ and hence results in $A^{(\lambda)}_\mu$

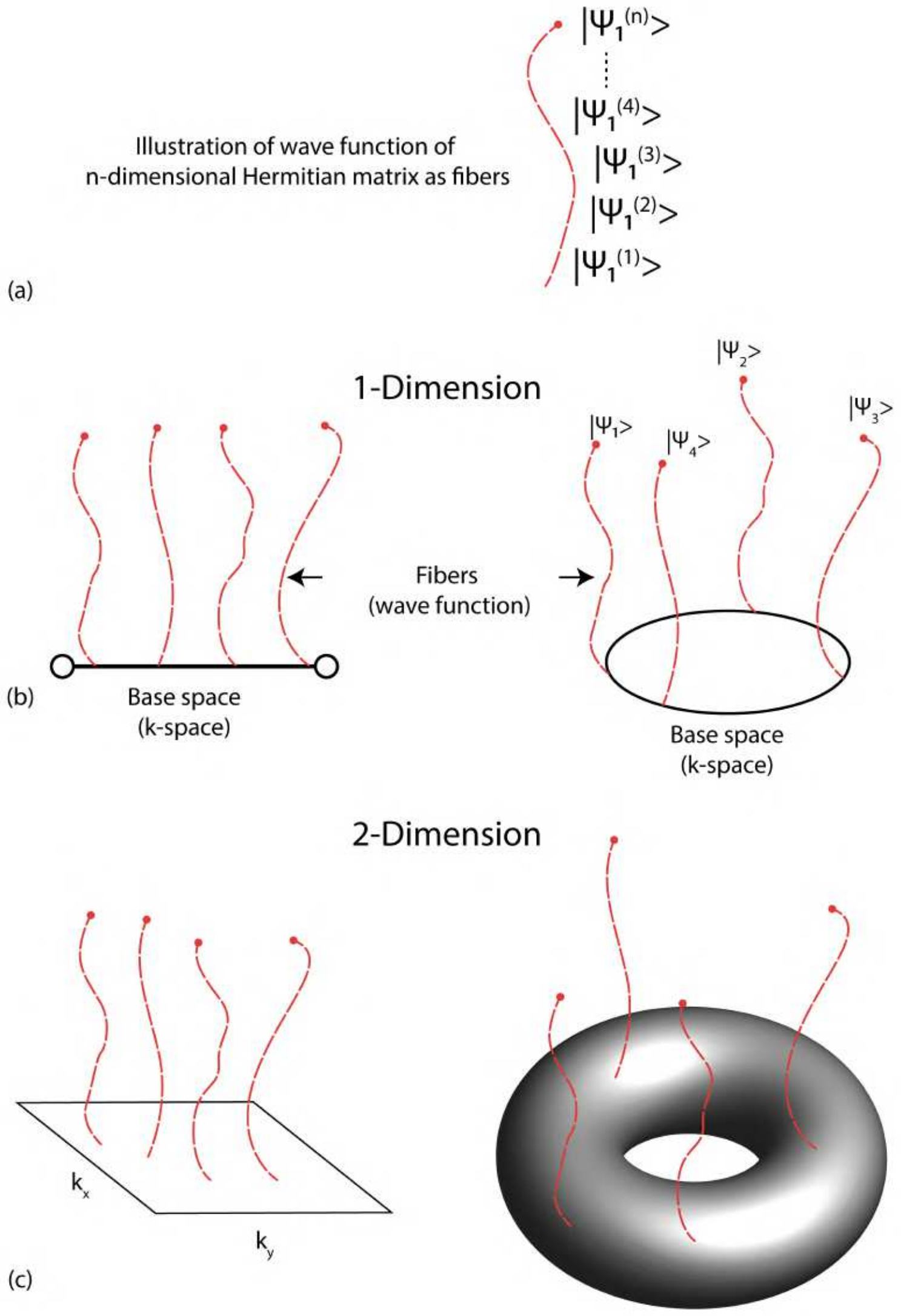

**Figure 2.4:** (a) illustration of wave function as fiber. Here $\left|\Psi_i^{(n)}\right\rangle$ is the wave function at the k point $i$ having a eigen value $E^n$ such that $E^0 < E^1 < E^2 \ldots < E^n$. Next we show the fibers attached to base space in (b) 1D k space and (c) 2D k space

$$\mathscr{F}_{\mu\nu}^{(\lambda)} = \partial_\mu A_\nu^{(\lambda)} - \partial_\nu A_\mu^{(\lambda)} \tag{2.11}$$

This now connects us to the ideas we had previously used, namely Equation 2.3 and Equation 2.4, where now for a finite closed loop $S$, we can calculate the acquired/accumulated Berry phase

$$\int_S \mathscr{F}_{\mu\nu}^{(\lambda)} dS \tag{2.12}$$

and if $S$ is the entire manifold, we get an object that uniquely counts the number of "holes" that the system is made up of. Infact when multiplied by $\frac{i}{2\pi}$, we get what is called the first Chern number for the manifold.

Now, it is time to address the fact that, while constructing this framework, we forced the eigen spectrum of $H$ to be non-degenerate. The Chern number is in general invariant under continuous deformations of $H$ (or equivalently $\left|\Psi^{(\lambda)}\right\rangle$). However, when $H$ becomes degenerate at some point, our construction of parallel transport breaks down. This is because, it becomes ambiguous which state to parallel transport to, as it is impossible to order the states according to their eigenvalue. This is at the core of the non-trivial topologies discussed in later chapters for specific materials in this thesis. We can therefore conclude that the Chern number is a topological invariant that is invariant under continuous deformations of $H$, as long as $H$ remains non-degenerate. Thus a continuous deformation that takes $H$ from one non-degenerate state to another non-degenerate state by crossing through a degenerate state, *has the possibility* of changing both the topological structure of the fiber bundle and the Chern numbers calculated on it.

Before moving further, we will try to get a mental picture of the above made construct of fiber bundles. Being a very abstract construct, having a visual caricature helps us in understanding how and why the structure we imposed makes sense.

Figure 2.4 shows the idea of fibers. For a given hermitian matrix, we have wave functions $\left|\Psi^{(n)}\right\rangle$ . Where we order the wave functions in increasing order of their eigen values such that if $\left|\Psi^{(n)}\right\rangle$ has the eigen value $E^n$, then we have $E^0 < E^1 < E^2 \ldots < E^n$. Next we see that our hermitian matrix is actually $k-$ dependent, thus giving us $\left|\Psi_k^{(n)}\right\rangle$, where we now attach $\left|\Psi^{(n)}\right\rangle$ to each point in $k-$space (which is now the base space). Thus we now have a construction of $k-$base space along with fibers $\left|\Psi_k^{(n)}\right\rangle$ which make our "fiber bundle". This is shown in (b) and (c) part of Figure 2.4. It is important to note that we not just have a simple base space, but a periodic base space which is shown in the right side. Now our entire discussion above was to understand how $\left|\Psi_k^{(n)}\right\rangle$ and $\left|\Psi_{k+dk}^{(n)}\right\rangle$ are related to each other, for which we introduced the notion of "Berry connection"

## 2.2 Semimetals and singularity

In the last part of the previous section, we mentioned why it was important that we consider Hermitian matrices with non-degenerate eigenvalues. Here we discuss what happens when we indeed do have degeneracies. In Equation 2.11, we defined the idea of curvature for a wave function. It turns out that when one has degeneracies, the curvature becomes undefined by having a singularity. This situation indeed leads to this singular point being the source of the "curvature vectors" or a source of Berry flux. There is a beautiful reformulation of this entire construct we have done above using magnetic fields, where now a point of "band crossing" would map to being a point where a monopole is placed. We can understand the origin of this singularity intuitively. When we have a degeneracy, we see that in Equation 2.11, $A_\nu^{(\lambda)} = A_\nu^{(\lambda-1)}$. Thus when one thinks of connecting this to the next fiber in $\nu + d\nu$, one fails to know which one in this fiber connects to which one in the next fiber. And thus, calculating $\partial_\nu A_\nu^{(\lambda)}$ becomes ill-defined.

If the reader is familiar with the idea of having "Dirac"(linear) like crossing being important to having topological systems, then one can easily see the importance of linearity from this argument. In the presence of systems that have linear dependence in the vicinity $d\nu$, there is no way for Equation 2.11 to vanish, which can happen if there is only quadratic dependence on $dv$. And in fact, one can have "trivial" crossings in systems where two eigenvalues are degenerate by just accident and thus we precisely know how they are related to the next fibers.

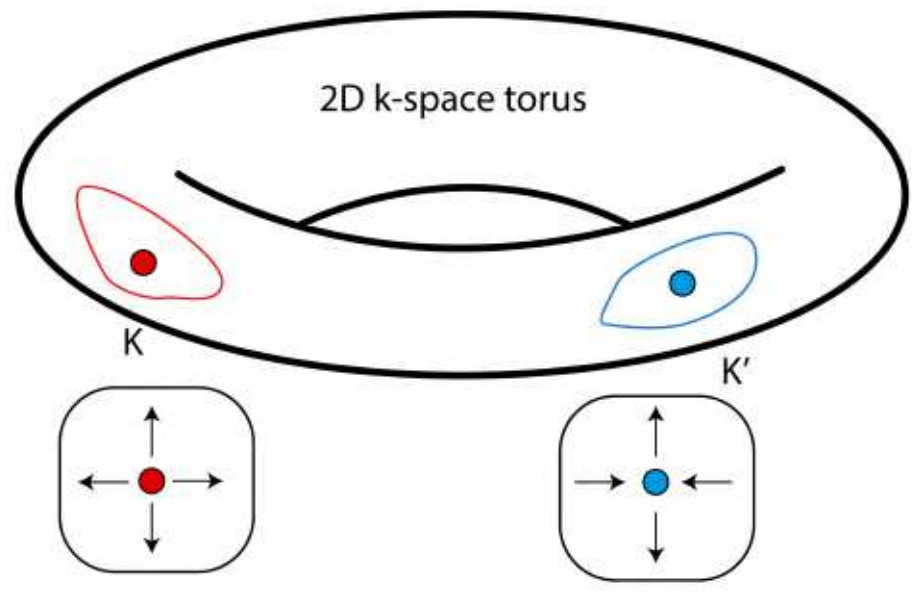

**Figure 2.5:** 2D k space torus of graphene with $K$ and $K'$ singular points taken out. Here around $K$ is the source while $K'$ drain.

Getting back to the singular nature of the point, we now try to work around this singularity. The obvious way to talk about the topology of a system with singularity is to just throw away the point in the base space where the singularity occurs. Upon closer look we see that this type of fiber bundle[14] is always topologically non-trivial (either locally or globally), to be defined below. For example, let's say we have an odd number of such singularities, then we see that we have we have a base space where, when we calculate a integral of the curvature around a closed loop that encloses the singularity (thrown out point), we get a non zero value. This is because of the fact that the singular point acts as a point of source of "flux". This might seem like a anomalous behavior where we have a system that keeps on producing (is a source of ) flux, but indeed this is what happens when one applies a magnetic field through a 2D system. The result of which is seen as a non-trivial edge state such as for example the quantum Hall effect.

14: Remember, fiber bundle is fibers with the base space

But let us now say we have an even number of singularities. In this case, we might have a situation where there are an equal number of singularities that act like sources and drains. Thus a closed loop over the entire system might not produce any net flux. However, we notice that there are other loops that one can draw locally that have non zero flux. These loops indeed represent a local non triviality in the system, the effect of which is the presence of edge state that connects these two source and drain points but does not connect throughout the system.

The reader might already have noticed another consequence of these non-trivial degeneracies. If one has a closed system, because of the fact that we cannot keep producing flux (or have a flux source alone), we always need to have these singularities in pairs. This is so because there is no net flux leaving a closed system. Thus, non-trivial crossings in a closed system must always occur in pairs. For instance, in graphene as we will later see, the Dirac cone (singularity) at $K$ is always accompanied by a Dirac cone at $K'$, its time reversed partner.

Figure 2.5 shows the singular $K$ and $K'$ points in the 2D BZ which are the source and drain. One would get a non-trivial winding number when integrating the curvature around a loop enclosing the singular points (shown by red and blue loops).

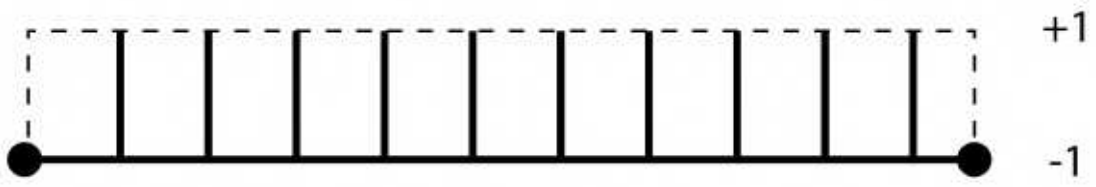

**Figure 2.6:** Line as base space that has lines attached to each point from $(-1, 1)$

## 2.3 Symmetry Indicators

Before we introduce the connection between group theory and topology in band theory, it is instructive to look at a concrete example of a system and its fiber bundle. We will take the prototypical example of the Su-Schrieffer-Heeger (SSH) Model[56]. But let us pause and first think about a very absurdly simple fiber bundle. Let us assume we have a base space which is a line segment. Let us also say that this line can be closed to form a circle (i.e. it is periodic). However, before closing the circle we attach yet another line segment, say $(-1, 1)$, to each point of the first line, such that we arrive at a two dimensional strip. This is shown in Figure 2.6.

It is important to note that although the two dimensions are similar to each other, there is an important conceptual difference between them: we view one dimension as a base space manifold, and the other as a fiber. Further, we assume the trivial connection on the strip, which parallel transports one value in one fiber to the same value in the nearby fiber. We now proceed to glue the base space together into a circle. When doing so, we need to identify not only the two end points of the base space with each other. We also need to identify each value in the fiber at one end point with a corresponding value in the fiber at the other end point. There are two qualitatively different ways to make this identification: in one case we glue them together by assuming that the connection transfers a value $x$ at one end point to the value x in the fiber at the other end. Alternatively, we can choose to instead connect x in one fiber to -x in the other. The difference between these two choices is demonstrated in Figure 2.7

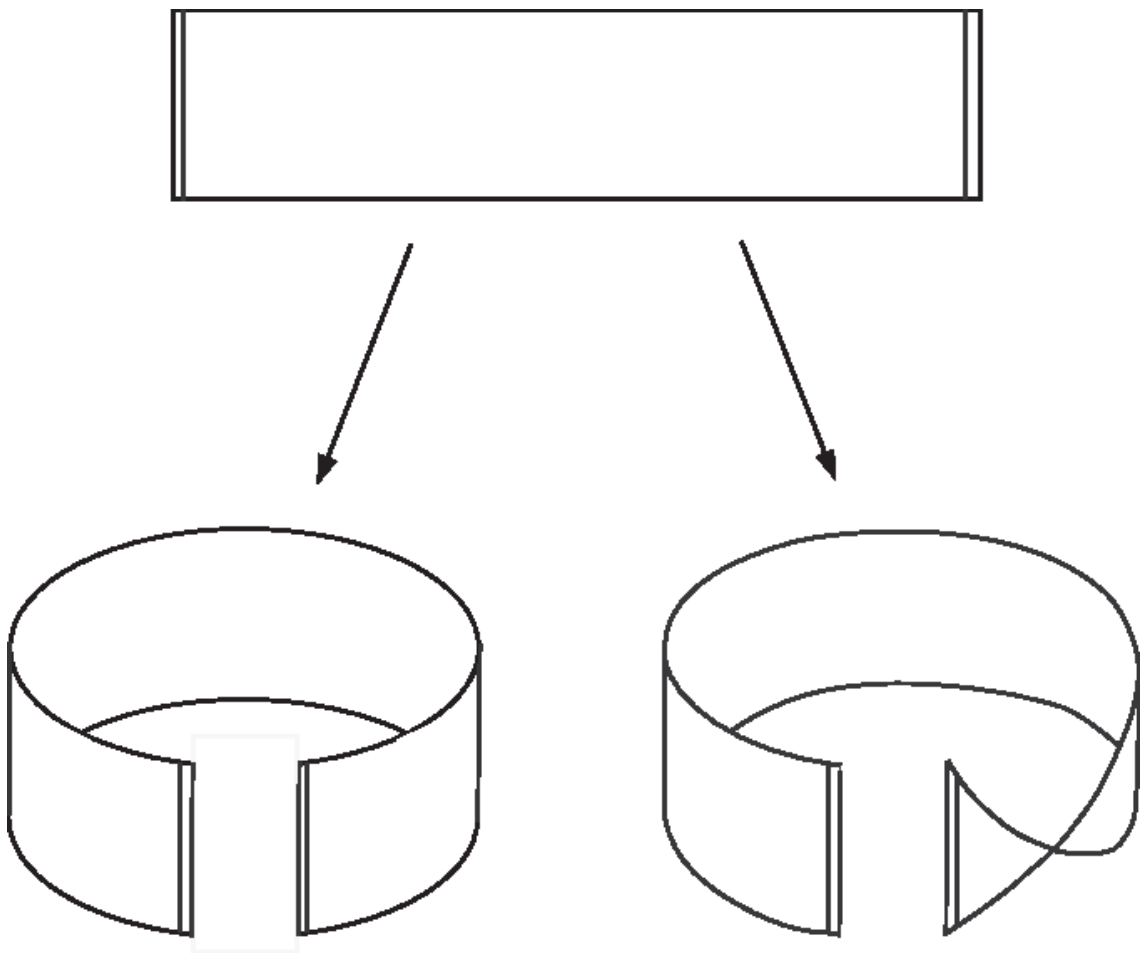

**Figure 2.7:** Different ways to join Figure 2.6 left-trivial right-Non-trivial

It is clear that in both cases the base space, the circle, is the same. The topological distinction is therefore not clear until we consider the whole fiber bundle. In the first case the fiber bundle is an ordinary circular strip, while the second one is a Möbius strip. This stands in contrast to our experience from differential geometry, where the topology has to do with the base space manifold alone, not the whole fiber bundle which consists of manifold plus tangent spaces.

We note the close relation between connection, topology, and parallel transport in this example.

- The connection at the end points is responsible for determining the topology of the whole fiber bundle, by either making it a normal strip or a Möbius strip.
- if we perform parallel transport of $x$ once around the base space, the result is x for the normal strip, while it is $-x$ for the Möbius strip.

This example here might seem quite a trivial or arbitrary construct, but we will use the two major outcome we saw above to understand a physical example i.e. the famous SSH model.

SSH model is a 1D chain of atoms with two atom in the unit cell with a hopping anisotropy between intra and inter unit cell interaction. This can be written as

$$\mathcal{H}_{SSH} = (t_1 + t_2 \cos(k))\, \sigma_x + t_2 \sin(k) \sigma_y \tag{2.13}$$

where $t_1$ and $t_2$ are the two anisotropic hopping terms. The energy spectrum of this system is gapped whenever $t_1 \neq t_2$. Let us try to visualize a more accurate picture of the fiber bundle and see how and where the non triviality of the topology comes in. The wave function of $\mathcal{H}_{SSH}$ is given by

$$\left|\Psi^{(1/2)}(k)\right\rangle = \begin{pmatrix} \pm e^{-i\phi(k)} \\ 1 \end{pmatrix} \tag{2.14}$$

where $\tan(\phi) = \frac{t_2 \sin(k)}{t_1 + t_2 \cos(k)}$, which is nothing but the ratio of the $\sigma_y$ and $\sigma_x$ components of $\mathcal{H}_{SSH}$. It is clear that the wave function at a given k point is purely determined by $\phi(k)$. This tremendously helps us in visualizing the wave functions as we now just need a way to plot the angle.

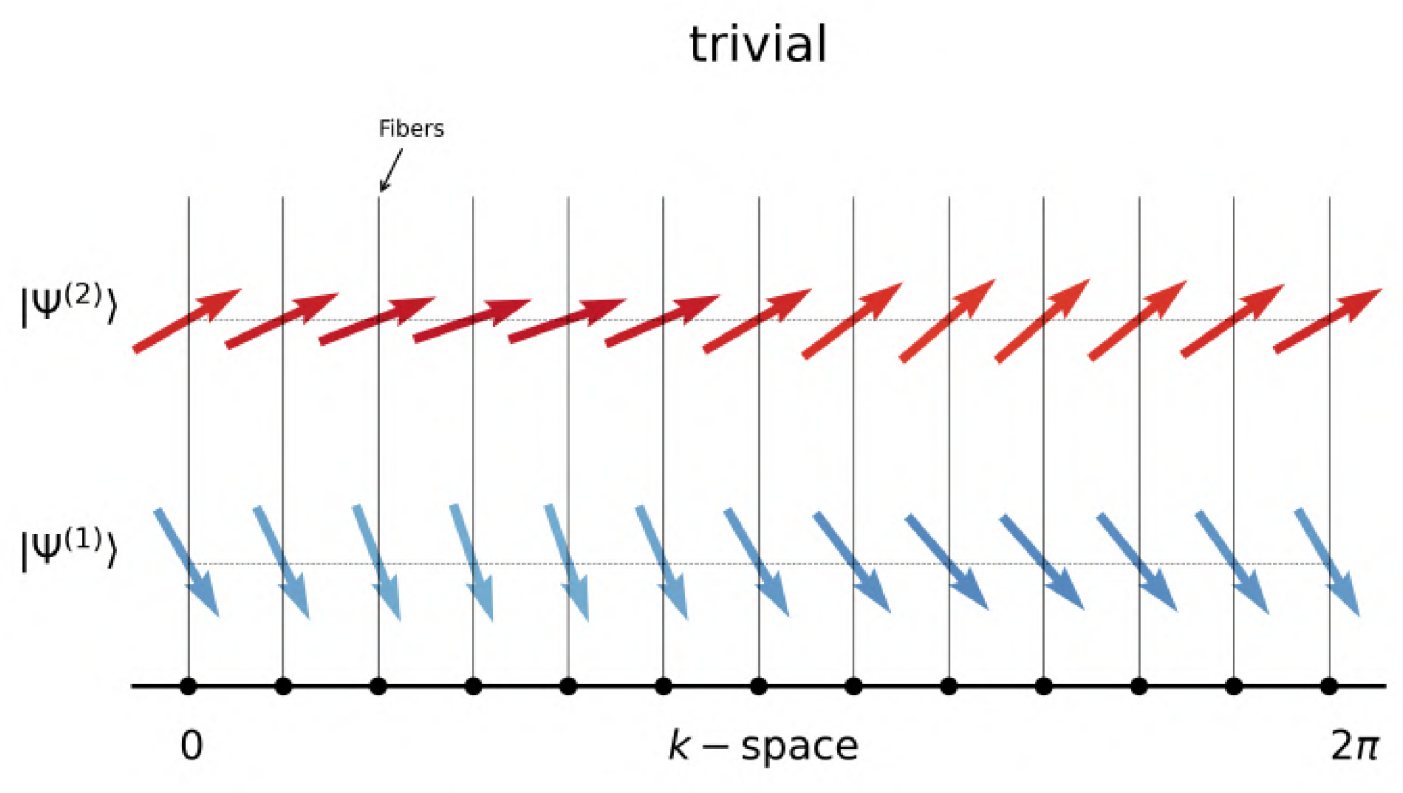

**Figure 2.8:** Fiber of SSH hameltonian in the limit $t_1 > t_2$

Similar to Figure 2.4, we will plot the base space (which here is a periodic 1D k space) and the fibers here are nothing but $\left|\Psi^{(i)}(k)\right\rangle$ where $i = 1, 2$ at each $k$ in the base space. We will use an arrow to point to the angle of the phase. Let us first look at the limit $t_1 > t_2$.

Figure 2.8 shows the above setup. Now, let us concentrate on the bottom part of the fiber $|\Psi^{(1)}(k)\rangle$, which is the wave function of the lower band.[15] First we see that the above mentioned pseudo construction of a fiber bundle shown in Figure 2.7 is in fact the same type of object we have here! Second, we verify that since the system is periodic in k-space, the angle of the vector at $k = 0$ is indeed the angle at $k = 2\pi$. Now to decipher the topology, all we do is look at how the nearby fibers at $k$ and $k + dk$ are related. We see that there is no change in the fibers as we move from 0 to $2\pi$. This indeed gives us the trivial construction that we mentioned above.

15: at half filled lattice, this is the wave function of the occupied electron

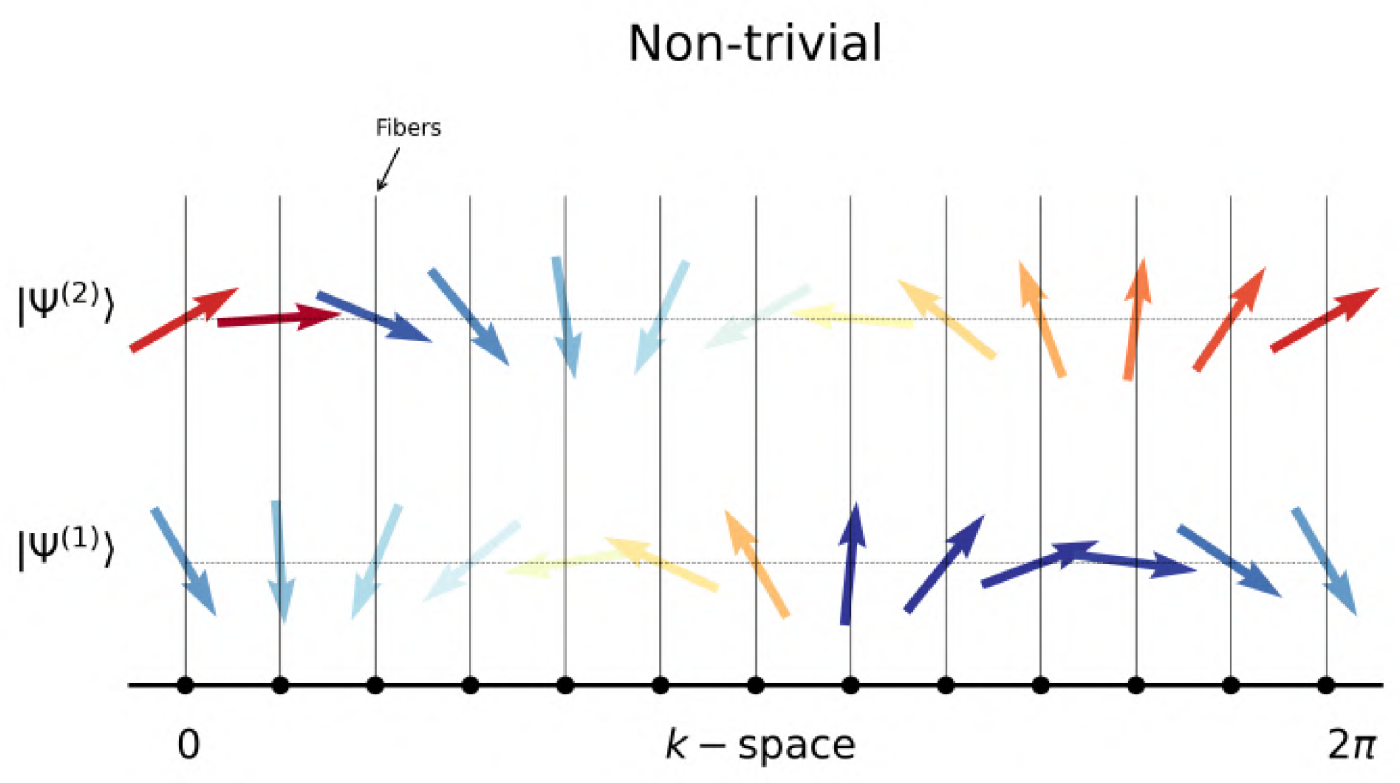

**Figure 2.9:** Fiber of SSH Hamiltonian in the limit $t_1 < t_2$

Now let us look at the case where $t_1 < t_2$., which is shown in Figure 2.9. Immediately, one sees the major difference in this system, that the vector at $k = 0$ does a complete rotation and then joins on to the other end $k = 2\pi$. This thus shows that the fiber in our system has a non-trivial topology (which actually can be deduced from a construct called "winding number") akin to the Möbius strip we discussed previously in Figure 2.7. This illustrates the first point we deduced from the previous Möbius strip example - *The connection at the end points is responsible for determining the topology of the whole fiber bundle, by either making it a normal strip or a Möbius strip.*

Further, we shall now look at the use of symmetry for detecting this topological transition. We know that our system can either be in a state of trivial winding or non-trivial winding as shown in Figure 2.8 and Figure 2.9. But because of the fact that the system is periodic, we are guaranteed that *if we perform parallel transport of $x$ once around the base space, the result is $x$ for the normal strip, while it is $-x$ for the Möbius strip*. This is shown in Figure 2.10, where we have plotted the same fibers (wavefunction), but now in a close loop representing the periodic $k-$space. It is easy to see that any vector at $k$ and $-k$ (connected by dotted lines), are always in the opposite direction in the non-trivial regime while in the same direction for the trivial one.

The above mentioned idea is indeed true for any general non-trivial topological system protected by symmetries. In the case of SSH, we have Chiral/Inversion symmetry that dictates that when we are in the non-trivial topology, the inversion value at $k$ and $-k$ are opposite to each other. Thus, to identify the topology of a SSH, all one needs to do is to calculate the value $\chi$ where

$$(-1)^{\chi} = \frac{\eta_{k=0}}{\eta_{k=\pi}} \tag{2.15}$$

where $\eta_k$ is the inversion eigenvalue at $k$. When $\chi = 0$, we have that the inversion eigenvalue at the TRIM[16] points are opposite of each other. $\chi = 1$ would correspond to the trivial case where we have the same inversion eigenvalues at points 0 and $\pi$ and hence no "inversion of bands" has occurred. A more rigorous proof of Equation 2.15 will be derived using "sewing matrix" theory in later chapters.

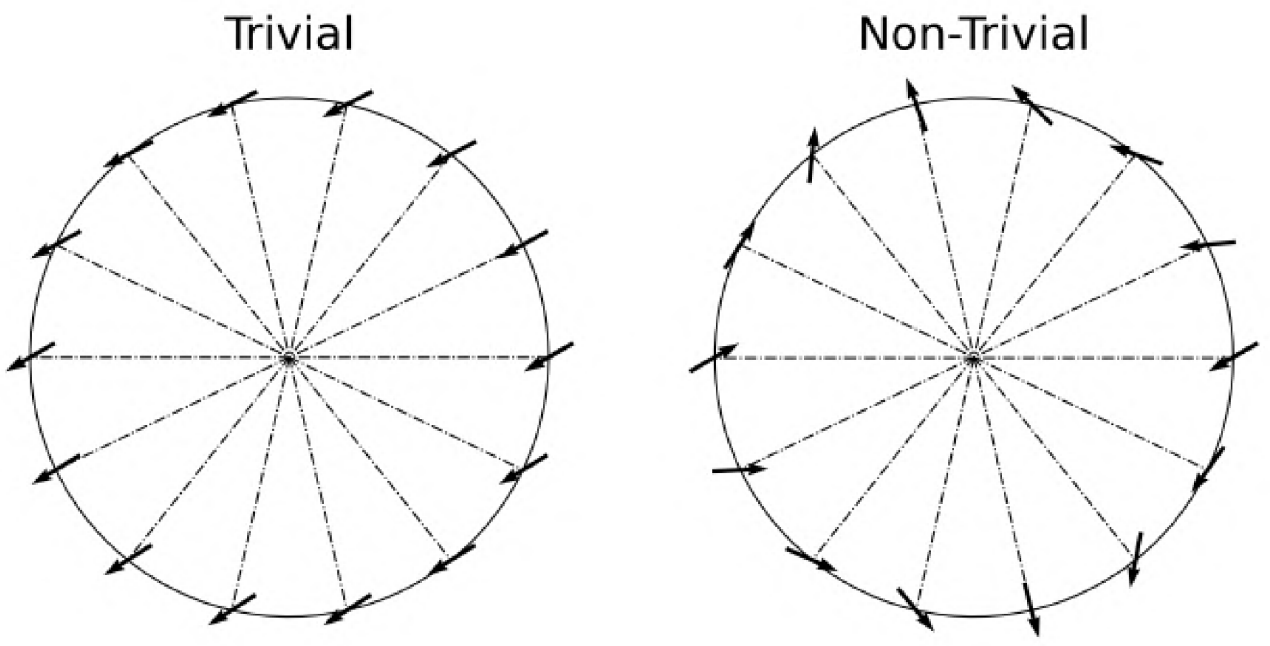

**Figure 2.10:** Trivial and Non-trivial limits of SSH's wave function. Note the direction of wave function at $k$ and $-k$ (diagonal opposite).

16: TRIM are Time reversal invariant momenta, which are points in $k-$space that map back to themselves under time reversal operation

## 2.4 Symmetries, orbitals, topology and obstruction

To understand the role of symmetries in a system, we need to be familiar with a few more terms. Instead of giving a formal definition of them, we will illustrate and give a intuitive feel for them with the help of an example - our trusted 1D chain. We will start with analyzing a 1D chain of atoms that at all time, respects inversion symmetry ($I$). Without loss of generality, we will assume that the origin is at the point 0 shown in Figure 2.11, and thus would be our center of inversion. Our space group $G$ is made up of $\{\{E|T\}, \{I|T\}|T = na; n \in Z\}$[17] , where we have labeled the group elements such that we have {Rotation/Inversion | Translation} and $E$ is the identity.

17: Since we consider spinless case, we have ignored Time Reversal Symmetry, but one indeed has the 1D space group $G$ as $\{\{E, T\}, \{I, 0\}, K\}$ where $K$ is the time reversal operator.

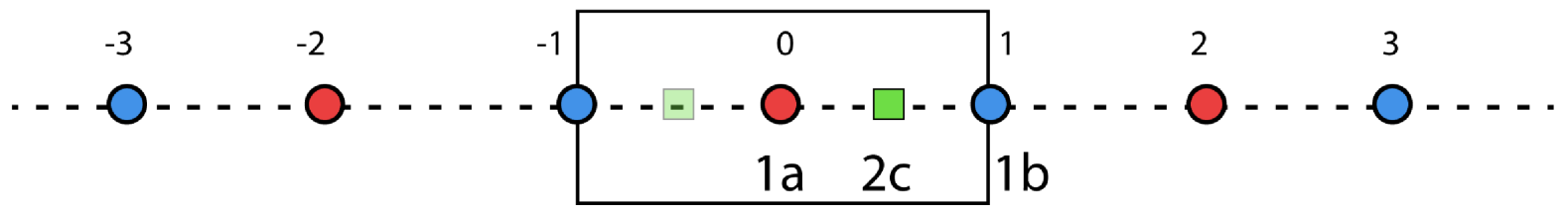

**Figure 2.11:** 1D chain of atoms with various Wyckoff positions.

Now, having fixed a symmetry, we can have 3 unique points that are differentiated by their *site symmetry group* (Definition 2.4.1)[18] .These are shown in the figure. First, the origin point is the point which under inversion maps back to it-self, thus the *site symmetry group* of this site

18: In crystallography the points 1a, 1b, 2c are known as Wyckoff positions.

is given by just $I$ and this site is called $1a$. Then there is another site which maps back to itself with inversion, but needs to be pulled back by a translation to do so. Site 1, when inverted goes to site -1 which needs to be translated back by a lattice vector to get it back to 1. Thus the site symmetry group of this site is $\{I|T\}$ and this is called site $1b$ and is at position $\frac{1}{2}$ lattice vector. And finally we have site $2c$ (placed at $x$ and $-x$) which is a unique site which has no symmetry whatsoever and the site symmetry group consists just of the identity. This site is labelled $2c$ because of the fact that when one places it on the right side of 0 (marked by bright green), one needs to place equivalently another site to the left (marked by light green) of 0 to respect inversion symmetry of the system. Thus one always needs 2 such sites if inversion must be respected, and hence the nomenclature.

**Definition 2.4.1** *The stabilizer group or site-symmetry group of a position q is the set of symmetry operations $g \in G$ that leave q fixed. Where G is the space group of the system*

Now we have 2 different representations that $I$ can have, namely even and odd functions with respect to inversion, we will label them |s>, |p> to show their connection with the spherical harmonics of the isolated atoms which obey full spherical rotation symmetry as given by the SO(3) Lie group, the group of all special orthogonal $3x3$ matrices ( i.e. the ones with determinant=+1). Because of the 1D nature, only the p-orbital (ie. spherical harmonic with l=1) along the 1D axis, say the z-axis is needed. All other spherical harmonics with l even are equivalent to s and all spherical harmonics with odd angular momentum l are equivalent to p as far as the inversion goes.

Thus $|s\rangle$ and $|p\rangle$ satisfy

$$I\,|s\rangle = +\,|s\rangle \tag{2.16}$$

$$I\,|p\rangle = -\,|p\rangle\,. \tag{2.17}$$

What this physically means is that we have at most two different type of orbitals that can be put at any of the above mentioned sites, giving us $|i\rangle_j$ where $i \in \{s, p\}$ and $j \in \{1a, 1b, 2c\}$.

From here, we jump off to the momentum space and now consider points in momentum space that are invariant under inversion. For out 1D case, that would correspond to points $k = 0, \pi$. We do this because, this gives us a relation/connection between symmetries in real space (which are tied to atomic orbitals) to symmetries in $k-$space (which are tied to k-space wave functions). We can thus uniquely label the wave functions at these inversion symmetric points in k space by their inversion eigenvalues. This is at the heart of "band representations", which is nothing but set of bands linked to a localized orbital respecting all the crystal symmetries. They relate electrons site to momentum space description. In the above case, for the points $1a$, we see that we have no translation in their corresponding site symmetry groups. Thus the representations $|s\rangle_a$ and $|p\rangle_a$ both do not pick up a phase when they move from 0 to $\pi$ in $k-$space. Thus they remain true to their eigenvalue at both the momenta. In contrast, $|s\rangle_b$ and $|p\rangle_b$ have a translation by lattice vector, which adds a phase to them at $\pi$ thus switching their representation at $\pi$. This is shown in Table 2.1. (For discussion, refer Remark 2.4.1[19] ).

| Rep. | 0 | $\pi$ |
|---|---|---|
| $\lvert s\rangle_{1a}$ | + | + |
| $\lvert p\rangle_{1a}$ | - | - |
| $\lvert s\rangle_{1b}$ | + | - |
| $\lvert p\rangle_{1b}$ | - | + |

Table 2.1: Band representations of 1D chain.

19: This is only given for intuitive and semi-rigorous purposes, For a detailed discussion of EBR and different ways to look at it, readers are guided towards Ref[57, 58]

**Remark 2.4.1** Let us start by describing our tight binding basis made up of $\phi_{\mathbf{R},j}$ of a wave-function of atom $j$ at unit cell $\mathbf{R}$ which satisfies

$$\langle \phi_{\mathbf{R}i} | \phi_{\mathbf{R}'j} \rangle = \delta_{\mathbf{R}\mathbf{R}'}\delta_{ij} \tag{2.18}$$

Then, we have our Hamiltonian $H(\mathbf{R})$

$$H_{ij}(\mathbf{R}) = \langle \phi_{0i} | H | \phi_{\mathbf{R},j} \rangle \tag{2.19}$$

Now, we define our Bloch basis function as

$$\left|\chi_j^{\mathbf{k}}\right\rangle = \sum_{\mathbf{R}} e^{i\mathbf{k}\cdot(\mathbf{R}+\mathbf{t}_j)} \left|\phi_{\mathbf{R}j}\right\rangle \tag{2.20}$$

where $t_j$ is the position of the atom $j$ inside the unit cell at $\mathbf{R} = 0$. Note that one can equivalently make the choice to define a Bloch basis as $\left|\tilde{\chi}_j^{\mathbf{k}}\right\rangle = \sum_{\mathbf{R}} e^{i\mathbf{k}\cdot\mathbf{R}} \left|\phi_{\mathbf{R}j}\right\rangle$ too.

We now can write the hamiltonian in this basis as

$$H_{ij}^{\mathbf{k}} = \left\langle \chi_i^{\mathbf{k}} | H | \chi_j^{\mathbf{k}} \right\rangle = \sum_{\mathbf{R}} e^{i\mathbf{k}\cdot(\mathbf{R}+\mathbf{t}_j-\mathbf{t}_i)} H_{ij}(\mathbf{R}) \tag{2.21}$$

which has Bloch Eigen states expanded as

$$\left|\psi_{n\mathbf{k}}\right\rangle = \sum_j C_j^{n\mathbf{k}} \left|\chi_j^{\mathbf{k}}\right\rangle \tag{2.22}$$

And it is indeed these $C_j^{n\mathbf{k}}$ column vectors that form the cell-periodic functions $|u_n(\mathbf{k})\rangle$ *i.e.* for a given band $n$, we have

$$|u_n(\mathbf{k})\rangle \quad \Rightarrow \quad \begin{pmatrix} C_1^{n\mathbf{k}} \\ \vdots \\ C_d^{n\mathbf{k}} \end{pmatrix} \tag{2.23}$$

To understand the values in Table 2.1 and to understand the concept of EBR, we now see how the crystal symmetry acts on these states. Lets start with a crystal symmetry operation given by $g = \{\mathscr{C}|\mathbf{T}\}$, which acts as $\mathbf{r} \longrightarrow \mathscr{C}\mathbf{r} + \mathbf{T}$. Now when $g$ acts on a generic $R + r_\alpha$, we get $r_\beta$ which satisfies the relation $\mathscr{C}(\mathbf{R}+\mathbf{r}_\alpha)+\mathbf{T}=\mathbf{R}'+\mathbf{r}_\beta$. We also define matrix $\mathscr{G}$ such that $g\left|\phi_{\mathbf{R},i}\right\rangle=[\mathscr{G}]_{j,i}\left|\phi_{\mathbf{R}',j}\right\rangle$ This is a matrix that acts on our TB basis that takes the atom/orbital at $i$ in the unit cell $\mathbf{R}$ to $j$ in unit cell $\mathbf{R}'$. Having these, now we can act $g$ on our basis $\left|\chi^{\mathrm{k}}_{\alpha i}\right\rangle$.

$$g\left|\chi^{\mathbf{k}}_{\alpha}\right\rangle=\sum_{\mathbf{R}} g e^{i\mathbf{k}\cdot(\mathbf{R}+\mathbf{r}_\alpha)}\left|\phi_{\mathbf{R},\alpha}\right\rangle \tag{2.24}$$

$$=\sum_{\mathbf{R}} e^{i(\mathscr{C}\mathbf{k})\cdot(\mathscr{C}(\mathbf{R}+\mathbf{r}_\alpha))}\left|\phi_{\mathbf{R}',\beta}\right\rangle[\mathscr{G}]_{\beta,\alpha} \tag{2.25}$$

$$=\sum_{\mathbf{R}} e^{i(\mathscr{C}\mathbf{k})\cdot\left(\mathbf{R}'+\mathbf{r}_\beta-\mathbf{T}\right)}\left|\phi_{\mathbf{R}',\beta}\right\rangle[\mathscr{G}]_{\beta,\alpha} \tag{2.26}$$

Thus, we have

$$g\left|\chi^{\mathbf{k}}_{\alpha}\right\rangle=\left|\chi^{\mathscr{C}\mathbf{k}}_{\beta}\right\rangle[\mathscr{G}]_{\beta,\alpha}\, e^{-i(\mathscr{C}\mathbf{k})\cdot\mathbf{T}} \tag{2.27}$$

This shows that the action of $g$ on $\left|\chi^{\mathbf{k}}_{i}\right\rangle$ just relates the orbitals to each other, except for when the symmetry involves translation ($\mathbf{T}$), where one picks up a $k$ dependent phase of $e^{-i(\mathscr{C}\mathbf{k})\cdot\mathbf{T}}$.

For the above 1D case, we saw that the *site symmetry group* at positions $1a$ was $g=\{I|0\}$ while for $1b$ we had $g=\{I|T\}$. Since we just have one atom/orbital per unit cell, we trivially map back to the same atom after inversion ($\mathscr{G}=1$). But in case of $1b$, we map back to an atom that is translated by $\mathbf{T}$. Thus for $\mathscr{C}=I$, we have

$$g\left|p/s\right\rangle^{k}_{1b}=\left|p/s\right\rangle^{(\mathscr{C}k)}_{1b}\mathscr{C}e^{-i(\mathscr{C}k)T} \tag{2.28}$$

$$=\left|p/s\right\rangle^{-k}_{1b}e^{ikT} \tag{2.29}$$

Hence for $1b$ the wave functions at $k=0$ and $k=\pi$ have opposite eigenvalue for $g$. In contrast, for $1a$ position, we have

$$g\left|p/s\right\rangle^{k}_{1a}=\left|p/s\right\rangle^{(\mathscr{C}k)}_{1a}\mathscr{C}e^{-i(\mathscr{C}k)T}|_{T=0} \tag{2.30}$$

$$=\left|p/s\right\rangle^{-k}_{1a} \tag{2.31}$$

Thus for $1a$ the wave functions at $k=0$ and $k=\pi$ have same parity.

Thus any orbital, put into this crystal structure that respects the inversion symmetry must be made up of one of the above following representations. Lets us construct a test bed example that has $s$ and $p$ orbitals together put at site $1a$.

The tight binding form of that Hamiltonian reads as

$$H(k)=-\left[e+\left(V_{ss\sigma}+V_{pp\sigma}\right)\cos(k)\right]\sigma_z-2V_{sp\sigma}\sin(k)\sigma_y \tag{2.32}$$

where we have taken the lattice constant to be unity and here $V_{ss\sigma}$,

$V_{pp\sigma}$, $V_{sp\sigma}$ are the Slaster Koster two-center interactions with interactions between same orbitals and interactions between s and p. We also have $e$ which is the on-site difference between the two orbitals. This Hamiltonian is the same as SSH one, that we previously saw, but with a rotation that moves the Pauli $z$ vector to $x$ ($\sigma_x \longrightarrow \sigma_z$). Let us first talk about the limit where $V_{sp} = 0$ and $e \neq 0$ where we have two bands that do not touch each other(to make it easier, think about the limit where along with $V_{sp} = 0$, we have $V_{ss} = 0$, $V_{pp} = 0$). In this case, when the system is half filled, we have an insulator whose occupied wave functions is nothing but [20]

20: To make it easier, we are assuming $V_{sp} = 0$, we have $V_{ss} = 0$, $V_{pp} = 0$

$$|1\rangle = \begin{pmatrix} 1 \\ 0 \end{pmatrix} \tag{2.33}$$

It is trivial to see that this wave functions is completely made up of s orbital. Realizing that the inversion operator is $\sigma_z$, we can see that our lower energy band is nothing but a band which is made up of the representation $|s\rangle_a$. This is not surprising as that is what we made the system to be in the first place. Now let us turn on $V_{sp} \neq 0$. We immediately see that there is a phase where we again have an insulator, but now the lowest energy eigen function is made up of

$$|1\rangle = e^{ik/2} \begin{pmatrix} \cos \frac{k}{2} \\ i \sin \frac{k}{2} \end{pmatrix} \tag{2.34}$$

Using the inversion operator, we can see that this limit corresponds to having a value of $+, -$ at the momenta $0, \pi$ for the occupied band. Looking at Table 2.1, this tells us that the occupied electrons act as if there it is a localized $s$ like state at position $1b$ even though our actual orbitals sit at $1a$.

This brings us to the classification of two kinds of *Atomic Insulators* [59]

- **Trivial atomic limit insulator (TAL)** : A set of bands is in the TAL when they possess symmetric, localized Wannier functions that reside on a Wyckoff position that is same as the Wyckoff position of the underlying ions
- **Obstructed atomic limit insulators (OAL)** : A set of bands is in the OAL when they possess symmetric, localized Wannier functions that reside on a Wyckoff position distinct from the Wyckoff position of the underlying ions and which cannot be smoothly deformed to the ionic position.

The first limit we saw, the ionic system($V_{sp} = 0$) corresponds to TAL while the second limit corresponds to OAL. The nomenclature that the first is called trivial implies that there is something non-trivial in OAL, and indeed OAL, in this case corresponds to having edge mode in case of 1D. But it is should be noted that although an edge mode is not guaranteed in OAL, there always needs to be a metallic edge mode on the boundary between trivial and OAL part of the system. This is because, these two systems are topologically different. This difference stems from the fact that the band representation of $(+, +)$ , $(-, -)$ can never smoothly be deformed to $(+.-)$, $(-, +)$ and thus needs to have a crossing.

This effect can also be verified by calculating the Wanier Charge Center *WCC* ( or center of their wanner functions) of the orbitals. This is nothing but the Zak phase ($\gamma$)[60, 61] over $2\pi$. The Zak phase of the system is given by,

$$\gamma = i \oint \langle 1(k) | \nabla_k | 1(k) \rangle \tag{2.35}$$

where $|1(k)\rangle$ is the occupied wave function. From Equation 2.33, we have a trivial Zak phase $\gamma = 0$. Thus, the WCC of the first TAL system is at the origin where the atoms indeed lie. But for Equation 2.34 we get $\gamma = \pm\pi$ , giving us a WCC of $\pm\frac{1}{2}$. This says that the charge center in this system is at an atomic position of $\pm\frac{1}{2}$ of the unit cell, which is the same conclusion we got from the band representations. From the above discussion, it seems like atomic obstruction is closely related to dimerization. In later chapters, we will use this fact to study the entanglement between spatially separated parts of the systems to probe this obstructed state. This deep connection between the Zak phase and atomic obstruction will be used latter in Chapter 7 to show that annihilation of Dirac fermions would necessarily lead the resulting insulator to OAL and not TAL.

These same band representations can be used to detect the usual topological insulators that are protected by time reversal symmetry and which are classified 10 fold using K-theory[62]. The advantage of using a more general idea of band representations is that, we are now not restricted to just k-space symmetries, but can easily well extend the ideas to real space symmetries. A more generalized definition of a topological system from their bands is given in Definition 2.4.2

**Definition 2.4.2** *A set of bands are in the atomic limit of a space group if they can be induced from localized Wannier functions consistent with the crystalline symmetry of that space group. Otherwise, they are topological.*

# 3 Computational methods

Throughout the thesis, since we are dealing with relatively weakly-correlated systems and the fact that we are studying topological properties, band structures at the level of Generalized Gradient Approximation (GGA)[63] within Density Functional Theory (DFT) suffice. Nonetheless, to obtain well defined energy bands that have meaning as excitation energies of the many-body system, i.e. quasiparticle energies instead of Kohn-Sham energies (which strictly speaking are only a set of Lagrange parameters to guarantee orthonormality of the one-electron orbitals in the DFT minimization process of the total energy) , we use also the more computationally expensive Many Body Perturbation Theory (MBPT). Furthermore, in order to get a more intuitive picture of transport properties and more qualitative idea of band structure, we develop tight binding models based on the physics of these first-principles band structures

DFT is based on the Hohenberg Kohn theorems and provides an alternative viewpoint avoiding the N-particle Schrödinger equation wave functions. Here, the system's total energy is entirely expressed in terms of the density distribution of the ground state $\rho_{GS}(r)$ which is expressed in terms of single particle wave function $\phi_i$. DFT reduces the calculations of the ground state properties of systems of interacting particles exactly to the solution of single-particle Hartree-type equations. For detailed description of the computational theory of DFT, a good read would be Ref.[64, 65]. All calculations were performed using the full-potential linearized muffin-tin orbital (FP-LMTO) method[66, 67] using the questaal package, which is fully described in Ref.[68] and available at [64].

To put it concisely, the single particle wave functions are eigenstates of a one-particle Schrödinger equation with an effective potential, which is expressed itself in terms of the density and obtained from the total energy functional as functional of density as a functional derivative. Meanwhile the density is expressed in terms of one-electron eigenstates by means of the "Aufbau" principle, meaning filling the states from the lowest one upward until the Fermi level or the number of electrons. The self-consistent solution of this set of equations, provides a minimum of the total energy as functional of the density. Everything that is not classic Hartree like Coulomb energy, or interaction with nuclei or one-electron kinetic energy is dubbed exchange + correlation and the different types of DFT are determined by their choice of the exchange correlation functional. For example, the total energy $E$ is given by

$$E = T + \frac{1}{2}\iint dr^3 dr'^3 \frac{n(\mathbf{r})n(\mathbf{r}')}{|\mathbf{r}-\mathbf{r}'|} + \int dr^3 v_{ext}(\mathbf{r})n(\mathbf{r}) + E_{xc}[n(\mathbf{r})] \quad (3.1)$$

where $n(r)$ is the electron density, exchange correlation is given by $E_{xc}[n(\mathbf{r})]$ and kinetic energy by $T$. Here one often approximates the exchange correlation energy locally by the same functional dependence on density as occurs in a homogeneous electron gas. In other words, the equation for exchange correlation in a homogeneous electron gas depends only on the constant density of the gas as a simple function (rather than a functional) and this function is now applied locally at each point. which is

$$E_{xc}[n(\mathbf{r})] = \int \epsilon_{xc}^{\text{hom}}[n(\mathbf{r})]n(\mathbf{r})d^3r \tag{3.2}$$

and the corresponding exchange correlation potential is given by

$$v_{xc} = \frac{\delta E_{xc}[n(\mathbf{r})]}{\delta n(\mathbf{r})} = \frac{\partial\left(\epsilon_{xc}(n)n\right)}{\partial n} \tag{3.3}$$

This approximation is called the Local Density Approximation (LDA).

We will now describe the idea of MBPT we use in this thesis, namely the Quasi Particle Self-consistent GW (QSGW) method. Stemming from the famous set of Hedin's equations, QSGW adds a layer of sophistication to the already existing GW theoryRemark 3.0.1. GW is a approximation of the full set of Hedin equations where the vertex is neglected. It is the first term in a series expansion of the self-energy in terms of the screened Coulomb interaction .The GW approximation delivers high quality band structures of many band insulators and simple metals and automatically includes many physical effects such as exact Fock exchange, localized Coulomb repulsion, dynamic screening, and dispersion forces. In fact one can show that the LDA+U method is a static and localized approximation to GW[69]. Most GW calculations are performed perturbatively by computing corrections to a DFT-like electronic structure. The final result thus depends on the initial starting DFT electronic state. This now makes GW a parametrized theory as we pick the starting wave functions to perturb on[1] . One overcomes this by constructing optimum static one-body Hamiltonian $H_0$ describing the independent particle system (or Quasi Particle, hence the QS in QSGW). Usually, GWA is computed as a one-shot approximation with LDA wave functions as input with the approximation that $\Sigma = iG_{LDA}W_{LDA}$ where $\Sigma$ is the self energy, $G_{LDA}$ is the bare particle propagator and $W_{LDA}$ is the screened Coulomb interaction constructed from $G_{LDA}$ in the Random Phase Approximation (RPA). But in QSGW, we start with a set of trial wave functions and QP energies $\{\psi_{\mathbf{q}n}, \epsilon_{\mathbf{q}n}\}$, from which we calculate the bare one-particle Greens function $G$. We then use this $G$ to compute the dynamical self energy $\Sigma^{\mathbf{q}}_{nn'}(\omega)$ using the GWA scheme given by

1: One might think that the input LDA, being a parameter free theory, should imply GW is parameter free too. As it turns out, for relative correlates systems like some oxides, starting with LDA+U (parametrized theory) is needed than pure LDA, hence the parameter dependence

$$\Sigma\left(\mathbf{r}, \mathbf{r}', \omega\right) = \frac{i}{2\pi} \int d\omega' G^0\left(\mathbf{r}, \mathbf{r}', \omega - \omega'\right) W\left(\mathbf{r}, \mathbf{r}', \omega'\right) e^{-i\delta\omega'} \tag{3.4}$$

Up till now, we have just describe the usual GWA scheme and nothing new, but once we have $\Sigma^{\mathbf{q}}_{nn'}(\omega)$, we generate an energy-independent, Hermitian $\Sigma^{\mathbf{q}}_{nn'}$ by adding $V^{\text{xc}}$ to the Hamiltonian where the latter is given by

$$V^{\text{xc}} = \frac{1}{2}\sum_{ij} |\Psi_i\rangle \left\{\text{Re}\left[\Sigma\left(\varepsilon_i\right)\right]_{ij} + \text{Re}\left[\Sigma\left(\varepsilon_j\right)\right]_{ij}\right\} \left\langle\Psi_j\right| \tag{3.5}$$

Thus we have obtained a mapping $V^{\text{eff}} \longrightarrow V^{GW}(\omega) \longrightarrow V^{\text{eff}}$, which can now be used to solve at the level of DFT, a new single particle eigenfunction and then again fed to the self consistent mechanism.

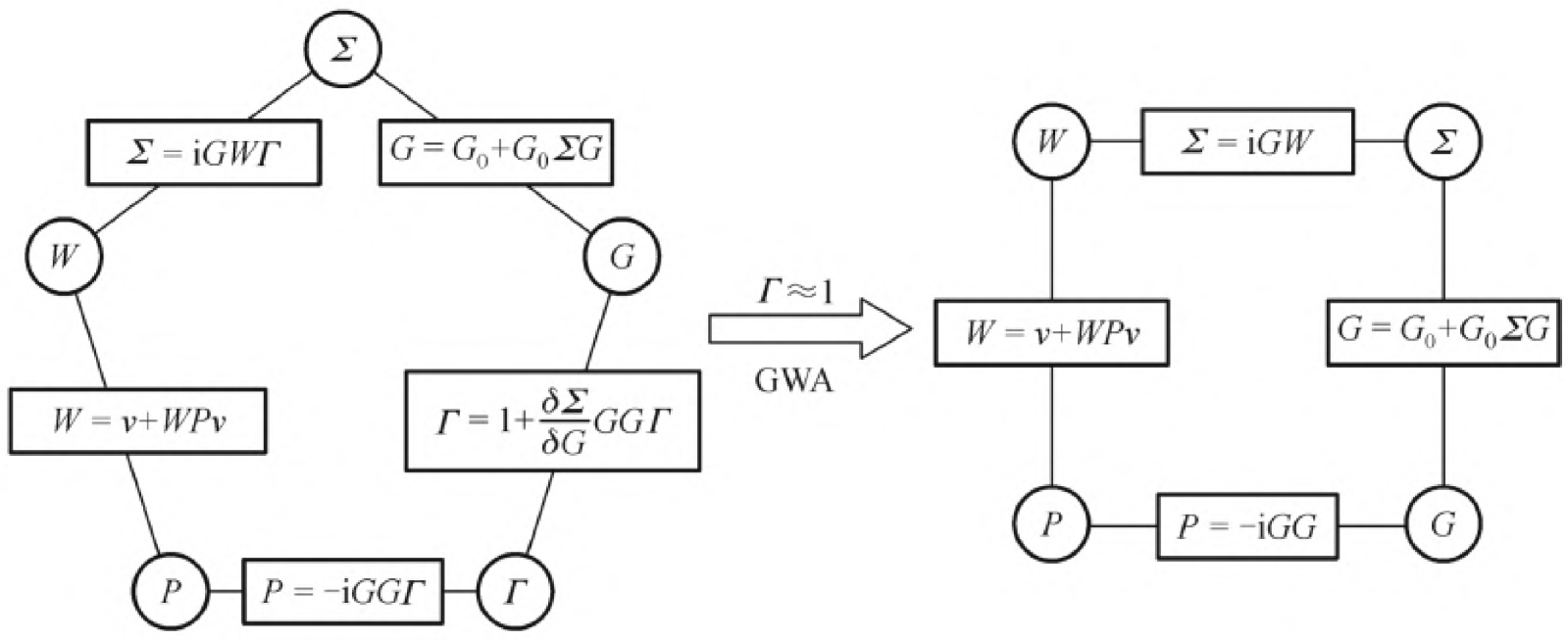

**Figure 3.1:** Flow chart for solving Hedin's equation

**Remark 3.0.1** Given bare particle propagator $G_0$, screen coloumb interaction $W$, vertex $\Gamma$, self energy $\Sigma$ and $P$, the polarizability, complete set of Hedin's equations are given by

$$G = G_0 + G_0\Sigma G \tag{3.6}$$
$$\Sigma = iGW\Gamma \tag{3.7}$$
$$W = V + WPV \tag{3.8}$$
$$P = -iGG\Gamma \tag{3.9}$$
$$\Gamma = 1 + \frac{\delta\Sigma}{\delta G}GG\Gamma \tag{3.10}$$

where $G$ is the dressed propagator. Solving these set of equations iteratively reaching a self-consistency, one would get the quantity $G$, which is the interacting Greens function. The $GW$ approximation corresponds to turning the last equation into $\Gamma \approx 1$, an immediate consequence of which is $\Sigma = iGW$, from where the name comes. Thus the resulting set of equations after making the $GW$ approximation amounts to solving,

$$G = G_0 + G_0\Sigma G \tag{3.11}$$
$$\Sigma = iGW \tag{3.12}$$
$$W = V + WPV \tag{3.13}$$
$$P = -iGG \tag{3.14}$$

This approximation basically reduces a step in the big picture of exactly solving the many body interaction problem shown by the self consistent flowchart in Figure 3.1 taken from Ref.[70]. This is a better approximation because the screened Coulomb interaction $W$ takes into account the major effect of correlation, namely the screening.

# Flat and low-buckled Sb | 4

**Abstract**

The electronic band topology of monolayer $\beta$-Sb (antimonene) is studied from the flat honeycomb to the equilibrium buckled structure using first-principles calculations and analyzed using a tight-binding model and low energy Hamiltonians. In flat monolayer Sb, the Fermi level occurs near the intersection of two warped Dirac cones, one associated with the $p_z$-orbitals, and one with the $\{p_x, p_y\}$-orbitals. The differently oriented threefold warping of these two cones leads to an unusually shaped nodal line, which leads to anisotropic in-plane transport properties and goniopolarity. A slight buckling opens a gap along the nodal line except at six remaining Dirac points, protected by symmetry. Under increasing buckling, pairs of Dirac points of opposite winding number annihilate at a critical buckling angle. At a second critical angle, the remaining Dirac points disappear when band structure opens a gap. Spin-orbit coupling and edge states are discussed.

## 4.1 Symmetry and Band dispersion

We start our discussion with the band structure of flat and slightly buckled monolayer Sb, as shown in Figure 4.1 and obtained in the generalized gradient approximation (GGA). The symmetry labeling of the bands is crucial to our understanding of the protection of the Dirac cones to be discussed. To make this symmetry labeling unambiguous, it is necessary to describe the symmetry operations and character tables in detail, which for keen readers is given in Section 4.10. Model Hamiltonian used throughout this section is described in Section 4.9.

### Band symmetry

The point groups of **k** applying along each symmetry line in the flat and buckled (labels in parentheses) case are given at the bottom of Figure 4.1.

Before proceeding with our study of the topological features of interest, we first point out some similarities and differences of the Sb band structure with the well known band structure of graphene. We immediately recognize the Dirac point at $K$, here labeled $E''$ corresponding to the $p_z$-derived bands (shown in blue) with $z$ perpendicular to the layer. In graphene, the Fermi level falls at this point but here, because of the additional valence electron, it lies higher in energy shown as the dash-dotted line and chosen as our reference energy. Another important difference with graphene is that the $s$-orbitals form a separate set of bands at lower energy rather than forming strongly hybridized sp$^2$ $\sigma$-bands. This results from a larger $E_p - E_s$ atomic energy difference relative to the hopping interactions between the sites. Nonetheless, we can see a little bit of $p$ contribution in the upper $s$-band from its slightly reddish color.

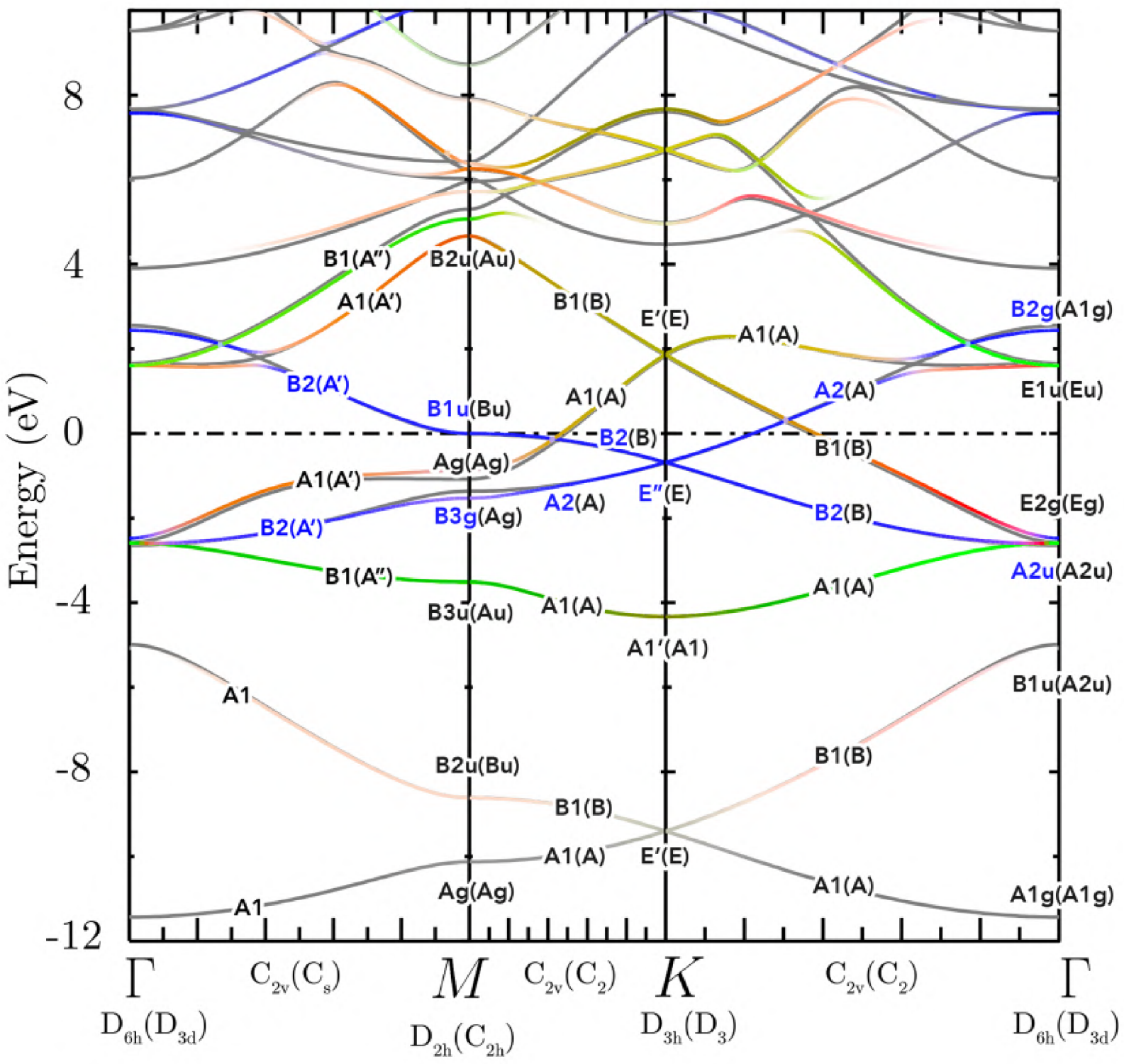

**Figure 4.1:** Symmetry labeled GGA band structure (in eV) of slightly buckled flat monolayer Sb; blue bands indicate bands dominated by $p_z$, red $p_x$, green $p_y$ (or their mixture), light grey in bottom two bands $s$; the bands of completely flat Sb (dark grey) differ only near the points were we see avoided crossings for the buckled case notably between bands 6-7 along $\Gamma - K$ and $\Gamma - M$ and the slight deviations of bands 4-5 near $M$;

The $\{p_x, p_y\}$ derived bands form a separate set of bands (indicated in red ($p_x$), green ($p_y$) and their mixture) with another Dirac cone $E'$ at $K$ above the Fermi level. The band structure of $\{p_x, p_y\}$ derived bands on the honeycomb lattice was discussed by Wu and Das Sarma[71] in a tight-binding approximation relevant to optical lattices where only the $\sigma$ interaction is non zero. Here, both the $V_\sigma$ and $V_\pi$ matter. While in a tight-binding model of each set of bands separately, the energy band derived from $s$, $p_z$ and $\{p_x, p_y\}$ are symmetric in energy with respect to their atomic energy band center, a feature usually referred to as particle-hole symmetry, the interaction with the higher lying Sb-$d$ bands here breaks this simplification.

The important point is that the Fermi level lies close to the intersection points of the $p_z$ derived and $\{p_x, p_y\}$ derived Dirac cones, one of which lies a little above $E_F$ along $\Gamma - K$ and the other a little below $E_F$ along $\Gamma - M$. We can see that the bands crossing at these points have different symmetry both in the flat and in the slightly buckled case and are thus protected by symmetry. Our symmetry analysis shows that it is a twofold rotation along the $\Gamma - K$ direction which is maintained both along $\Gamma - K$ and along $K - M$ even after buckling and protects the existence of these band crossings on the high symmetry lines. On the other hand, it is the horizontal mirror plane symmetry that protects the nodal line in the flat case.

## 4.2 Dirac cones and nodal line

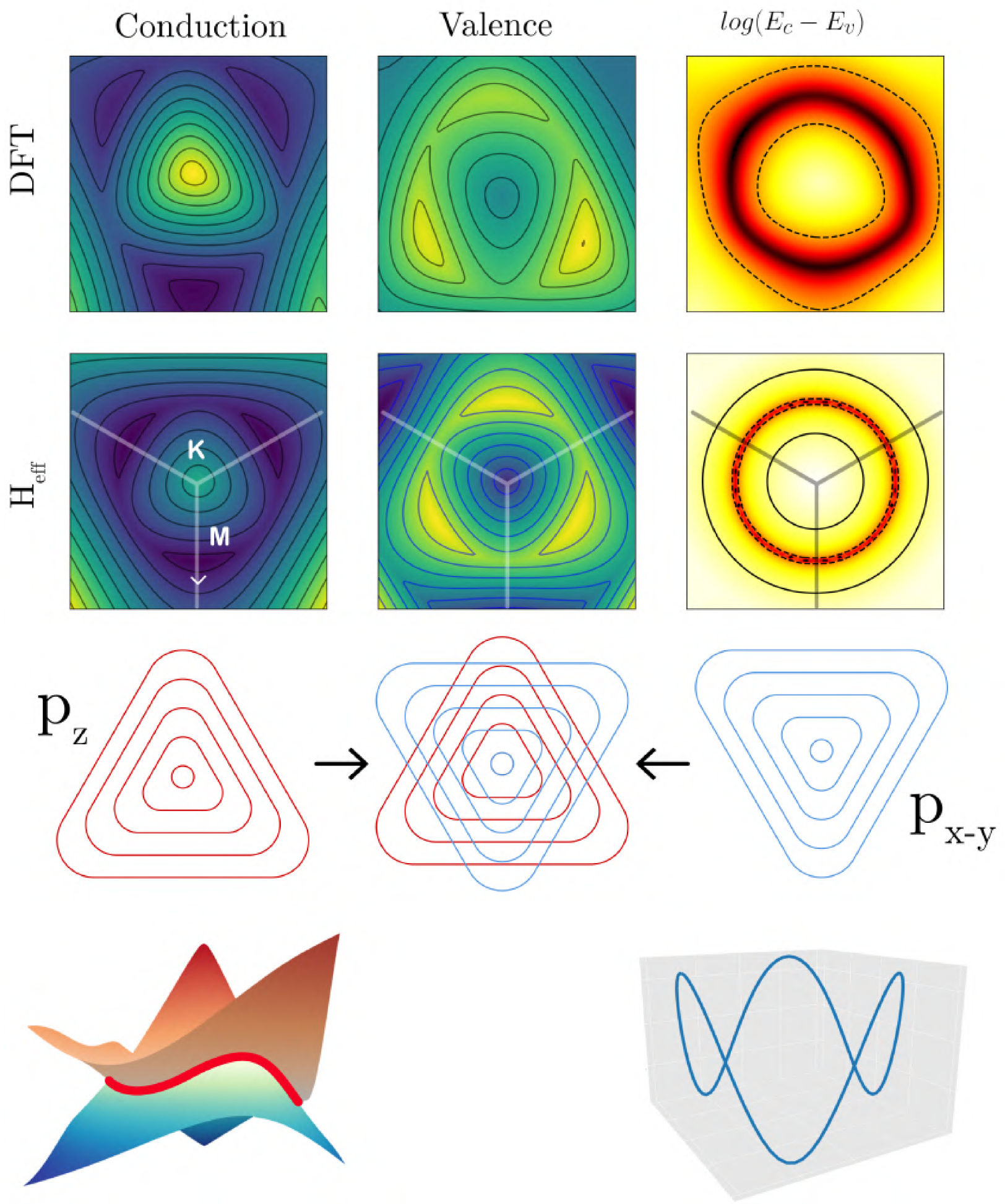

**Figure 4.2:** Contour plots of two Dirac cones and their intersection (on log scale) in DFT (top) and low energy effective hamiltonian (bottom); yellow is high, blue is low, explanation in text
Schematic of cone intersecton
3D view of (DFT) intersecting Dirac cones and (e) nodal line

The Dirac cones around $K$ are shown in Figure 4.2. In the top row of the contour plots, we present the DFT results and below it the corresponding results from an effective low-energy Hamiltonian based on symmetry and a nearest neighbor tight-binding model. In Figure 4.2 by valence (conduction) band we mean bands 5 (6) at each **k**-point numbered in order of increasing energy without regard to band crossings or the nature of the band. Thus close to $K$, the conduction band is the empty $E'$ $\{p_x, p_y\}$ derived band, whereas the valence band is the $E'$ $p_z$ derived band. These respective cones have trigonal symmetry and are seen as the triangular contours in the center of each figure. Beyond the crossing of the cones, the roles of conduction and valence band are reversed. The yellow (high value) regions in the "valence band" are at the intersection of the bands along $K - \Gamma$ while the dark blue (low value) in the CB correspond to the flat $p_z$ derived band along $K - M$. We can see that in the center the contours have a triangular shape but are rotated by 30° from each other as is further shown in the schematic sketch (c). This is also shown in a 3D view in part (d) and leads to a nodal line (e) with the Lissajous like shape, where again the blue surface corresponds to the valence band and the brown one to the conduction band. The triangular warping of the energy surfaces results from the terms of order $q^2$ in an expansion around the point $K$ and can be derived fully analytically from the tight-binding

Hamiltonian for the $p_z$ derived bands as is shown before. The linear terms of the Dirac cones are isotropic. Because $p_z$ ($\{p_x, p_y\}$) orbitals are odd (even) with respect to the horizontal mirror plane, they are derived from a separate $2\times2$ and $4\times4$ Hamiltonian matrix. Both of these can be further reduced to the eigenvalues of a $1\times1$ and $2\times2$ matrix because of the "particle-hole" symmetry within this model and analytical expressions can be derived for them at $\Gamma$, $K$ and near $K$. The threefold symmetry around $K$ is expected from the pointgroup of $K$ which is $D_{3h}$. The Dirac cone states can thus be written

$$\begin{aligned} E_z &= \Delta_z \pm v_z q \pm \frac{q^2}{m_z}\cos(3\phi), \\ E_{x,y} &= \Delta_{x,y} \pm v_{x,y} q \pm \frac{q^2}{m_{x,y}}\cos(3\phi) \end{aligned} \tag{4.1}$$

Here $\Delta_z$, $\Delta_{x,y}$ are the centers of the $(E', E')$ Dirac cones at $K$, $v_z$, $v_{x,y}$ are the Dirac linear dispersion velocities a and $m_z$, $m_{x,y}$ are effective mass parameters. The actual effective masses depend on the direction of $\mathbf{q}$ represented by its azimuthal angle $\phi$ from the $x$-axis ($\Gamma - K$ direction) leading to a warping of the constant energy lines with three-fold symmetry, while the velocity is isotropic. The opposite sign of the mass parameter for the two cones leads to their relative rotation by 30° and is found to be responsible for the interestingly shaped nodal line. Within the tight-binding model the sign of $m_{x,y}$ is found to be controlled by the ratio of the $V_\sigma$ and $V_\pi$ interactions. The situation here is reminiscent of that in AA bilayer graphene but with the difference that here the two Dirac cones have different velocities and warping terms. As a result their intersection is not a simple circular nodal line.

## 4.3 Consequences for transport

This unique shape of nodal ring gives rise to an equally interestingly shaped 2D Fermi surface as shown in Figure 4.3(a) (obtained from the effective mass Hamiltonian). The Fermi surface can be seen to consist of electron and hole pockets at 60° from each other and exhibits electron hole contact points (EHCP) where discontinuities occur in the band velocity. This figure shows that the carrier type changes from electron to hole type every 60° as we go around the Fermi surface, an effect that has been named goniopolarity.[46] Furthermore, depending on the precise location of the Fermi energy, which could in principle be varied by doping or gating, the electron or hole transport could be larger or smaller as the direction is changed in-plane. This situation is somewhat similar to the case of a tilted nodal line in a 3D $\mathbf{k}$-space, where interesting effects on the frequency dependent conductivity result from the cyclide geometry of the resulting Fermi surface.[72]

Here, we have calculated the static diffuse longitudinal conductivity $\sigma(E_F, \phi)/\tau$ apart from the, at this point unknown, relaxation time $\tau$ as function of azimuthal angle $\phi$ and the thermopower (or Seebeck coefficient) which is proportional to $d\ln\sigma(E,\phi)/dE|_{E=E_F}$ and whose sign reflects the charge of the carriers. These are shown in Figure 4.3(c) and (d) and show that the conductivity has modest anisotropy at the ~14 % level but the thermopower changes discontinuously from positive to negative at the EHCPs. Interestingly, it is positive for the directions corresponding to electron transport because $d\sigma/dE|_{E_F} < 0$. This is because the conductivity varies rapidly with energy near $E_F$ and the Fermi energy occurs just above a peak in the $\sigma(E, \phi)$ related to a logarithmic singularity in the density of states (DOS) resulting (see Figure 4.3(b)) from the saddle-point band structure at the point $B_{1u}$ at $M$. Inserting an order of magnitude estimate of $\tau = 10^{-13}$s would give a resistivity of order 0.2 $\mu\Omega$cm and a thermopower of order 5 $\mu$V/K which is relatively high due to the Fermi level's occurrence near a peak in the DOS. The unique feature here however is the discontinuous angular dependence of the thermopower. Numerous other opportunities in the optical conductivity and magnetotransport related to this unique 2D nodal line remain to be explored. While we pointed out the goniopolarity here for the completely flat system which hosts a nodal line, this property is expected to hold up even after somewhat buckling the system, so for less strong in-plane tensile strains, as long as some thermal excitation or doping provides carriers in the corresponding slightly gapped bands.

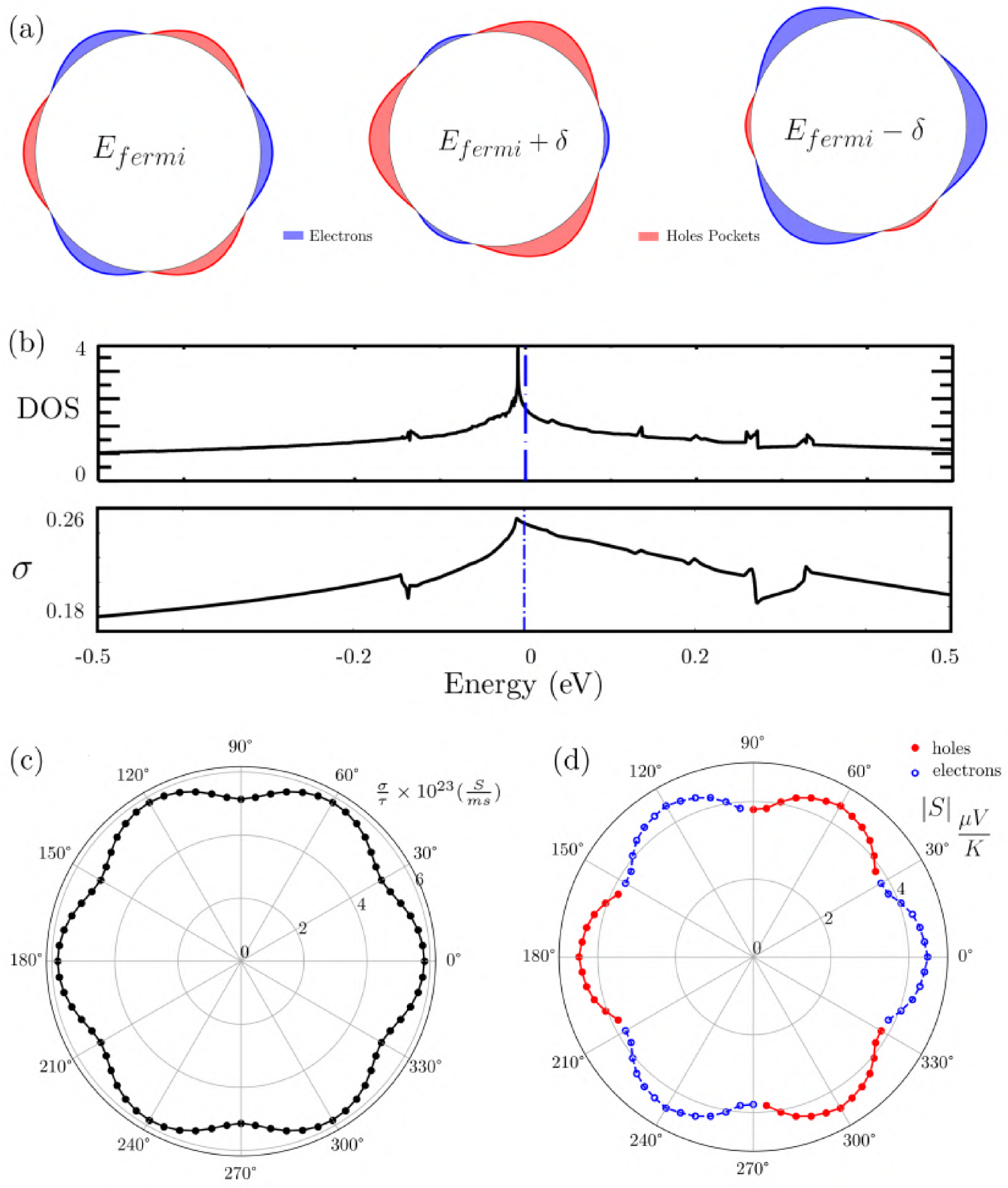

**Figure 4.3:** a) Fermi surface portions around each point $K$, (b) DOS and $\sigma(E)/\tau$ (arb. units) (c) Angular dependence of diffuse conductivity $\sigma(E_F)/\tau$ and (d) absolute value of thermopower with sign indicated by color (red < 0, blue > 0)

## 4.4 Changes in topology due to buckling

Next, we address the changes in band structure due to buckling. The buckling leads to an interaction between the $p_z$ and $\{p_x, p_y\}$ derived energy bands because the horizontal mirror plane symmetry no longer applies. We assume here that the bond-lengths between Sb atoms stays the same but the in-plane lattice constant shrinks as the buckling is increased. This is qualitatively consistent with the relaxation results[41] showing a decrease in vertical distance between the Sb atoms $d$ as function of in-plane lattice constant $a$. Thus in our tight-binding model the $V_\sigma$ and $V_\pi$ interactions stay the same but their relative contribution to the hopping integrals changes. Within the tight-binding model the interaction terms between $p_z$ on the one hand and $\{p_x, p_y\}$ on the other hand are proportional to $(V_\sigma - V_\pi)$ and to $\sin(2\theta)$ where $\theta$ is the buckling angle. For small buckling the coupling is thus linear in $\theta$. The symmetry labeled band structure shows that in the six high symmetry directions around $K$ the crossing is protected by symmetry and thus the interaction needs to go to zero every 60°. Hence by symmetry, the low energy Hamiltonian describing the behavior near these Dirac cones can be written

$$H^{buckled}_{\mathbf{k}+\mathbf{q}} = \begin{bmatrix} \Delta_{x,y} + \frac{q^2\cos(3\phi)}{m_\pi} - qv_{x,y} & A\theta\sin(3\phi) \\ A\theta\sin(3\phi) & \Delta_z + \frac{q^2\cos(3\phi)}{m_z} + qv_z \end{bmatrix} \tag{4.2}$$

with $A$ some constant and a $\sin(3\phi)$ behavior of the off-diagonal coupling. **??** shows the effect of buckling on the nodal line around $K$ by plotting the difference between conduction and valence band in a log plot and verifies the existence of six Dirac points.

Increasing the buckling either in the DFT or in the tight-binding model we find that the Dirac touching points around $K$ move closer toward $M$ along the $K - M - K'$ line and closer to $\Gamma$ along the $K - \Gamma$-line. At some critical angle the two Dirac points along $K - K'$ annihilate each other when they reach $M$. This is shown in Figure 4.4(b) In the DFT results, this occurs for about $\theta_c \approx 7°$.

The reason why they can annihilate is that they have opposite winding number +1 and −1. The winding number was calculated either from the effective low energy Hamiltonian or by calculating the Berry curvature in the tight-binding model The winding number, which can be thought of as a topological charge, is a conserved quantity [73]. Thus the only possible way to remove Dirac cones is to merge and annihilate them. The reason for the merging of the Dirac cones is related to time reversal symmetry, which guarantees that there is an equivalent and oppositely charged Dirac cone at $-\mathbf{k}$ for every Dirac cone at $\mathbf{k}$. Thus by symmetry, these Dirac cones of opposite sign come together and annihilate at the two types of time reversal invariant points $\Gamma$ and $M$.

[74] have analyzed the merging of 2D Dirac points in terms of a universal Hamiltonian, which shows that near such a point the bands correspond to a massive dispersion in one of the in-plane directions and a massless one in the orthogonal in-plane direction, which leads to an unusual $\sqrt{E}$ onset of the density of states.[75] which leads also to interesting changes in Landau levels[76] The behavior of the energy surfaces near the merging is shown in Figure 4.4(d).

After the merging of the Dirac points at $M$, upon further buckling, the Dirac points along $K - \Gamma$ keep moving closer to $\Gamma$. When they reach $\Gamma$ at a second critical angle of about 27°, they annihilate in pairs and the gap at $\Gamma$ beyond at this buckling first closes and then reopens, indicating a band inversion between bands of different symmetry label. The $E'$ and

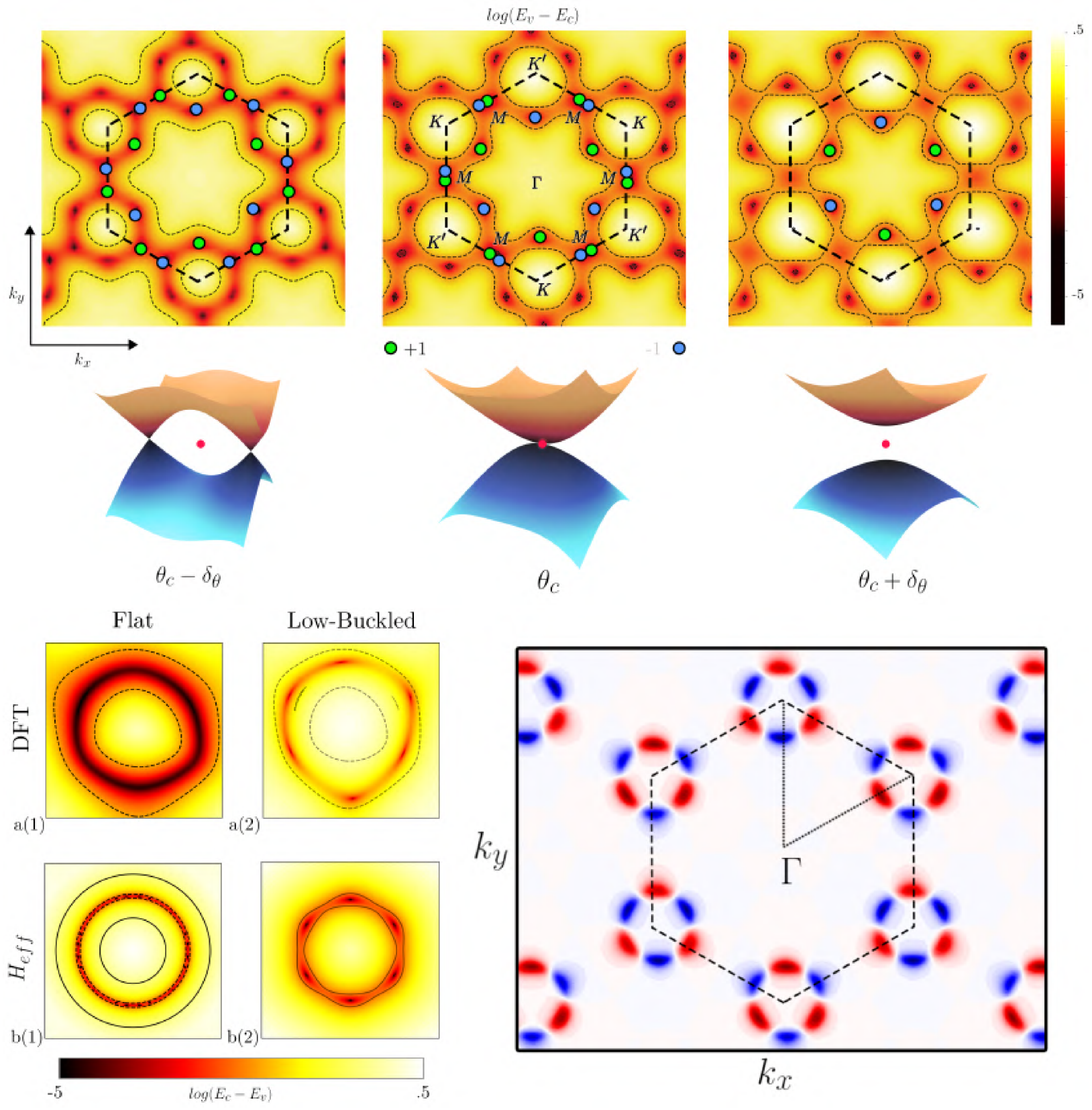

**Figure 4.4:**

(a) Nodal line gap opening and formation of six Dirac points in slightly buckled honeycomb; (b) Berry flux around each of the Dirac cones surrounding each $K$-point; (c) movement and merging of Dirac cones as buckling angle increases; (d) corresponding behavior of constant energy surfaces around $M$ (red dot).

$E'$ points splitting increases significantly with increasing buckling. With increasing buckling the distinction between $p_z$ and $\{p_x, p_y\}$ becomes less and less meaningful and a new type of hybridization between all three bands forming a bonding set of bands and antibonding set of bands emerges for the fully buckled ground state of the system. The opening of the gap corresponds to the transition from a topologically non-trivial to a trivial gap at $\Gamma$, which was studied earlier in literature[41] starting from the equilibrium large buckling by reducing the buckling under a tensile in-plane strain. We should clarify here that by non-trivial we mean in the sense of a topoligical insulator (TI) protected by time reversal. The system even in its fully buckled form has non-trivial topoligal crystal insulator (TCI) character.[77]

Later in Chapter 5, we will see that what was previously thought to be a trivially gapped insulator, is indeed not "trivial". The final transition once all the dirac cones meet at $\Gamma$ is indeed a higher order topological transition.

## 4.5 Spin-orbit coupling effects

Turning on spin orbit coupling (SOC), a gap opens up at each of the Dirac points. We can see that it is larger for the upper $E'$ point (0.56 eV) than at the lower $E''$ point (0.17 eV) and intermediate at the Dirac points near $E_F$. The non-trivial nature of the band crossings leads to topologically required edge states when the gap is opened by SOC. For the nearly flat case, these were studied in Ref. [45]. However, we need to keep in mind how the gaps at the $K-M$ and $\Gamma-K$ Dirac points are placed energetically

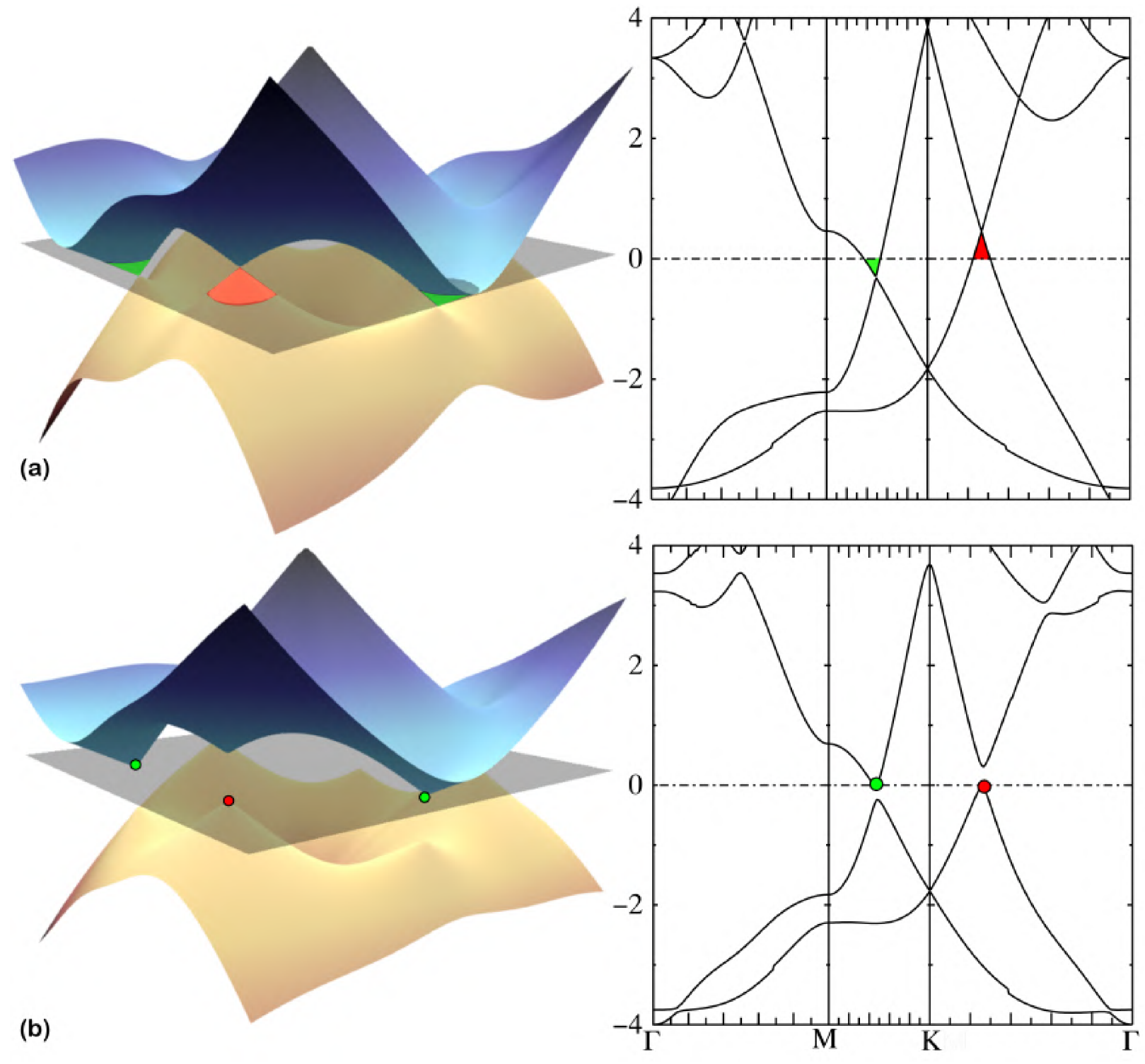

**Figure 4.5:**

QS*GW* band dispersion for the low-buckled system without (top) and with (bottom) SOC and the corresponding conical energy surfaces. Energies in eV.

relative to each other. While up to this point, we considered mostly generic properties which are topologically invariant, we now need to worry about the accuracy of the band structure and in particular the correct slope and placement of the different Dirac points relative to each other. To this end we perform the band structures in the QS*GW* approach which is known to give much more accurate single particle excitations than DFT in a semilocal approximation. We can see that this affects the band velocities of the Dirac cones and the energy difference of the $E'$, $E''$ Dirac points at $K$. For the fully buckled equilibrium system QS*GW* gives a gap of 2.9 eV, significantly larger than the 1.3 eV obtained in GGA and somewhat larger than the 2.28 eV from hybrid functional calculations.[78] In Figure 4.5 we can see that the highest VBM at the Dirac point along $\Gamma - K$ lies at the same energy as the lowest conduction band at the Dirac point along $K - M$. So, the system is an indirect zero gap semiconductor. Because the SOC is weaker in arsenene, there is then a non-zero indirect overlap between the occupied and empty bands at different Dirac points. The unique feature of this band structure is that it should have a topologically spin polarized edge state associated with these SOC induced spin-texture inverted gaps at specific **k**-points, even though the overall gap of the system is zero or slightly negative. Such a situation has been labeled a gapless topological insulator (GTI). It has been proposed to possibly occur due to electron-electron interaction effects[79] but is here found even in the independent particle approximation.

## 4.6 Topological edge states

As is well known, topological features in the band structure are closely related to protected edge states. The tight-binding model allows us to study the formation of these edge states explicitly in a finite nanoribbon.

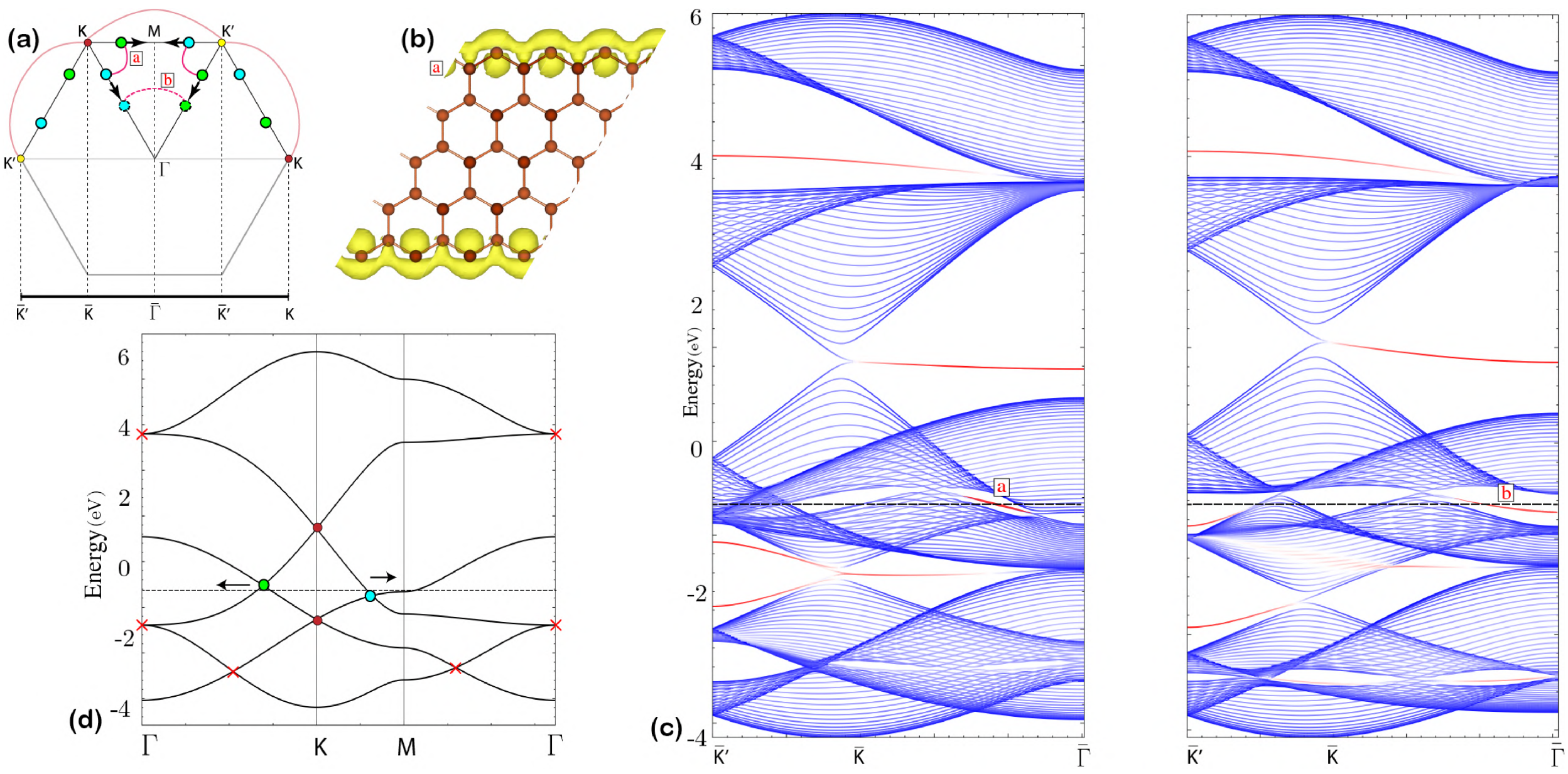

**Figure 4.6:** (a) folding of 2D Brillouin zone onto 1D Brillouin zone in nanoribbon indicating schematically the Dirac points linked by edge states and their motion under increased buckling; (b) 6 unit cell side nanoribbon with surface state from DFT; (c) band structure of nanoribbon in tight-binding model indicating the topologically protected edge states (in red) associated with Dirac point pairs labeled as in (a) for two buckling angles, one before and one after the merging of Dirac cones at $M$; (d) 2D tight-binding band structure for the low buckling case.

We choose to cut the honeycomb lattice along the zigzag direction and keep it periodic in the direction perpendicular to it but with a width of 60 unit cells. As further confirmation, Figure 4.6(b) shows a smaller 6 unit cell nanoribbon, calculated at the DFT level with the resulting edge state. The Brillouin zone folding is shown in Figure 4.6(a). The Dirac points are color coded according to their winding number and linked by lines for the pairs that will be connected by corresponding edge states. These connections found here explicitly from the numerical calculations are consistent with the theoretical considerations of Ryu and Hatsugai[80] and with explicit calculation of the Zak phase along different sections of $\mathbf{k}_{\|}$ showing that the edge states occur in the region where the Zak phase is $\pi$ and hence non-trivial. Note that this depends not only on the type of edge but also on the symmetry type of orbital involved in these bands. For instance, the edge states emanating from the $p_z$ derived Dirac cone at $K$ at about $-2$ eV extends between $\bar{K}$ and $\bar{K}'$ as is the case in graphene. However, the edge state in the gap near the Fermi level (marked (b) in the right panel of Figure 4.6(c) connects the Dirac point with $\bar{\Gamma}$ and continues in the next Brillouin zone back to the periodically repeated Dirac point. This difference from graphene results from the different type of orbitals involved and the region where the Zak-phase is non-trivial

The main part of the figure shows the 1D band structures indicating the edge states in red labeled to identify them with particular Dirac point pairings in the Brillouin zone figure following the same labeling. We here focus mainly on the Dirac cones near the Fermi level but other ones can be seen at energies farther away from the Fermi level. These are related to other linear band crossings and the $\Gamma$-point degenerate levels (marked by $\times$) which can be seen to occur in the tight-binding band structure for the 2D periodic band structure at energies farther removed from the Fermi surface. In the low-buckling case, from top to bottom, the edge states are related to the $E_{1u}$ state at $\Gamma$, the $E'$ state, the connection between $\Gamma - K$ and $K - M$ Dirac points, the $E''$ Dirac point and the $E_{2g}$ state at $\Gamma$, with the latter two interacting along $\bar{K} - \bar{K}'$. Finally, an edge state connected to the lowest energy Dirac crossings along $\Gamma - K$ can be seen along $\bar{K} - \bar{\Gamma}$. The edge states associated with the low energy crossing along $\Gamma - M$ cannot be seen in this nanoribbon because the $\Gamma - M$ direction is the one along which we fold the bands. Because various of these edge states connect Dirac points not along a high symmetry direction, one expects them to be present for other cut-outs of the 2D lattice in arbitrary directions.

## 4.7 Stability and Metastability of flat and buckled 2D monolayers

Here we discuss the stability of flat *vs.* buckled forms of 2D monolayer Sb and As in the $\beta$-structure. In Fig. Figure 4.7 we show the results of first-principles density functional calculations using the PBE-GGA functional for the total energy as function of in-plane lattice constant of the honeycomb lattice. In the buckled case, the structure for a given $a$ is allowed to fully relax, leading to a relative high buckling angle of $\sim 33°$ defined by $\tan\theta = \sqrt{3}d/a$ with $d$ the vertical distance between Sb atoms along the $z$-axis and $a$ the in plane lattice constant. We can see that a nearly flat structure, with higher in-plane lattice constant, exists as a second metastable minimum.

The dynamical stability of monolayer honeycomb Sb was studied by Zhao *et al.* [44] by means of phonon calculations in the harmonic approximation. It shows that up to a 19% in-plane tensile strain, the system is dynamically stable while beyond this critical strain, imaginary phonon modes appear around $K$ and near $\Gamma$. We have confirmed these results using the Quantum Espresso code[81] and also found that the flat form shows imaginary modes. However, at finite temperature anharmonic terms in the potential could still keep the system from becoming unstable. Typically a dynamic symmetry breaking instability of this type only occurs below a critical temperature because, at higher displacement, fourth order terms become dominant. Secondly, what one needs to consider here is not the stability of the free-standing flat form of Sb but the flat form under an applied tensile in-plane strain. This would actually require one to check the positive definiteness of $\partial\mathcal{H}/\partial u_i \partial u_j$ where $\mathcal{H} = \mathcal{U} - V\eta \cdot \sigma$ with $\mathcal{U}$, the internal energy, $\eta$ the strain and $\sigma$ the stress and $V$ the volume, as function of atomic displacements in the cell $u_i$. Such anahrmonic calculations and including the strain explicitly in the formulation are not at all trivial.[**Bianco17**] Fortunately, they are not needed because the experiment has already confirmed the stability of the material in flat form.[38]

Although free standing monolayer Sb in the $\beta$-structure clearly has a high buckling angle, it has unambiguously been demonstrated experimentally that the flat or nearly flat structure can be stabilized by epitaxial in-plane tensile strain by putting the Sb on a Ag(111) structure.[38] The band structure of free standing flat monolayer Sb is discussed in the main section. For comparison, we here show the band structure of a 10 layers thick Ag layer with a monolayer of Sb on top in Fig. Figure 4.8. The

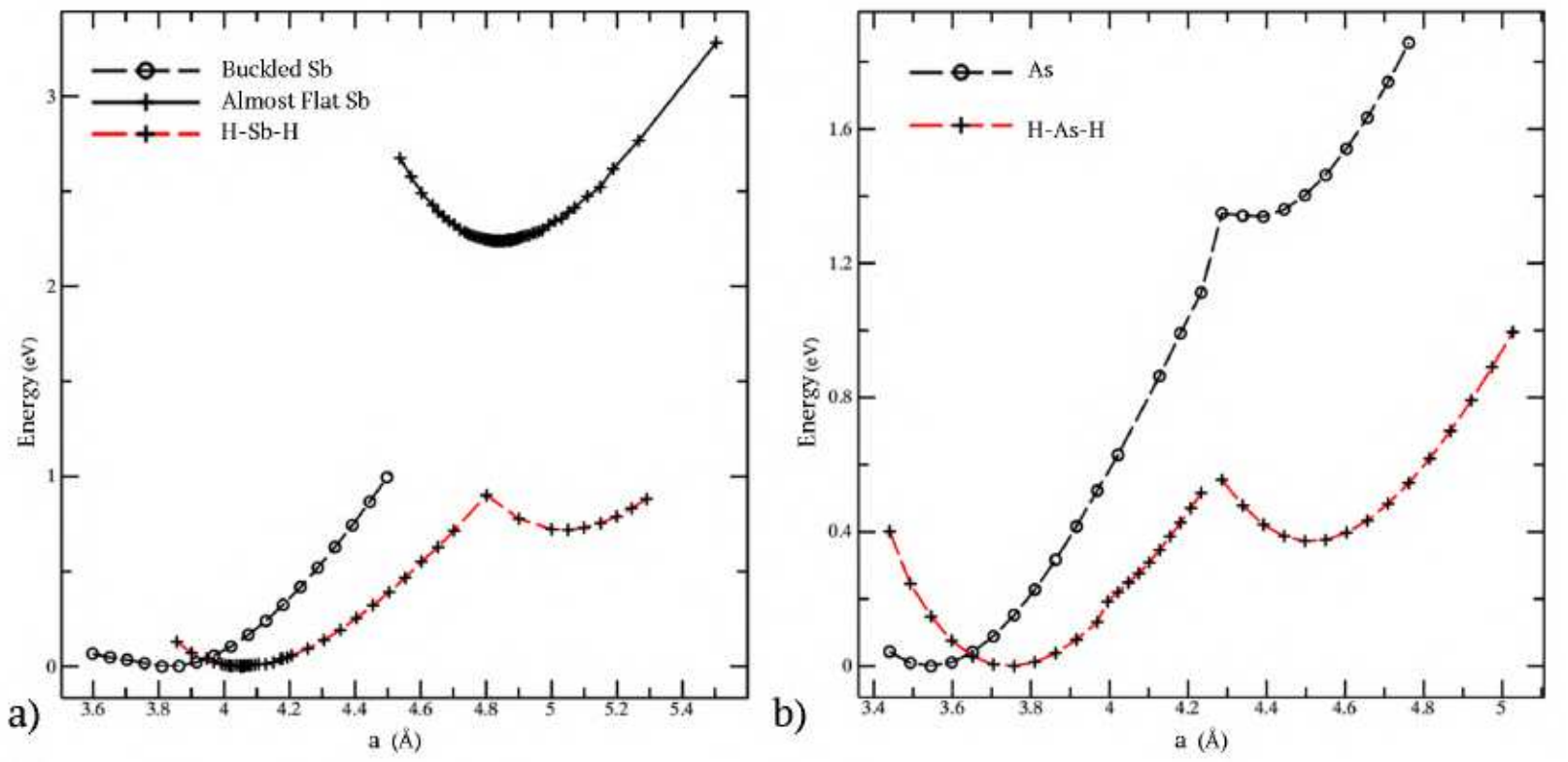

**Figure 4.7:**

Fully relaxed total energy per 2-atom unit cell as function of in-plane lattice constants for Sb (left) and As (right)

structure was relaxed with DFT before calculating the band structure. The bands weighted by their Sb contribution are shown in red. This shows that the features of monolayer Sb can still be recognized clearly on top of the Ag background, especially in the important region near the Dirac crossings and Fermi energy where the Ag density of states is low. This confirms the X-ray photemission (XPS) investigations by Shao *et al.* [38] indicating a weak interaction between substrate and monolayer, which could maintain at least approximately its relevant properties while providing the required in-plane tensile strain to stabilize the flat form.

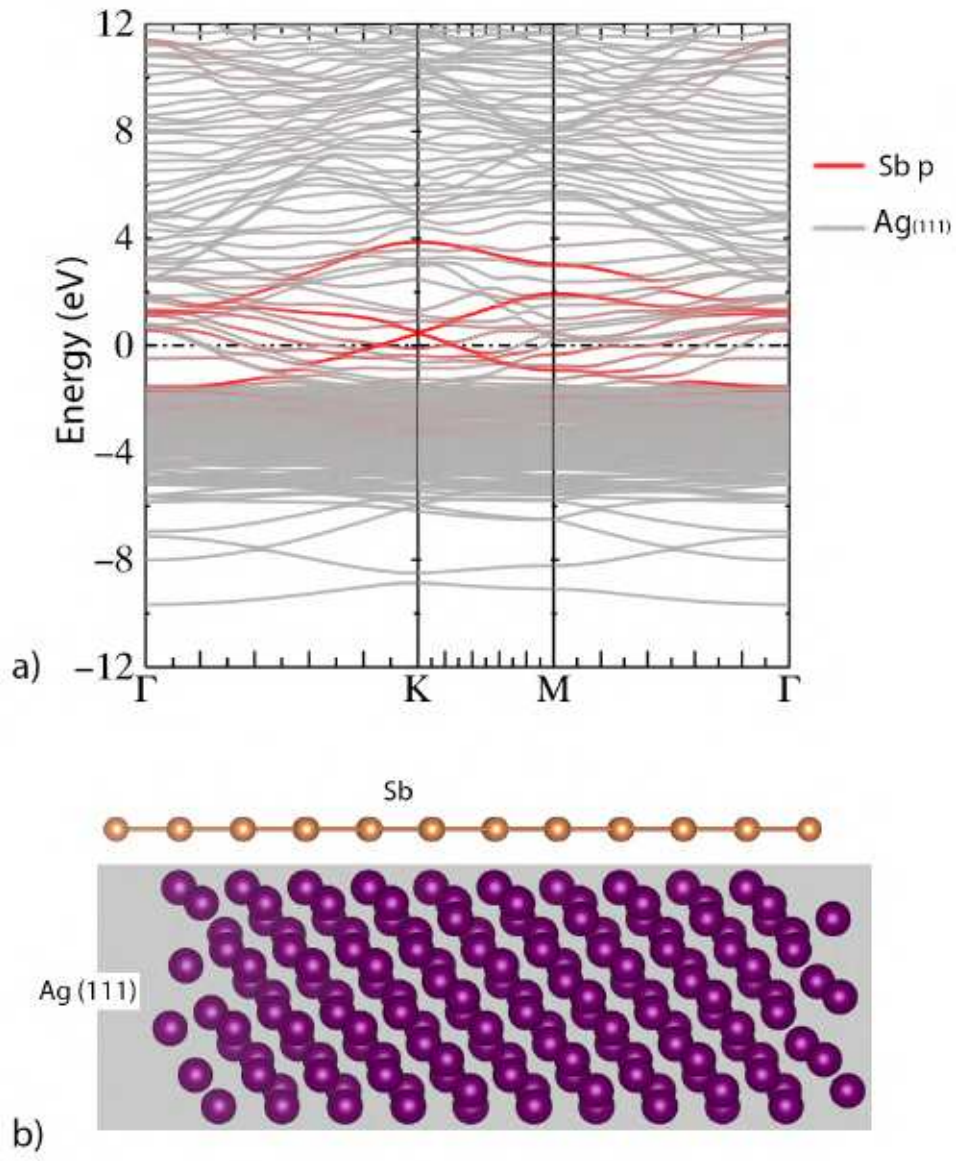

**Figure 4.8:**

Band structure of a Ag (111) slab with Sb monolayer adsorbed.

This shows that there is a route forward to experimentally investigate the band structure aspects studied in the above section by means of epitaxial stabilization and the investigation of monolayer Sb under high tensile strain in a flat or nearly flat form is not just a theoretical exercise but could potentially be realized experimentally by adjusting the coupling to the underlying substrate or varying the lattice constant of the substrate. Of course, for topologically induced effects on the transport, one would then also have to consider scattering to and from the underlying Ag band structure and the effects of the substrate symmetry breaking. Nanoscale mechanical systems can potentially also be designed for free standing membranes suspended over a hole in the substrate as is currently done for graphene, black phosphorous etc. What level of strain can be sustained in such systems is not yet clear but 2D systems are known to be able to sustain larger strains than 3D systems. Alternative substrates, like h-BN with van der Waals type interactions might also be considered as a way to maintain the required tensile strain while at the same time minimizing the interactions so as to preserve the properties of flat monolayer Sb.[45] In any case the properties studied here for free-standing flat mono-layer Sb should serve as a well-defined theoretical limit of the physically realizable flat Sb on a suitable substrate with sufficiently weak interactions. The perturbations caused by such interactions would have to be considered in future work.

Finally hydrogenating the structure with H on top or below the Sb on alternating Sb, is seen to lower the energy of the flat Sb structure significantly although it still has higher energy than the equilibrium structure. Similar results are found for As. In the this paper we have not studied the hydrogenated form. In that case the Fermi level lies at the $\{p_x, p_y\}$-derived $E'$ Dirac point at $K$. This case was studied in Refs. [82, 83]

## 4.8 Arsenene

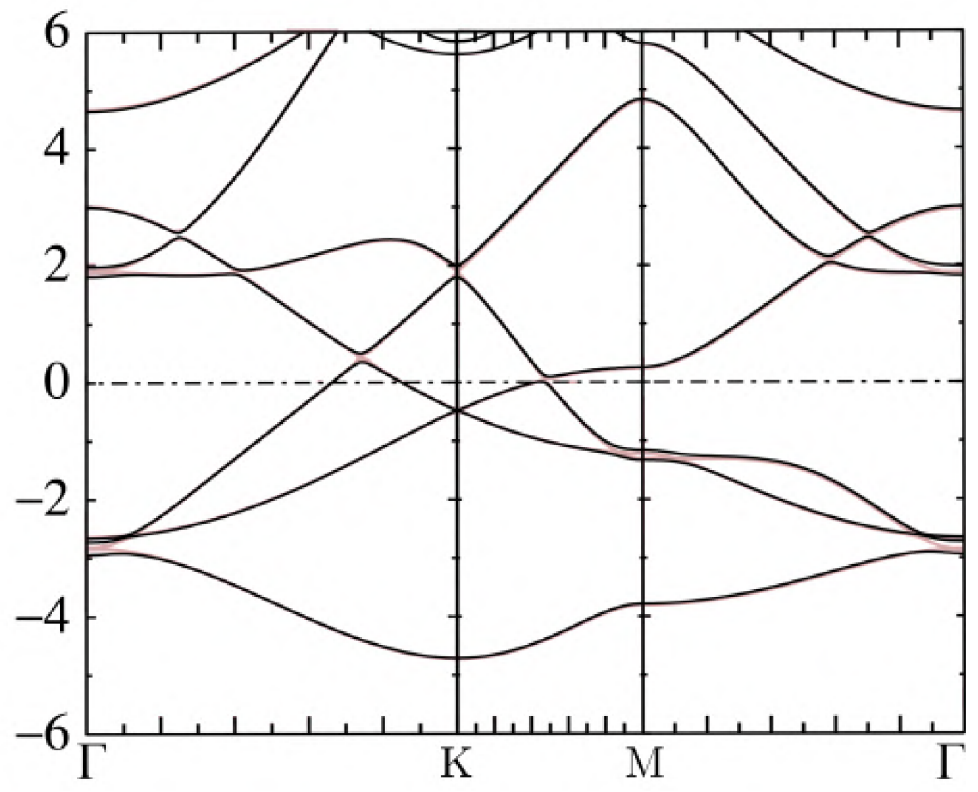

**Figure 4.9:**

Band structure of arsenene (As) with (black) and without (red) SOC at the GGA level.

While we mostly focused on antimonene (Sb), very similar results hold for arsenene (As). The relative stability of the flat and buckled honeycomb structure was already shown in Fig. Figure 4.7. The band structure at the GGA level is shown in Fig. Figure 4.9 both with and without spin-orbit coupling. We can see that the Fermi level again occurs near the crossing of the two Dirac cones centered at $K$. The crossings of these cones along $\Gamma - K$ and $K - M$ are again tilted with respect to each other. The spin-orbit coupling opens a gap at all the Dirac points. We can see that as in Sb, the spin-orbit opened gap is larger for the upper $E'$ $\{p_x, p_y\}$ derived Dirac point at $K$ than for the lower $E''$ $p_z$ derived Dirac point. At the new Dirac points along $\Gamma - K$ and $K - M$, the splitting is intermediate. The main point is that the gaps opened here at the Dirac points near $E_F$ are straddled with respect to each other so that the system remains overall metallic, in contrast with the Sb case in the above section Fig. 3, where a zero indirect gap situation emerges.

## 4.9 $p_x - p_y - p_z$ Model Hamiltonian

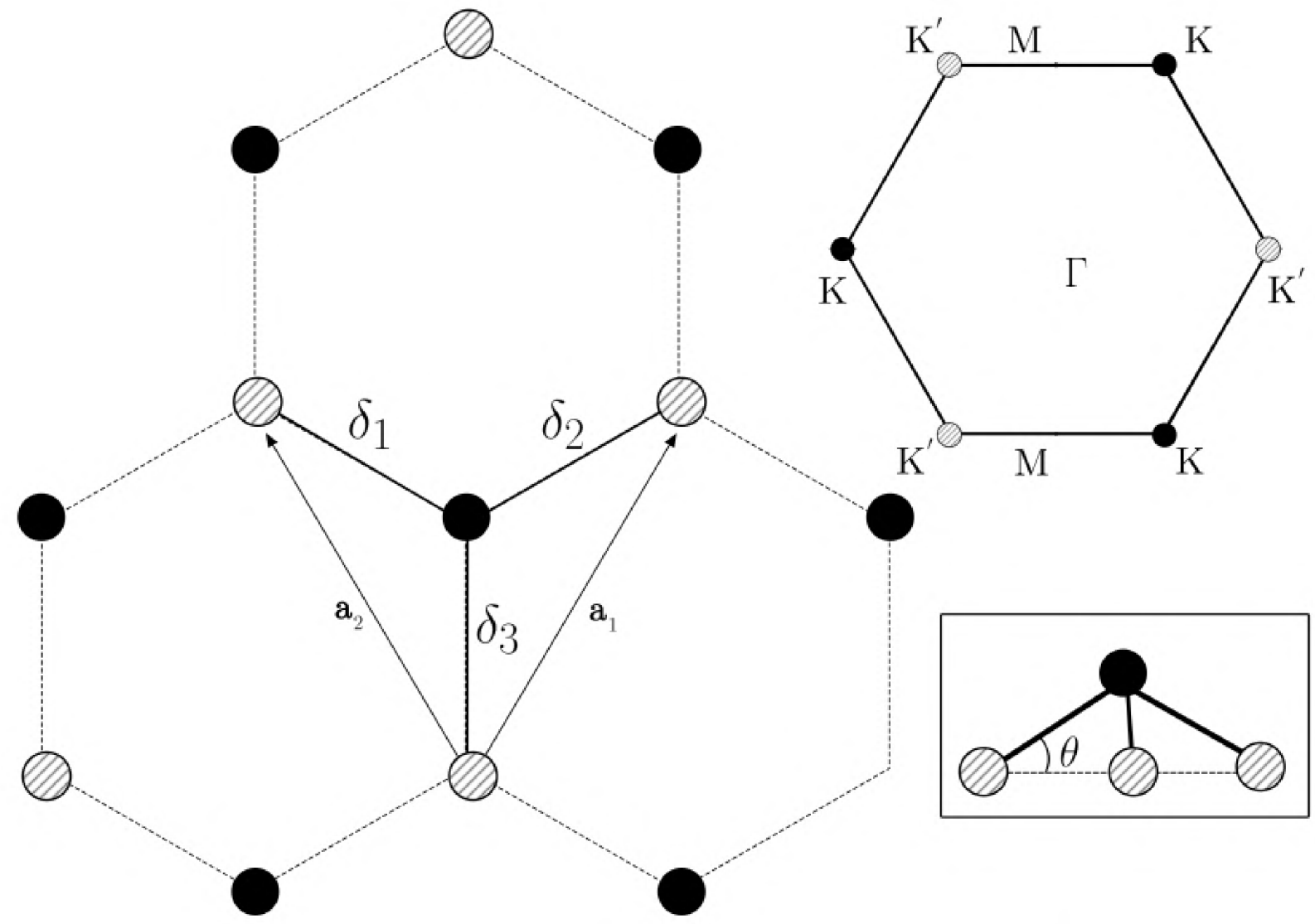

**Figure 4.10:** Lattice of low buckled antimonene. Shaded and dark circles represent the Sb at different height. Inset shows the definition of buckling used in the paper along with the corresponding $1^{st}$ $BZ$.

We start by constructing the nearest neighbor tight-binding Hamiltonian for the Sb-$p$ derived orbitals. The structure of the 2D system is shown in Fig. Figure 4.13. For the bands near the Fermi level, one can ignore the contribution of $s$ states as they occur at much lower energy as seen in the previous section. Separating the orbitals according to their sublattice $A$, $B$, the Hamiltonian takes the block form:

$$H = \begin{matrix} & A & B \\ A & \bar{\Delta} + \mu_A I_{3\times3} & \bar{H} \\ B & \bar{H}^* & \bar{\Delta} + \mu_B I_{3\times3} \end{matrix} \tag{4.3}$$

Here

$$\bar{\Delta} = \begin{bmatrix} \Delta_{x,y} & & \\ & \Delta_{x,y} & \\ & & \Delta_z \end{bmatrix} \tag{4.4}$$

gives the energy shift of the atomic $p$-orbitals from the zero reference energy. We included here the fact that the on-site diagonal energy for $p_z$ orbitals may be different from that of $\{p_x, p_y\}$ orbitals. We can also switch on a different potential $(\mu_A, \mu_B)$ on each sublattice and examine which band crossings open up as a gap in response. The off-diagonal $AB$-blocks involve the Bloch sum over the three nearest neighbors:

$$\bar{H} = T_1(V_\pi, V_\sigma, \theta)e^{i\mathbf{k}\cdot\boldsymbol{\delta}_1} + T_2(V_\pi, V_\sigma, \theta)e^{i\mathbf{k}\cdot\boldsymbol{\delta}_2} + T_3(V_\pi, V_\sigma, \theta)e^{i\mathbf{k}\cdot\boldsymbol{\delta}_3} \tag{4.5}$$

where

$$\begin{aligned} \boldsymbol{\delta}_1 &= (-\cos\theta\sqrt{3}/2, \cos\theta/2, \sin\theta)a/\sqrt{3}, \\ \boldsymbol{\delta}_2 &= (\cos\theta\sqrt{3}/2, \cos\theta/2, \sin\theta)a/\sqrt{3}, \\ \boldsymbol{\delta}_3 &= (0, -\cos\theta, \sin\theta)a/\sqrt{3} \end{aligned} \tag{4.6}$$

in terms of the lattice constant $a$ as shown in **??**. The side of the hexagon is $a/\sqrt{3}$. Here $T_1$, $T_2$, $T_3$ can be expressed in terms of the $V_\pi$ and $V_\sigma$ components of the nearest neighbor hopping interaction using the Koster-Slater two-center approximation.[84] We assume here that under buckling, the bond distance $a$ is kept fixed and the buckling angle just changes the relative contributions of $V_\pi$ and $V_\sigma$ to the hopping integrals. We chose the $V_\pi/V_\sigma$ ratio to be −0.275. This value is close to Walter Harrison's universal ratio.[85] Together with the $\Delta_{x,y} - \Delta_\sigma$ parameters, they were chosen to present a similar band structure to the DFT results shown in Figure 4.1 on page 28. However, a detailed fit was not attempted because the actual band structure is affected by interaction with nearby $d$ and $s$ bands and the main purpose of our tight-binding model is to study the generic behavior of the bands as function of buckling angle and mostly the topological features.

Each $T_i$ is a $3 \times 3$ matrix in the basis of the $p_m$, $m = (x, y, z)$ orbitals.

$$T_1(V_\pi, V_\sigma, \theta) = \begin{bmatrix} V_\pi\left(-\frac{3\cos^2\theta}{4}+1\right) + V_\sigma\frac{3\cos^2\theta}{4} & (V_\pi - V_\sigma)\frac{\sqrt{3}\cos^2\theta}{4} & (V_\pi - V_\sigma)\frac{\sqrt{3}\sin\theta\cos\theta}{2} \\ (V_\pi - V_\sigma)\frac{\sqrt{3}\cos^2\theta}{4} & V_\pi\left(-\frac{\cos^2\theta}{4}+1\right) + V_\sigma\frac{\cos^2\theta}{4} & (V_\sigma - V_\pi)\frac{\sin\theta\cos\theta}{2} \\ (V_\pi - V_\sigma)\frac{\sqrt{3}\sin\theta\cos\theta}{2} & (V_\sigma - V_\pi)\frac{\sin\theta\cos\theta}{2} & V_\pi\cos^2\theta + V_\sigma\sin^2\theta \end{bmatrix} \tag{4.7}$$

$$T_2(V_\pi, V_\sigma, \theta) = \begin{bmatrix} V_\pi\left(-\frac{3\cos^2\theta}{4}+1\right) + V_\sigma\frac{3\cos^2\theta}{4} & (V_\sigma - V_\pi)\frac{\sqrt{3}\cos^2\theta}{4} & (V_\sigma - V_\pi)\frac{\sqrt{3}\sin\theta\cos\theta}{2} \\ (V_\sigma - V_\pi)\frac{\sqrt{3}\cos^2\theta}{4} & V_\pi\left(-\frac{\cos^2\theta}{4}+1\right) + V_\sigma\frac{\cos^2\theta}{4} & (V_\sigma - V_\pi)\frac{\sin\theta\cos\theta}{2} \\ (V_\sigma - V_\pi)\frac{\sqrt{3}\sin\theta\cos\theta}{2} & (V_\sigma - V_\pi)\frac{\sin\theta\cos\theta}{2} & V_\pi\cos^2\theta + V_\sigma\sin^2\theta \end{bmatrix} \tag{4.8}$$

$$T_3(V_\pi, V_\sigma, \theta) = \begin{bmatrix} V_\pi & 0 & 0 \\ 0 & V_\pi\sin^2\theta + V_\sigma\cos^2\theta & (V_\pi - V_\sigma)\sin\theta\cos\theta \\ 0 & (V_\pi - V_\sigma)\sin\theta\cos\theta & V_\pi\cos^2\theta + V_\sigma\sin^2\theta \end{bmatrix} \tag{4.9}$$

It is insightful to first understand the unbuckled case ($\theta = 0$). In that case, the $p_z$ orbitals are decoupled from the $p_x, p_y$ orbitals because the former are odd with respect to the horizontal mirror plane and the latter are even.

The $z$-part of the Hamiltonian becomes the well-known

$$H_z(\mathbf{k}) = \begin{bmatrix} \Delta_z & V_\pi g_0(\mathbf{k}) \\ V_\pi g_0^*(\mathbf{k}) & \Delta_z \end{bmatrix} \tag{4.10}$$

with

$$g_0(\mathbf{k}) = \sum_{j=1}^{3} e^{i\mathbf{k}\cdot\boldsymbol{\delta}_j} \tag{4.11}$$

giving the purely $\pi$-bonded bands. At $\Gamma$ we find $E(\mathbf{k}_\Gamma) = \Delta_z \pm 3V_\pi$ with eigenvectors $\frac{p_{zA} \mp p_{zB}}{\sqrt{2}}$. At $K$, we have $e^{i\mathbf{k}\cdot\boldsymbol{\delta}_1} = e^{-i2\pi/3}$, $e^{i\mathbf{k}\cdot\boldsymbol{\delta}_2} = e^{i2\pi/3}$ $e^{i\mathbf{k}\cdot\boldsymbol{\delta}_1} = 1$ and $g_0(\mathbf{k}) = 0$ giving the doubly degenerate eigenvalue $\Delta_z$. This is the Dirac point $E''$.

The $xy$ part of the Hamiltonian has the off-diagonal part

$$\bar{H}_{xy} = \begin{bmatrix} \frac{3V_\sigma+V_\pi}{4} g_+(\mathbf{k}) + V_\pi e^{i\mathbf{k}\cdot\boldsymbol{\delta}_3} & \frac{\sqrt{3}(V_\pi-V_\sigma)}{4} g_-(\mathbf{k}) \\ \frac{\sqrt{3}(V_\pi-V_\sigma)}{4} g_-(\mathbf{k}) & \frac{3V_\pi+V_\sigma}{4} g_+(\mathbf{k}) + V_\sigma e^{i\mathbf{k}\cdot\boldsymbol{\delta}_3} \end{bmatrix} \tag{4.12}$$

with $g_\pm(\mathbf{k}) = e^{i\mathbf{k}\cdot\boldsymbol{\delta}_1} \pm e^{i\mathbf{k}\cdot\boldsymbol{\delta}_2}$. At $\Gamma$, the $g_-(\mathbf{k}_\Gamma) = 0$ and we obtain two degenerate eigenvalues $\Delta_{x,y} \pm (V_\sigma + V_\pi)3/2$ with eigenstates $(p_{xA} \mp p_{xB})/\sqrt{2}$ and $(p_{yA} \mp p_{yB})/\sqrt{2}$. At $K$, the matrix can still be diagonalized analytically. The off-diagonal part here takes the form

$$\bar{H}_{xy}(\mathbf{k}_K) = \begin{bmatrix} -\frac{3}{4}(V_\sigma - V_\pi) & i\frac{3}{4}(V_\sigma - V_\pi) \\ i\frac{3}{4}(V_\sigma - V_\pi) & \frac{3}{4}(V_\sigma - V_\pi) \end{bmatrix} \tag{4.13}$$

The Hamiltonian then has a double degenerate eigenvalue $E = \Delta_\pi$ with eigenvectors $\pi_A = (p_{xA} + ip_{yA})/\sqrt{2}$ and $\pi_B^* = (p_{xB} - ip_{yB})/\sqrt{2}$, the Dirac point $E'$, and two non-degenerate eigenvalues $\Delta_{x,y} + (V_\pi - 3V_\sigma)/2$ with eigenvector $(\pi_A^* - \pi_B)/\sqrt{2}$ $\Delta_{x,y} - (V_\pi - 3V_\sigma)/2)$ with eigenvector $(\pi_A^* + \pi_B)/\sqrt{2}$. In other words, it can be diagonalized in the basis of the $\pi_A$, $\pi_A^*$, $\pi_B$ and $\pi_B^*$ orbitals.[71]

We now examine the band surfaces in 3D, in particular the intersection of the down pointing Dirac cone derived from the $p_x, p_y$ orbitals and the upward pointing Dirac cone derived from the $p_z$ orbitals. To further study this crossing analytically we expand the tight-binding Hamiltonian around $K$, *i.e.* for $\mathbf{k} = \mathbf{k}_K + \mathbf{q}$ for small $\mathbf{q} = (q\cos\phi, q\sin\phi)$. The azimuthal angle $\phi$ of the $\mathbf{q}$ is measured from the $X$-direction for the $K$-point along $x$ and $q = |\mathbf{q}|$.

The eigenvalues are symmetric about the $\Delta_z$ and $\Delta_{x,y}$ and given by

$$\begin{aligned} E_z &= \Delta_z \pm v_z q \pm \frac{q^2}{m_z}\cos(3\phi), \\ E_{x,y} &= \Delta_{x,y} \pm v_{x,y} q \pm \frac{q^2}{m_{x,y}}\cos(3\phi) \end{aligned} \tag{4.14}$$

Here $v_z$, $v_{x,y}$ are the Dirac linear dispersion velocities a and $m_z$, $m_{x,y}$ are an effective mass parameter. The actual effective masses depend on the direction of $\mathbf{q}$ leading to a warping of the constant energy lines with three-fold symmetry, while the velocity is isotropic. For the $p_z$ Hamiltonian, one finds in our nearest neighbor tight binding Hamiltonian, $v_z = \sqrt{3}\pi V_\pi$ and $m_z^{-1} = -V_\pi \pi^2/2$. Thus both are completely determined by the $V_\pi$-interaction between $p_z$ orbitals. On the other hand, one may also keep $m_z$ and $v_z$ as independent parameters to make the effective low energy Hamiltonian and eigenvalues applicable beyond the tight-binding model. Their form is dictated by symmetry.

For the $\{p_x, p_y\}$ derived bands, the eigenvalues of the tight-binding Hamiltonian are also found to be symmetric about $\Delta_\pi$ and are found by diagonalizing $H_{eff} = \bar{H}_{xy}\bar{H}_{xy}^\dagger$ and taking $\pm\sqrt{\lambda}$ of its eigenvalues $\lambda$. These eigenvalues were worked out by Wu *et al.* [71] for the special case that $V_\pi = 0$, which applies to optical lattices. In that case, two bands are found to be flat, and two are Dirac cone like and have exactly the same shape as for the $p_z$ orbitals. With a mixture of both $V_\sigma$ and $V_\pi$, the expressions of the expansion in $q$ become too complex to be useful but the important finding is that by choosing the $m_{x,y}$ of opposite sign as the $m_z$ orbitals the warping is found to be rotated by 30°. In the DFT band structure in Figure 4.1 on page 28 we can see that the particle-hole symmetry about $\Delta_{x,y}$ which is the $E'$ value above the Fermi level, no longer holds. This is because of the interactions with the higher lying bands which are derived from the Sb-$d$ orbitals and not included in the model. Therefore, it is not that useful to find expressions for the velocity and mass parameter of this Dirac cone in terms of the tight-binding model because the latter has only limited validity. What matters is that both Dirac cone bands near $K$ and extending up to the region of their intersection can be described by Eq. 4.14 which represents two trigonally warped Dirac cones and that in the DFT results these cones are found to be rotated 30° with respect to each other.

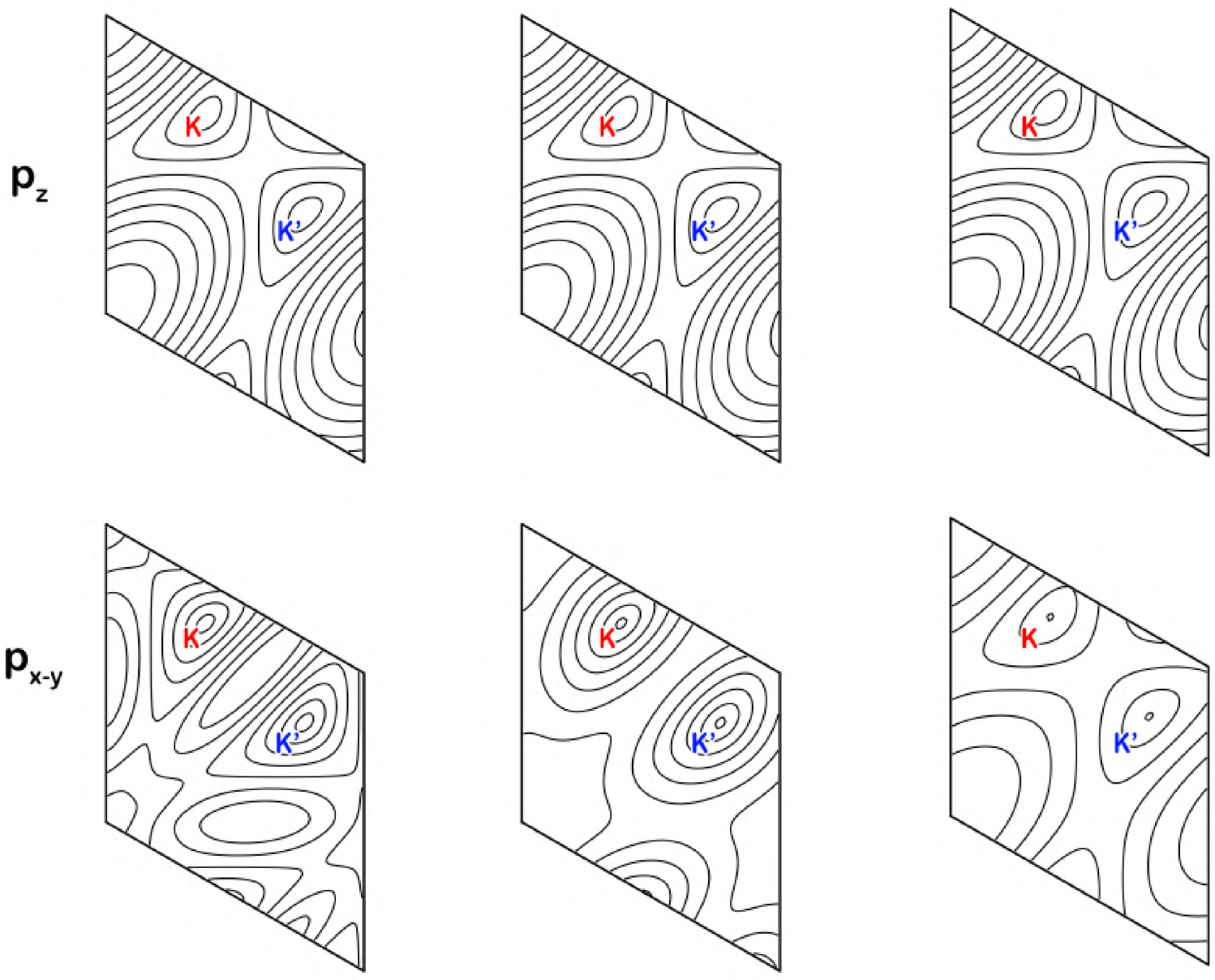

**Figure 4.11:** Contour plot of valence $p_z$ (top) and conduction $\{p_x, p_y\}$ (bottom) bands for various values of $(V_\pi, V_\sigma)$. From left to right, $(V_\pi, V_\sigma)$ =(−0.66,2.4),(−0.66,3.6),(−0.66,4.44) eV. $K$ and $K'$ points are marked. The Γ-point occurs at each of the corners of the reciprocal unit cell spanned by $\mathbf{b}_1, \mathbf{b}_2$.

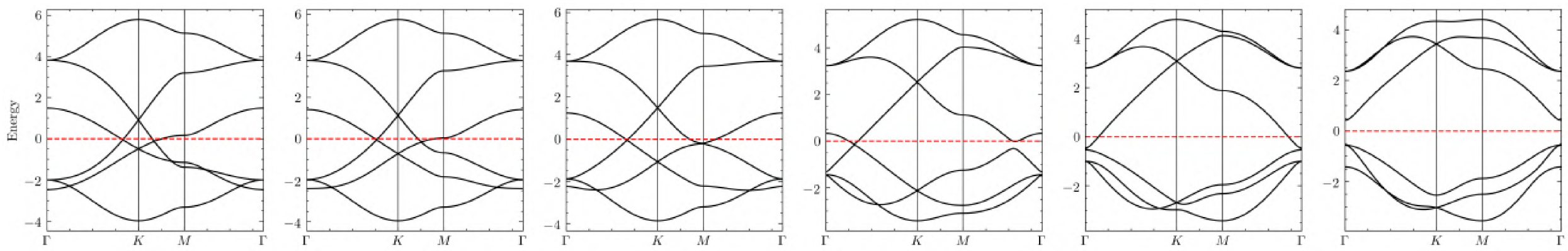

**Figure 4.12:**

Evolution of band structure ( in eV) at different buckling angles $\theta$. From left to right) 0°, 5°, 7.34°, 20°, 36°. In this figure the energy units are eV and $V_\sigma$ = 2.4 eV, $V_\pi$ = −0.66 eV. The red-dashed line at 0 energy is the Fermi energy in the flat case but is kept fixed in the later figures without adjusting the Fermi energy. In the before last panel, the Fermi energy lies exactly at the degenerate band touching at Γ.

Examining the bands in Figure 4.1 on page 28, one can see that the upper band of the $p_z$ derived Dirac cone deviates upward from the linear behavior and thus has positive $m_z$ which agrees with the $V_\pi < 0$. The lower band of the $p_{x,y}$ derived Dirac cone can be seen to bend down and thus has the opposite mass $m_{x,y} < 0$. Note that the direction $K - \Gamma$ corresponds to $\phi = \pi$. and the direction $K - M$ corresponds to $\phi = 2\pi/3$.

Both of the Dirac cones can be obtained from an effective low Hamiltonian

of the form

$$H_{eff} = \begin{bmatrix} \Delta_z & qv_z + q^2\frac{\cos(3\phi)}{m_z} \\ qv_z + q^2\frac{\cos(3\phi)}{m_z} & \Delta_z \end{bmatrix} \tag{4.15}$$

with a similar one for the $p_{x,y}$ case.

To illustrate the behavior of the Dirac cone warping, we show in Fig.Figure 4.11 contour plots obtained in the tight-binding Hamiltonian for different choices of $V_\sigma, V_\pi$ for the $p_z$ derived and $\{p_x, p_y\}$ derived cones. One can see that while for the $p_z$ derived ones the corners of the triangular contours around $K$ are point in the $K - M - K'$ direction and the shape does not depend on $V_\sigma$ because these bands only involve $V_\pi$. In contrast the $\{p_x, p_y\}$ derived cones for the first choice of $V_\sigma, V_\pi$ parameters which best matches the DFT bands and are used in Figure 4.1 on page 28, the cones are rotated by 30° with respect to the $p_z$-cone. The flat edge of the triangle is now along the $K - M - K'$ direction. However, as we change the $V_\sigma, V_\pi$ these cones can become almost circular (middle case) or their warping rotated the same as for $p_z$ as $V_\sigma$ is increased. Thus the rotated warping of the $p_z$ relative to the $\{p_x, p_y\}$ derived cones is sensitive to the relative values of $V_\sigma$ and $V_\pi$ and this is what is ultimately responsible for the shape of the nodal line and the occurrence of six Dirac points after buckling.

The effects of buckling in our model are incorporated through the $\theta$ dependence. When $\theta \neq 0$ the off-diagonal terms between the $p_z$ block and $\{p_x, p_y\}$ blocks are turned on. The resulting $6 \times 6$ matrix is readily diagonalized numerically but analytic expressions are no longer possible. Instead in Figure 4.1 on page 28 we then focus on the effective low energy Hamiltonian near the nodal line, which explains its breaking up into six separate Dirac points.

In Fig. Figure 4.12 we show the tight-binding bands of the $6 \times 6$ $p$-orbital derived Hamiltonian for various buckling angles. The leftmost figure for the flat case can be compared with the DFT results given in Figure 4.1 on page 28. The $V_\sigma$, $V_\pi$ and $\Delta_{x,y}$, $\Delta_z$ parameters were chosen to give about the correct ratios of the splitting of the two Dirac points $E'$, $E''$ at $K$, the splitting of the outermost eigenvalues $A_2'$, $A_1'$ of the $\{p_x, p_y\}$ derived bands at $K$ and the splitting of the $p_z$ derived bands at $\Gamma$. Note that both the $\{p_x, p_y\}$ derived bands and $p_z$ derived bands are 'particle-hole' symmetric about their center of gravity, the $E'$ and $E''$ Dirac points at $K$, which are displaced from each other by $\Delta_{x,y} - \Delta_z$ for the flat case. We can see that the lower band levels $A_{2u}$, $E_{2g}$ at $\Gamma$ are then close to each other while the upper band levels at $\Gamma$, $B_{2g}$ and $E_{1u}$ are more separated and inverted from the DFT bands. We are not trying to reproduce the DFT bands exactly because these upper bands are influenced by the interaction with the Sb-$d$ bands in the DFT results. Our main goal here is just to see the qualitative evolution of the bands under buckling.

We can see the Dirac points move toward $M$, disappear at $M$ and then the remaining ones move closer to $\Gamma$ and finally the full gap opens.

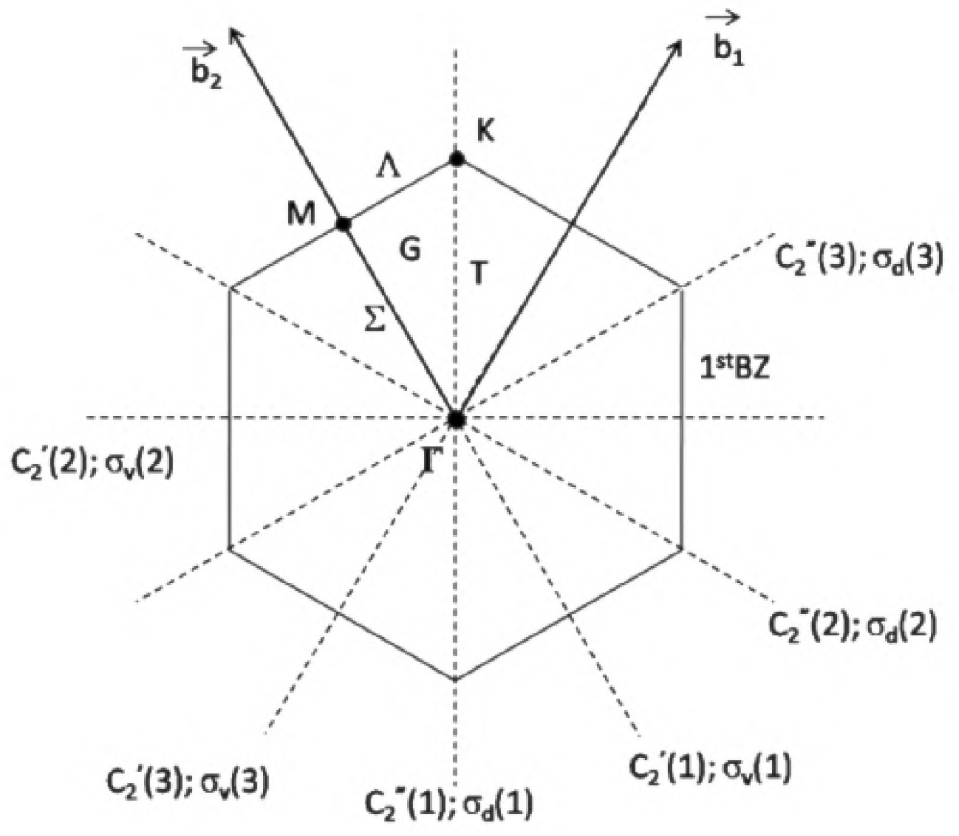

**Figure 4.13:** $1^{st}$ Brillouin zone and symmetry elements of the honeycomb lattice

## 4.10 Symmetry Analysis

$1^{st}$ Brillouin zone of flat honeycomb Sb is shown in Figure 4.13. Flat monolayer Sb in the honeycomb structure has the same spacegroup and pointgroup as graphene. Several prior papers have addressed the symmetry labeling of the bands in graphene[86–88] but still some confusion exists because of the non-uniqueness in specifying the symmetry operations and irreducible representations. The point group of the crystal is $D_{6h}$ and this is also the point group at Γ. The character table is given in Table Table 4.1 on the following page. Note that in $D_{6h}$ there are two sets of 2-fold rotations lying in the plane. We choose $U_2$ to pass through the atoms, while the $U_2'$ pass through the bond centers. In reciprocal space this implies that $U_2$ lies along the $\Gamma - M$ lines and $U_2'$ lies along the $\Gamma - K$ lines. The corresponding mirror planes $i * U_2 = \sigma_v$ and $i * U_2' = \sigma_v'$ are perpendicular to these axes, so $\sigma_v$ goes through the bond centers and $\sigma_v'$ goes through the atoms. We choose the lattice vectors as $\mathbf{a}_1 = a\hat{\mathbf{x}}$, $\mathbf{a}_2 = -\frac{1}{2}a\hat{\mathbf{x}} + \frac{\sqrt{3}}{2}a\hat{\mathbf{y}}$ and one of the points $K$ in the Brillouin zone thus lies along $x$ and one of the points $M$ lies along $y$. The $K$-points rotated by 60° we label $K'$. We note points $K$ rotated by 120° are equivalent in that they are related by a reciprocal lattice vector whereas $K$ and $K'$ are not. Likewise we denote the $M$-points rotated by 60°, $M'$ and rotated by 120° as $M''$. These are nonequivalent but rotating them by 180° gives equivalent $M$-points.

The group $D_{6h}$ can be viewed as the direct product $D_6 \otimes C_i$ where $C_i$ is the group consisting of the identity and the inversion. (It could also be viewed as $C_{6v} \otimes C_i$ or $C_{6v} \otimes C_s$ or $D_6 \otimes C_s$ with $C_s$ the group formed by $\{E, \sigma_h\}$ and this is the reason behind some of the discrepancies between previous symmetry labelings.) The irreducible representations follow the usual notation in which subscript $g$ means even with respect to inversion and $u$ means odd with respect to inversion. The corresponding labels of the Koster notation[89] are also included.

Although the symmetry aspects for graphene and Sb are the same, a difference is that the $s$-$p$ splitting is larger in Sb relative to the hopping interactions and hence the $s$ states form separate bands from the atomic $p$-state derived bands. The $s$-states are even in the horizontal mirror plane and form bonding an antibonding combinations on atoms $A$ and $B$ in the unit cell: $(s_A + s_B)/\sqrt{2}$ and $(s_A - s_B)/\sqrt{2}$. The operations $C_2, C_6, U_2', i, S_6, \sigma_v$ change sublattices $A$ to $B$ and vice versa. Thus it is clear that the bonding state belongs to $A_{1g}$ and the antibonding state to $B_{1u}$. The $p_z$ states are odd *vs.* the $\sigma_h$ and are thus decoupled by symmetry from the $s$ and $p_x, p_y$. Again they form bonding and antibonding states $(p_{zA} + p_{zB})/\sqrt{2}$ and $(p_{zA} - p_{zB})/\sqrt{2}$ which are now respectively of symmetry $A_{2u}$ and $B_{2g}$. The $p_x$ and $p_y$ build the representations $E_{2g}$ and $E_{1u}$.

**Table 4.1:** Character table of $D_{6h}$

| $D_{6h}$ | | $E$ | $C_2$ | $2C_3$ | $2C_6$ | $3U_2$ | $3U_2'$ | $i$ | $\sigma_h$ | $2S_6$ | $2S_3$ | $3\sigma_v$ | $3\sigma_v'$ |
|---|---|---|---|---|---|---|---|---|---|---|---|---|---|
| $\Gamma_1^+$ | $A_{1g}$ | 1 | 1 | 1 | 1 | 1 | 1 | 1 | 1 | 1 | 1 | 1 | 1 |
| $\Gamma_2^+$ | $A_{2g}$ | 1 | 1 | 1 | 1 | −1 | −1 | 1 | 1 | 1 | 1 | −1 | −1 |
| $\Gamma_4^+$ | $B_{1g}$ | 1 | −1 | 1 | −1 | 1 | −1 | 1 | −1 | 1 | −1 | 1 | −1 |
| $\Gamma_3^+$ | $B_{2g}$ | 1 | −1 | 1 | −1 | −1 | 1 | 1 | −1 | 1 | −1 | −1 | 1 |
| $\Gamma_6^+$ | $E_{2g}$ | 2 | 2 | −1 | 1 | 0 | 0 | 2 | 2 | −1 | 1 | 0 | 0 |
| $\Gamma_5^+$ | $E_{1g}$ | 2 | −2 | −1 | 1 | 0 | 0 | 2 | −2 | −1 | 1 | 0 | 0 |
| $\Gamma_1^-$ | $A_{1u}$ | 1 | 1 | 1 | 1 | 1 | 1 | −1 | −1 | −1 | −1 | −1 | −1 |
| $\Gamma_2^-$ | $A_{2u}$ | 1 | 1 | 1 | 1 | −1 | −1 | −1 | −1 | −1 | −1 | 1 | 1 |
| $\Gamma_4^-$ | $B_{1u}$ | 1 | −1 | 1 | −1 | 1 | −1 | −1 | 1 | −1 | 1 | −1 | 1 |
| $\Gamma_3^-$ | $B_{2u}$ | 1 | −1 | 1 | −1 | −1 | 1 | −1 | 1 | −1 | 1 | 1 | −1 |
| $\Gamma_6^-$ | $E_{2u}$ | 2 | 2 | −1 | 1 | 0 | 0 | −2 | −2 | 1 | −1 | 0 | 0 |
| $\Gamma_5^-$ | $E_{1u}$ | 2 | −2 | −1 | 1 | 0 | 0 | −2 | 2 | 1 | −1 | 0 | 0 |

Now, the group of $\mathbf{k}$, $\mathscr{G}_\mathbf{k}$, consists of those operations that turn $\mathbf{k}$ into itself up to a reciprocal lattice vector. For point $K$, which we choose along the $x$-axis, these are $\{E, 2C_3, 3U_2', \sigma_h, 2S_3, 3\sigma_v\}$, which build the group $D_{3h}$. At $M$ the group $\mathscr{G}_{\mathbf{k}_M}$ consists of $\{E, C_2, U_2, U_2', i, \sigma_h, \sigma_v, \sigma_v'\}$ building the group $D_{2h}$. The character tables of these groups are given in Tables Table 4.2, Table 4.3 on the following page.

| $D_{3h}$ | | $E$ | $2C_3$ | $3U_2'$ | $\sigma_h$ | $2S_3$ | $3\sigma_v$ |
|---|---|---|---|---|---|---|---|
| $K_1$ | $A_1'$ | 1 | 1 | 1 | 1 | 1 | 1 |
| $K_2$ | $A_2'$ | 1 | 1 | −1 | 1 | 1 | −1 |
| $K_6$ | $E'$ | 2 | −1 | 0 | 2 | −1 | 0 |
| $K_3$ | $A_1''$ | 1 | 1 | 1 | −1 | −1 | −1 |
| $K_4$ | $A_2''$ | 1 | 1 | −1 | −1 | −1 | 1 |
| $K_5$ | $E''$ | 2 | −1 | 0 | −2 | 1 | 0 |

**Table 4.2:** Character table of $D_{3h}$.

Note that in the group $D_{3h}$ representations labeled by superscript $'$ are even with respect to the $\sigma_h$ and $''$ are odd. In $D_{2h}$ note that which irrep is called $B_{2g}$ or $B_{3g}$ depends on the choice of $U_2$ or $U_2'$ being chosen as first or second set of 2-fold axes in the $xy$-plane. We added another label to indicate specifically which axes are chosen for the $M$ lying along $y$.

Along the $\Gamma - K = T$ axis the symmetry operations remaining are $\{E, U'_{2(x)}, \sigma_h, \sigma_{v(xz)}\}$ building the group $C_{2v}$. The irreps of this group depend on which of the mirror planes one chooses as first. We here choose $\sigma_h$ as first mirror plane. Along $\Gamma - M = \Sigma$ the group is also $C_{2v}$ but now the symmetry elements remaining are $\{E, U_{2(y)}, \sigma_h, \sigma'_{v(yz)}\}$. Finally along the line $M - K$ the group is the same as along $\Gamma - K$. The character tables used are given in Table Table 4.4 on the next page.

Next, we consider the modifications that occur upon buckling. In this case, the mirror planes through the atoms remain but the 2-fold rotation through the atoms is no longer valid. Similarly, the rotation axis through the bond center remains but the mirror plane through the bond axis is no longer valid. The six-fold rotations also are no longer valid but the inversion remains. The corresponding changes from flat to buckled in the groups $\mathcal{G}_{\mathbf{k}}$ are given in Table Table 4.5. The resulting group is $D_{3d}$ and its character table is given in Table Table 4.6.

**Table 4.3:** Character table of $D_{2h}$.

| $D_{2h}$ | | $E$ | $C2$ | $U_{2(y)}$ | $U'_{2(x)}$ | $i$ | $\sigma_h$ | $\sigma_{v(xz)}$ | $\sigma'_{v(yz)}$ |
|---|---|---|---|---|---|---|---|---|---|
| $M_1^+$ | $A_g$ | 1 | 1 | 1 | 1 | 1 | 1 | 1 | 1 |
| $M_3^+$ | $B_{1g}$ | 1 | 1 | −1 | −1 | 1 | 1 | −1 | −1 |
| $M_2^+$ | $B_{2g}$ | 1 | −1 | 1 | −1 | 1 | −1 | 1 | −1 |
| $M_4^+$ | $B_{3g}$ | 1 | −1 | −1 | 1 | 1 | −1 | −1 | 1 |
| $M_1^-$ | $A_u$ | 1 | 1 | 1 | 1 | −1 | −1 | −1 | −1 |
| $M_3^-$ | $B_{1u}$ | 1 | 1 | −1 | −1 | −1 | −1 | 1 | 1 |
| $M_2^-$ | $B_{2u}$ | 1 | −1 | 1 | −1 | −1 | 1 | −1 | 1 |
| $M_4^-$ | $B_{3u}$ | 1 | −1 | −1 | 1 | −1 | 1 | 1 | −1 |

| $C_{2v}$ | $\Gamma - K$ | $E$ | $U'_{2(x)}$ | $\sigma_h$ | $\sigma_{v(xz)}$ |
|---|---|---|---|---|---|
| | $\Gamma - M$ | $E$ | $U_{2(y)}$ | $\sigma_h$ | $\sigma'_{v(yz)}$ |
| $(T,\Sigma)_1$ | $A_1$ | 1 | 1 | 1 | 1 |
| $(T,\Sigma)_3$ | $A_2$ | 1 | 1 | −1 | −1 |
| $(T,\Sigma)_2$ | $B_1$ | 1 | −1 | 1 | −1 |
| $(T,\Sigma)_4$ | $B_2$ | 1 | −1 | −1 | 1 |

**Table 4.4:** Character table of $C_{2v}$.

**Table 4.5:** Groups $\mathcal{G}_{\mathbf{k}}$ for flat and buckled cases

| | Γ | Γ-M | M | M-K | K | K-Γ |
|---|---|---|---|---|---|---|
| Flat | $D_{6h}$ | $C_{2v}$ $\{E, U_{2(y)}, \sigma_h, \sigma'_{v(yz)}\}$ | $D_{2h}$ | $C_{2v}$ $\{E, U'_{2(x)}, \sigma_h, \sigma_{v(xz)}\}$ | $D_{3h}$ | $C_{2v}$ $\{E, U'_{2(x)}, \sigma_h, \sigma_{v(xz)}\}$ |
| Buckled | $D_{3d}$ | $C_s$ $\{E, \sigma'_{v(yz)}\}$ | $C_{2h}$ | $C_2$ $\{E, U'_{2(x)}\}$ | $D_3$ | $C_2$ $\{E, U'_{2(x)}\}$ |

Next, we consider the modifications that occur upon buckling. In this case, the mirror planes through the atoms remain but the 2-fold rotation through the atoms is no longer valid. Similarly, the rotation axis through the bond center remains but the mirror plane through the bond axis is no longer valid. The six-fold rotations also are no longer valid but the inversion remains. The corresponding changes from flat to buckled in the groups $\mathcal{G}_{\mathbf{k}}$ are given in Table Table 4.5. The resulting group is $D_{3d}$ and its character table is given in Table Table 4.6.

| $D_{3d}$ | | $E$ | $2C_3$ | $3U'_2$ | $i$ | $2S_6$ | $3\sigma'_v$ |
|---|---|---|---|---|---|---|---|
| $\Gamma_1^+$ | $A_{1g}$ | 1 | 1 | 1 | 1 | 1 | 1 |
| $\Gamma_2^+$ | $A_{2g}$ | 1 | 1 | −1 | 1 | 1 | −1 |
| $\Gamma_3^+$ | $E_g$ | 2 | −1 | 0 | 2 | −1 | 0 |
| $\Gamma_1^-$ | $A_{1u}$ | 1 | 1 | 1 | −1 | −1 −1 | |
| $\Gamma_2^-$ | $A_{2u}$ | 1 | 1 | −1 | −1 | −1 | 1 |
| $\Gamma_3^-$ | $E_u$ | 2 | −1 | 0 | −2 | 1 | 0 |

**Table 4.6:** Character table for $D_{3d}$.

We can thus easily convert the labels from the $D_{6h}$ case to the $D_{3d}$ case according to Table **??**. The same holds for the odd *vs.* inversion irreps labeled by the $u$ subscript.

| $D_{6h}$ | $A_{1g}$ | $A_{2g}$ | $B_{1g}$ | $B_{2g}$ | $E_{2g}$ | $E_{1g}$ |
|---|---|---|---|---|---|---|
| $D_{3d}$ | $A_{1g}$ | $A_{2g}$ | $A_{2g}$ | $A_{1g}$ | $E_g$ | $E_g$ |

**Table 4.7:** Compatibility between $D_{6h}$ and $D_{3d}$ group irreps.

Now at $K$, the group becomes $D_3$ because we loose the inversion. The character table thus consist just of the upper left block in Table Table 4.6 on the preceding page. The irreps stay the same as at $\Gamma$ but without the $g, u$ subscripts. At $M$ we now have the group $C_{2h}$ consisting of $\{E, U'_{2(x)}, i, \sigma'_{v(yz)}\}$. Its character table is given in Table Table 4.8.

| $C_{2h}$ | | $E$ | $U'_{2(x)}$ | $i$ | $\sigma'_{v(yz)}$ |
|---|---|---|---|---|---|
| $M_1^+$ | $A_g$ | 1 | 1 | 1 | 1 |
| $M_2^+$ | $B_g$ | 1 | −1 | 1 | −1 |
| $M_1^-$ | $A_u$ | 1 | 1 | −1 | −1 |
| $M_2^-$ | $B_u$ | 1 | −1 | −1 | 1 |

**Table 4.8:** Character table of $C_{2h}$.

Along the line $\Gamma - K$ the group is $C_2$ consisting of $\{E, U'_{2(x)}\}$. The same applies to the line $M - K$. The character table is given in Table Table 4.9. Along $\Gamma - M$, the group is $C_s$ consisting of $\{E, \sigma'_{v(yz)}\}$ with characters given in Table Table 4.10.

| $C_2$ | | $E$ | $U'_{2(x)}$ |
|---|---|---|---|
| $T_1$ | $A$ | 1 | 1 |
| $T_2$ | $B$ | 1 | −1 |

**Table 4.9:** Character table of $C_2$.

We may also consider a flat structure but making the $A$ and $B$ atoms different. This would apply to the case of monolayer h-BN. Then starting from $D_{6h}$ we loose the inversion but we keep the horizontal mirror plane. The group at $\Gamma$ in that case is $D_{3h}$ consisting of $\{E, 2C_3, 3U_2, \sigma_h, 2S_3, 3\sigma_v'\}$. The group at $K$ in that case is the group $C_{3h}$ consisting of $\{E, 2C_3, \sigma_h, 2S_3\}$. At $M$ the group becomes $C_{2v}$ consisting of $\{E, U_2, \sigma_h, \sigma_v'\}$. Along the lines $\Gamma - K$ and $M - K$ the group is $C_s$ consisting of $\{E, \sigma_h\}$ and along $\Gamma - M$ it is the same as at $M$.

| $C_s$ | | $E$ | $\sigma'_{v(yz)}$ |
|---|---|---|---|
| $\Sigma_1$ | $A'$ | 1 | 1 |
| $\Sigma_2$ | $A''$ | 1 | −1 |

**Table 4.10:** Character table of $C_s$.

Finally, making the system both buckled and breaking the inversion. Then the group at $\Gamma$ is only $D_3$, containing $\{E, 2C_3, 3U_2\}$. At $K$ it becomes $C_3$ at $M$ and along $\Gamma - M$ it becomes $C_2$ with only elements $\{E, U_2\}$ along $\Gamma - K$ and $M - K$ there is no symmetry left at all.

The character table of $C_3$ is given in Table Table 4.11

| $C_3$ | | $E$ | $C_3$ | $C_3^{-1}$ |
|---|---|---|---|---|
| $K_1$ | $A$ | 1 | 1 | 1 |
| $K_2$ | $E$ | 1 | $\omega$ | $\omega^*$ |
| $K_3$ | $E$ | 1 | $\omega^*$ | $\omega$ |

**Table 4.11:** Character table of $C_3$, $\omega = e^{2i\pi/3}$.

We note that in the group $C_3$, Koster *et al.* [89] labels the two irreps which cannot be made real as two separate irreps while in the 'chemical' notation, they are both labeled $E$. This is because if one ignores spin these two irreps are each other's complex conjugate and become degenerate by time reversal. They form a Kramers doublet. However, taking into account the spin 1/2 no degeneracy between the two occurs because time reversal takes spin up into spin down. Adding the horizontal mirror plane just adds another label ′ or ″ for even or odd under that operation. Thus we can see that that the degenerate levels $E$ in the buckled case or $E'$ and $E''$ would be allowed to split and open a gap. This is well known to open the gap in the honeycomb BN case.

# $(d-2)$ Higher Order Topology of buckled Group - V

# 5

**Abstract**

Higher Order Topological Insulators (HOTI) are $d$-spatial dimensional systems featuring topologically protected gap-less states at their $(d-n)$-dimensional boundaries. With the help of *ab-initio* calculations and tight binding models along with symmetry considerations we show that monolayer buckled honeycomb structures of group-V elements (Sb,As), which have already been synthesized, belong in this category and have a spinless charge fractionalization of $\frac{e}{2}$ at the corner states as well as weak topological edge states, protected by $S_6$ symmetry, which classify this system as a quadrupole topological insulator. The robustness of these edge and corner states to perturbations is explicitly demonstrated.

In Chapter 4, we saw the non trivial properties of Sb as function of buckling turning from Nodal line to Dirac points and finally an insulator. In this chapter, we will see that this final state of insulator is in no way a trivial insulator, but rather a higher order topological insulator. We will start the discussion by observing the band structure of a 1D nano ribbon of Sb in its completely buckled form (insulator form).

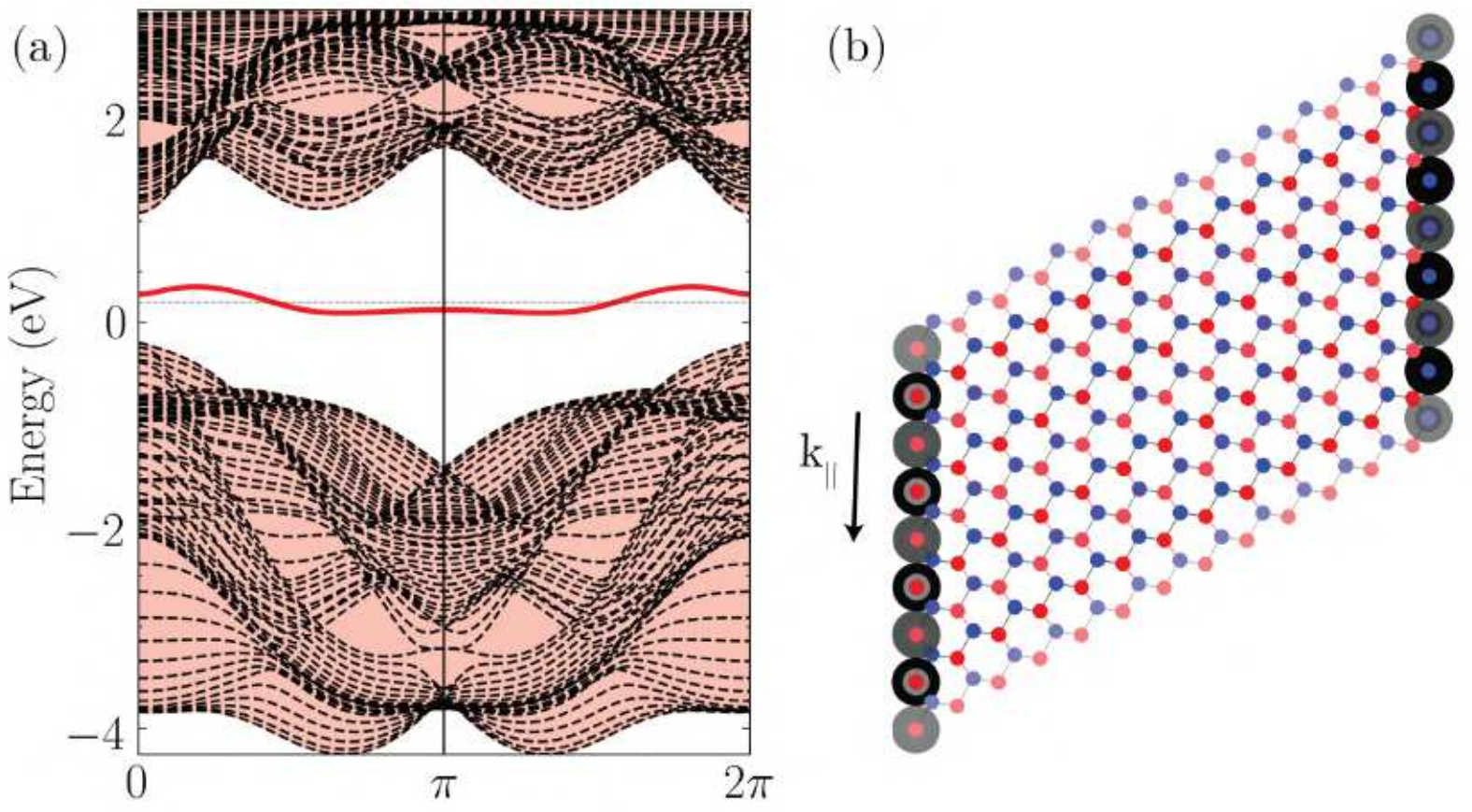

**Figure 5.1:** (a) Band structure of nanoribbon of buckled monolayer Sb in local density approximation along the zigzag edge $\mathbf{k}_{\|}$, showing edge states(red); (b) structure and edge state wave function modulo squared.

Figure 5.1 shows the 1D band structure as obtained from a tight-binding model parametrized with maximally localized Wannier function extracted hopping integrals from a projector augmented wave (PAW)[90] calculation of the 2D periodic system using the Quantum Espresso [81] and Wannier90 codes.[91].

## 5.1 Origin of edge states

In a broad sense the origin of these edge states is related to the "obstructed atomic limit" (OAL), a concept introduced by Bradlyn *et al.* [92, 93] Accordingly, a set of bands is in the OAL when they possess symmetric, localized Wannier functions that reside on Wyckoff positions distinct from the atomic positions which cannot be smoothly deformed to the latter. This corresponds to a weak topological phase.

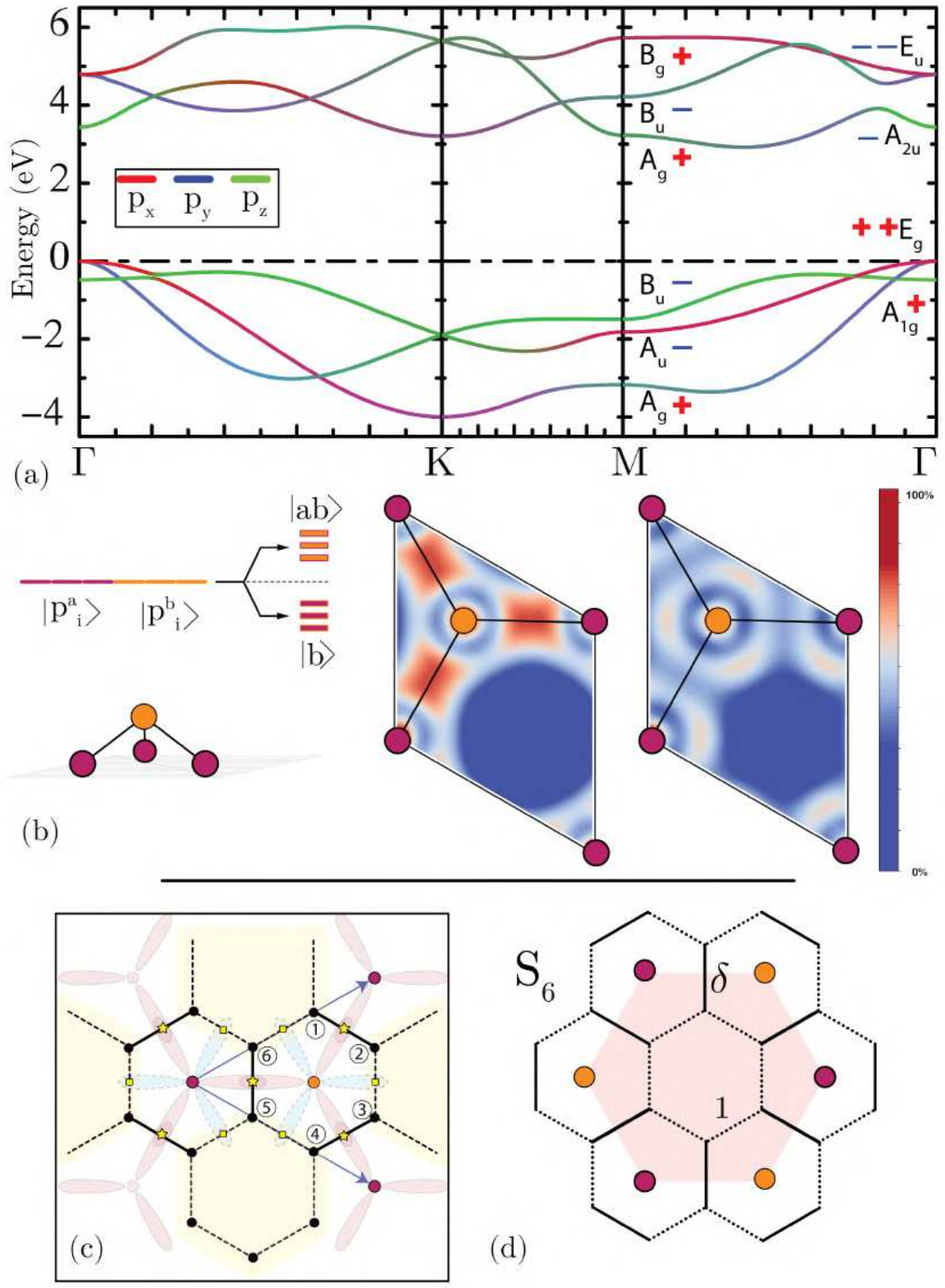

Figure 5.2: *(a)* QSGW band structure of honeycomb-buckled Sb; *(b)* schematic of $p$-levels from atomic limit to buckled Sb and contour plot of total electron density from the 3 $p$−derived valence ($|b\rangle$) and conduction ($|ab\rangle$) bands; *(c)* relation of bond centers to Kekulé model: see text; *(d)* Kekulé model.

In the present case, the bands of interest result from the Sb-$\{p_x, p_y, p_z\}$ orbitals which form three sets of bonding (occupied, valence) and three sets of antibonding (empty,conduction) bands. These bands are shown in Figure 5.2 along with the relevant symmetry labeling and indicating the atomic orbital character of each band. This band structure is actually obtained at the quasiparticle self-consistent (QS)$GW$ level[94] which guarantees accurate band gaps but for the remainder of this paper, this is not important and a DFT or even simpler tight-binding (TB) models have the same set of irreducible representations present in the valence (VB) and conduction (CB) band manifolds. Significant hybridization between all three $p$-orbitals is apparent. To better understand the origin of the gap and hybridization, we plot in Figure 5.2(b) the $\sum_{n\mathbf{k}} |\psi_{n\mathbf{k}}|^2$ separately for $n \in$ VB and CB manifolds. These show clearly that the Wannier function corresponding to the VB are localized at the bond centers, while the ones of the CB are centered at the "antibond centers", obtained from the bond center by inversion about the atomic sites. This in itself is already a clear

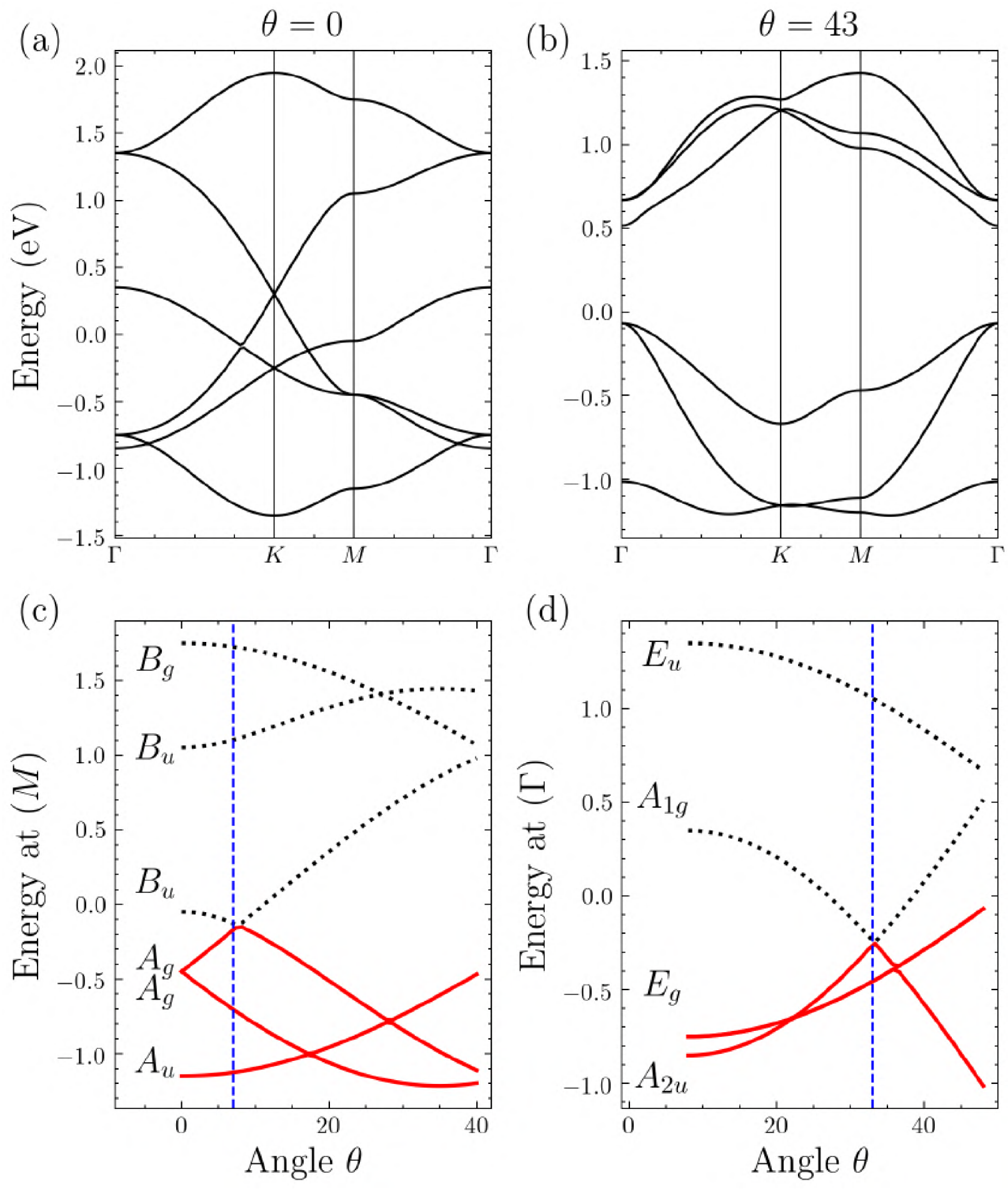

**Figure 5.3:** Bulk band structure at the limit where (a) buckling ($\theta$)=0 and (b) $\theta$=33$^o$ bands at (c) M and (d) Γ as function of buckling angle and their corresponding irreducible representations. Blue lines mark the point of two non trivial transitions. red and black dotted lines show the occupied and unoccupied energies

indicator of the non-trivial nature of the band structure. Intuitively, if one cuts the system along these bonds, dangling bond like edge states are expected.

Another way to realize the origin of non triviality is to look at non trivial band crossings that exchange irreducible representations as function of buckling.

Figure 5.3 shows the switch in irreps as function of buckling. As one can see, around 7$^o$ (shown by blue lines), we have the first transition that was discussed in Chapter 4 around M. Upon further buckling, around 33$^o$ we see a switch in irreps at Γ where essentially $A_{2u}$ and $A_{1g}$ irreps switch. This switch is accompanied by change in the inversion symmetry charecter of the occupied bands which will be of importance when discussing about the higher order invariant.

## 5.2 Relation to Kagome and Kekulé lattice

Remarkably the bond centers and antibond centers from respectively a Kagome and *anti-Kagome* lattice . More interestingly, a further transformation relates this band structure to the Kekulé lattice. As shown in Figure 5.2(c) the bond centers (yellow stars) of the (pink) bonding orbitals of Sb atoms (orange and red circles) and the corresponding antibond (blue) centers (yellow squares) can alternatively be viewed as the bond centers of yet another honeycomb lattice (black circles) with alternating bond strength. We may recognize this as the bond modulated honeycomb lattice, known as the Kekulé lattice. We can see in Figure 5.2(d) that this lattice consists of hexagon shaped molecules connected by (intracell) bonds of strength 1 within the Wigner-Seitz unit cell and connected by intercell bonds of strength $\delta$. It is useful to note that the bond centers of the original buckled Sb lattice, which coincide with the bond centers of the final Kekulé lattice all lie in the same horizontal plane. The Kekulé lattice is thus strictly 2D and the inversion operator in 2D is identical with a two-fold rotation $C_2$ about the $z$-axis. Thus the point group of the Kekulé lattice is $C_6$ while that of the Sb lattice is $S_6$ but the two are simply related by replacing the $C_2$ operation by the inversion operation, which we will denote $\mathcal{I}$.

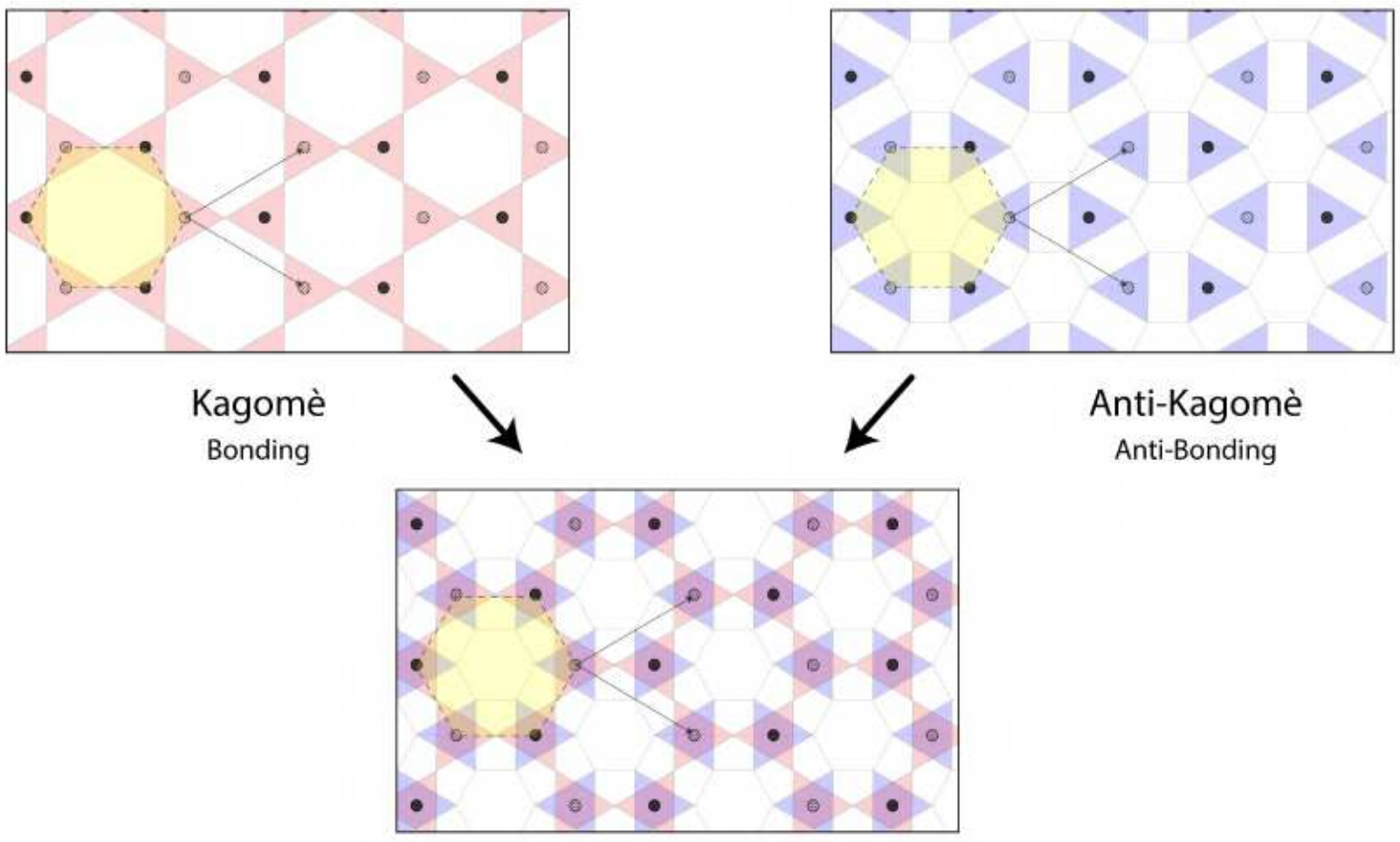

**Figure 5.4:** Bonding and Anti bonding states forming Kagome and Anti kagome

The topological properties of the Kekulé lattice have been discussed in a number of recent papers.[95–99] These papers have not only shown the presence of topological edge states but also topological corner states in the Kekulé lattice in the nontrivial condition $\delta > 1$. Not surprisingly, one can think of Kekulé as a hexagonal 2D generalization of the Su-Schrieffer-Heeger (SSH) model[100] or a set of SSH models arranged next to each other.

Figure 5.4 show how the bond centers form a Kagome lattice and the antibond centers form what we call here an anti-Kagome lattice. The filled and open circles are the Sb atoms point up or down the plane of the projection plane the corners of the pink triangles are the bond centers and form a Kagome lattice. On the right, the blue triangle corners are at the anti-bond centers and form a distinct lattice which we call here anti-Kagome. Both are shown superposed on each other in the bottom figure.

In Figure 5.5 we show a further comparison of the DFT, Wannier NN and two other models. The one labeled $H_p$ corresponds to the NN-TB model used in [101]. This model uses $\{p_x, p_y, p_z\}$ orbitals and the standard Slater-Koster two-center approximation to write the hopping integrals in terms of $V_{pp\sigma}$ and $V_{pp\pi}$ interactions. This allowed us to follow the behavior as function of buckling angle, assuming the relative amount of $V_{pp\sigma}$ $V_{pp\pi}$ depend only on angle but the bond lengths and hence the $V_{pp\sigma}$ and $V_{pp\pi}$ two-center integrals themselves stayed fixed. In this model, also the $p_x, p_y$ orbital energies are slightly displaced in energy from the $p_z$ orbitals thereby breaking the particle-hole symmetry. Although there are changes in the ordering of bands within the occupied manifold, and within the unoccupied manifold, the set of irreducible representations comprising the VB and CB manifolds stay the same and this is the only feature that matters for the topological aspects discussed in the main paper. Finally, we show in this figure the results of the band structure of the Kekulé model, whose tight-binding Hamiltonian can be written.

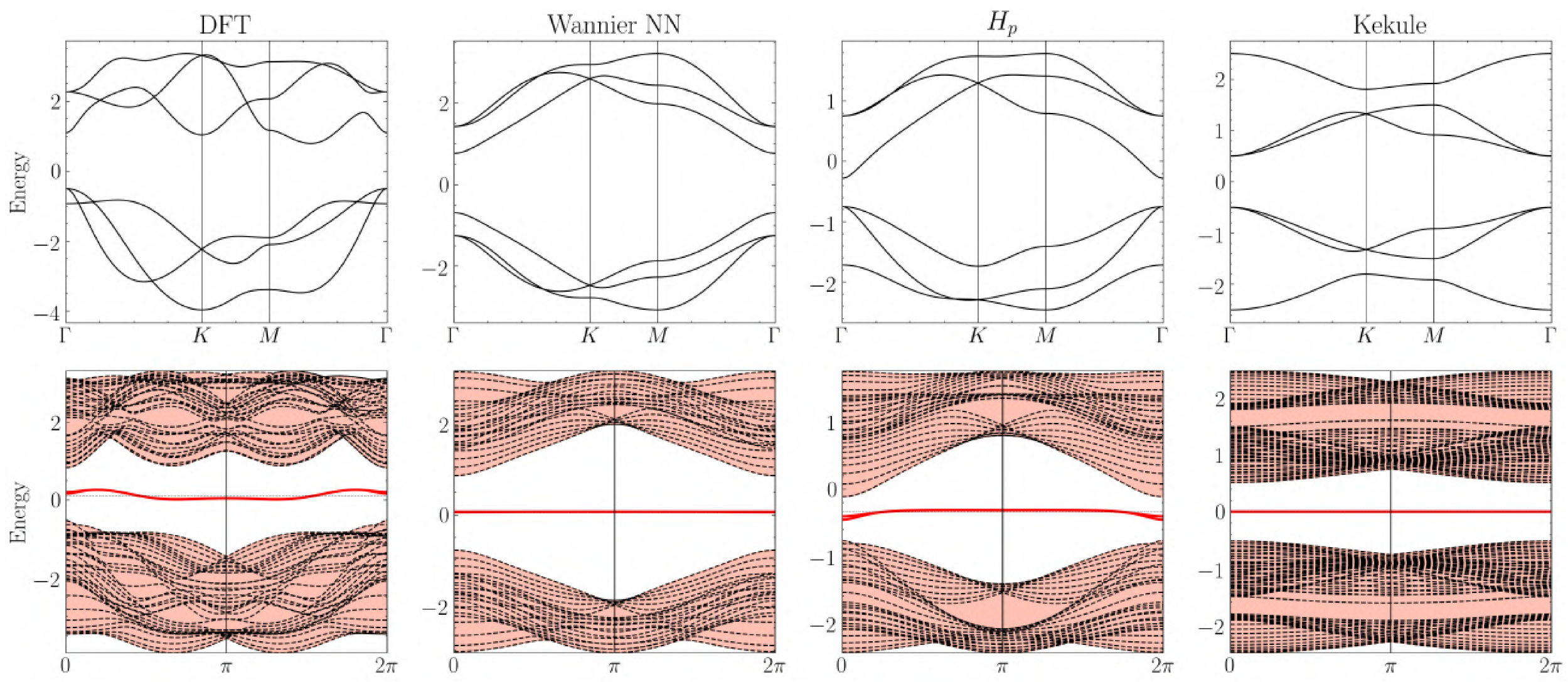

**Figure 5.5:** Bulk and edge band structures at different level of theories

$$H = \begin{pmatrix} 0 & \delta & 0 & e^{i\mathbf{k}.\mathbf{a}_3} & 0 & 1 \\ \delta & 0 & 1 & 0 & e^{i\mathbf{k}.\mathbf{a}_2} & 0 \\ 0 & 1 & 0 & \delta & 0 & e^{i\mathbf{k}.\mathbf{a}_1} \\ e^{-i\mathbf{k}.\mathbf{a}_3} & 0 & \delta & 0 & 1 & 0 \\ 0 & e^{-i\mathbf{k}.\mathbf{a}_2} & 0 & 1 & 0 & \delta \\ 1 & 0 & e^{-i\mathbf{k}.\mathbf{a}_1} & 0 & \delta & 0 \end{pmatrix} \tag{5.1}$$

where $\mathbf{a}_1 = (1, 0)$, $\mathbf{a}_2 = (\frac{1}{2}, \frac{\sqrt{3}}{2})$ The Kekulé orbital centers are numbered 1-6 in Fig.2(c) in the main text and the lattice vectors are indicated by blue arrows.

## 5.3 Corner states

We now turn our attention to the higher order topologically required corner states in a 0D system that preserves the rotational symmetries of the periodic lattice, in other words, a finite hexagonal portion cut out from the buckled buckled honeycomb lattice. First, let us mention that such hexagonal nanoflakes have already been fabricated[102]. Benalcazar *et al.* [103] have described topological invariants and the related occurrence of corner-states and their charge fractionalization for $C_n$ groups.

The conditions under which corner-states occur depend on the symmetries of the states at high-symmetry $\mathbf{k}$-points. While the system under study here has $S_6$ instead of $C_6$ symmetry, we can easily generalize their procedure. According to Benalcazar *et al.* [103] for the $C_6$ group, the topological invariant is $\chi^{(6)} = ([M_1^{(2)}], [K_1^{(3)}])$, which, for example at $M$ indicates the difference in number of eigenstates in the occupied bands manifold of the $C_2$ operation indicated by the superscript corresponding to a given eigenvalues (1, the subscript) at $M$ and $\Gamma$. A 6-fold rotation can be viewed as the product of a 3-fold and 2-fold rotation. Similarly, a 3-fold rotation-inversion is the product of a 3-fold rotation and an inversion. Thus we merely have to replace the two fold rotation by inversion for our case. Thus the important indicator becomes $[M_\pm^{(i)}]$ which is the difference in number of bands with even/odd character under the inversion operation at $M$ and at $\Gamma$. We can see from the symmetry labeling in Figure 5.2(a) that the $[M_\pm^{(i)}] = \pm 2$. According to Table I in [103], the invariants at $K$ and $M$ are either (2,0),(0,2) or (0,0) where the last case is the trivial topology. Thus, our system corresponds to the $h_{3c}^{(6)}$ *primitive generator* class in Benalcazar *et al.*'s notation. This indeed means that there are 3 filled bands and that the generator is obtained from orbitals centered at the Wyckoff position $c$. In the buckled Sb system, the space group is $P\bar{3}c1$ or $D_{3d}^3$ or #164 and the Wyckoff position is $d$. Thus, generalizing their notation our primitive generator would be $h_{3d}^{(\bar{3})}$ but in terms of charge fractionalization and polarization would have the same invariants as $h_{3c}^{(6)}$. (For the Kekulé case, the plane space group is $p3m1$ and the Wyckoff position is $c$.) It implies that there is no net dipole in the plane and the corner charge fractionalization will be $e/2$ in each $\pi/6$ sector.

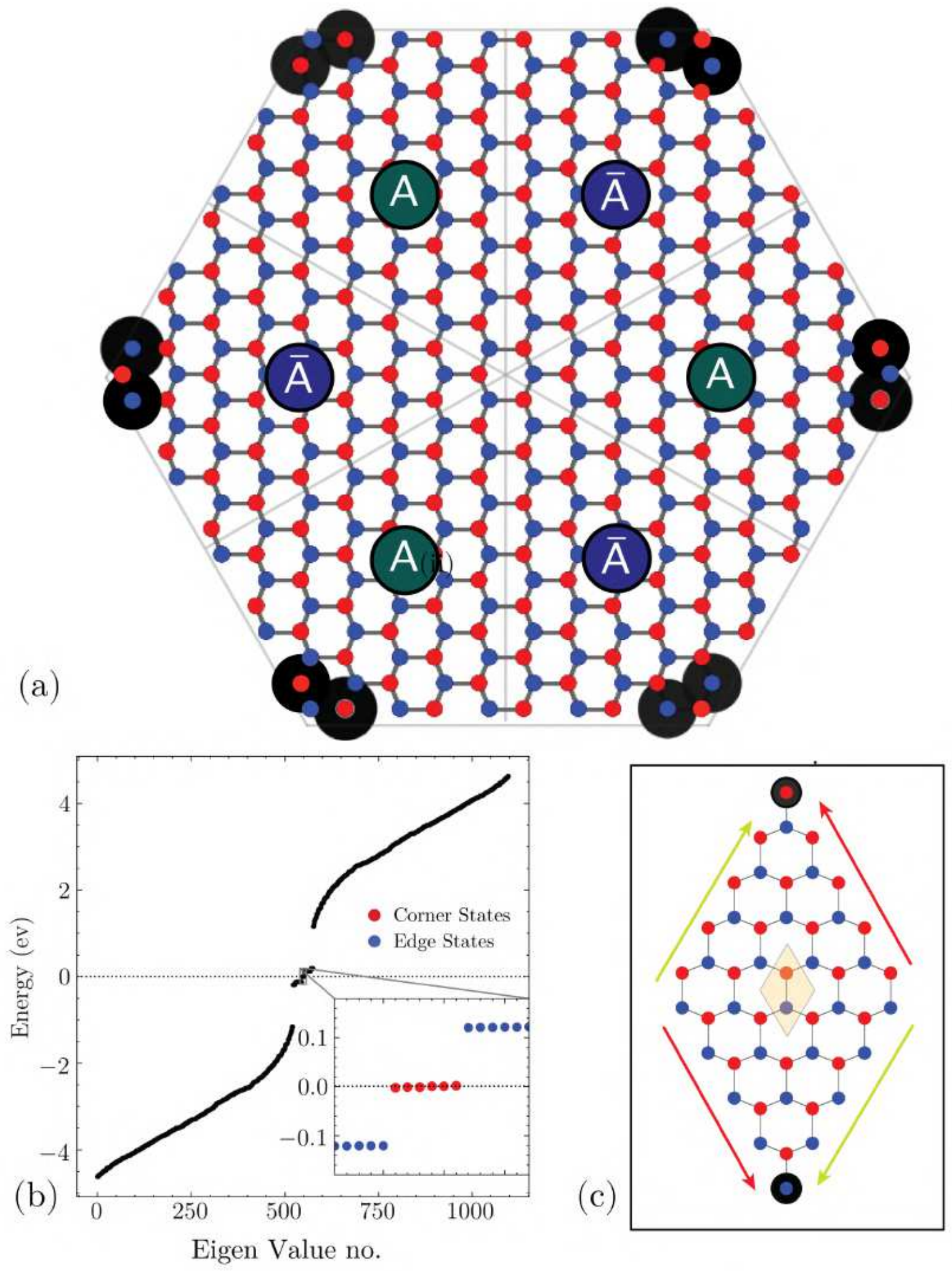

**Figure 5.6:** (a) Finite size fragment of honeycomb (red and blue circles show Sb atoms below and above the plane), large black circes represent $|\psi|^2$ for the localized corner states indicated in in the eigenvalue spectrum in (b).

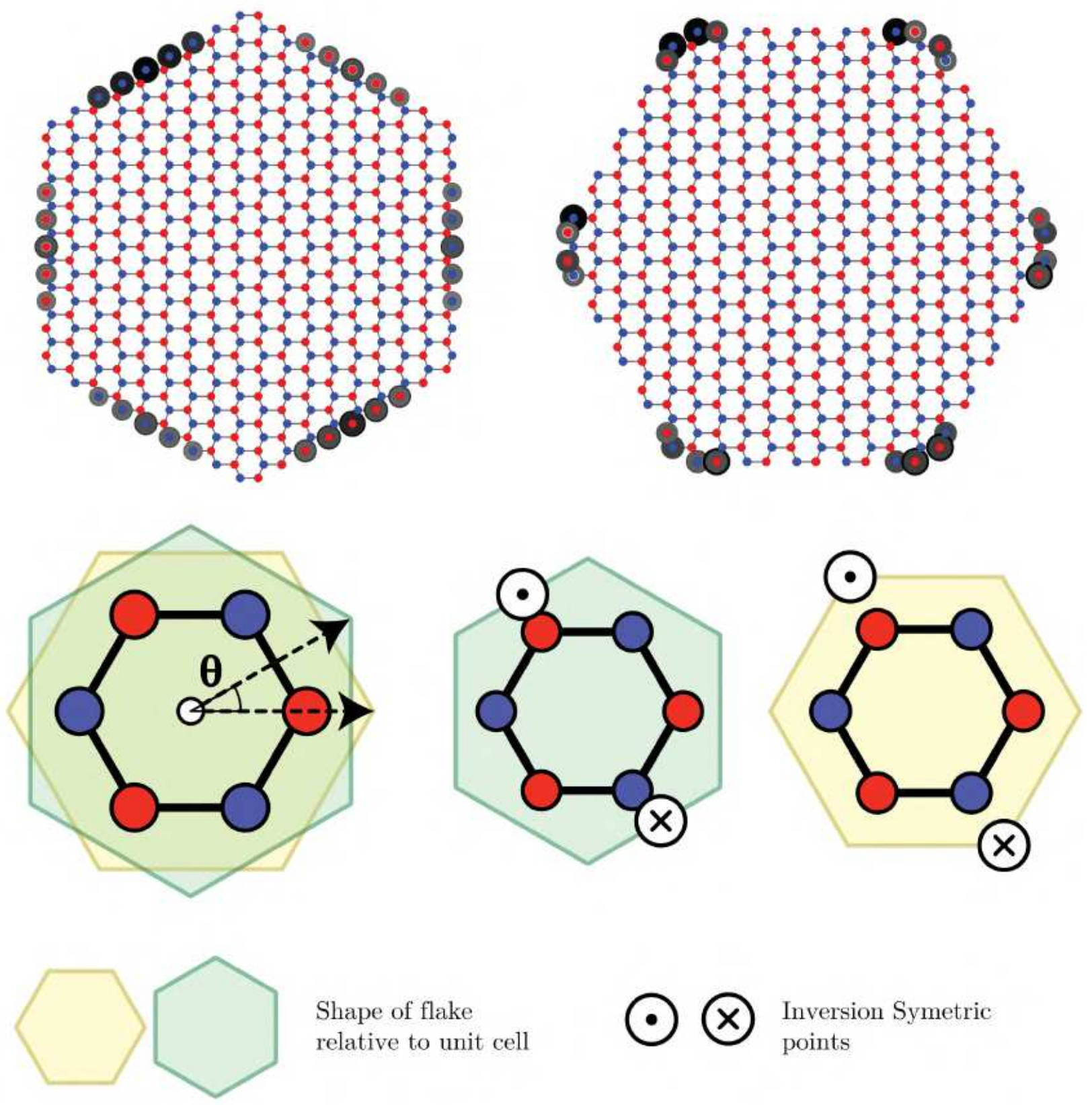

**Figure 5.7:** Different $C_6$ symmetric flake geometry and the mid gap state wave-function

We verify this prediction in Figure 5.6. The calculation is performed on a hexagonal 0D flake within the $\{p_x, p_y, p_z\}$ NN-TB Hamiltonian (rather than the simplified Kekulé model, which is however topologically equivalent). The eigenvalues of this Hamiltonian are shown in Fig.Figure 5.6(b). One can see that near zero energy, in the band gap there occurs both edge states (blue dots in inset of (b)) as well as mid-gap corner states (red). Their wave function modulo squared is indicated by the black circles in Fig. Figure 5.6(a) clearly showing that these states are localized on the corner atoms. The $\pi/6$ sectors are labeled alternating $A$ and $\bar{A}$, which are related by the inversion symmetry. One should note that though the existence of such corner states is deeply tied to the choice of symmetric unit cell, $\pi/6$ charge-fractionalization is not. In any general $S_6$ symmetric $0D$ fragment, one is guaranteed $e/2$ charge-fractionalization. For example, in Figure 5.6, we used either a perfectly hexagonal 0D model or an integer multiple of the 2D primitive unit cell given by the $\mathbf{a}_1$, $\mathbf{a}_2$ lattice vectors. This lead to symmetrically distributed corner localized states within each $\pi/6$ sector or a quadrupole symmetry showing charge localization However more generally localized states at the edges can be obtained from other 0D fragments as illustrated here in Figure 5.7. These states generally occur at edge points related to another point on the circumference of the 0D system by inversion symmetry In the figure on the left we see edge states localized on zigzag edges but not on the corners joining them, while on the right we see states localized at the corners joining arm-chair edges. The figures below illustrate the relation between the overall 0D nanoscale fragment and the lattice unit cell and which points are related by inversion which is the key crystallographic symmetry protecting these topological features.

## 5.4 Relation to multi-pole insulators

A closely related point of view on the origin of the corner states arises in the context of quantized multipole insulators. Although the net polarization in the plane is zero for our case, the system has a non-zero quadrupole insulator character.

For a system with inversion symmetry, the contribution to the polarization projected on direction $i$ from band $n$ can be obtained from the the eigenvalues of the inversion operator at Time Reversal Invariant Momentum (TRIM) points $M$ and $\Gamma$ and is given by [97, 99, 104]

$$P_i^n = \frac{e}{2}(q_i^n \bmod 2) \quad \text{with} \quad (-1)^{q_i^n} = \frac{\eta_n(M_i)}{\eta_n(\Gamma)} \tag{5.2}$$

where $\eta^n(\boldsymbol{k})$ is the eigenvalue of the inversion operation. The quadrupole moment is then given by

$$\mathbb{Q}_{ij} = \sum_{n}^{N_{occ}} \frac{P_i^n P_j^n}{e} \tag{5.3}$$

For our system $(P_1^n, P_2^n) = (0,0), (\frac{e}{2}, \frac{e}{2}), (\frac{e}{2}, \frac{e}{2})$ for bands $n = 1, 2, 3$ numbered from bottom to top. Thus the net dipole moment is trivially $(e, e)$ as it should be for a $C_6(S_6)$ symmetric system but the net quadrupole moment has magnitude $e/2$, the effect of which can be seen in a $d-2$ system cut along the lattice vectors as shown in Fig. Figure 5.6(c). In this figure we see the dipoles on opposite edges canceling each other but leading to a net quadrupole with charge accumulation at the two corners indicated by the black circles showing $|\Psi|^2$ of the corner localized state in the TB model. Not surprisingly, given the close connection pointed out earlier, similar corner states and quadrupole character are also found for the Kekulé lattice in the non-trivial limit.[99] Furthermore we note that not only regularly shaped fragments as considered here host such corner localized states, but more generally shaped 0D objects can host localized states on the perimeter when points on opposite sides are related by inversion symmetry.

## 5.5 Robustness against disorder

It is important to note that the even though our system is a Topological Crystal Insulator (TCI), which are strictly speaking only weakly protected, the corner states are fairly robust to bulk disorder. We show this numerically by adding uniformly distributed random on-site bulk terms in a range of the magnitude of the band gap. This does confirm that approximate symmetries that are preserved on average are enough to host the fractional charges. But as soon as one perturbs the edges, the corner fractionalization gets destroyed. However, because of the existence of edge states in the corresponding higher dimensional $1D$ system, one still preserves the edge modes in the disordered system with perturbed edge states.

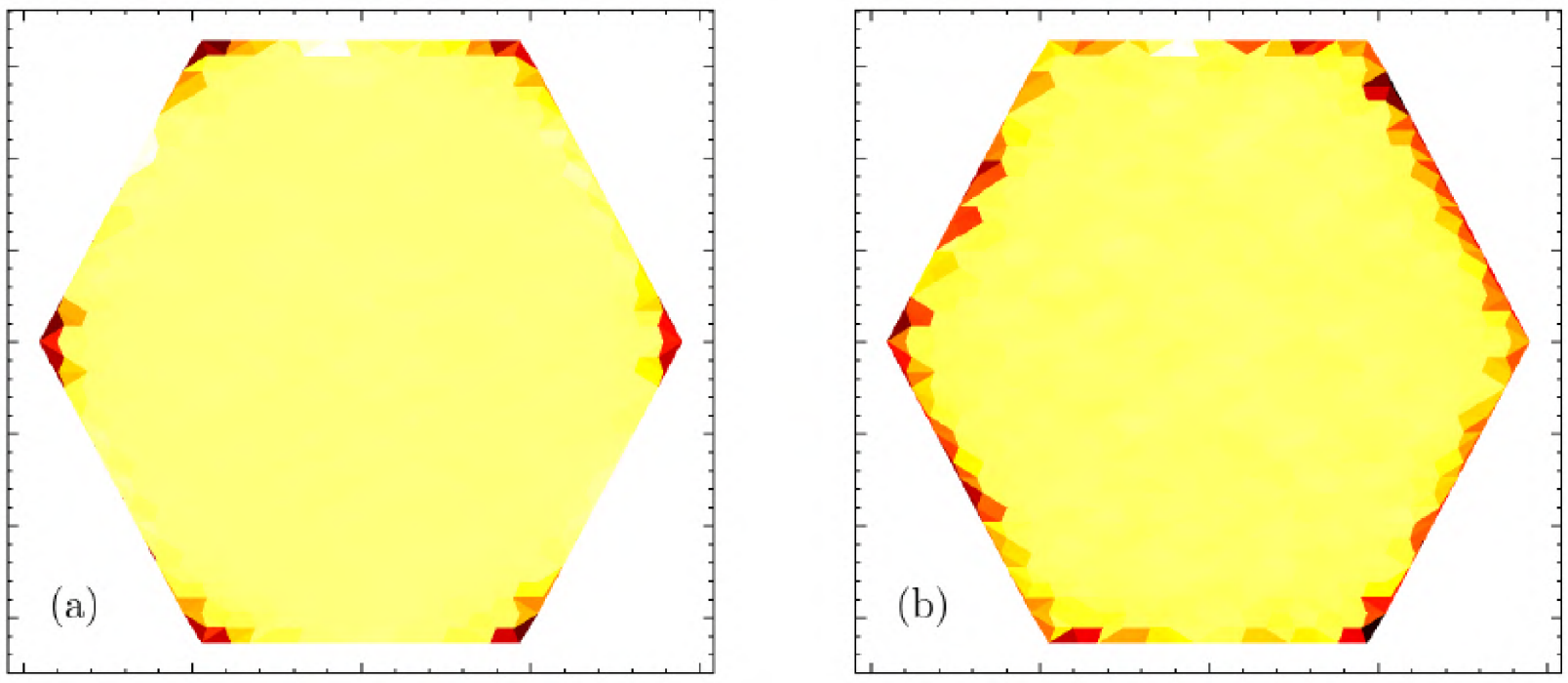

**Figure 5.8:** Total charge distribution of occupied states with on-site potential added stochastically from a uniform distribution of $[-\frac{E_g}{2}, \frac{E_g}{2}]$ *(a)* everywhere except edge atoms *(b)* everywhere

Figure 5.8 shows the total charge distribution in a $C_6$ symmetric D0 system made up of 4200 sites ×3 = 12600 orbitals. Random on-site potential sampled from a uniform distribution in the interval $[-\frac{E_g}{2}, \frac{E_g}{2}]$ (where $E_g$ is the band gap) was added on (a) all atoms other than the edges (b) everywhere uniformly. Both simulations were run 500 times and the average charge density is plotted. This shows that bulk disorder does not destroy the corner states while disorder also at the edges does destroy the corner states but still shows localized states along the edges.

# Topological quantum switch and 1D wires in Sb

# 6

**Abstract**

Based on the recently found non-trivial topology of buckled antimonene, we propose the conceptual design of a quantized switch that is protected by topology and a mechanism to create configurable 1D wire channels. We show that the topologically required edge states in this system can be turned on and off by breaking the inversion symmetry (reducing the symmetry from $S_6$ to $C_3$), which can be achieved by gating the system. This is shown to create a field-effect quantum switch projected by topology. Secondly we show that by locally gating the system with different polarity in different areas, a soliton-like domain wall is created at their interface, which hosts a protected electronic state, in which transport could be accessed by gated doping.

## 6.1 Introduction

Field effect transistors (FET) are the key elements in modern electronics and function by turning on and off the conductivity through a channel between source and drain by means of a gate. Typically, the channel is a 2D electron gas between two semiconductors and the gate voltage is applied through an insulating layer, such as an oxide in a MOSFET, which thereby controls the carrier concentration in the 2D gas. An important quantity is the ratio between on and off state resistance. Here we propose a switch based on turning off the conduction in topological edge states of a 2D topological crystal insulator by means of an electric field gate. In the 2D Sb system under consideration, the conducting edge states are topologically protected by inversion symmetry and an electric field breaks the inversion symmetry thereby turning the conducting channel off in an abrupt way. Furthermore the conductance itself can be in the ballistic quantized regime. This is illustrated in Figure 6.1 *(left)*.

A somewhat related but not identical idea for a topology based switching device was proposed by Liu *et al.* [105]. However that device is based on breaking the mirror plane symmetry, which protects a topological state, by means of an electric field in the 3D material SnTe. Strong effects on magnetotransport due to mirror symmetry breaking were also reported by Wei *et al.* [106].

A second intriguing idea in advanced electronics is that of reconfigurable circuits. Cheng *et al.* [107–109] demonstrated this idea by creating a conducting local 2D electron gas at the LAO/STO ($LaAlO_3/SrTiO_3$) interface induced by removing surface adsorbed OH groups using a scanning tunneling microscope tip along selected lines drawn on the surface of LAO a few layers above the LAO/STO interface. Here we propose a different mechanism for creating reconfigurable quasi 1D semi-conducting channels in the monolayer Sb system. Furthermore the

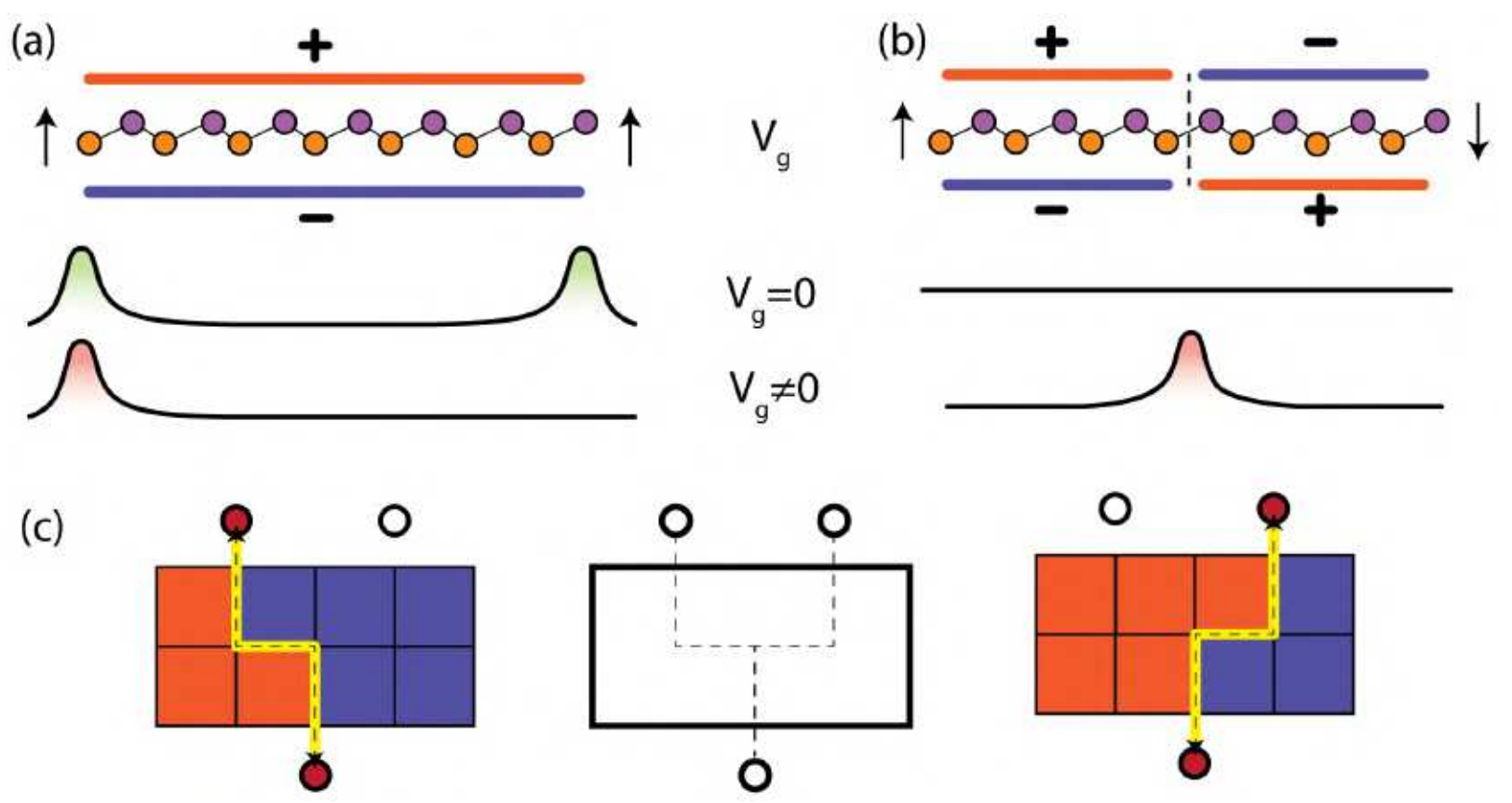

**Figure 6.1:** Illustration of the two devices *(a)* topological quantum switch *(b)*quasi 1D domain wall channels *(c)* Configurable 1D wire in a pixilated local gated system

conductance in this 1D system could show quantized conductance in the low-temperature ballistic regime.[110] This is also based on breaking the inversion symmetry. But instead of breaking it in the entire system, we show that by locally breaking inversion with opposite polarity one can form domain walls hosting localized 1D dispersing states in the gap which can become semi-conducting channels for electron or holes. Unlike the drawing mechanism used in the LAO/STO system, which requires nanoscale scanning probe type manipulations, the process to reconfigure the path in our proposed scheme is much simpler. First, it is a low energy process as one needs a very small amount of bias to open and close the channel because its existence is based on topology. Second it would only involve applying bias voltages to a static configuration of localized pixelated gates without the need for mechanical motion. A schematic illustration of this device is shown in Figure 6.1 *(right)*

The main idea behind both types of devices presented here is based on applying an electric field normal to the layers which breaks the inversion symmetry between the two sublattices because of the alread present buckling which exposes each sublattice to a different potential in the presence of an electric field. The use of an electric field normal to the layers of a 2D system to tune the electronic structure has been considered before, in particular for flat monolayer Sb.[45] However, in that case the electric field's role is to break the horizontal mirror plane rather than the inversion symmetry. The topological edge states in that regime are different and result from the spin-orbit coupling.

## 6.2 Symmetry and topology

In monolayer honeycomb group-V systems, unlike in graphene, the atomic $p$-orbital derived bands are disentangled from the $s$-orbitals.[101] In the completely flat limit, interaction between $p_z$ and $p_x, p_y$ is absent because of mirror symmetry. Buckling breaks the mirror symmetry and becomes a tuning parameter for the interaction and after a critical angle ($\approx 33^\circ$ for Sb) or equivalently reduction of tensile in-plane strain, the system gaps up and attains the lowest energy stable structure that has been synthesized[111]. In this structure, all three $\{p_x, p_y, p_z\}$ orbitals are hybridized and determine the low-energy physics near the gap, by forming a set of bonding and antibonding combinations in the valence and conduction band respectively. As described in Ref. [112] the valence band manifold has bond-centered Wannier functions obeying $S_6$ symmetry. This places the system in the "obstructed atomic limit" (OAL) as defined by Bradlyn *et al.* [92, 93] which has non-trivial crystalline topological (TCI) properties, leading to weakly protected edge-states as well as corner states. In Ref. [112] we showed the system can be mapped to the bond-alternating honeycomb or Kekulé lattice, which can be viewed as a 2D generalization of the iconical Su-Schrieffer-Heeger (SSH) model whose alternating weak and strong bonds are well known to lead to protected end states when the unit cell is chosen so that the strong bonds are broken at the ends of a finite chain. The $S_6$ group has 3D inversion $\mathcal{I}$ as well as $C_3$ symmetry operations and it is the inversion symmetry which protects the edge and corner states.[103, 112, 113] This inversion is directly connected with the identical Sb atoms in this bipartite system, which leads to mid-gap 1D edge bands and corner states of 0D symmetry preserving flakes.

Unlike for a flat honeycomb lattice, because of buckling, one can easily break this sub-lattice (inversion) symmetry by an applied electric field (or gating) in the direction perpendicular to the layers, thereby creating a modulated on-site potential. This is obvious from Figure 6.1. This then leads to a $C_3$ point group as the reduced symmetry. As a result the edge states gap up and hence the fractionalized electrons previously spread over the two identical edges will now reside on the one edge (corner) with the lower energy. This, in principle, leads from partially filled edge (corner) states on both edges to a completely filled and empty edge (corner) state on opposite sides and hence a metal-insulator transition in the case of dispersing edge states or a switch in which half of the corner states switch from half occupied to fully occupied and the other ones to empty as detailed next.

## 6.3 Topological switch

At low temperature, graphene is known to have quantized low bias conductance in 1D armchair (zigzag) edge modes given by[114] $G = 4n\frac{e^2}{h}\tilde{t}$ ($G = 2(2n+1)\frac{e^2}{h}\tilde{t}$) with $\tilde{t}$ the transmission coefficients to the leads, whereby the conductance occurs in steps but which are not exact multiples of the quantum of conductance. In this nanoscale ballistic transport regime, the conductance is determined by the number of 1D transverse modes contributing to the transport.[110] In the present case of a 1D Sb nanoribbon and for zero bias normal to the plane of the ribbon, we have two zero energy edge modes (one for each edge) with half occupation for each spin channel, therefore giving $G = 4\frac{e^2}{h}\tilde{t}$ conductance, when the Fermi level lies within this band. In contrast to graphene, where the edge bands connect the host valence and conduction bands, the large gap in the 2D Sb host system ($\sim$ 3eV) implies a huge barrier before carriers can access the high/low lying 2D host bands in the ribbon giving us zero conductance for any Fermi level location in the gap outside the range $(E_{min}, E_{max})$ set by the 1D dispersion of the edge bands. Note that the dispersive nature of the 1D edge bands and their bandwidth stem from the next nearest neighbor interaction along the parallel direction and are hence small (of order 0.3 eV). Breaking the inversion symmetry between the sublattices by a gate voltage, as already described in the previous

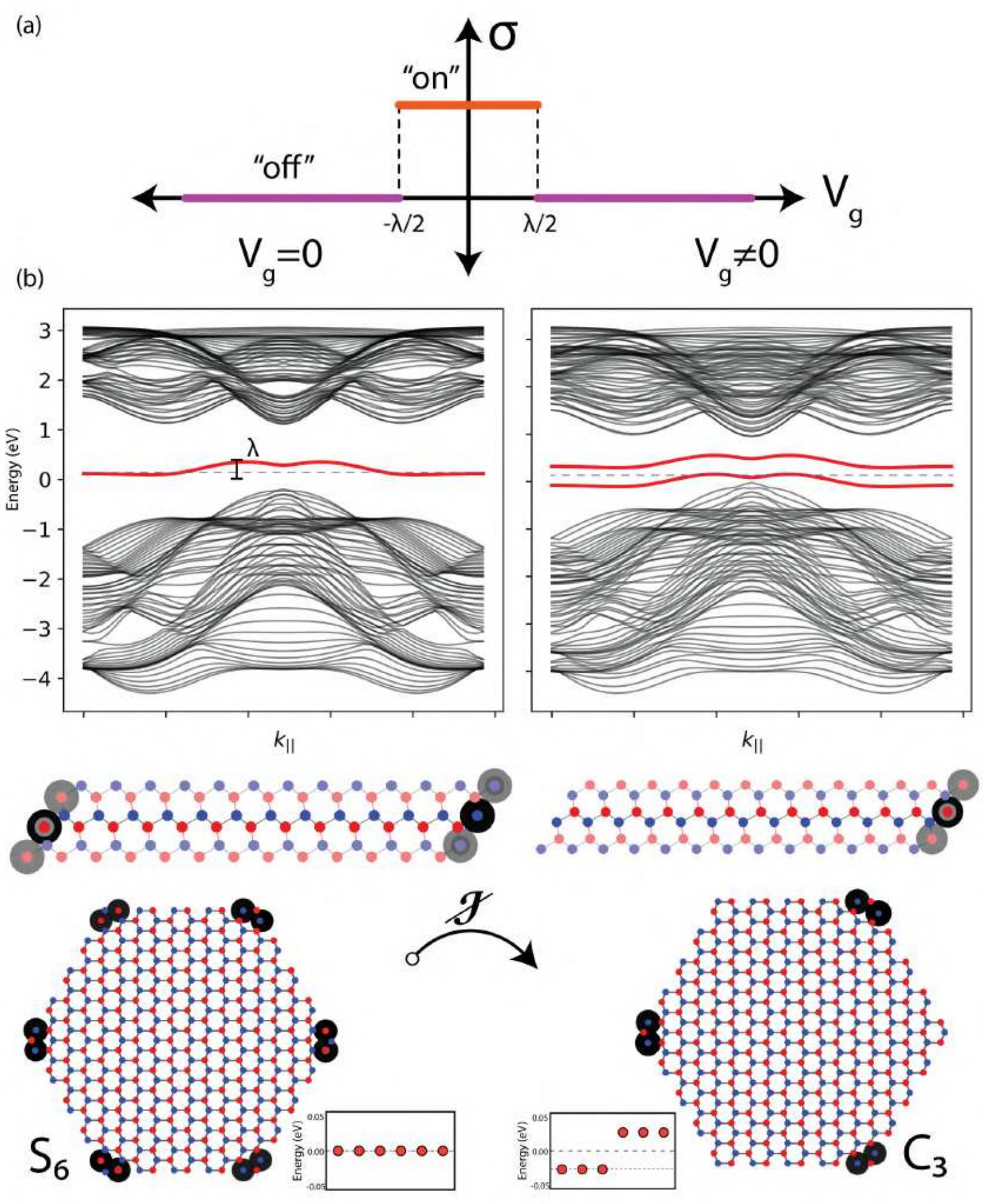

**Figure 6.2:** *(a)* Operation of "on" and "off" state using the device *(b)* band structure with gate voltage turned on and off and currosponding $|\Psi|^2$ *(c)* Effect of Inversion breaking in corner states of 0D system

section, now opens a gap between the edge bands as soon as the splitting becomes larger than the 1D edge band dispersion. Assuming that the gate voltage changes the on-site potentials on the two sub-lattices by $\pm\Delta/2$ and thus opens a splitting $\Delta$ between the energy centers of the two edge bands and a dispersion in the edge modes of width $\lambda$, a gap between the two edge bands occurs when $|\Delta| > \lambda$ and one enters the insulating limit with zero conductance. Because $\lambda$ is small compared to the gap of the 2D system, only a small gate voltage will be required to switch the device from its "on" (conducting) to its "off" or insulating state.

There are three major advantages to device of this type compared to traditional switches: (i) this switch has topological protection as the *on* state is protected from perturbations including defects in the lattice because of topological non-triviality (as shown in [112]); (ii) unlike traditional switches, we have quantized binary states; and finally (iii) the barrier to turn off the switch requires only low power with a value set by the dispersion of the edge modes.

Another possible device that uses the inversion symmetry as switch is to use the same mechanism in the corresponding finite size symmetric hexagonal flakes of the system. With inversion symmetry preserving the $S_6$, one is guaranteed (spinless) charge fractionalization of $e/2$ in each $\pi/6$ sector. But as soon as one reduces the$S_6$ symmetry to $C_3$, the edge modes can interact and open a gap. This leads to the $e/2$ charges being transferred to the respective previously inversion symmetric point. (Including spin the charge trasfer occurs for each spin so becomes $e$.) Thus at zero bias voltage, where inversion symmetry is preserved, one has no potential difference between each $\pi/6$ sectors while a inversion breaking non-zero bias potential difference between $C_3$ symmetric inversion points as shown in Figure 6.2(c).

## 6.4 1D Quantum wires

We now look at selectively breaking inversion symmetry in the system with opposite polarity which we will show leads to a domain boundary localized state in the gap. To gain qualitative insight, note that breaking inversion by biasing the system essentially acts like adding a mass gap of $m\sigma_z$ to the 1D SSH system. This breaks the chiral/inversion symmetry (in the 1D case) which can be easily seen in the finite 1D chain with one end being A $(+m)$ site and the other end being B $(-m)$ site. Thus the previously degenerate edge modes trivially gap with the occupied state being at the low energy $B$ side. The same physics holds here with one edge being occupied by the low potential part of the inversion broken sub-lattice and other side being empty when an inversion breaking bias is applied.

For creating a 1D quantum wire, we split the device into two regions, where in one region we apply a bias voltage of $+V$ and in the other $-V$. This has the effect of creating alternating onsite terms on one side and the opposite onsite terms on the other side with the junction between them having a domain wall of two $V$.

### Non-chiral SSH

To understand the energetics of this system, we start by looking at the same effect in a simpler 1D SSH, which is nothing but the 1D projection of the buckled Sb 2D system. The SSH Hamiltonian can be written as

$$H(\delta t, \alpha) = \sum_i (t + \delta t) c^{\dagger}_{i,A} c_{i,B} + (t - \delta t) c^{\dagger}_{i+1,A} c_{i,B} + \tag{6.1}$$

$$\left(-1^{\theta[i]}\right) \alpha \sum_i (c^{\dagger}_{i,A} c_{i,A} - c^{\dagger}_{i,B} c_{i,B}) + h.c. \tag{6.2}$$

where $c^{(\dagger)}_{i,j}$ creates (annihilates) an electron in unit cell $i$ on sublattice site $j$ and $\theta[x]$ is the Heaviside step function. The first terms is the usual SSH Hamiltonian with $\delta t$ controlling the hopping anisotropy. For $\delta t = 0$, we are at the critical point between the trivial/non-trivial regimes of the SSH model and the system would be metallic with zero gap. However, the effect we are concerned with here would still be valid because it depends on the second term. We just want a gap to more clearly see the edge states but whether the SSH gap is of the trivial or non-trivial type is irrelevant for the following discussion. The second term here adds a mass gap of $\alpha\sigma_z$ for $i < 0$ and $-\alpha\sigma_z$ for $i > 0$. It is easy to see that this system has inversion w.r.t. origin, the sites to the left and right of origin (A and B) gain the same value of $\alpha$ to their on-site term.

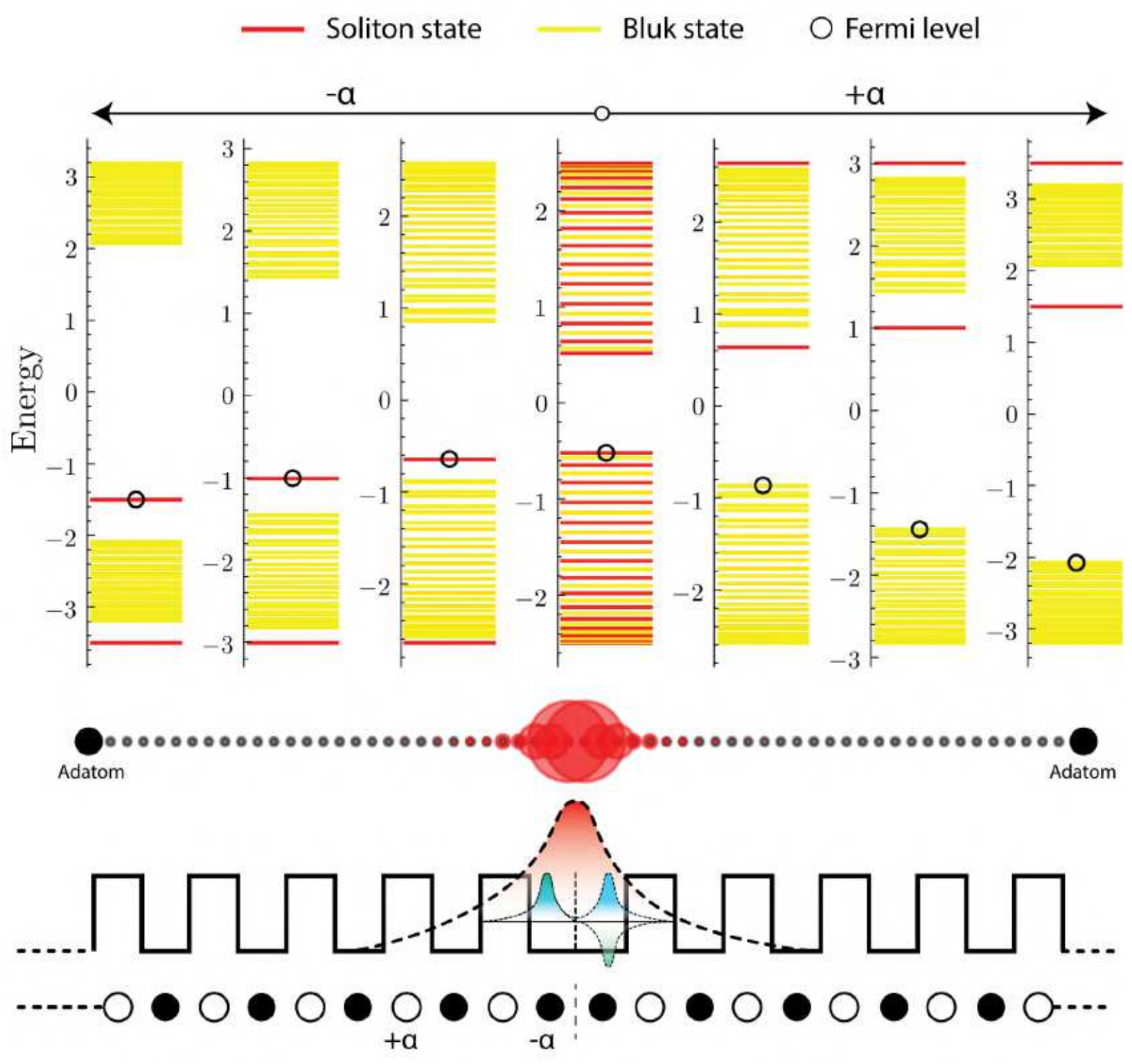

**Figure 6.3:** *(top)* Energy spectrum of (6.2) as function of $\alpha$ with $\delta t \neq 0$ red levels are projected onto two sites near origin and yellow on bulk. (Energy scales are different)*(middle)* $|\Psi|^2$ of the localized state. *(bottom)* potential profile and the illustration of localized state and corresponding symmetric and antisymmetric wave-function

Figure 6.3 shows the eigenvalue spectrum of the system. The states colored red are the ones localized at the origin where the switch occurs between the two polarities. Since the system has lost its periodicity because of the step function, we treat the system as a large but finite number of sites. Whether it is trivial or not depends on the sign of $\delta t$. We consider a system with $\delta t > 0$, so that there is indeed a gap as in the SSH. In the non-trivial limit we would have an edge state which is not essential for the following discussion and hence is avoided by adding an adatom to the edges as shown in Figure 6.3. For the effect we consider, it is not important whether the SSH is in the trivial or non-trivial regime as long as there is a gap. Indeed the effect is related to the potential shifts on the atoms rather than on the alternation of strong and weak bonds. Alternatively, we may also envision a periodic Kronig-Penney (KP) type model with alternating wells and barriers. Starting from a finite chain, what the gate voltage does is to flip the wells into barriers in the right half of the system, which creates a *well* of double width at the center as two atoms with on-site shift $-\alpha$ and $-\alpha$ become adjacent at the boundary between the two regions. For the opposite gate voltage, $+\alpha$ occurs at both center atoms, or in the KP model, a *barrier* of double width occurs.

For $\alpha = 0$ (shown in middle of the top plot), one can see that the spectrum is gapped because of the interaction anisotropy from $\delta t \neq 0$ but the energy states are all having comparable contribution at the origin. In other words, the states are extended and not localized at the domain boundary. Next, applying a negative bias $-\alpha$ to the atoms next to the origin, as illustrated at the bottom, we see that states localized at the origin (indicated in red) form below and above the occupied bands. They split farther and farther apart from the continuum of levels as the magnitude of the bias $|\alpha|$ is increased.

For $\alpha = 0$ as in the case of SSH, we have bonding valence and anti-bonding conduction bands in the lattice with respect to the strong bond. Within the valence band, the weak bonds will form the most bonding combination of the bond-centered Wannier orbitals (same sign on each site) at the bottom of the valence band and the most antibonding arrangement, (alternating signs) at the top of the band. But with the $-\alpha$ on-site shift, at the domain boundary we create a pair of local bonding and anti-bonding states from the domain boundary adjacent atoms. Thus a state with even stronger bonding character forms below and another one with more antibonding character above the VBM. In the KP model, one may think of the lowest state in each well as corresponding to a strong bond of the SSH model and these bond orbitals are now weakly coupled by tunneling through the barriers and forming a band. In this context, one might reason that the center well being wider will have a lower level and hence when it interacts with the other wells, a new level will appear below the band. But there will now also be a second antisymmetric band in the central well of double width and this one will occur slightly above the band but well below the CBM formed by broadening the antisymmetric states of the narrow wells into a band. The electron count does not change, so this level becomes the highest occupied level of the system and is fully occupied. Thus, one can see that we have essentially created a potential domain wall hosting a localized state at the origin, as shown in the bottom part of Figure 6.3. If the system is slightly doped p-type, say by removing one electron from the whole system, the split-off top level localized at the center will become partially filled, or in other words, a hole becomes preferentially trapped at this site.

For opposite polarity, where the two central atoms experience an upward potential shift $+\alpha$ or a *barrier*, we can see in the right panels of energy levels, that now states localized on the origin appear above and below the conduction band. In the KP model one may think of this bound state in the gap as resulting from having a weaker tunneling trough the double well at the center. The highest occupied band stays at the top of the valence band but a bound state localized near the domain wall now appears below the conduction band. So, for slight n-type doping the electron will become localized at the boundary.

We may call this domain wall boundary localized state a soliton state. In fact, the boundary could be moved in principle by shifting the region where positive and negative bias is applied, which would move the localized state around like a solitary wave. Note, however, that it does not originate from having two strong bonds or two weak bonds next to each other as in the SSH soliton like states but rather by having two adjacent atoms with downward or upward potential (or mass shift).

Another interesting perspective for the occurrence of this state can be deduced from topology. Breaking chiral symmetry breaks the notion of topological protection for the system, but still guarantees one localized occupied and one unoccupied edge mode (as long as SSH non triviality is satisfied)[115]. As mentioned before, we essentially have created two copies of non-chiral SSH with one being the mirror symmetric reflection of the other (Figure 6.3 lowest panel). Thus, at the origin, the ends of these two chains meet which then topologically guarantees us a localized state. The left system consists of a lattice with $+\alpha\sigma_z$ while right $-\alpha\sigma_z$ and thus there is a domain boundary between the two ionic lattices creating a protected bound state.

## Solitons in Buckled Sb

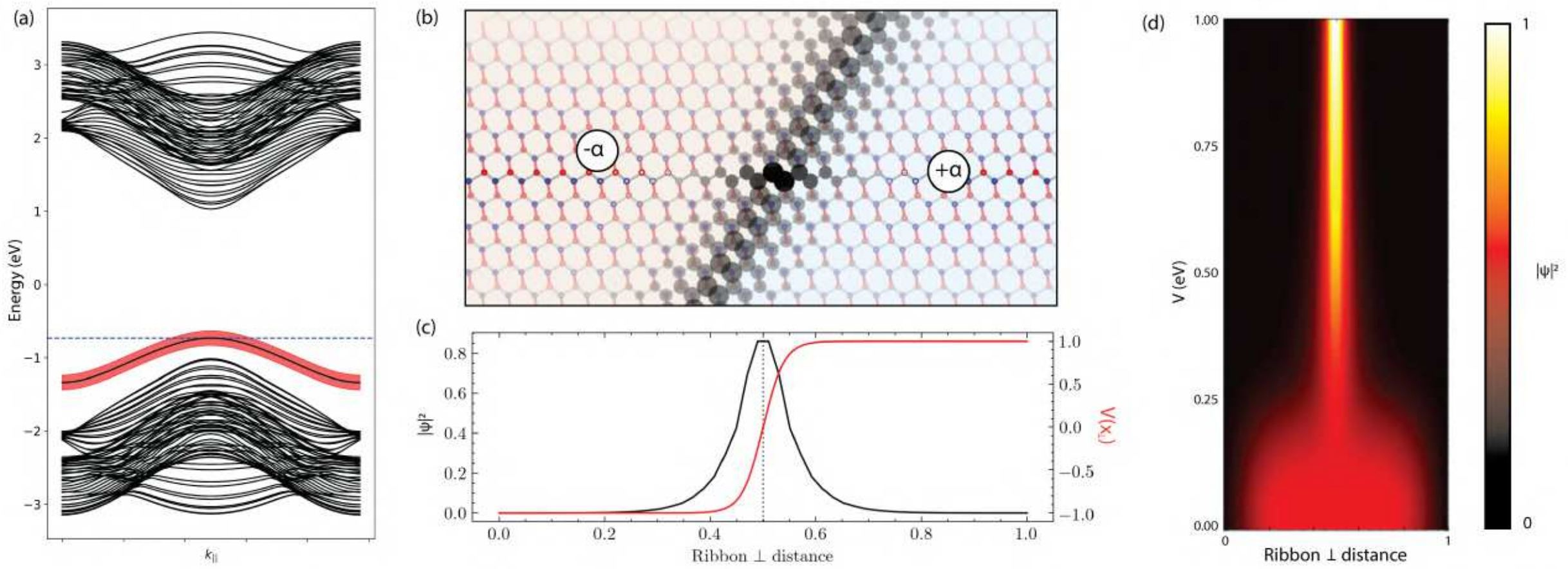

**Figure 6.4:** *(a)* Band structure of the nano-wire when $\alpha = 0.5eV$ with the domain wall state marked in red *(b)* Real space distribution of the domain wall wavefunction *(c)* Gate potential profile and $|\Psi|^2$ of domain wall state. *(d)* $|\Psi(\alpha, x_\perp)|^2$ showing the localization

Now moving on to our actual system, we can follow the same procedure to create this domain wall state. But since we have the added dimension $k_{\|}$, we have a dispersive band at the 1 dimensional domain wall. To simulate a realistic device, we use $V_{gate} = \tan(wx)$ as shown in Figure 6.4(c). This essentially ensures that the domain wall is smooth and the smoothness factor is controlled by $w$.Figure 6.4(a)shows the band structure of this system (for $\alpha = 0.6$ eV). Again, as in the case of SSH, we have added adatoms at the edges of the nanoribbon to remove the outer edge modes. Alternatively, one could use two semi-infinite systems and use a Green's function technique. Marked in red is the localized domain wall band and its localization is plotted as $|\psi|^2$ by the gray scale in Figure 6.4(b). To numerically check the localization we also look at the $\int dk|\psi(k,x)|^2$ which is exponentially localized normal to the domain boundary as shown by the applied potential profile. We also show the degree of localization as function of applied gate voltage in Figure 6.4(d). Clearly the higher the gate voltage, the narrower the soliton like bound state becomes. Strictly speaking what we mean by gate voltage here is the applied on-site terms on the Sb atoms. How these are related to an actual applied voltage would depend on the technical implementation of the device and the screening or self-consistent potential in the Sb layer. Modeling that in detail is beyond the scope of this qualitative paper which focuses on the concept.

These domain wall states remain occupied/unoccupied and become the highest valence/ lowest conduction bands depending on the sign of $\alpha$ just like in the SSH toy model. Thus by electron or hole doping the system (possibly by global gating), one could achieve localized conduction because electrons or holes would thermalize to these domain boundary bound states. Interestingly, since the polarity of the channel dependents on the sign of applied voltage, one can turn a hole doped system conducting and insulating along this channel by simple reversing the global gate polarity.

Although we have here discussed the principle of the creation of a 1D wire only, it doesn't take much imagination to see that one could apply such opposite voltages in patches creating 1D wires at their boundaries. By creating a pixelated gate on top and below the Sb layer, one could imagine just changing the pattern where the voltages are applied and hence where the wires are created by which pixels are addressed. This is in principle a design for a reconfigurable network or circuit of connecting wires as shown in Figure 6.1(c). As mentioned above, this localization is guaranteed by topology and thus defects/perturbations in the bulk part away from the domain wall would not interfere with this localization, hence creating robust 1D quantum channels. In order to make the 1D wires at the boundaries conducting the system needs to be doped in just the right way that free carriers, holes or electrons will tend to be confined to the 1D localized domain wall states. This doping itself can conceivably also be realized by a global gating of the system rather than by adding chemical dopant atoms, whereby the overall carrier density in the system would also be controllable. We emphasize that domain walls of this type can be made along several directions in the 2D Sb. Their origin depends on breaking the inversion symmetry and is thus not confined to specific edges, such as zigzag or arm-chair edges at which the domains meet.

# Annihilating Dirac fermions | 7

**Abstract**

We show that annihilating pairs of Dirac fermions necessitate topological transition from the critical semimetallic phase to an Obstructed Atomic Limit (OAL) insulator phase instead of a trivial insulator. This is shown to happen because of *branch cuts* in the phase of the wave functions, leading to a non-trivial Zak phase along certain directions. To this end, we develop $\mathbb{Z}_2 \times \mathbb{Z}_2$ invariant and also study the phase transition using Entanglement Entropy. We use low energy Hamiltonians and numerical results from model systems to show this effect. These transitions are observed in realistic materials including strained graphene and buckled honeycomb group-V (Sb/As).

## 7.1 Introduction

There has been much progress in the past years improving our understanding of topological phases. This has extended the possibility of having topologically non-trivial systems which were previously restricted to internal symmetries[116], to systems that are protected by crystalline symmetries[117–120]. Many new possibilities of Symmetry Protected Topological phases (SPT) were realized including multipole insulators[20], fragile insulators[26] and boundary obstructed topological phases [22] .Ref. [121] showed that, by combining crystalline symmetries with underlying orbital symmetries, one can obtain a class of topologically non-trivial insulators called "Obstructed Atomic Limit" (OAL) insulators which are distinct from trivial "Atomic insulators" (AI). OAL insulators are systems where Wannier charge centers (WCC) of the occupied band manifold are exponentially localized on Wyckoff positions that are different from the atomic positions. They thus cannot be adiabatically connected to a phase where the WCC are located on-top of the atoms (AI). This results in a *dimerized* insulating state. These phases of matter often lead to charge fractionalization in co-dimensions[1] $\geq 1$ because of the *filling anomaly*[122]. This may, in general lead, to *extrinsic* higher-order topologically insulators[123, 124] *i.e.*, systems where the occurrence of zero-energy corner states or gap-less hinge modes may depend on properties/symmetries of the boundary.

1: Co-dimension is a term used to indicate the difference between the dimension of certain objects and the dimension of a smaller object contained in it. For ex, co-dimension 1 of a 3D system is its 2D counterpart while co-dimension 2 of a 3D system is a 1D system.

Meanwhile, Dirac systems are the classic example of a symmetry-protected gap-less topological phase. These are systems which have topological charge around nodal points stabilized by having Time Reversal Symmetry (TRS) along with a spatial symmetry. Apart from the ubiquitous example of graphene, we have recently showed that buckled mono-layer As and Sb have Dirac nodes protected by $C_2$ located along high symmetry lines[101]. There can also be systems where the nodes are located on a general point in the BZ[125]. These *unpinned* Dirac nodes can be gapped out without breaking TRS and the crystal symmetry protecting them by annihilating them in pairs of opposite topological charges at the Time Reversal Invariant Momenta (TRIM). There have been various artificial systems where this kind of merging can/has been accessed including the honeycomb Fermi gas[126, 127], microwave experiments [128] and mechanical graphene[129]. It has also been shown that electronic 2D graphene[130], $\alpha$-(BEDT-TTF)$_2$I$_3$ [131], Phosophorene[132] also has this feature upon applying strain. More recently, we showed that 2D group-V buckled honeycomb systems (Sb,As) undergo the same transitions[101, 112] and interestingly we found that they undergo 2 different merging transitions. In this chapter, we show that systems which are topological Dirac semimetals necessarily undergo *transition to a topologically non-trivial OAL upon annihilating Dirac cones of opposite winding number*. We start by looking at the low energy Hamiltonian describing the annihilation physics, after which we use the Zak phase to understand the topology of the system. We will then proceed to use symmetry indicators to find the topological classification and finally study the phases using the idea of Entanglement Entropy (EE).

## 7.2 Complex phase and annihilation

It was shown by G. Montambaux *et al.* [130] that the Hamiltonian of a annihilating Dirac system takes the universal form

$$H(k_x, k_y) = \left(d + \frac{k_x^2}{2m_x}\right)\sigma_x + p_y k_y \sigma_y \tag{7.1}$$

Here $m_x, p_y$ controls the effective mass (along $k_x$) and velocity (along $k_y$) directions. Close to annihilation, the system takes massless and massive dispersion along the two directions. It is clear that $H$ obeys TRS giving us $H(\boldsymbol{k}) = H^*(-\boldsymbol{k})$, has inversion symmetry ($\mathcal{I}$) given by $\sigma_x H(\boldsymbol{k})\sigma_x = H(-\boldsymbol{k})$ and Chiral symmetry ($\mathcal{C}$) as $\sigma_z H(\boldsymbol{k})\sigma_z = -H(\boldsymbol{k})$. Finally, the system also has a mirror/$C_2$ ($\mathcal{M}$) symmetry that leaves the entire $k_y = 0$ line invariant given by $\sigma_x H(k_x, k_y)\sigma_x = H(k_x, -k_y)$. This symmetry is the key that protects and allows the Dirac points to move $k_x$. It has been shown[130] that, based on the sign of $(m_x d)$, the ground state solution can either be in the (i) semimetallic limit with two sets of oppositely wound Dirac cones at $k_x = \pm\sqrt{-2m_x d}$ or (ii) insulating limit . We show here that the phase diagram of this Hamiltonian is much richer than realized in Montambaux *et al.* [130] as seen in Figure 7.1. Phases **I** and **III**, which were previously thought to be trivial insulators both, belong to the class of the OAL. This effect can be seen in the surface of their respective co-dimension $\geq 1$ limit with protected boundary states based on the symmetries of termination.

Figure 7.1 shows the phase diagram of $H(\boldsymbol{k}; m_x, d)$. Phases **I** (**III**) correspond to an insulating state in bulk without (with) boundary states in the co-dimension=1 system. Phases **II** (**IV**) are semimetallic systems with two Dirac cones and edge states propagating outside (in-between) in their co-dimension=1 system. The difference between these phases are encoded in the phase of their complex wave function. The wave functions of Equation 7.1 are given by $|\pm\rangle = \frac{1}{\sqrt{2}}\left(1, \pm e^{-i\theta(k)}\right)$ where[2]

2: Note that the two eigenstates differ by a sign in the second component, in other words, they differ by a phase $\pi$.

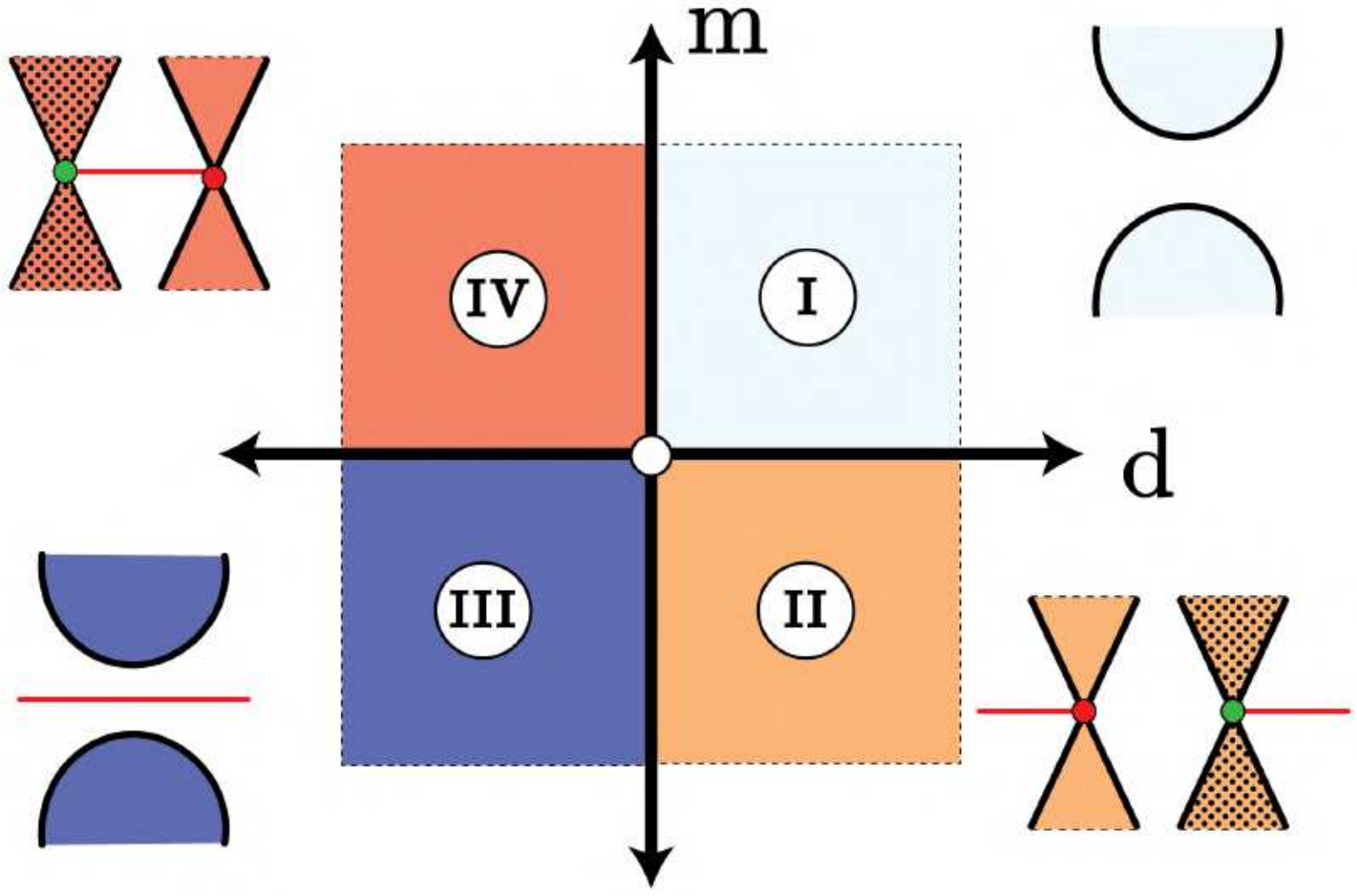

**Figure 7.1:** Phase Diagram of Equation 7.1 and the corresponding lower dimensional eigenstates where the origin is at the center. Green and Red dots in **II** & **IV** indicate the winding number of ±1. Red band denotes the surface state

$$\theta(\boldsymbol{k}) = \tan^{-1}\left(\frac{p_y k_y}{d + \frac{k_x^2}{2m_x}}\right) \tag{7.2}$$

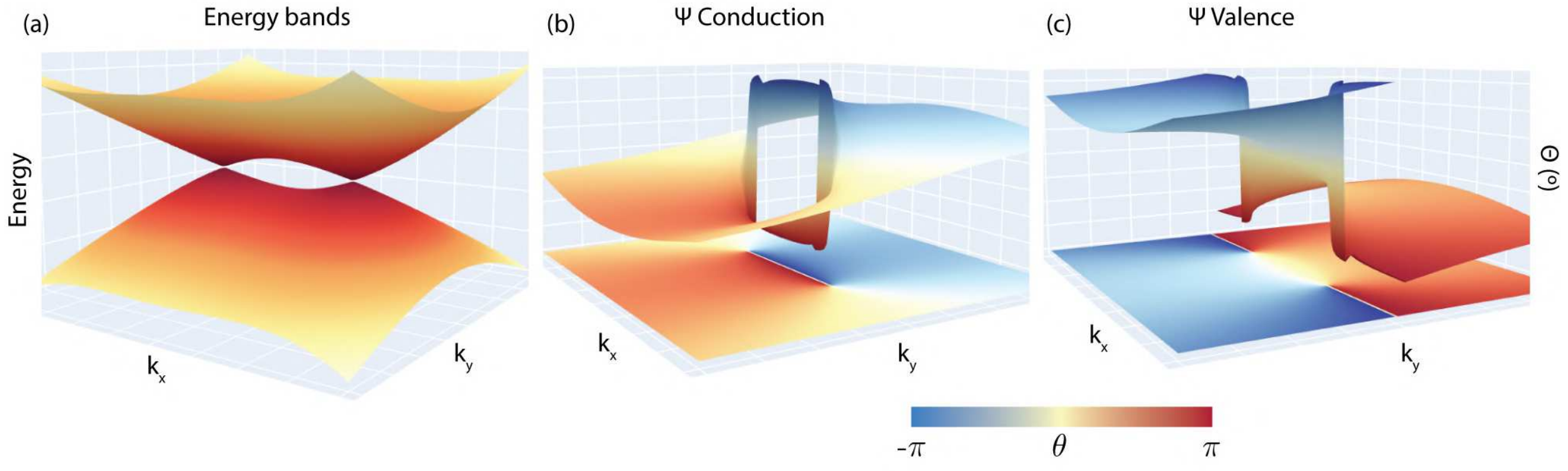

**Figure 7.2:** (a) 3D band dispersion showing Dirac cones of opposite winding number and Wave function's Riemann sheet (complex phase) for (b) conduction (c) valence band with the phase plotted again in the base for clarity showing the singularity.

Figure 7.2 shows the phase at each $\boldsymbol{k}$ for the conduction (b) and valence bands (c) along with the eigenvalues (a) of Equation 7.1 for $m_x d < 0$ with two Dirac cones along with the phase overlaid on top (and projected on bottom for clarity). The complex wave functions form sheets in $\boldsymbol{k}$ space which are nothing but Riemann sheets[133]. Close to $\pm k_0 = \sqrt{-2m_x d}$, Equation 7.1 can be expanded to $\pm\sqrt{\frac{2d}{m_y}} q_x \sigma_x + p_y k_y \sigma_y$ where $q_x = k_x \mp \sqrt{-2m_x d}$, corresponding to the dispersion of two Dirac cones of opposite winding number. To understand the mechanism better, we now consider a complex holomorphic function $f(z) = (z + k_0)^{-1} - (z^* + k_0)^{-1}$. In the phases **II** and **IV** (Figure 7.2) corresponding to the Dirac cone limit, it is easy to see that the complex phase wave functions of Equation 7.1 are the same as $\pm f(z)$ defined in the complex plain $\boldsymbol{k} = k_x + ik_y$. As seen from Figure 7.2, The two wave functions in the semimetallic phase consist of branch cuts from $k_x \in (-k_0, +k_0)$ and $k_x \in (-k_0, -\infty) \bigcup (k_0, +\infty)$. Both the wave function are related by a phase shift of $\pi$ and the choice of branch cut picked up by each wave function is determined by the sign of $m_x$ and $d$. Similarly, as one tunes the parameters to get to the insulating state, $k_0 \longrightarrow 0$. This completely detaches the Riemann surface of one of the wave function and connects it to the next branch. This results in extending the branch cut from $(-\infty, \infty)$ for one of the wave function and closing the branch cut on other. This is shown in Figure 7.3

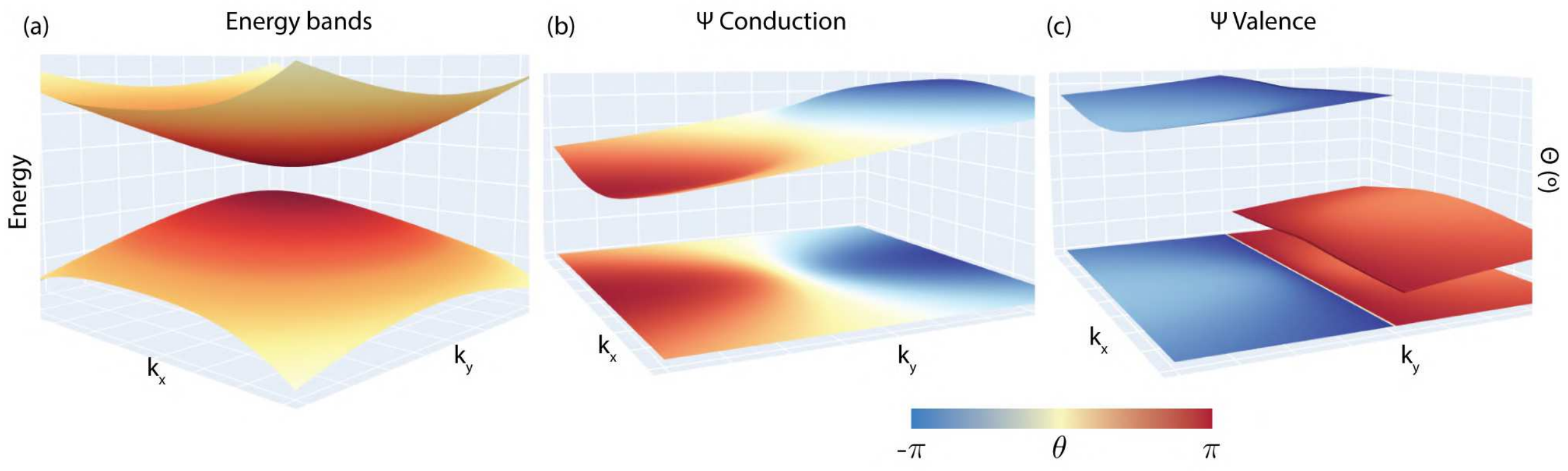

**Figure 7.3:** a) 3D band dispersion showing Insulating state and Wave function's Riemann sheet (complex phase) for (b) conduction (c) valence band with the phase plotted again in the base for clarity showing the singularity.

To quantify the topology, one can calculate the Zak phase[60, 134], which is given by $\gamma = i \oint \langle u_{n\mathbf{k}} | \nabla_{k_\perp} | u_{n\mathbf{k}} \rangle$ where $|u_{n\mathbf{k}}\rangle$ is the lattice-periodic part of the wavefunction. Here $k_\perp$ is the perpendicular direction along which the integration is being done[3] . Or equivalently the Wannier Charge Center (WCC), which for the $n^{th}$ band along the $i^{th}$ direction is given by

3: It is denoted by $k_\perp$ as the corresponding $k_\parallel$ is the direction in which $k$ vector survives and thus the system is periodic in the $\parallel$ direction.

$$x_i^n = \frac{1}{2\pi} \int_{-\pi}^{\pi} dk_i \mathcal{A}_n(k_i), \tag{7.3}$$

where $\mathcal{A}_n(\mathbf{k}) = i \langle u_{n\mathbf{k}} | \nabla_{\mathbf{k}} | u_{n\mathbf{k}} \rangle$. It is easy to see that the Zak phase along a given direction becomes necessarily non-vanishing when it crosses through a phase discontinuity. This can happen when the phase continues to the next complex branch by having a *branch point/cut* as we discussed above. From the previous discussion, $\gamma^x(k_y)$ is 0 everywhere except for the points passing through the branch cut, where

$$\gamma^x(k_y) = i \oint dk_x \left\langle u_n(k_x, k_y) \middle| \partial_{k_x} \middle| u_n(k_x, k_y) \right\rangle. \tag{7.4}$$

For example, in the semimetallic case, based on the values of $m_x$ and $d$, one gets a non-trivial Zak phase either inside or outside the Dirac cones. Once the parameters are tuned to reach either of the insulating phases **I**/**III**, one identically gets a Zak phase of $\pi$ or 0. Non-trivial Zak phase along this direction indicates a non-trivial 1D topology along this path which results in a surface spectrum on co-dimension $\geq 1$ (attached to the respective $k_\parallel$). It should be noted that although the existence of surface states in the OAL limit is topologically guaranteed, the protection is not. The protection stems from the underlying symmetry of the surface termination[4] .

4: By protection, we mean the degeneracy of the states belonging to the surface spectrum.

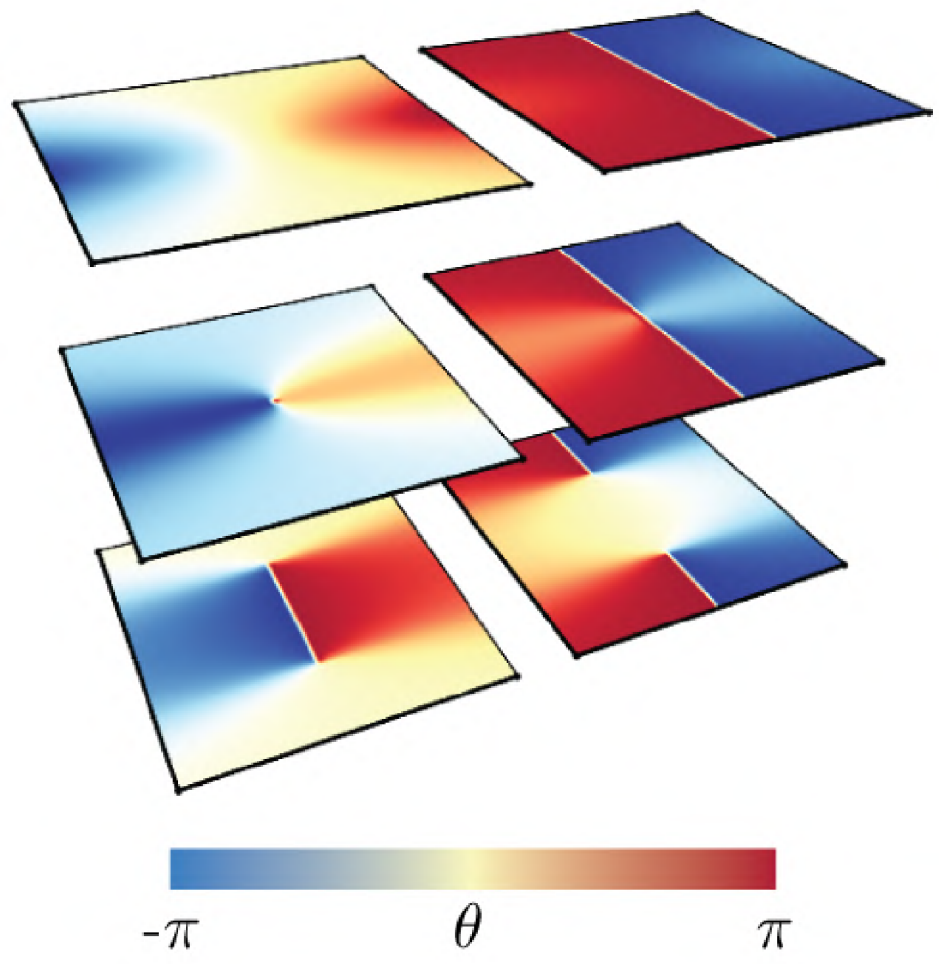

**Figure 7.4:** Phase in the 2D BZ as as function of transition from bottom to top

## 7.3 Symmetry indicators and annihilation

Analogous to the Su–Schrieffer–Heeger (SSH) model [56], the effect of a non-trivial Zak phase can be captured using the symmetry indicators at High Symmetry Points (HSP). For the present case, we start by looking at the different symmetry groups that are preserved throughout the BZ. A band crossing at a generic point along high symmetry line is protected by a symmetry ($g$) which is preserved throughout the high symmetry line (in this case it is given by $\mathscr{M}$ that makes the line $k_y = 0$ invariant). Hence, the only way to gap this kind of system is to annihilate the opposite winding band crossing points with each other at a TRIM HSP. This point would have a higher symmetry group ($G$) such that $g \in G$. Before the annihilation, at HSP the bands are made up of two distinct irreducible representations or *irreps*, say $\mathscr{A}, \mathscr{B}$ and along the high symmetry line, by $a$ and $b$. Because of compatibility, $a$ and $b$ need to have different parity eigen value given by $+, -$[5] . Another way to see this is to realize that, at the transition point occurring at HSP, the two bands have a massive dispersion along one direction and thus need to have different parity eigenvalue in order for the bands to become degenerate at the HSP. This is illustrated in Figure 7.5(a), where the path $\Gamma - X_1$ has a symmetry that protects the crossing of the bands made of different irreps[6] . As one travels from $\Gamma - X_1$, the parity eigenvalue of the bands necessarily change because of the crossing. Now there can be several scenarios of parity value ordering on the other HSPs based on the various phases we are at. For instance, in Figure 7.5(a), the parity eigenvalues are such that a switch happens along the path $X_1 - M$ and thus one gets a non-trivial SSH-like edge mode along that path emerging from the Dirac crossing (as shown in inset). This thus indicates that the edge state is of type **II**. Once the points annihilate at $\Gamma$ , one is ensured a parity change at $\Gamma$. Because of the massless nature of the transition, we now have switched the parity not just along the path $X_1 - M$, but also along $\Gamma - X_2$. This leads to a surface state throughout the BZ in the corresponding 1D system taking us to the phase **III**.

5: Here, we mean parity w.r.t inversion operator. As mentioned in the previous text, inversion is a symmetry of our system and hence one can always talk about inversion eigen values at high symmetry TRIM points.

6: Note that in 2D there are always 4 TRIM points and thus Figure 7.5 describes the most general BZ. The $, -$ signs here are the parity with respect to inversion at the TRIM points but along the HSL, the symmetry operation that allows for the crossing could be a different operation that however must be compatible with the inversion parities at the HSP at the ends of each line.

This switching is not accidental and is deeply connected to the polarization (or Zak phase) of the system. For an inversion symmetric system, one can re-write the Zak phase along a given direction $i$ connecting $X_i$

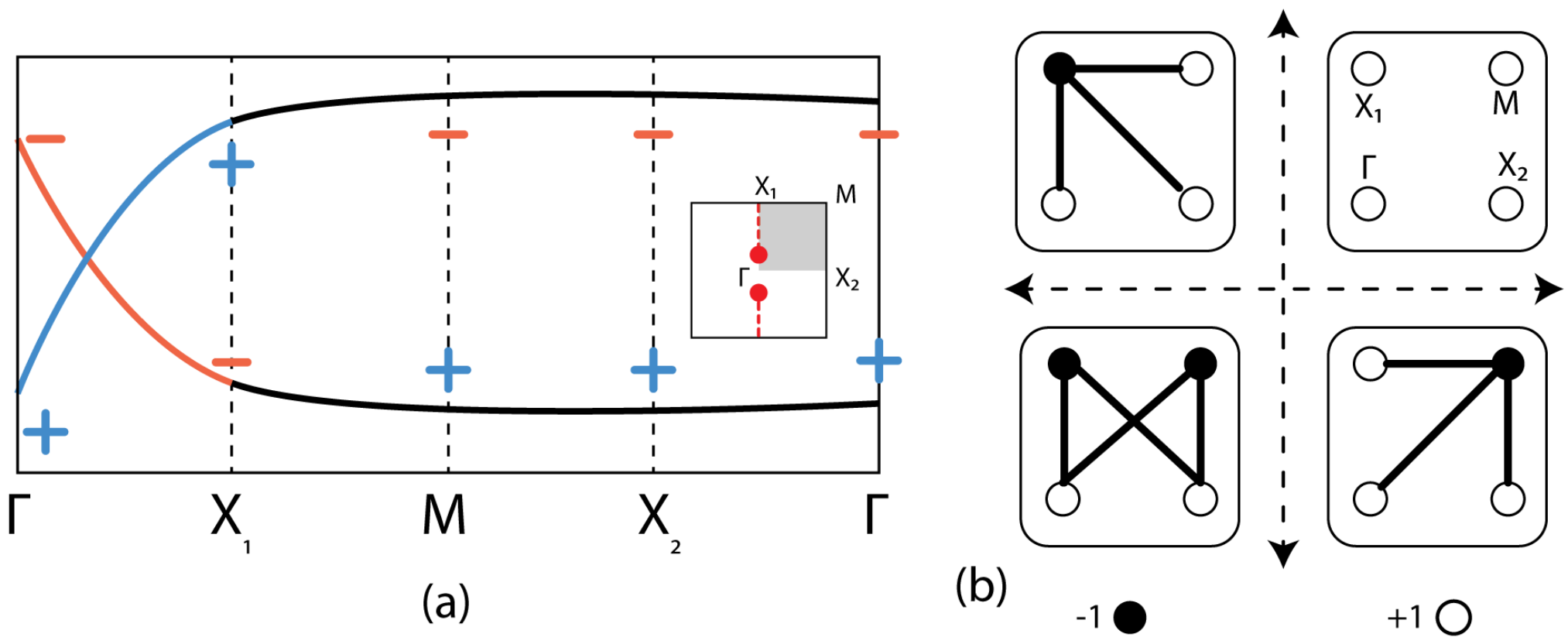

**Figure 7.5:** *(a)* Band structure and parity eigen values of a system in phase **II** *(b)* 2D BZ and TRIM points shaded according to their parity eigenvalue with graph connecting points of opposite parity

and $Y_i$ by $\pi\chi_i$ (for proof see Subsection 6 (Zak quantization)) where $\chi_i$ is[118]

$$(-1)^{\chi_j} = \prod_{j \in occ.} \frac{\eta_j(\mathrm{X_i})}{\eta_j(Y_i)} \tag{7.5}$$

where $\eta_j(X_i)$ is the parity eigenvalue of $j^{th}$ band at $X_i$. $\chi_i$ is a quantized $\mathbb{Z}_2$ invariant. Figure 7.5(b) shows $\prod_{j \in occ.} \eta_j(\mathrm{X_i})$ denoted by $\circ = 1$ and $\bullet = -1$ at TRIM points and their corresponding phases. Starting with phase **I**, we have all the eigen values trivially equal to each other. Sans any inversion, this gapped system is in the *"trivial OAL"* limit with no edge states. For phase **II**(**IV**), we have created a symmetry protected crossing along $X_1 - M$ which changes the parity at $M$ ($X_1$) and hence a non-trivial Zak phase along $X_2 - M$ ($\Gamma - X_1$). As we proceed to move the formed crossing and annihilate at $X_1$, we switch the parities at $X_1$, leading to **III** where now there is a non-trivial Zak phase along both $X_2 - M$ and $\Gamma - X_1$. To intuitively derive an invariant, we consider a graph with $(\Gamma, X_1, M, X_2)$ as shown and connect the points with different parity values. It is easy to see that an even number of connections indicate a insulating phase while a odd number of connections, a semimetallic phase. Secondly, within even (odd) connections, a connection between $\Gamma - M$ indicates a surface state connecting throughout BZ (outside the Dirac point). Thus we have a $\mathbb{Z}_2 \times \mathbb{Z}_2$ invariant given by $(\zeta_1, \zeta_2)$ in Equation 7.6 and Equation 7.7. Here, $\zeta_1$ determines if the system is gapped or not while $\zeta_2$ gives us information about the edge modes.

$$(-1)^{\zeta_1} = \prod_{j \in occ} \eta_j(\Gamma)\eta_j(\mathrm{X_1})\eta_j(\mathrm{M})\eta_j(\mathrm{X_2}) \tag{7.6}$$

$$(-1)^{\zeta_2} = \prod_{j \in occ} \eta_j(\Gamma)\eta_j(\mathrm{M}) \tag{7.7}$$

To make the idea concrete, we shall look at some examples in Subsection 7.4.

## Zak quantization

In the presence of inversion symmetry, one can calculate the quantized polarization/Zak phase by looking at the parity eigenvalue at the TRIM points[135]. With the Berry connection defined as $\mathcal{A}^l_{ij}(\boldsymbol{k}) = -i \left\langle u_i(\boldsymbol{k}) \middle| \partial_{k_l} \middle| u_j(\boldsymbol{k}) \right\rangle$, The polarization along the $i^{th}$ direction is defined as

$$P_i = \frac{1}{2\pi} \sum_{j=1}^{occ} \int_{-\pi}^{\pi} dk_1 dk_2 \mathcal{A}^i_{jj}(\boldsymbol{k}) \tag{7.8}$$

$$= \frac{1}{2\pi} \int_{-\pi}^{\pi} dk_1 dk_2 \,\mathrm{Tr}\left[\mathcal{A}^i_{lm}(\boldsymbol{k})\right] \tag{7.9}$$

Now we use the fact that we have a symmetry that leaves a line in $\boldsymbol{k}$ space invariant (as our Dirac point moves to annihilate) without loss of generality, we call it mirror symmetry (although it could be $C_2$ as well) and mirror operator is denoted by $M$ (and $\hat{\mathcal{M}}$ in $\boldsymbol{k}$ space), we construct the sewing matrix $\mathcal{B}_{\mathcal{M}}$ as

$$(\mathcal{B}_{\mathcal{M}})_{ij}(\boldsymbol{k}) := \langle u_i(\mathcal{M}\boldsymbol{k})| \hat{\mathcal{M}} |u_j(\boldsymbol{k})\rangle \tag{7.10}$$

$$= \langle u_i(-\boldsymbol{k})| \hat{\mathcal{M}} |u_j(\boldsymbol{k})\rangle \tag{7.11}$$

Applying the Mirror to Equation 7.9 we get

$$P_i = \frac{i}{2\pi} \int_{-\pi}^{\pi} dk_1 dk_2 \,\mathrm{Tr}\left[\left\langle u_l(\boldsymbol{k}) \middle| \hat{\mathcal{M}}^\dagger \partial_{k_i} \hat{\mathcal{M}} \middle| u_m(\boldsymbol{k}) \right\rangle\right] \tag{7.12}$$

$$= \frac{i}{2\pi} \int_{-\pi}^{\pi} dk_1 dk_2 \,\mathrm{Tr}\left[\left\langle u_l(\mathcal{M}\boldsymbol{k}) \middle| \mathcal{B}^\dagger_{\mathcal{M}}(\boldsymbol{k}) \partial_{k_i} \mathcal{B}_{\mathcal{M}}(\boldsymbol{k}) \middle| u_m(\mathcal{M}\boldsymbol{k}) \right\rangle\right] \tag{7.13}$$

$$= \underbrace{\frac{i}{2\pi} \int_{-\pi}^{\pi} dk_1 dk_2 \,\mathrm{Tr}\left[\left\langle u_l(-k_1,k_2) \middle| \partial_{k_i} \middle| u_m(-k_1,k_2) \right\rangle\right]}_{-P_i} + \underbrace{\frac{i}{2\pi} \int_{-\pi}^{\pi} dk_1 dk_2 \,\mathrm{Tr}\left[\mathcal{B}_{\mathcal{M}}(\boldsymbol{k})^\dagger \partial_{k_i} \mathcal{B}_{\mathcal{M}}(\boldsymbol{k})\right]}_{\chi_i \to \text{Winding of } \mathcal{B}_{\mathcal{M}}(\boldsymbol{k}) \text{ along } i} \tag{7.14}$$

$$P_i = \frac{1}{2}\chi_i \tag{7.15}$$

Where $\chi_i = \frac{i}{2\pi} \int_{-\pi}^{\pi} dk_1 dk_2 \,\mathrm{Tr}\left[\mathcal{B}_{\mathcal{M}}(\boldsymbol{k})^\dagger \partial_{k_i} \mathcal{B}_{\mathcal{M}}(\boldsymbol{k})\right]$. Since $(\mathcal{B}_{\mathcal{M}})_{ij}$ is unitary, we can rewrite

$$\mathrm{Tr}\left[\mathcal{B}_{\mathcal{M}}(\boldsymbol{k})^\dagger \partial_{k_i} \mathcal{B}_{\mathcal{M}}(\boldsymbol{k})\right] = \partial_{k_i} \log(\det\left[(\mathcal{B}_{\mathcal{M}})_{ij}(\boldsymbol{k})\right]) \tag{7.16}$$

Using the fact $\partial_{k_i} \log(\det\left[(\mathcal{B}_{\mathcal{M}})_{ij}(k_1,0)\right]) = \partial_{k_i} \log(\det\left[(\mathcal{B}_{\mathcal{M}})_{ij}(-k_1,0)\right])$, we can now calculate the winding number of the sewing matrix along direction $k_1$ at $k_2 = 0$

$$\implies \chi_1 = \frac{i}{2\pi}\int_{-\pi}^{\pi} dk_1 \partial_{k_1} \log(\det[\mathcal{B}_{\mathcal{M}}(k_1,0)]) \tag{7.17}$$
$$= \frac{i}{\pi}\int_{0}^{\pi} dk_1 \partial_{k_1} \log(\det[\mathcal{B}_{\mathcal{M}}(k_1,0)]) \tag{7.18}$$
$$= \frac{i}{\pi}\log\left(\frac{\det(\mathcal{B}_{\mathcal{M}}(\pi,0))}{\det(\mathcal{B}_{\mathcal{M}}(0,0))}\right) \tag{7.19}$$
$$= \frac{i}{\pi}\ln\prod_{n\in\mathrm{occ}}\frac{\eta_n(\mathrm{X}_1)}{\eta_n(\Gamma)} \tag{7.20}$$

Last step is trivial when one notices that the eigenvalue of $\mathcal{M}$ operator is $\pm 1$ and thus $\det(\mathcal{B}_{\mathcal{M}}(\boldsymbol{k})) = \prod_{j\in occ.}\eta_j(\boldsymbol{k})$. Finally as $\chi_1$ is constant under $k_2$ over a smooth gauge[136], Equation 7.20 is true $\forall k_2$ can be rewritten to get us back to equation in the main paper given by

$$(-1)^{\chi_i} = \prod_{j\in occ.}\frac{\eta_j(\mathrm{X_i})}{\eta_j(\Gamma)} \tag{7.21}$$

The same procedure can be used to get $\chi_2$. As noted in Ref.[136] this can be seen intuitively too as follows. In the presence of Mirror/$C_2$ symmetry, we have $\mathcal{M}\mathcal{H}(k_x,k_y)\mathcal{M} = \mathcal{H}(-k_x,k_y)$. It is easy to check that at TRIM points both inversion and $\mathcal{M}$ symmetry maps the points back to themselves and thus they both make a good quantum number for the system at TRIM points[7] . This means that we have $[\mathcal{H}(\boldsymbol{k_i}),\mathcal{M}/\mathcal{I}] = 0$ $\forall \boldsymbol{k_i} \in$ TRIM, thus as described in the previous sections, when the Dirac points meet at one of these TRIM points, they exchange their parity eigenvalue creating an inversion or lack thereof. This entire proof, though was specific for 2D systems, can be generalized for 3D systems and indeed gives more interesting results as one can have 2 kinds of symmetries that causes protected semimetal

7: Both $\mathcal{M}$ and $I$ leaves the TRIM point invariant

1. Line-invariant symmetries i.e. symmetry that leaves a line in k space invariant ($C_2$)
   (or)
2. plane invariant symmetries (ex. Mirror) and thus one could have nodal loops closing in and Dirac/Weyl semimetals annihilating. Further work to study these systems are in need.

## 7.4 Phase Transitions and Entanglement entropy

The most distinctive difference in the phases described above are the idea of *dimerization* which can be probed using entanglement. Entanglement has been used to understand topological phases of system including QSHI[137], chern insulators [138] and topological insulators[139–141] and recently even for higher-order topological systems [142] and topological phase transitions[143, 144] in such systems. This is done by calculating the entanglement spectrum (ES) of the system.

Given a ground state wavefunction of a system $\Psi_{GS}$, one can spatially separate it into two parts $A$ and $B$. Then the entanglement spectrum is the eigenspectrum of the the correlation matrix

$$C_{ij}^{A} = \mathrm{Tr}\left(\rho_A c_i^{\dagger} c_j\right) \tag{7.22}$$

where $c_i/c_i^{\dagger}$ are the annihilation/creation operator for sites $\in A$. The entanglement eigenspectrum $\xi_i$ of $\hat{C}$, for our system, is going to be dependent on $k_{\perp}$ for a cut along $k_{\parallel}$. In the thermodynamic limit, most of the eigenvalues lie exponentially close to 0 and 1[145] while a value of $\frac{1}{2}$ captures the physics of the maximally entangled zero mode. Using $\xi_i(k_{\perp})$, one can define the entanglement entropy (EE) as

$$\mathcal{S} = -\int_{BZ} dk_{\perp} \sum_i [\xi_{i,k_{\perp}} \log \xi_{i,k_{\perp}} + \left(1 - \xi_{i,k_{\perp}}\right) \log\left(1 - \xi_{i,k_{\perp}}\right)]$$

$\mathcal{S}$ measures the mean degree of entanglement between the given spatial split with value ranging from 0 to 1 with 1 being maximally entangled. One can see from the co-dimension=1 spectrum that the phase **II**, **IV** have contribution to entangled states for some values of $k_{\perp}$, while **III** is maximally entangled throughout the BZ. From Equation 7.23, this results in

$$\mathcal{S}_{\alpha} = \begin{cases} 0 & \alpha = \mathbf{I} \\ 0 < S_{\alpha} < 1 & \alpha \in \{\mathbf{II}, \mathbf{IV}\} \\ 1 & \alpha = \mathbf{III} \end{cases} \tag{7.23}$$

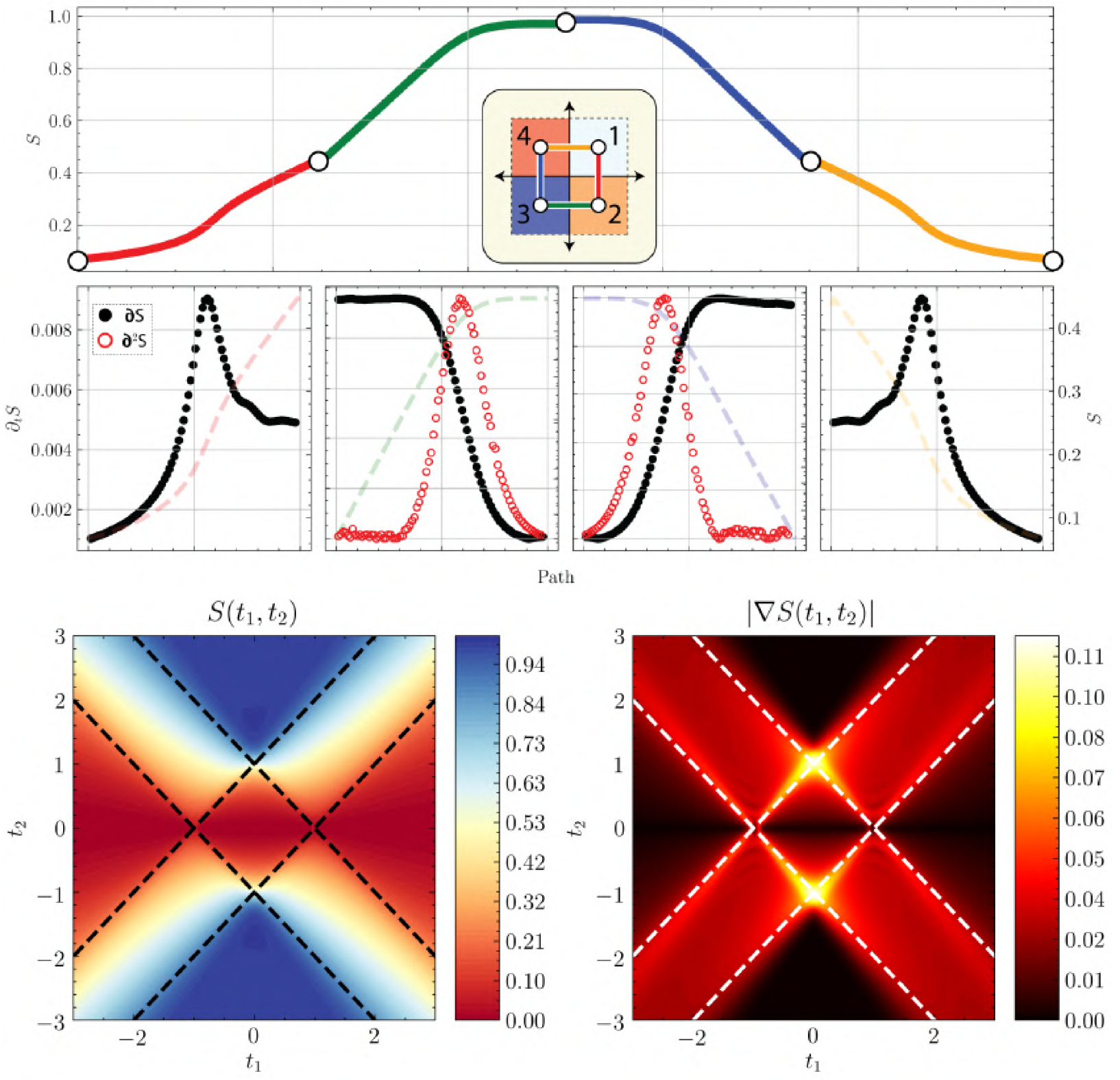

Figure 7.6: *(top)* Entanglement entropy $S$ through the path sown in the inset for *Strained* graphene *(middle)* $\partial S$ (black) and $\partial^2 S$ (red) for the corresponding path in top pannel ($S$ is shown in background for clarity). *(bottom)* $S$ and $\partial S$ throughout the parameter space for the strained graphene, black and white lines in left and right panel shows the transition points from semi-metal to insulator in bulk.

We choose anisotropic graphene with the 3 nearest neighbor interactions given by $t_1, t_2, t_3$ as an example to illustrate $\mathcal{S}$ (details given in [146]). Figure 7.6 *(top)* shows the numerically calculated EE with parameters traversing the space between phases. Red and yellow paths correspond to transitions from insulating to semi-metallic phase in bulk. End states where chosen such that the Dirac cones end up mid way in the BZ, thus leading to surface state spanning half of the lower dimensional system. This reflects in the EE being exactly one half at the end point of the path. Green and blue paths represent transitions from the semi-metalic to the insulating phase. This insulating phase has a ground state wave function with a phase discontinuity. This final insulating phase corresponds to the system where the surface state is present throughout the lower dimensional k space, and is thus maximally entangled, giving us EE of 1. It is interesting to calculate the change in entropy as one traverses the paths. The bottom part of the figure shows $\partial S$ in black indicating that the entropy has a first order divergence when it transits from the insulating phase with a non singular ground state to a semi-metallic state. These kind of divergences are also observed in Ref.[147]. This behavior is also observed with a path connecting **III** and **I** traversing through the origin. Contrastingly, if we consider the other two transitions, they show divergence not in their first order, but rather the second order derivative $\partial^2 S$, as shown by the red curve. The lower part of the figure shows $\mathcal{S}$ and $|\nabla\mathcal{S}|$ calculated throughout the space $t_1, t_2$ with $t_3 = 1$. Again, one can see the first order divergence when crossing from **II**,**III**,**IV** (yellow,blue,yellow in $\mathcal{S}$) to **I** (red)

## Examples

**1D Example:** We start by looking at the 1D SSH system given by

$$\mathcal{H}_{SSH} = (t_1 + t_2\cos(k))\,\sigma_x + t_2\sin(k)\sigma_y \tag{7.24}$$

This system, at half filling is metallic for $t_1 - t_2 = \delta = 0$ with a Dirac dispersion and has an insulating phase on either side of the $\delta$. Based on the sign of $\delta$ the system can either be in OAL or not and this choice precisely describes if the system *dimerizes* within or outside the unit-cell. This system thus undergoes a phase transition from **I** to **III** crossing the origin at $\delta = 0$. This can be seen by expanding $\mathcal{H}_{SSH}$ near the vicinity of the Dirac point, where it reduces to $(t_1 - t_2)\sigma_x + t_2 k\sigma_y$ which is nothing but the projection of Equation 7.1 onto the $k_y$ axis.

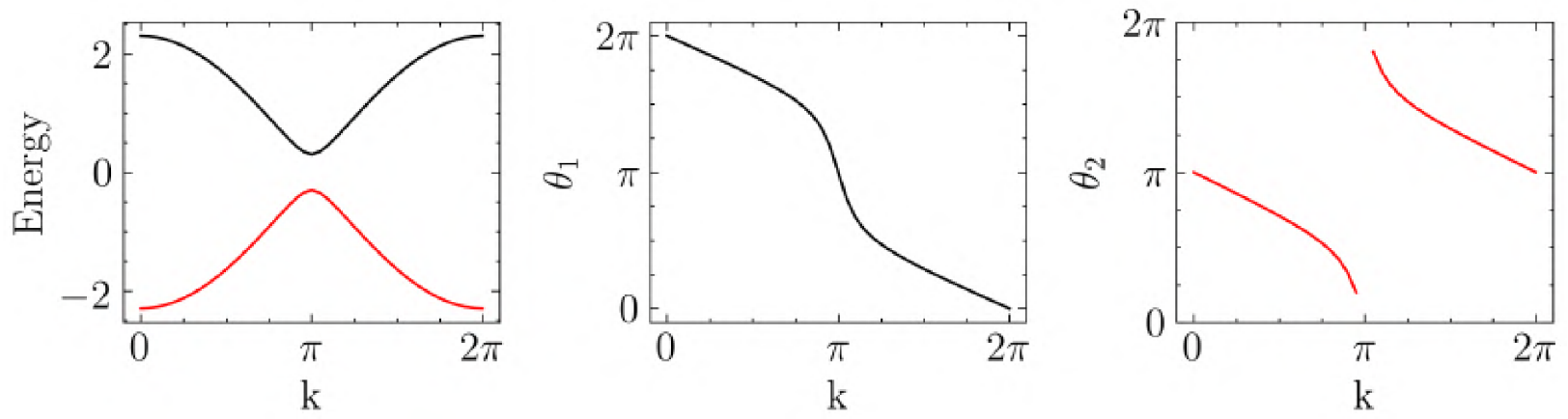

**Figure 7.7:** Eigen-value of SSH and the respective complex phase ($\theta$ in precious sections) of the wave function (red lower band, black top band) in the non-trivial regime (OAL)

Figure 7.7 shows the band structure of the system in the non-trivial regime where, due to branch point in the periodic k space, one gets a non-trivial Zak phase. One can easily see the evolution of the phase from non-trivial to trivial phase in Figure 7.8(left). The right part of the figure shows the evolution of the 2D version of annihilating Dirac cones from the semimetallic phase to the non-trivial phase for comparison.

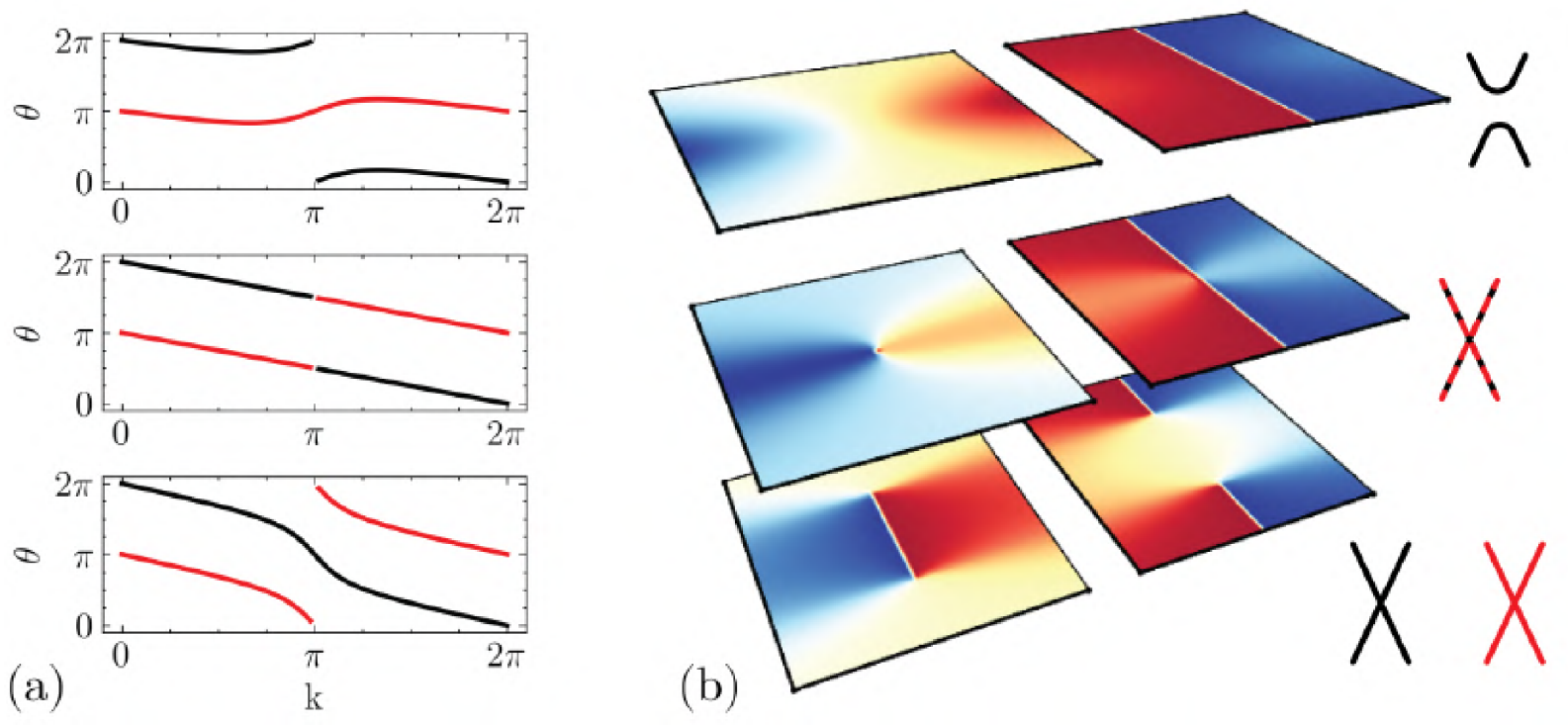

**Figure 7.8:** (left) Evolution of phase of SSH bands from trivial to non-trivial regime (right) compared to the 2 dimensional case of the same as discussed earlier. In left figure, red and black indicate the two wave functions. In the right figure left and right planes indicate the two wave function while the color indicates maximum and minimum values.

The can also be seen by computing the Entanglement entropy, where one sees a first order divergence with $\delta$ being the order parameter as shown in Figure 7.9. This is similar to the transition seen in main text from phase **I** to **III**.

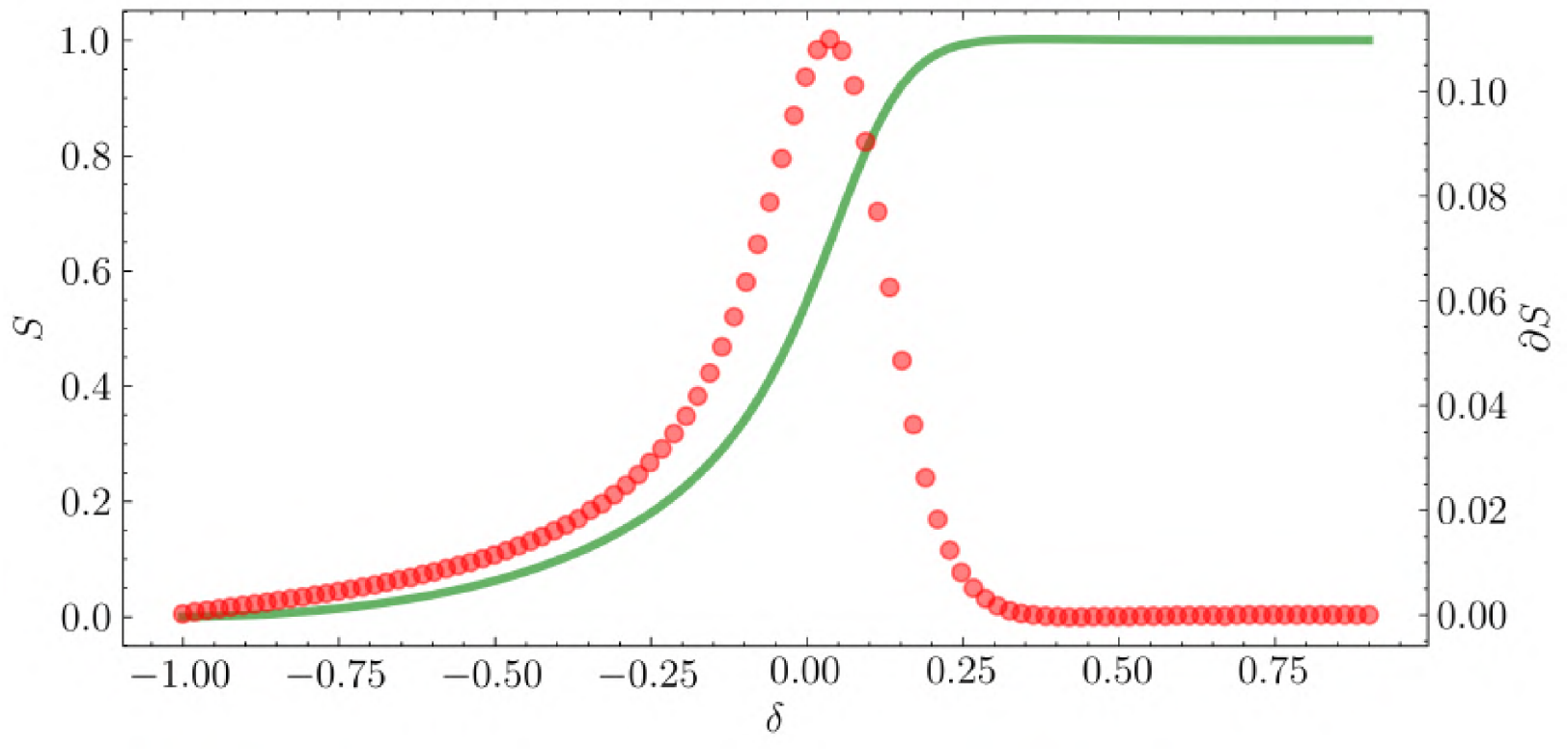

**Figure 7.9:** $S$ (green) and $\partial_\delta S$ (red) for $\mathscr{H}_{SSH}$

Though this is not an example of *moving* annihilation where the Dirac fermions move in k-space as a function of tuning parameters, this annihilation happens by mixing Dirac fermions from the $1^{st}$ BZ and $2^{nd}$ BZ by doubling a single atom unit cell and folding the BZ. This indeed happens in real systems, for example graphene with Kekule distorted interactions involve folding the $3^{rd}$ BZ which essentially puts the Dirac cones at K and K′ together at Γ in the new smaller BZ. Thus any small anisotropy in interaction opens up interaction between the two Dirac cones and gaps the system. We have already explored this physics in Chapter 5 (($d-2$) Higher Order Topology of buckled Group - V) on page 48.

## 2D Examples

**CCSSH:** Moving to 2D, apart from the example discussed in the main section with anisotropic graphene (details of which follow), we introduce yet another example where we can move the Dirac points. It is a simple extension of SSH that we call $CCSSH$(criss-cross SSH) and is shown in Figure 7.10 (left).

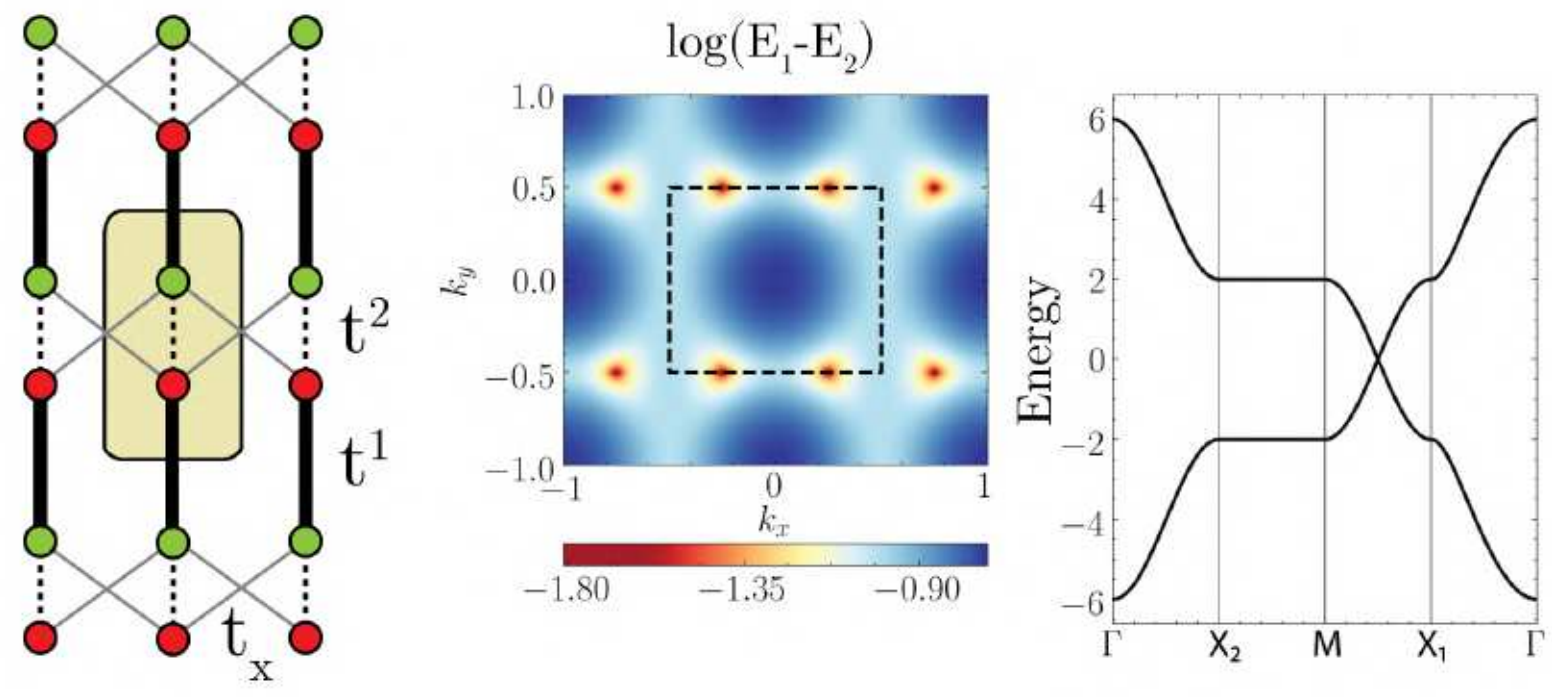

**Figure 7.10:** *left to right* Lattice model, $\log(E_1 - E_2)$ in $\boldsymbol{k}$ space of the bulk band and bulk band structure

Because of the bipartite nature of CCSSH have,

$$\mathcal{H}(\mathbf{k}) = h_x(\mathbf{k})\sigma_x + h_y(\mathbf{k})\sigma_y \quad (7.25)$$

$$(7.26)$$

where

$$h_x(\mathbf{k}) = t_1 + t_2\cos(k_y) + 2t_x\cos(k_x) \quad (7.27)$$

$$h_y(\mathbf{k}) = t_2\sin(k_y) \quad (7.28)$$

Equation 7.27, is SSH like along $k_y$ with an added $t_x$ interaction that is patterned *criss-cross*. It is easy to see that the system has Time Reversal Symmetry (TRS) giving us $\mathcal{H}(\boldsymbol{k}) = H^*(-\boldsymbol{k})$, has inversion symmetry ($\mathcal{I}$) given by $\sigma_x\mathcal{H}(\boldsymbol{k})\sigma_x = \mathcal{H}(-\boldsymbol{k})$ and Chiral symmetry ($\mathcal{C}$) as $\sigma_z\mathcal{H}(\boldsymbol{k})\sigma_z = -\mathcal{H}(\boldsymbol{k})$. Finally, the system also has a Mirror/$C_2$ ($\mathcal{M}$) symmetry that leaves the $k_y = 0$ line invariant and is given by $\sigma_x\mathcal{H}(k_x, k_y)\sigma_x = \mathcal{H}(k_x, -ky)$.

Figure 7.10 shows the band structure for values of the parameters $t_i$ chosen ad hoc to represent various phases mentioned in the main text (values of which are given in Table 7.1), along with $\log(E_1 - E_2)$ to show the occurrence of unpinned Dirac points in the BZ. As always because of Time Reversal Symmetry (TRS), we are guaranteed a Dirac cone of opposite winding number at $-\boldsymbol{k}$ for each $\boldsymbol{k}$. This system can be tuned to realize various phases mentioned in the text as shown in Figure 7.11 with $t$'s chosen to provide an example of the discussed phases (values in SM). We show the bulk band structure with the Dirac cone and the gapped system along with the 1D band structure.

Phases **II** and **IV** represent the protected semi metallic regime, where the branch cut occurs either inside(**IV**) or outside(**II**) the Dirac cones. They differ either by *(i)* sign flip of $t_x$ or *(ii)* by the choice of the bulk unit cell used to make the boundary, just as in the case of SSH. Phases **I** and **III** which are formed upon annihilating the above Dirac points, form the OAL limit. This can be seen from their respective surface bands. These two phases can again be tuned by either the relative size of $|t_1|$ and $|t_2|$ or the choice of bulk unit cell. This same effect can be deduced from their respective ground state wavefunction's phases and calculating the number of branch cuts along the way using $\gamma^x(k_y)$. The wavefunction phase of each of the phases in the phase-diagram are shown along side in figure where one can see the branch cuts. This brings us to the important fact that, though the 0 co-dimensional bulk ground states ($\Psi^0_{GS}$) of the insulating/metallic phases are related by a phase/unit cell shift, the corresponding co-dimension= $n$ ground state ($\Psi^n_{GS}$) clearly distinguishes them.

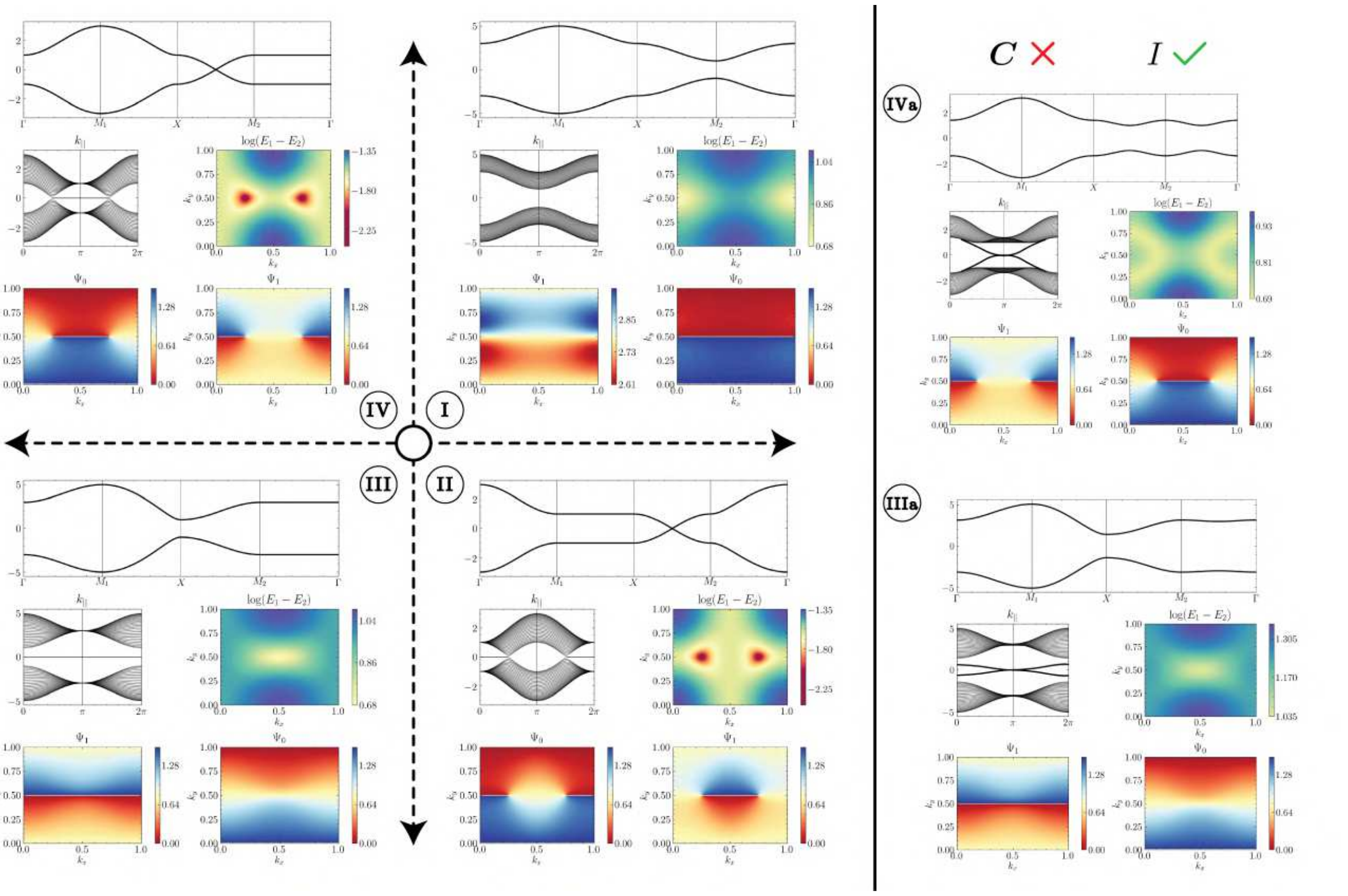

**Figure 7.11:** *(left)* Bulk Band structure, Edge spectrum, $\log(E_2 - E_1)$, and phase of $\Psi_1$ and $\Psi_2$ for phase **I,II,III,IV** given in text with parameters shown in Table 7.1. *(right)* phase **IV** $\rightarrow$ **IVa** and **III** $\rightarrow$ **IIIa** upon breaking Chiral symmetry but preserving inversion.

Now let us look at this as a concrete example of the symmetry indicators given in Equation 7.6. Using our inversion matrix given by $\sigma_x$, we can get the inversion eigenvalue of the half-filled system at the TRIM points by computing $\text{sgn}[t_1 + t_2\cos(k_y) + 2t_x\cos(k_x)]$ for $(k_x, k_y) = (\Gamma, X_1, M, X_2) = (0,0), (0,\pi), (\pi,\pi), (\pi,0)$. The values are shown in Table 7.1 along with the connected network as described in above section where points of opposite values are connected. It is easy to see that the net invariant of finding the semi-metal/insulator is just done by merely counting the number of lines in network being even or odd while the other two phases are determined by cross-connection. From the network graphs it is quite easy to see that rotating the system $\pi/2$ to place the non-trivial interaction along the other axis is given by , and . To make the transition clearer, we show in Figure 7.12 the dynamics of parity values close to trivial and non-trivial transitions (Note the linear and quadratic dispersion at the transition point in the middle panels).

Next, we elucidate the role of chiral symmetry in the system. All the examples mentioned above respect chiral symmetry with $\sigma_z$ being the chiral operator. This leads to the surfaces states formed being degenerate throughout the co-dimension $\geq 1$ system as seen in Figure 7.11. As it turns out, this is not a necessary condition for the surface protection. This can be easily seen by the fact that our invariants (and hence branch cuts) are unchanged with addition of a chiral breaking $\sigma_z$ term and solely depends on $\sigma_x$. To show this we explicitly break the chiral symmetry by adding term such as $2t_c\cos(k_x)\sigma_z$. Figure 7.11 (right) shows the surface spectrum where we have broken the chiral symmetry. It is easy to see that this does not nullify the existence of edge state, but rather breaks the degeneracy of edge state that is guaranteed by chiral symmetry. This breaks it everywhere except at the point $k = \pi(0)$ in case of odd (even) chiral term as the term goes back to 0 giving back locally the chiral symmetry. Thus this leads to an interesting strong topological crystalline insulator state protected by inversion symmetry. This model is extremely similar to the example proposed by Ref.[117].

**Table 7.1:** Parameters used in Figure 7.11 along with $\eta(\boldsymbol{k})\ \forall \boldsymbol{k} \in$ TRIM

| Phase | $t_x$ | $t_1$ | $t_2$ | $t_c$ | $Inversioneigenvalue(\Gamma, X_1, M, X_2)$ | Graph |
|---|---|---|---|---|---|---|
| **I** | -0.5 | 3.0 | 1.0 | 0 | (+, +, +, +) | 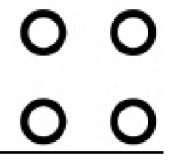 |
| **II** | 0.5 | 1.0 | 1.0 | 0 | (+, +, -, +) | 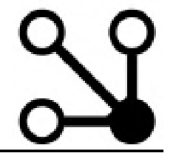 |
| **III** | -0.5 | 1.0 | 3.0 | 0 | (+, -, -, +) | 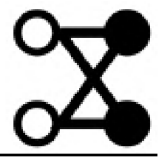 |
| **IV** | -0.5 | 1.0 | 1.0 | 0 | (+, -, +, +) | 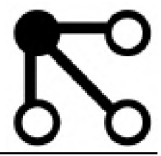 |
| **IIIa** | -0.5 | 1.0 | 3.0 | 0.5 | (+, -, -, +) | 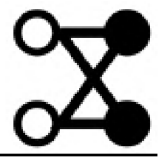 |
| **IVa** | -0.5 | 1.0 | 1.0 | 0.5 | (+, -, +, +) | 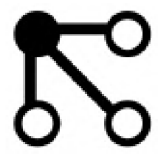 |

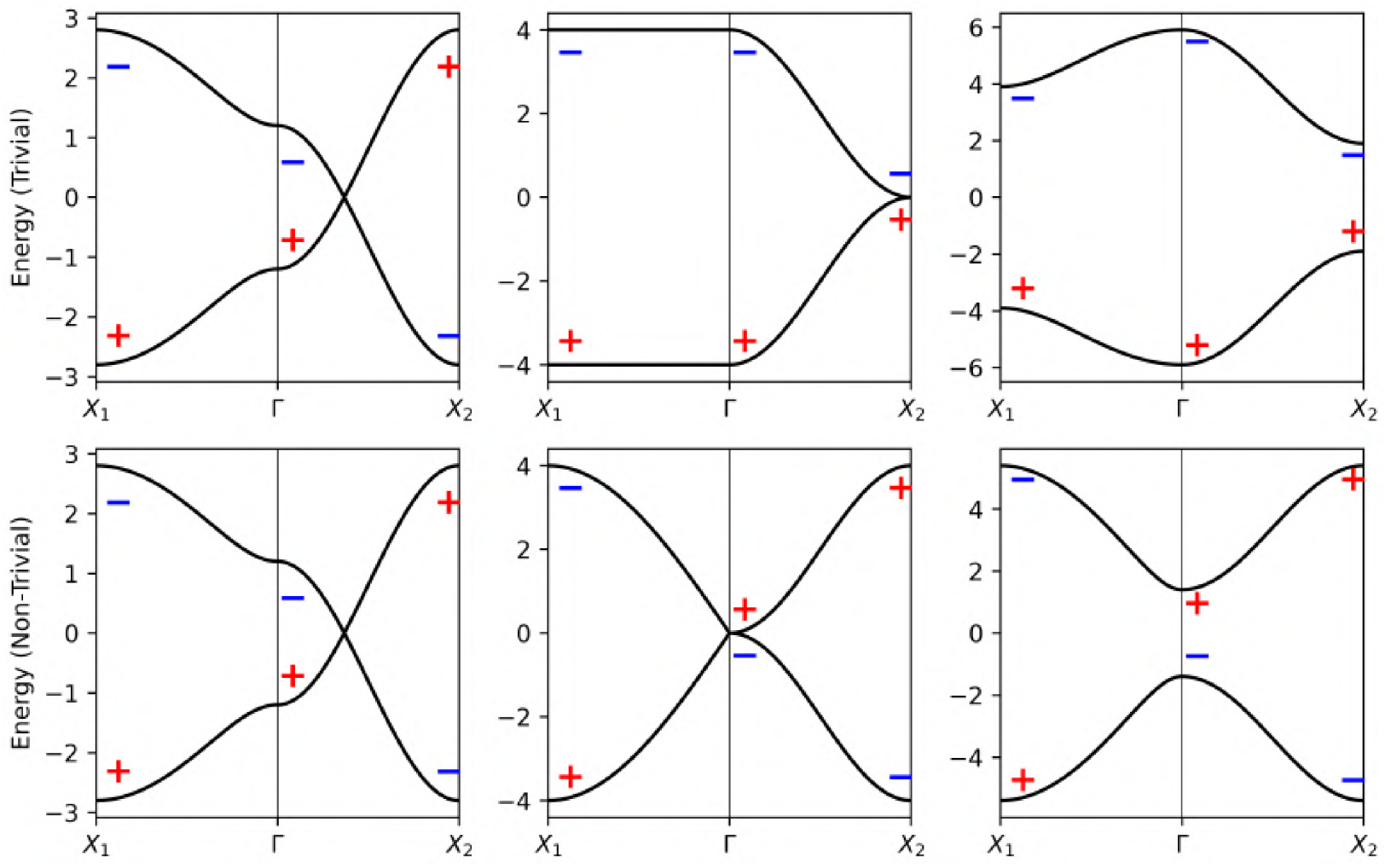

**Figure 7.12:** Transition to OAL by annihilating Dirac points. + and − represents the parity Eigen values*(top)* trivial phase (**I**) where the system annihilates the points at Γ *(bottom)* Non-trivial phase (**I**) where the system annihilates the points at $X_2$

**Anisotropic graphene:** This system is shown in Figure 7.13(a) and the Hamiltonian is given by $\mathcal{H} = h_x(\boldsymbol{k})\sigma_x + h_y(\boldsymbol{k})\sigma_y$ with

$$h_x(\mathbf{k}) = -\sum_{i=1}^{3} t_i \cos(\delta_i \cdot \boldsymbol{k}) \tag{7.29a}$$

$$h_y(\mathbf{k}) = -\sum_{i=1}^{3} t_i \sin(\delta_i \cdot \boldsymbol{k}) \tag{7.29b}$$

where $\delta_1 = \left(\frac{\sqrt{3}}{3}, 0\right), \delta_2 = \left(-\frac{\sqrt{3}}{6}, \frac{1}{2}\right), \delta_3 = \left(-\frac{\sqrt{3}}{6}, -\frac{1}{2}\right)$ and $t_i$ the anisotropic interaction parameters. The eigenspectrum of this system is given by

$$E_{\pm} = \pm\sqrt{h_x(\boldsymbol{k})^2 + h_y(\boldsymbol{k})^2} \tag{7.30}$$

It can be shown[148] that for $t_3 = 1$ when the condition

$$||t_1| - 1| \leq |t_2| \leq |t_1| + 1 \tag{7.31}$$

is satisfied, one is guaranteed a Dirac points at $\pm\boldsymbol{k}$ where $\pm\boldsymbol{k}$ is the solution of

$$\begin{aligned} \cos(\mathbf{a}_1 \cdot \mathbf{k}) &= \frac{1 - t_1^2 - t_2^2}{2t_1 t_2} \\ \cos(\mathbf{a}_2 \cdot \mathbf{k}) &= \frac{t_2^2 - t_1^2 - 1}{2t_1} \\ \cos((\mathbf{a}_1 - \mathbf{a}_2) \cdot \mathbf{k}) &= \frac{t_1^2 - t_2^2 - 1}{2t_2} \end{aligned} \tag{7.32}$$

where $\mathbf{a}_1, \mathbf{a}_2$ are the lattice vectors.

Figure 7.13 shows the band structure of the finite dimensional model afor different phases, determined by the choice of parameters given in Table 7.2

| Phase | $t_1$ | $t_2$ | $t_3$ |
|---|---|---|---|
| I | 1 | 1 | 3 |
| II | -1 | 1 | 1 |
| III | 1 | 3 | -1 |
| IV | 1 | 3 | -1 |

**Table 7.2:** Parameters used in Figure 7.13

In the above section, path chosen for Entanglement Entropy is an interpolation between the lines connecting these parameter values in hyper-parameter space. Once again, one can easily verify the validity of the symmetry indicator by calculating the inversion eigenvalue at $\Gamma, X_1, M, X_2$ marked in the figure. Similarly, annihilation transition in a honeycomb lattice made up of $p_x - p_y - p_z$ is observed when one buckles the honeycomb lattice by moving from $D_{6h} to D_{3d}$ where one still gets movable Dirac points preserved by $C_2$ symmetry which has two annihilation transitions at $M$ and $\Gamma$ leading to the phase **III**[101]

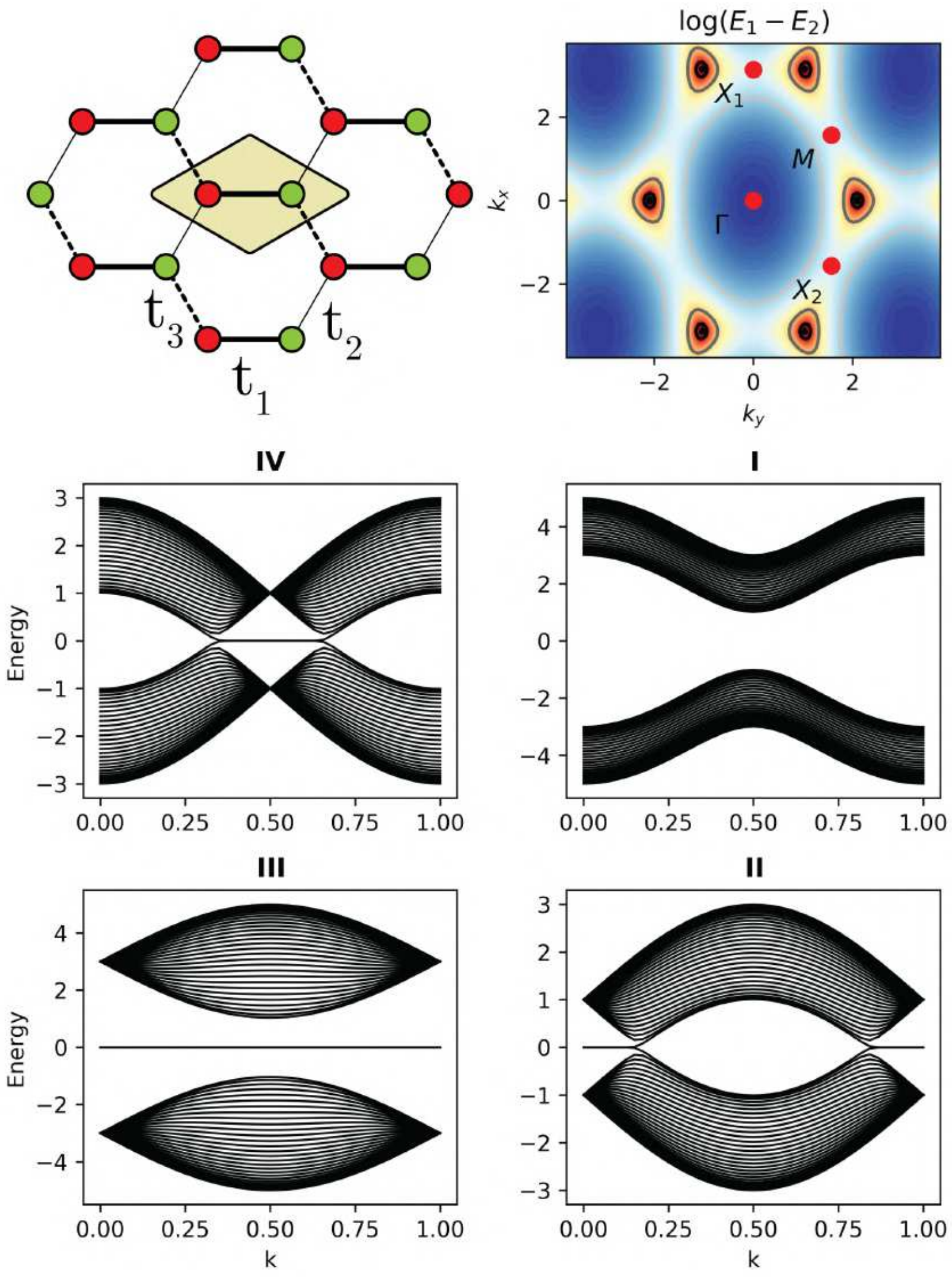

Figure 7.13: Anisotropic graphene lattice, $\log(E - 1 - E_2)$ at $t_1 = t_2 = t_3 = 1$, edge spectrum at $t$ values given in Table 7.2

## 7.5 Summary

We have shown that the annihilation of Dirac fermions, which follows a universal Hamiltonian leads to non-trivially gapped system. The annihilation dynamics has a rich phase diagram with phases distinguishable by the path dependent Zak phase. We showed that two of the phases correspond to the Obstructed Atomic Limit insulators while the other two are semimetallic. These phases have vastly different ground state wave functions in the corresponding co-dimension $\geq 1$ system. We showed the effect of transitions between these phases on their entanglement entropy. These transitions can be realized in realistic systems like buckled Sb/As, Strained graphene as well in artificial systems. Moreover, since this is a universal property of annihilation, this should in principle also be applicable for more exotic 2D Dirac systems like d-wave superconductors[149], though ideas extending atomic insulators to BdG systems needs to be extended.

# 2-D spin-polarized electron/hole gas in $LiCoO_2$

# 8

**Abstract**

First-principles calculations show the formation of a 2D spin polarized electron (hole) gas on the Li ($CoO_2$) terminated surfaces of finite slabs down to a monolayer of by Li donating its electron to the $CoO_2$ layer forming a Co-$d - t_{2g}^6$ insulator. By mapping the first-principles computational results to a minimal tight-binding models corresponding to a non-chiral 3D generalization of the quadripartite Su-Schriefer-Heeger (SSH4) model, we show that these surface states have topological origin.

## 8.1 Introduction

$LiCoO_2$ has been mostly studied as cathode material in Li-ion batteries.[150–152] However, its layered structure also lends itself to the possibility of extracting interesting ultrathin mono- or few layers nanoflakes. A chemical exfoliation procedure has recently been established by Pachuta *et al.* [50] and similar exfoliation studies have also been done on $Na_xCoO_2$.[153] The $R\bar{3}m$ structure of $LiCoO_2$ consists of alternating $CoO_2$ layers, which consist of edge sharing $CoO_6$ octahedra, and Li layers stacked in an ABC stacking. By replacing lithium by large organic ions, the distance between the layers swells and they can then be exfoliated in solution and redeposited on a substrate of choice by precipitation with different salts.

Inspired by these experiments we investigated the electronic structure of $LiCoO_2$ few layer systems with various Li and other ion terminations and as function of thickness of the layers using density functional theory (DFT) calculations.[**questaal-paper**, **compdetails**] We found, surprisingly that Li no longer fully donates its electron to the $CoO_2$ layer but instead a surface state appears above the Li and is occupied with a fraction of an electron per Li.

The Li bands in bulk $LiCoO_2$ lie at energies $E > 5$ eV above the Fermi level, consistent with the mostly ionic charge donation picture mentioned above. So, the fact that a Li related surface state comes down sufficiently close to the Fermi level to become partially occupied is truly surprising. Furthermore because it is accompanied by the opposite surface $CoO_2$ becoming spin-polarized it leads actually to a spin-polarized electron gas on the Li side which is located primarily above the Li atoms. The main question we address in this paper is: why does this happen? The answer is that this is a topological effect. We show that the DFT calculations can be explained by a minimal tight-binding (TB) model, closely related to the quadripartite Su-Schrieffer-Heeger (SSH4) model which has been shown to support topologically protected surface states for specific conditions on the interatomic hopping integrals.[154] However, while the original SSH4 model has chiral symmetry protecting the surface state at zero energy, in the present case, the $Li/CoO_2$ electronegativity difference leads to surface states which would tend to still place the surface electrons on the $CoO_2$ side. The crucial element that allows the Li surface state to become partially filled is the strong lateral interaction between Li atoms on the surface. The resulting band broadening leads the Li surface band to dip below the top of the $CoO_2$ localized surface band leading to a partial electron/hole occupation in these bands respectively. As a further proof of the importance of the lateral interaction, we find that when we place 1/2 Li per cell on opposite sides of the slab, whereby the Li occur along 1D rows, the Li surface band has then only 1/3 of the band width and no longer dips below the Fermi level.

## 8.2 Density Functional Results

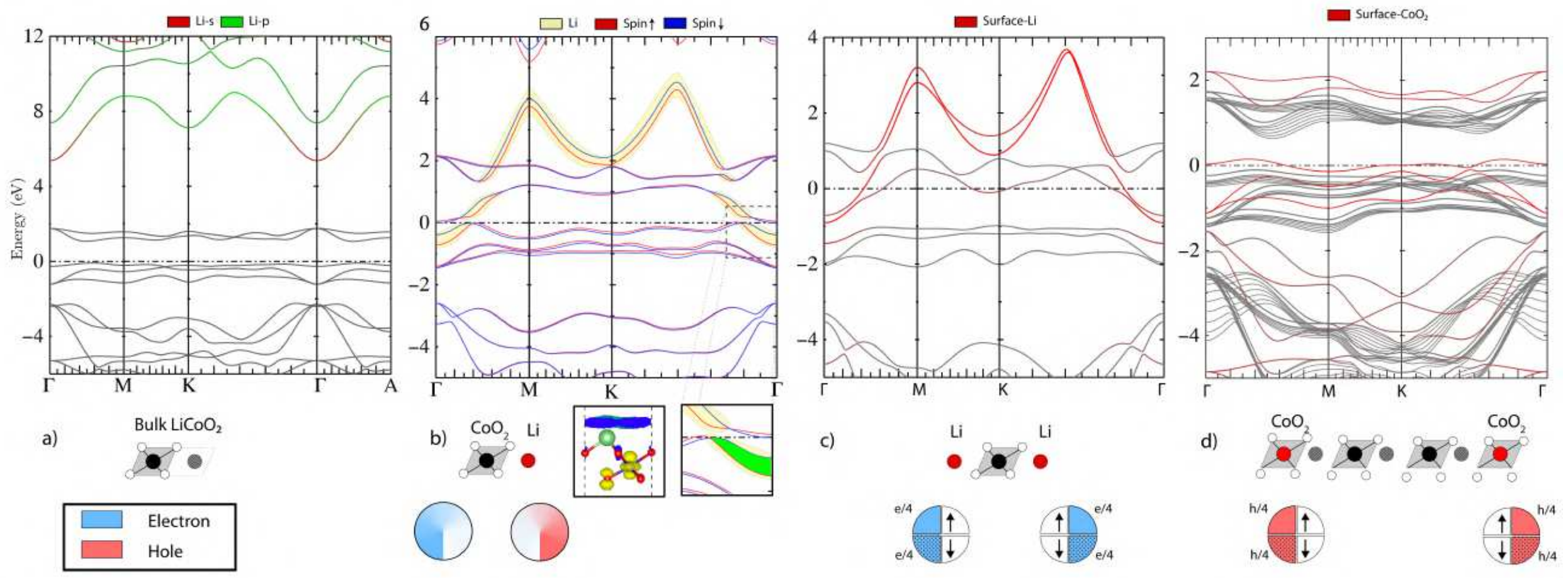

**Figure 8.1:** *(a)* Bulk band structure of $LiCoO_2$, *(b)* Spin polarized 2D $LiCoO_2$ mono-layer. Insets below, right: zoom in near Γ showing spin polarizaton in green shading, left: electron density of occupied surface state, *(c)* symmetric Li terminated $Li_{x+1}(CoO_2)_x$ with $x = 1$, *(d)* symmetric $CoO_2$ terminated $Li_{x-1}(CoO_2)_x$ with $x = 10$ neglecting spin-polarization. Below each panel: structural model and electron occupation of surface states.

We start by comparing the band structure in bulk $LiCoO_2$ with that of a monolayer $LiCoO_2$ in Fig. Figure 8.1 (a) and (b). In the bulk case, we find an insulating band structure with a gap between the filled $t_{2g}$ and $e_g$ Co-$d$ bands . The Li $s$ and $p$ derived bands, highlighted in color occur at high energy indicating that they donate their electron to the Co-$t_{2g}$ orbitals and support a mostly ionic picture of the bonding. In strong contrast, in the monolayer system, Fig. Figure 8.1(b), we find an additional set of spin-polarized bands, as highlighted in the figure by yellow shading. Orbital decomposition shows that this band is Li related and its 2D dispersion closely matches that of a hypothetical 2D monolayer of Li atoms. Importantly, it has not only Li-$s$ but also Li-$p_z$ character indicating the formation of Li-$sp_z$ hybrid states. This free-electron-like band has avoided crossings with the Co-$e_g$ bands around 2 eV. It dips down below the Fermi level with an electron pocket near Γ. Inspection of the corresponding wavefunction modulo squared shown in the inset below it shows clearly that it is a surface state hovering slightly above the Li atom. This band depends somewhat on the location chosen for the Li atom, as shown in Section 8.4 (Density functional theory results) but its general characteristics are robust. In the lowest energy structure it is found to be spin polarized. This results from the high density of states at the Fermi level due to the rather flat $CoO_2$ surface band which becomes partially filled, which may be viewed as inducing a Stoner instability. For such a partially filled $CoO_2$ layer, correlation effects stemming from the strong on-site Coulomb energies, might be important. Calculations at the LSDA+U[155] and quasi-particle self-consisent QS$GW$ levels[**Kotani07**] in SM, show that the results are robust with respect to these more accurate treatments of exchange and correlation. The fact that its maximum density lies above the Li atom is consistent with a $sp_z$ hybrid orbital derived band. Bader analysis[156, 157] shows that the surface Li accumulates about $0.4e$ while a planar average of the electron density integrated over the $z$ region just above the Li atom, indicates that $\sim 0.25e$ resides above the Li atom. The latter corresponds to a rather high density of 5-8×10$^{14}$ $e/\mathrm{cm}^2$. The precise value depends on the cut-off of the spatial region over which we integrate. Further calculations for thicker slabs with 3 and 4 layers with the same termination of one Li terminated and one $CoO_2$ terminated surface give very nearly the same surface charge densities. Orbital decomposition in these thicker slabs shows that a second surface state occurs near the Fermi level and is localized on the opposite $CoO_2$ terminated surface layer and crosses the Li related one. We find that there is a net hole concentration on that layer.

To rule out that this would be a supercell artifact arising from the polar nature of the structure, in which an artificial dipole could arise over the vacuum region, we also consider a symmetrical slab with both surfaces Li terminated. In this case, the system is in some sense overcompensated by having one additional Li. The band structure for this case is shown in Fig. Figure 8.1(c) and has no spin-polarization. A similar surface state is then found on both both Li terminated surfaces and, in fact, one can see that the occupied electron pocket in this band near Γ is larger. For symmetry reasons it must contain a fractionalized $e/4$ for each spin and on each surface, so a net charge of 1/2 electron per Li. Similarly, (Fig. Figure 8.1(d)) symmetric $CoO_2$ termination leads to a surface state on both surface layers with equal hole concentration. This case does prefer a spin-polarized state as shown in Section 8.4 (Density functional theory results)

We also considered a 1/2 Li per Co symmetrically at both terminating surfaces. In that case the Li is placed in 1D rows on the surface and a 1D electron gas is found above these Li rows, as shown in Section 8.4 (Density functional theory results). However, the density of electrons in this case is about 10 times smaller. We further inspect the dilute limit of 1 Li per 4 Co atoms and still find an even smaller residual small charge density in an orbital locally above that Li. However, no Li localized surface state dipping below the Fermi level is found in these cases with a reduced Li surface concentration. This indicates that sufficient lateral interaction between Li atoms is required to generate a significant occupation of the surface states with electrons. Replacing the Li terminating layer by Be (also overcompensating the system from the $CoO_2$ point of view) we find a higher electron density in a Be related surface band. Replacing Li by Na gives similar results but with different band widths of the surface band because of the stronger overlap of the Na orbitals (refer Section 8.4 (Density functional theory results)).

## 8.3 Minimal Topological SSH4 model

To explain these remarkable results, we now consider a minimal tight-binding model. First, it is clear that the Li needs to be represented by two $sp_z$ orbitals pointing toward the $CoO_2$ layer on either side. It is well known that an even number of orbitals is required in a 1D model to obtain topologically non-trivial band structures.Therefore we choose to represent the $CoO_2$ layer by two $s$-like Wannier orbitals. One could think of these as representing the $a_1$-symmetry of the $D_{3d}$ group or $d_{z^2}$ orbitals on Co with $z$ along the layer stacking **c**-axis making bonding orbitals with O-$p$ on either side of the Co. Of course, this does not represent the full set of $CoO_2$ layer derived bands but we will argue that it represents the relevant bands leading to the surface states. The important point is that the $CoO_2$ and Li each are represented by two Wannier type orbitals whose centers are not on the atoms but on the bonds in between atoms in the layer stacking direction.

This minimal model is a non-chiral version of the SSH4 model. Ordering the orbitals as $\{|Li^a\rangle, |CoO_2^a\rangle, |Li^b\rangle, |CoO_2^b\rangle\}$ the Hamiltonian for the above 1D system (with distance between the layers set to 1) is represented by the following $4\times4$ matrix:

$$
\begin{aligned}
H_{1d} &= \begin{pmatrix} \delta & 0 & \tau_1 & \tau_4 e^{ik_z} \\ 0 & -\delta & \tau_2 & \tau_3 \\ \tau_1 & \tau_2 & \delta & 0 \\ \tau_4 e^{-ik_z} & \tau_3 & 0 & -\delta \end{pmatrix} = \begin{pmatrix} \delta\sigma_z & \mathbf{s}^*(k_z) \\ \mathbf{s}(k_z) & \delta\sigma_z \end{pmatrix} \\
&= \sigma_x \otimes \mathbf{h}(k_z) - i\sigma_y \otimes \mathbf{a}(k_z) + \delta \mathbb{1}_2 \otimes \sigma_z
\end{aligned}
\tag{8.1}
$$

where $\tau_1$,$\tau_2$,$\tau_3$, are intra-unit cell interaction while $\tau_4 = \tau_2$ is the out of unit cell interaction, $\delta$ is the ionic on-site term for Li relative to $CoO_2$. The second form of the Hamiltonian focuses on its $2\times2$ block structure, in which $\mathbf{s}(k_z)$ is a 2×2 matrix which is split in its hermitian, $\mathbf{h}(k_z) = \mathbf{h}(k_z)^\dagger$, and anti-hermitian, $\mathbf{a}(k_z)^\dagger = -\mathbf{a}(k_z)$, parts, allowing us to finally write the block structure of the Hamiltonian in terms of the Pauli matrices and a $2\times2$ unit matrix $\mathbb{1}_2 = \sigma_0$. For our system, $\tau_2 = \tau_4 = t^z_{\mathrm{Li-CoO_2}}$ corresponds to the interaction between the Li and $CoO_2$ layers while $\tau_1$=$t^z_{Li} = (E^{\mathrm{Li}}_s - E^{\mathrm{Li}}_p)/2$ corresponds to the interaction between the two Li $sp_z$'s on the same Li atom, and $\tau_3 = t^z_{\mathrm{CoO_2}}$ to O-Co-O interaction within the layer.

This model, which is the SSH4 model for $\delta = 0$ corresponding to chiral symmetry, has been shown [154] to have non-trivial topology which requires zero-energy edge states when $\tau_1\tau_3 < \tau_2\tau_4$. In fact, in that case, the winding number, which characterizes the topology $\mathcal{W} = \oint \frac{dk_z}{2\pi}\partial_{k_z} \arg\det\{s(k_z)\}$ is 1, while in the other case it is 0. This condition in our case, indicates that the covalent Li-$sp_z$–$CoO_2$ interaction is stronger than the intra $CoO_2$ interaction or the Li-$sp_z$ interaction on the same Li atom.

When $\delta$ is not zero this model becomes non-chiral and the zero energy surface states move up and down in energy and become localized on opposite edges which would tend to localize the electrons on one side only, in the present case obviously the $CoO_2$ side because it would have lower energy $-\delta$ because of its electronegative character. Therefore, to explain the electron occupation of the Li-derived surface state, we need to generalize our model to include the lateral in-plane interactions.

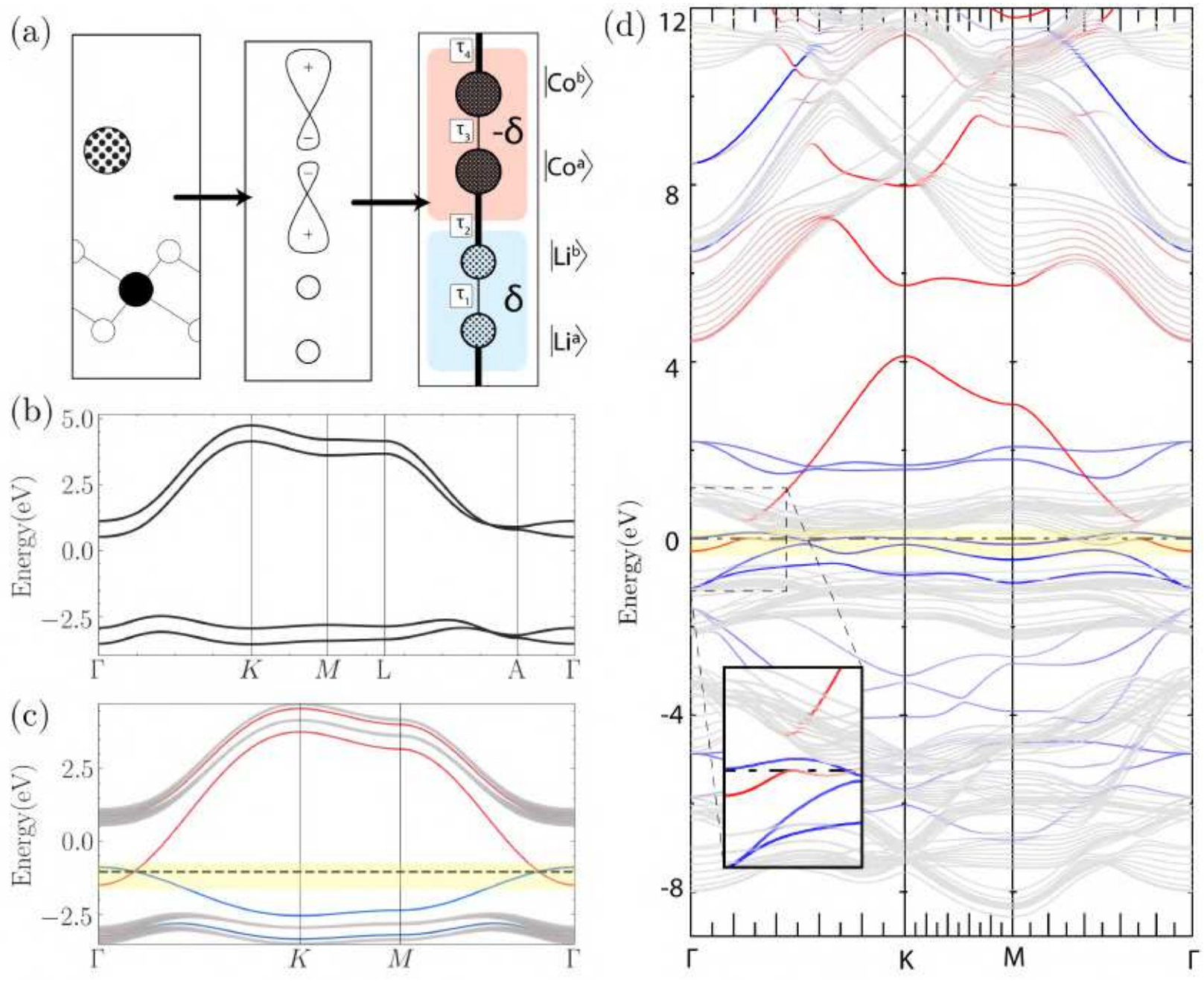

**Figure 8.2:** a) $LiCoO_2$ unit cell, 1D-TB model; b) Bulk band structure of 3D Hamiltonian; c) Band structure of finite slab *red* Li surface, *blue* opposite surface $CoO_2$; d) DFT band structure of 14nm slab. (Inset shows surface states near Fermi level.)

We introduce in-plane $t^{xy}_{\text{Li}}$ and $t^{xy}_{\text{CoO}_2}$ interactions on the planar trigonal lattice and define $f_{Li(Co)} = 2t^{xy}_{Li(Co)} \sum_{i=1}^{3} cos(\boldsymbol{k} \cdot \boldsymbol{\delta_i})$ where $\pm\boldsymbol{\delta}_i$ are the 6 vectors pointing toward the nearest neighbors and $\mathbf{k}_{\|} = k_1\, \mathbf{b}_1 + k_2\, \mathbf{b}_2$ is the in-pane 2D wave vector. This modifies only the diagonal terms in Equation 8.1 leading to

$$H_{3d}(\,\mathbf{k}_{\|}) = \sigma_x \otimes \mathbf{h}(k_z) - i\sigma_y \otimes \mathbf{a}(k_z) + \sigma_0 \otimes [\delta + \Delta(\,\mathbf{k}_{\|})]\sigma_z \tag{8.2}$$

after we drop out a constant from the Hamiltonian diagonal. Here, $\Delta(\,\mathbf{k}_{\|}) = \frac{f_{Li} - f_{Co}}{2}$, is added to the $\delta$ in the 1D model and can be thought of as a dimensional crossover parameter,[158] which tunes the influence of the in-plane dimensions. Physically, $\Delta(\,\mathbf{k}_{\|})$ is proportional to the width of the energy bands in $\mathbf{k}_{\|}$ space, which is $\Gamma = \Gamma_{Li} + \Gamma_{CoO_2} = 6(t^{xy}_{\text{Li}} + t^{xy}_{\text{CoO}_2})$.

Figure 8.2(c) shows the band structure of the above $3D$ Hamiltonian while (d) shows the energy levels of the $2D$ periodic system with a finite number of layers along the $z-$axis with top most layer having Li and bottom layer $CoO_2$. While the 3D periodic TB system is seen to have a wide gap between high-lying Li derived bands and low lying $CoO_2$-derived bands, two surface bands with significant $\mathbf{k}_{\|}$ dispersion are seen in part (c), which are respectively localized on the Li (red) and $CoO_2$ (blue) sides and are found to cross each other near the Fermi energy.

The interlayer parameters used in the TB Hamiltonian are chosen to satisfy the SSH4 non-triviality criterion. The in plane interactions determining $\Delta(\,\mathbf{k}_{\|})$ are chosen to resemble the DFT band structure (shown in part (d) for a 14 nm (30 layers) thick in-plane periodic layer) with $t^{xy}_{\text{CoO}_2} \approx -0.1 t^{xy}_{\text{Li}}$. The opposite dispersion of these surface bands is obvious from the DFT results and translates to these in-plane interactions having opposite sign. In the actual system, it is clear that $t^{xy}_{\text{Li}} < 0$ as it is a $\pi$-interaction between Li-$p_z$ states combined with $\sigma$-interaction between the $s$-part of the $sp_z$ orbitals and in absolute value is much larger than the $t^{xy}_{CoO_2} > 0$. This is important because it means that the Fermi level is pinned at the intersection of the two surface bands, which indicate an overall semimetallic case with as many holes in the $CoO_2$ surface band as there are electrons in the Li surface band when they overlap.

While we recognize that this model is representing only part of the bands of the actual physical system, the correspondence of the surface bands in the DFT and the TB-model Hamiltonian convincingly captures the essence of the relevant physics. We thus conclude that the surface states originate from the topologically nontrivial SSH4 character of the interlayer bonds which correspond to a non-zero winding number. To this is added a term in the Hamiltonian orthogonal to the space of the $\sigma_x$, $\sigma_y$ parts of the Hamiltonian which define a complex plane in which the winding number is defined. One can generally write such a Hamiltonian as $H = \mathbf{d} \cdot \boldsymbol{\sigma}$ where $d_{\|}$ corresponds to the $(x, y)$ components in a complex plane defining the winding number.[159, 160] The loop defining the winding number is now above or below the complex plane. Its projection on the complex plane encircling the origin or not defines the winding number and therefore topological non-trivial/trivial character. The component $d_\perp$ to this plane determines the energy position of the surface states $E_s = \pm|d_\perp|$,[160] away from zero by the chiral symmetry breaking. In the bipartite SSH model, the third component of the $\boldsymbol{\sigma}$ would be simply the third Pauli matrix $\sigma_z$ while in our case it is $\sigma_0 \otimes \sigma_z$. However, because of the in-plane band dispersion, such an SSH4 like model now applies at each $\mathbf{k}_{\|}$. This turns the loop outside the plane into a cylinder (shown in Fig. Figure 8.3(b) centered at energy $2\delta$ from the plane with height given by the 2D band width.

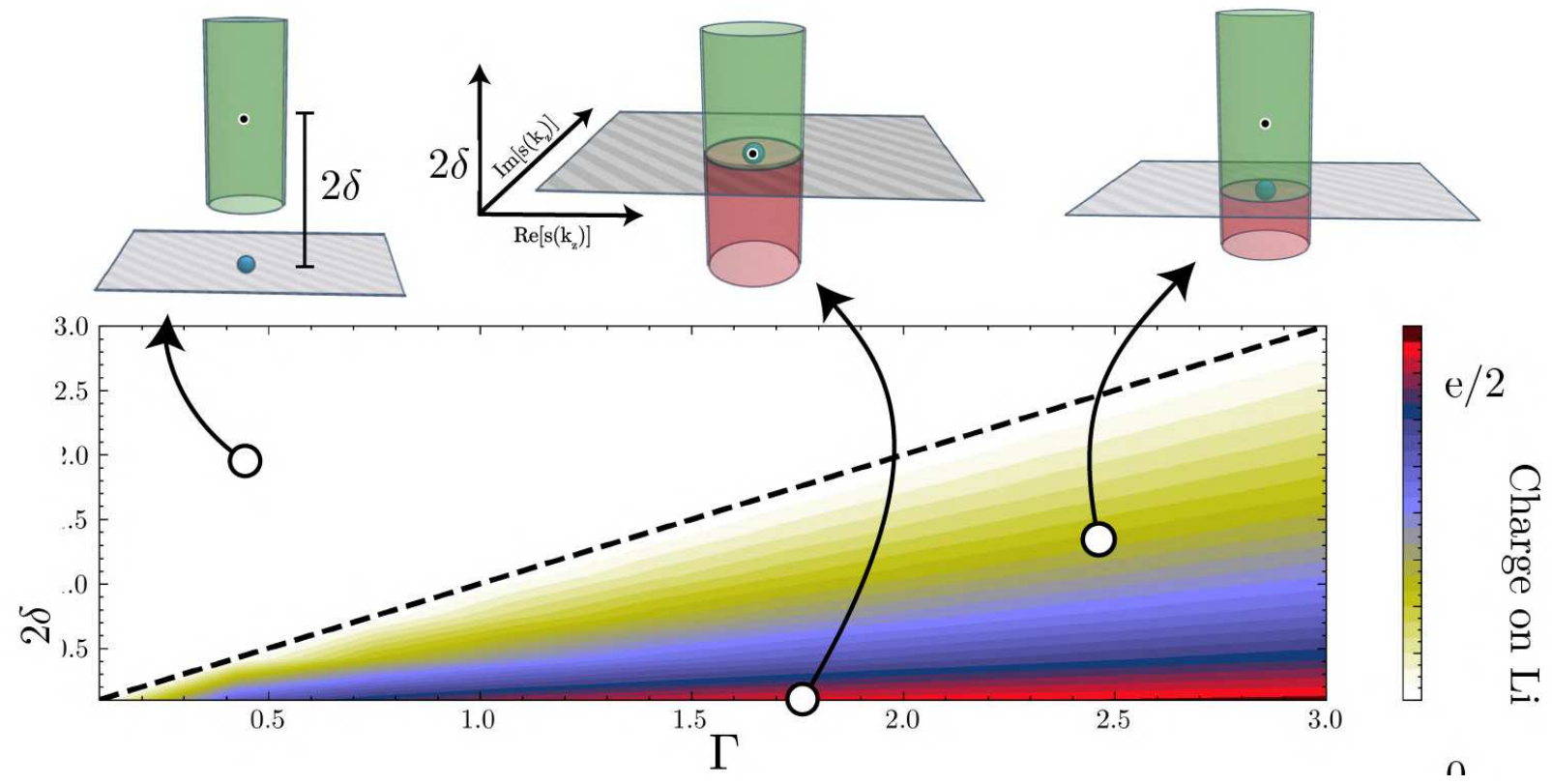

**Figure 8.3:** a) $LiCoO_2$ unit cell, 1D-TB model; b) Bulk band structure of 3D Hamiltonian; c) Band structure of finite slab *red* Li surface, *blue* opposite surface $CoO_2$; d) DFT band structure of 14nm slab. (Inset shows surface states near Fermi level.)

The amount of charge on the Li is determined by how much the bottom of the Li band overlaps with the top of the $CoO_2$ band. Assuming a steplike density of states (DOS) for each band near these band edges and parabolic free electron like bands, which makes sense near the band edges for a 2D system, we can easily determine the Fermi energy, located in the surface bands from the fact that the number of electrons in the upper band equals the number of holes in the lower band. Using the density of states for free electrons in 2D to be proportional to the effective mass in each band, and the proportionality of the inverse effective mass to the hopping parameter or bandwidth of the tight-binding model for that band, we find that $q_{Li}/q_{CoO_2} = (\Gamma - 2\delta)/(\Gamma + 2\delta)$ if $2\delta > \Gamma$ and zero otherwise.

One can easily see that if the splitting of the two surface band centers (which is $2\delta$) is larger than the sum of half their band widths then the charge will still all be localized on the $CoO_2$ and zero on Li. A full numerical calculation of the net surface charge density within the tight-binding model resulting from the overlapping bands is given in SM and shown to closely agree with the above approximate result (check Section 8.4 (Density functional theory results)). In the cylinder topological model, the part of the cylinder that dips below the surface corresponding to $\delta = 0$ gives the amount of charge on the Li side, again when assuming a constant DOS.

The above model is consistent with the facts from our DFT calculations presented in the SM, that for Na with larger in-plane interactions and hence larger 2D surface band width, a larger surface charge density is found than for Li on the surface. The same is true for Be which also has a smaller electronegativity difference and hence smaller $\delta$ in our model and, in fact gives an additional electron to the surface states. Finally, when two equal surface terminations are used then there is an overall inversion/mirror symmetry in the center of the slab and thus by symmetry requires that the additional electron in the surface states is spread equally over both sides. The same is true for holes for the case of two $CoO_2$ terminated surfaces. In the SM, we show furthermore that the occurrence of the surface states is related to the entanglement spectrum.[161–164]

In summary, we have shown that surfaces of $LiCoO_2$ finite slabs host topologically required surface states related to the SSH4 like nontrivial interlayer interactions of Li-$sp_z$ and $CoO_2$ bond-centered Wannier orbitals. As a result of strong lateral interactions, the Li related surface band can become partially occupied and host a spin-polarized 2DEG of fairly high electron density.

While several angular resolved electron spectroscopy (ARPES) and scanning tunneling microscopy studies (STM) have been published in the past [152, 165–168] for both $Li_xCoO_2$ and $Na_xCoO_2$ they were generally focused on the bulk rather than on the search for surface states, which may thus have been missed.

In Section 8.4, we start by discussing the various surface configuration of mono-layer $LiCoO_2$ including its energetics and band structures. We also show that this general phenomena is not restricted to just $LiCoO_2$ but can also be achieved in other layered $CoO_2$ systems including (but not exclusive to) (Na/Be)$CoO_2$. We then proceed to discuss the effect of in-plane interaction by removing the surface Li concentration and placing them in 1D rows. Finally we consider the effects of both local and non-local correlations beyond the local spin denstity functional approximation (LSDA).

Section 8.6 contains the information of the entanglement spectrum of the minimal SSH4 model introduced in the main text. We also discuss the quantitative agreement of the surface charge derived from topological arguments by numerically comparing it to a effective 2 band Hamiltonian. Finally, we elucidate the inversion symmetric termination case discussed in the main text where charge fictionalization of $\frac{e}{2}$ is guaranteed by symmetry

## 8.4 Density functional theory results.

**Li-location** In Fig. Figure 8.4 we show the changes in band structure for different surface locations of the Li. Their relative total energies are given in Table Table 8.1 . The lowest energy position **(1)** for Li is on top of the Co, position **(2)** is above the bottom O, **(4)** above the top layer O and **(3)** in the center of the 2D cell.

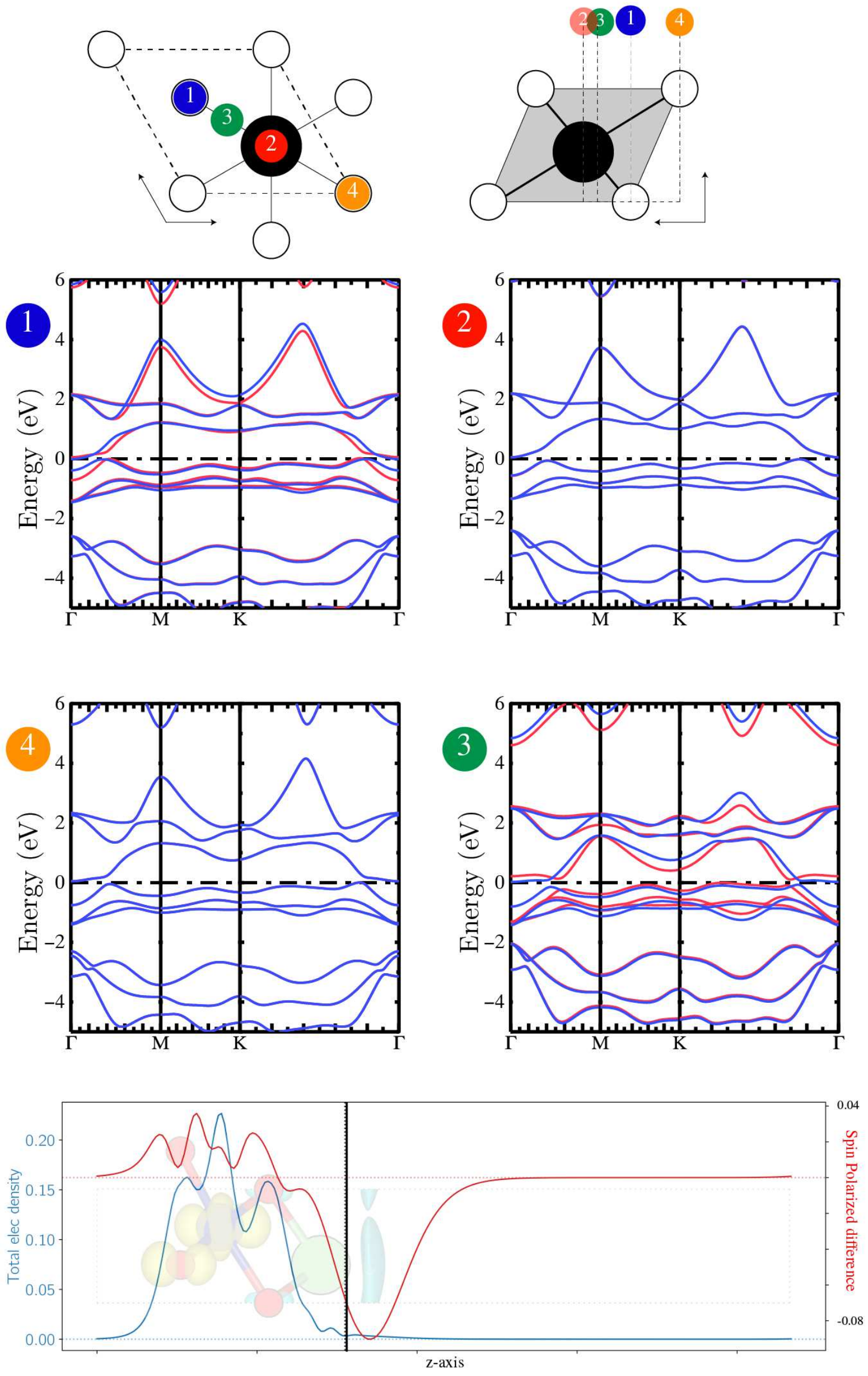

**Figure 8.4:** *(top)* top and side view of all 4 symmetric positions of Li on top of 2D $CoO_2$ lattice. *(middle)* Spin polarized LDA band structures for the corresponding structures. *(bottom)* Total electron density (blue) integrated along in-plane axis and spin difference $= n_{\downarrow} - n_{\uparrow}$ (red). Black line shows the real space limit for integrating the surface charge density

| Structure | 1 | 2 | 3 | 4 |
|---|---|---|---|---|
| Energy (eV/f.u.) | 0 | 0.09 | 0.32 | 0.81 |

**Table 8.1:** Relative energy of the structures in Figure 8.4

It is interesting to note that in structure **(3)** shown in Figure 8.4, the effective interaction is between Li and $CoO_2$ layer is mediated by O-$p$ instead of Co-$d$. This change is captured in the band structure by the lowering of the center of the *free-electron* like Li band compared to other cases. On the other hand, this is clearly not the lowest total energy.

Interestingly, we find negligible spin-polarization for both locations **(2)** and **(4)**.

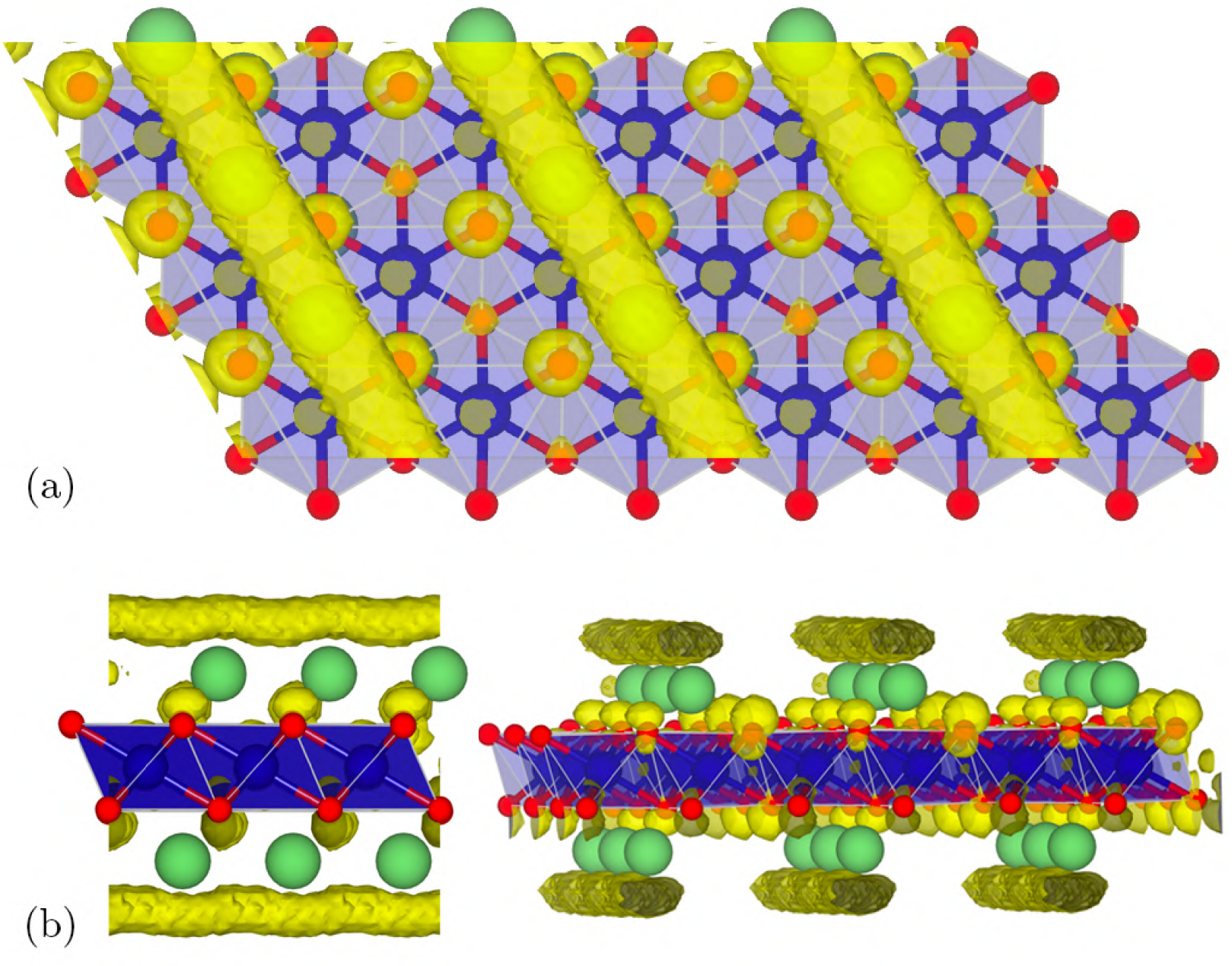

**Figure 8.5:** 2D electron density in case of $CoO_2$ monolayer with Li on either side but arranged in rows of Li, achieving one Li per $CoO_2$. yellow isosurface of the electron density correspond to $1.04 \times 10^{-5}$

At the bottom of Fig. Figure 8.4 we show the 2D planar averaged electron density and its spin polarization for location **(1)** set against the structure. The region over which we integrated the surface density is to the right of the vertical black line, which is placed just above the Li atom location. This integration gives about 0.25 $e$, while the total Bader charge associated with Li is 0.4 $e$. This shows that 62.5 % of it lies above the Li atom.

**Other systems** In Fig. Figure 8.6, we show the band structures for various other cases . Part (a) shows the case of a monolayer of $CoO_2$ compensated by one Li per Co but with Li arranged at half the surface density on each surface. The Li atoms then occur in rows. The corresponding electron density is shown in Fig. Figure 8.5. While showing some electron density just above the Li atom rows, it should be pointed out that this electron density is a factor 10 times smaller than for the full Li coverage. Correspondingly, we see that the surface bands related to Li do not dip below the Fermi level. Although centered at about the same energy, the band width of this surface band is now about 3 times smaller because the Li only have 2 neighbors (along the 1D rows) instead of 6 in the plane. This prevents this surface band to become occupied. However, from the color coding (red for the Li contribution to the band) we can see that the highest occupied band does contain some Li contribution and forms an electron pocket around Γ. The plot of the corresponding wave function modulo squared is what is shown in Fig. Figure 8.5. This Li row related 1DEG cannot be explained within the SSH4 based tight-binding model. It would require a more complete description of the Li-$CoO_2$ layer interactions.

Next, in Fig. Figure 8.6(b) we show the case of a fully Na covered $CoO_2$ monolayer with Co on one side. This is similar to the corresponding Li case discussed in the main part of the paper but show that with Na, the electron pocket around Γ is increased in size. This is consistent with the larger lateral interactions between Na. Finally, in Fig. Figure 8.6(c) we show the case of Be covered $CoO_2$. Compared to Li, we now overcompensate the $CoO_2$. In this case the electron density in the surface 2DEG is even larger but we also obtain a larger spin-splitting.

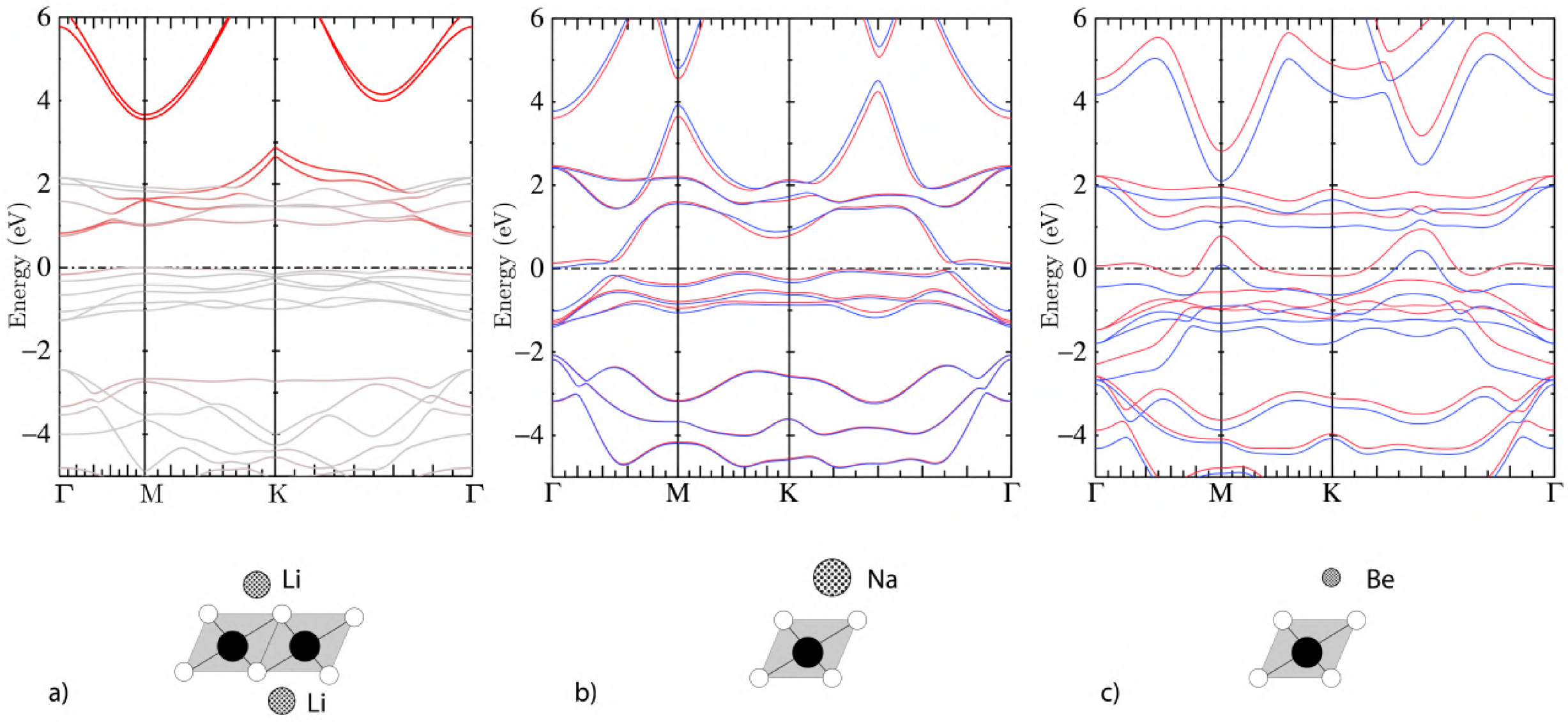

**Figure 8.6:** (a) $LiCoO_2$ mono-layer with the Li atoms forming 1D chain (Red color is the Li projected band); (b) $NaCoO_2$; (c) $BeCoO_2$ band structures (*red/blue* bands denote the majority and minority spins).

## 8.5 Correlation effects

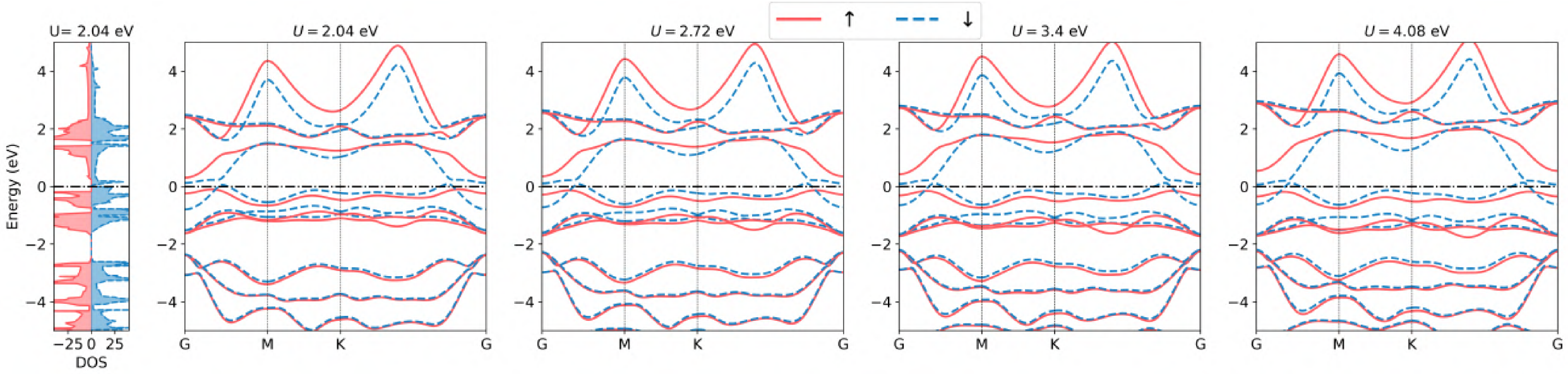

**Figure 8.7:** DOS and band structure of majority and minority spins for monolayer $LiCoO_2$ for various values of U

**Local correlations:** While pure bulk $LiCoO_2$, in which Li fully donates its electron to the $CoO_2$ layers, is clearly a non-magnetic band-insulator based on the electron count with a gap between octahedral filled $t_{2g}$ and empty $e_g$ bands, the incomplete electron donation from the surface Li to surface $CoO_2$ layers leads to a partially filled $d$-$t_{2g}$ band in the monolayer and $CoO_2$ surface regions of the finite slabs. Because of the strong on-site Coulomb interacton on Co-$d$ states, this may lead to correlation effects beyond those treated at the Local Spin Density Approximation (LSDA). In fact, the effect of partial $d$ filling is already manifested in LSDA leading to a spin-polarized solution. A complete treatment of spin correlations due to partial occupancy may require Dynamical Mean Field (DMFT), which is beyond the scope of this study. Here, we investigate whether the qualtitative band strucuture picture in terms of the occurrence of surface states holds up and to what extent the charge accumulated in the Li surface band 2DEG is robust to the inclusion of on-site Coulomb terms within the mean-field type LSDA+U method. To this end Figure 8.7 shows the DOS and band structure of monolayer $LiCoO_2$ for different values of $U$. We treat the LSDA+U in the simplest form [155] where a potential $V_{m\sigma} = U(\frac{1}{2} - n_{m\sigma})$ is added on orbital $d_{m\sigma}$ as function of its occupation number $n_{m\sigma}$. This shifts empty states up by $U/2$ and filled states down by $U/2$. Several initial density matrices or occupation numbers are tried out and the results give the one with the lowest total energy. First, one sees no major change in the qualitative picture regarding the non trivial surface bands in the presence of strong on-site Coulomb interaction. Secondly, it can be seen from the DOS that the cause for spin splitting arises from the avoidance of a high density of states at the Fermi level, in other words, the Stoner instability. A DOS peak occurs near the Fermi level because of the almost flat bands near Γ (here labeled G). It is important to note that the surface charge concentration is mainly determined by the overlap of the Li and $CoO_2$ surface bands, which in turn depends on the Li surface band width and the on-site potential difference between the effective Li and $CoO_2$ layers. The addition of an effective Hubbard $U$ to the local Co-$d$ states does not disrupt that potential difference (although there might be a slight effect because of Li-Co-O hybridization). This can be seen more quantitatively by calculating the Bader charge[156, 157] associated with the Li atom as function of the $U$ value. The Bader approach consists in partitioning the system into regions bounded by a "zero flux" surface, *i.e.* a 2D surface on which the real space charge density is a minimum in the direction perpendicular to the surface. The total charge inside the Bader surface surrounding each atom is then integrated and gives the Bader charge. While not a unique way of apportioning charge to atoms, it is a precisely defined way.

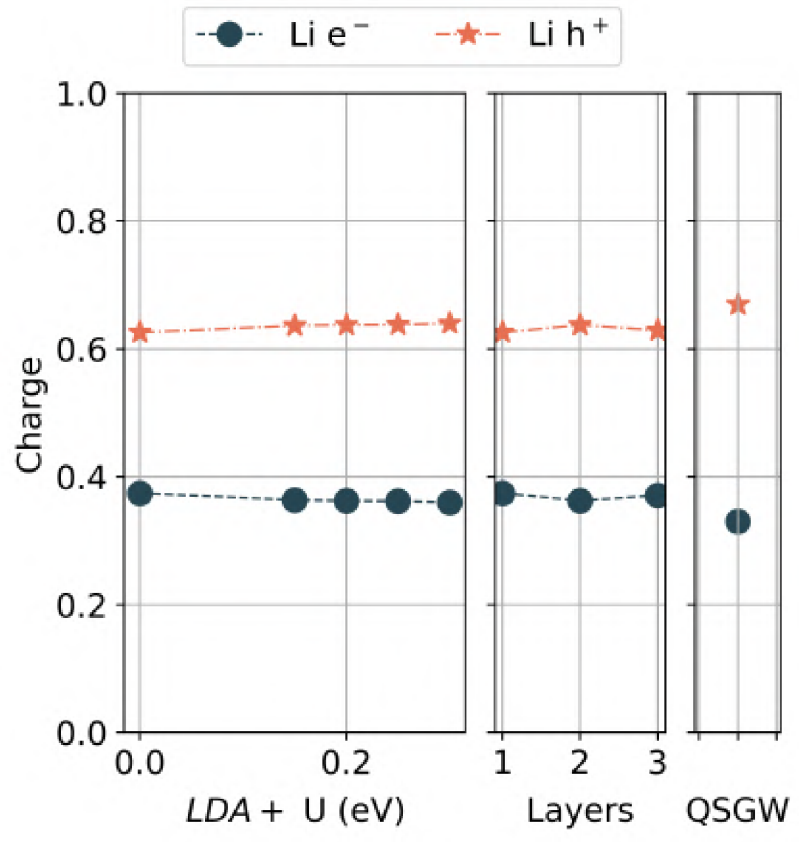

**Figure 8.8:** Electron and hole (1-electron) charge on Li atom for (a) various values of U at LDA+$U$ theory (b) LDA for various number of layers (c) QSGW

Figure 8.8(a) shows the electron and hole (1-electron) Bader charge for various values of $U$ on the Li atom. One can see that there is almost no change in the surface charge content on Li atom as function of $U$. Figure 8.8(b) also shows the surface charge dependence on the thickness (number of layers) of the finite size sample. Again, because this does not change the on-site potentials on Li vs. Co, no significant change in Li-surface 2DEG charge is observed.

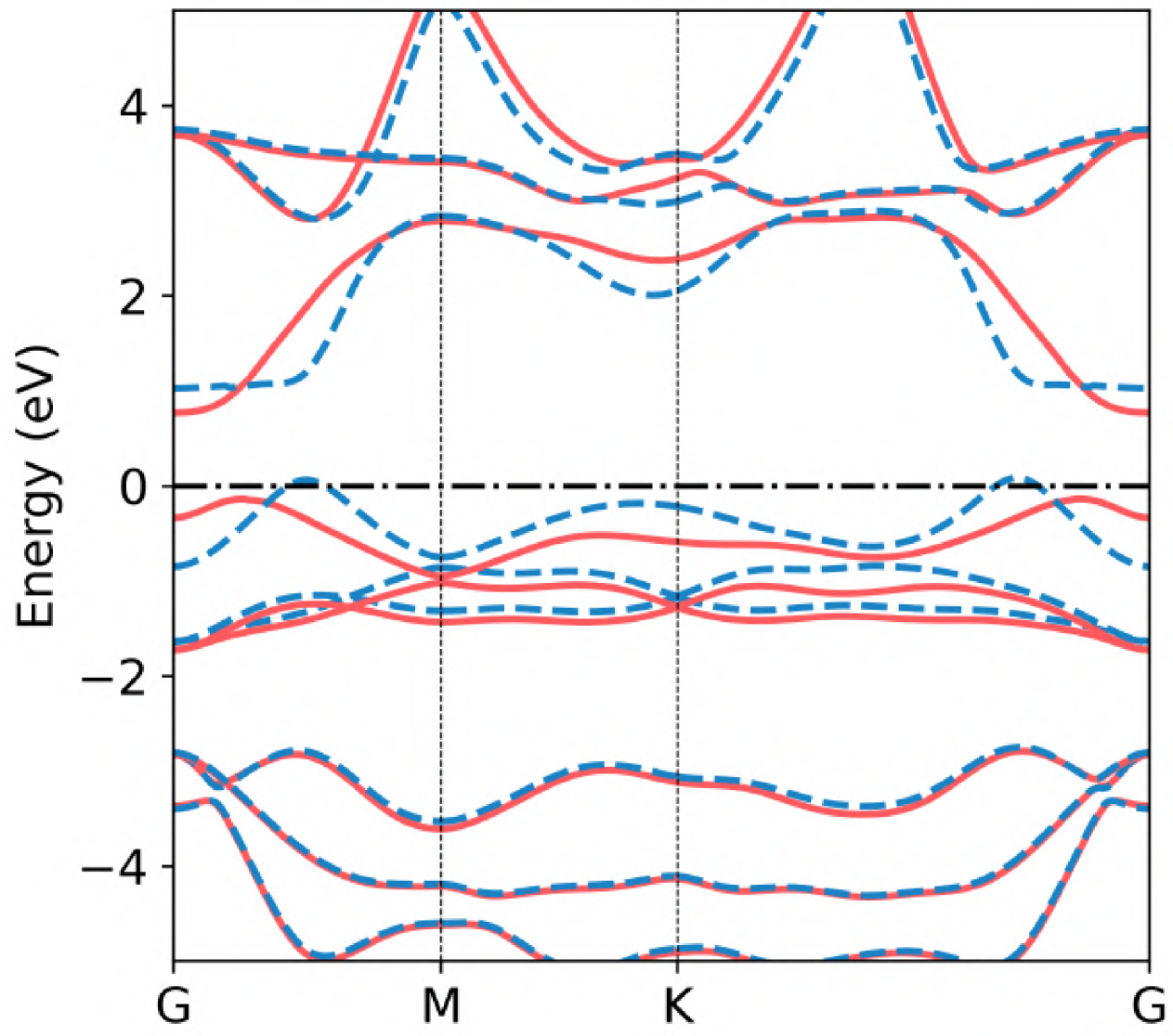

**Figure 8.9:** QSGW band structure of the monolayer $LiCoO_2$

**Non-local correlations:** In LSDA+U only onsite Coulomb terms are added on a particular orbital, the Co-$d$-orbitals and $U$ is essentially treated as an empirical parameter. To account for non-local screening effects, we also study the monolayer band structure using a many-body perturbation theory approach: the quasi-particle self-consistent (QS) $GW$ method,[**Kotani07**, 169] where $G$ and $W$ are the one-electron Green's function and screened Coulomb interaction respectively. In the QS$GW$ approach, the energy-dependent self-energy $\Sigma(\omega) = iG(\omega) \otimes W(\omega)$ ($\otimes$ means convolution) is replaced by a hermitian energy-averaged non-local exchange-correlation potential used in the independent particle Hamiltonian $H_0$ (defining the $G$ and $W$) and which is iterated to self-consistency. This means that the Kohn-Sham eigenvalues of the independent particle Hamiltonian $H_0$ become the same as the quasi-particle excitation energies of the many-body system and become independent of the starting independent particle Hamiltonian ($H_0$). We here choose the initial $H_0 = H_{LDA}$.

Again, qualitative features of the band structure (shown in Fig. Figure 8.9) remain intact as the occurrence of surface state is dictated by topology. Because of quasi-particle renormalization due to screening, slight changes in the surface charge is observed as shown in Figure 8.8(c).

**Spin-polarization for symmetric $CoO_2$ termination:** In Fig. Figure 8.10 we show the spin-polarized band structure of the symmetrically $CoO_2$ terminated monolayer $Li_9(CoO_2)_{10}$, which was shown without spin-polarization in the main text in Fig.1d. The origin of spin polarization is again that in this case partially filled Co-$d$ bands occur.

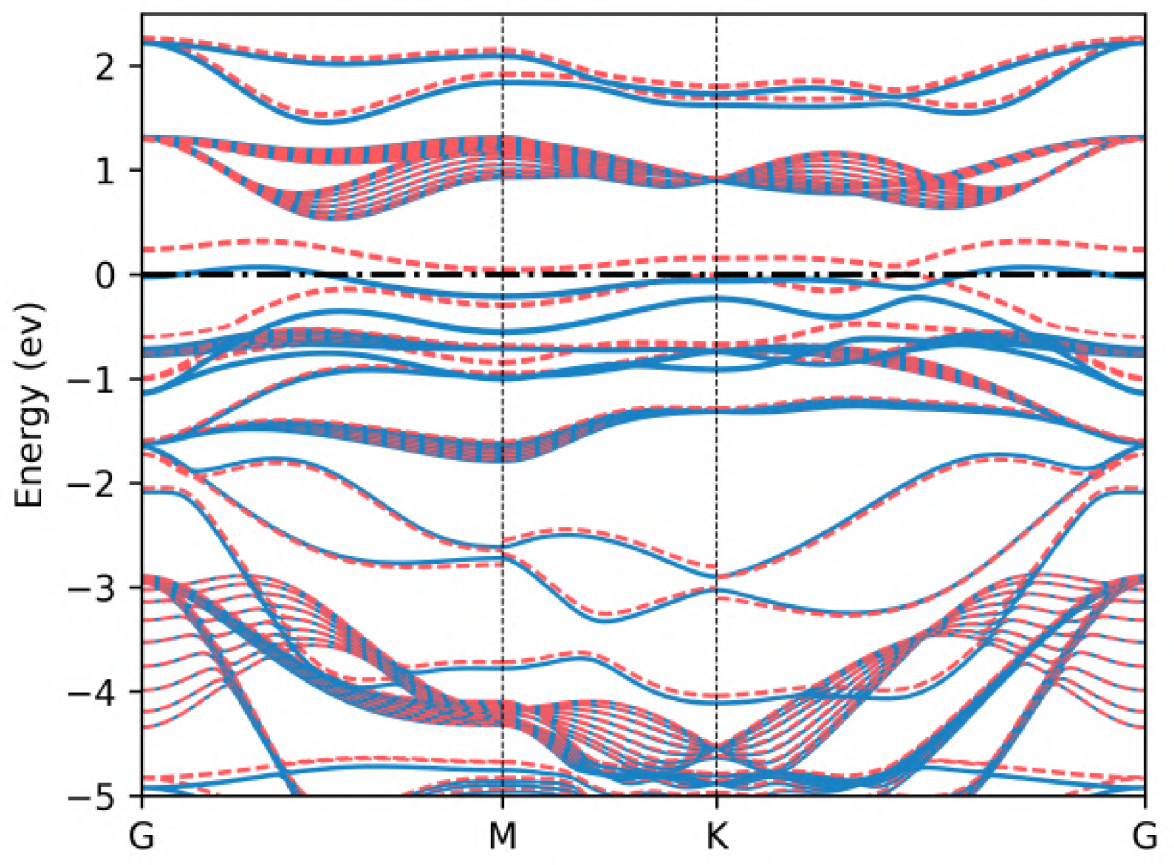

**Figure 8.10:** Band structure of $Li(CoO_2)_2$ symmetrically terminated slab, including spin-polarization.

## 8.6 Entanglement spectrum

To understand the surface states better, we use the idea of *entanglement spectrum* (ES)[170] which has been found to be a generally useful theoretical tool in investigations of topological states[161–164]. The main idea of the ES is that the eigenvalues of the hermitian correlation matrix of the occupied eigenstates, restricted to a subsystem A of the combined system (A+B), provide already information on the existence of surface states when the system would be split in separate A and B parts and of the topologically non-trivial nature of the system. In our case of non-interacting electrons, the correlation matrix is defined in terms of the Bloch functions expanded in the tight-binding basis set as follows. Although our system is periodic in $x$ and $y$ direction we here consider Bloch states only in one direction combined with the layer direction $z$ in which the non-trivial SSH4 topology applies. Let the eigenstates be $|\psi^n_{k_x}\rangle = e^{ik_x x}|u_{nk_x}\rangle = \sum_{j\alpha} e^{ik_x x}[u^n_{k_x}]_{j\alpha}|\phi_{j\alpha}\rangle$, where $i$ labels the sites, which can be either in the A or B part of the system and $\alpha$ labels orbitals per site, then the correlation matrix restricted to the A-subsystems is given by

$$C^A_{i\alpha,j\beta}(k_x) = \sum_n^{occ}[u^n_{k_x}]^*_{i\alpha}[u^n_{k_x}]_{j\beta}, \qquad \text{with } i \in A, j \in A \tag{8.3}$$

If we remove the restrictions on $i, j$ then we drop the superscript $A$. The eigenvalues of this correlation matrix $\xi(k_x)$ define the ES. If we would not include the restriction, this correlation matrix is built from idempotent projection operators and thus has eigenvalues 0 or 1 only. As shown in [164] and elsewhere, the existence of eigenvalues deviating strongly from 0 or 1, near 1/2 are an indicator of the entanglement of the states between its subparts and thus of the non-trivial topology and the existence of surface states.

The eigenvalues $\xi(k_x)$ of $C_{ij}(k_x)$ matrix at a given $k_x$ are shown in Figure 8.11(b) along with the 1D eigenvalues of our tight-binding model and clearly show the one-to-one correspondence between the ES containing eigenvalues near 1/2 with the existence of surface states. Furthermore we see that for a different choice of the SSH4 TB parameters, no surface

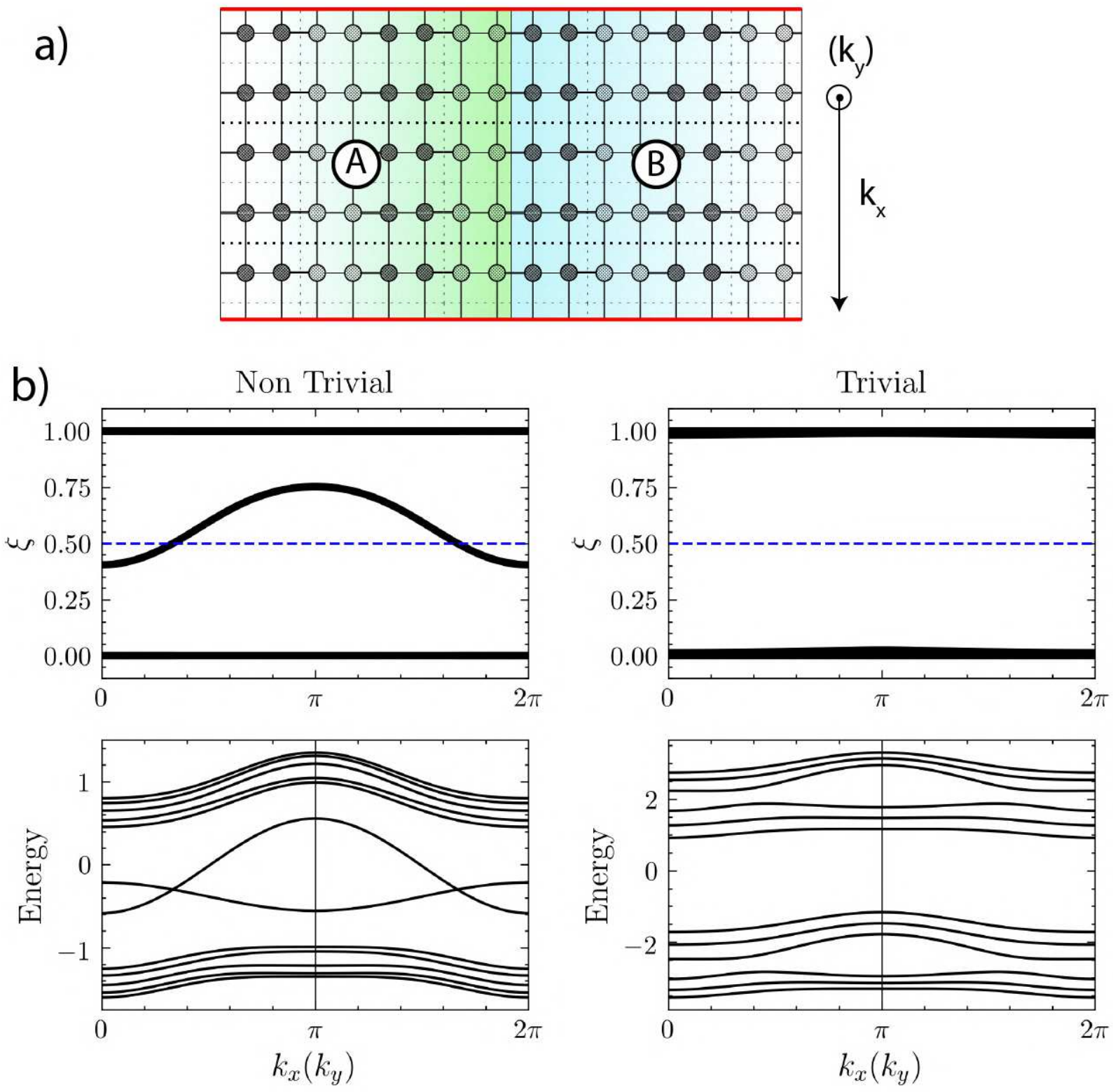

**Figure 8.11:** a) Partitioning of the system in two parts for calculating the entanglement spectrum of the reduced 2D system b) *(top)* entanglement spectrum of the 2D system along the cut *(bottom)* band structure of 1D ribbon of the corresponding 2D Hamiltonian showing the edge states for (left) topologically non-trivial and right trivial choice of SSH4 parameters.

states exist and correspondingly no ES with eigenvalues near 1/2.

The connection between $\xi$ and the previous topological picture in terms of cylinders in Fig. 3 of the main paper, is that for each $\mathbf{k}_{||}$ the eigenvalues $\xi > 0.5$ ($\xi < 0.5$) correspond to a loop above (below) the $z = 0$ plane in Fig. 3 of the main paper as indicated by the *green/red* color in the cylinders of Fig. 3(a). The $\xi = 0.5$ eigenvalue corresponds exactly to the $z = 0$ plane.

## 8.7 Surface charge calculation

In Fig. Figure 8.12 we show the results of a numerical calculation of the surface state occupancy as function of the energy separation of the two surface bands within the tight-binding approximation. We can see that it is in good agreement with the model results in the main paper. Note that beyond $\delta = 3$ here the overlap of the bands is zero and no charge occurs on the Li side. For $\delta = 0$ we are in the limit where the charge on the Li is 1/2 by symmetry.

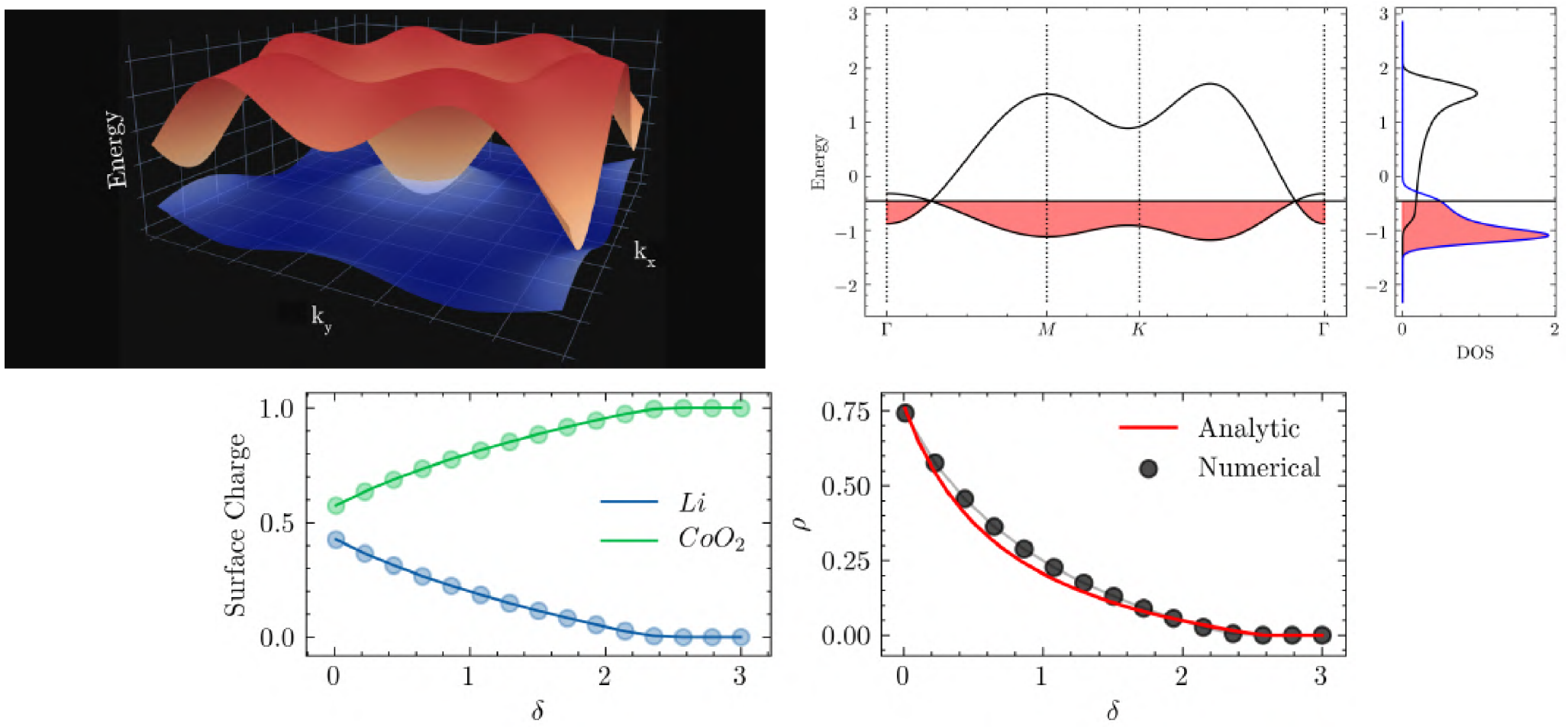

**Figure 8.12:** *(Top left)* Intersecting surface bands in TB model, *(Top right)* band structure of surface states and Density of States (DOS) contribution from each band (black on Li side, and blue on $CoO_2$ side) Surface states have a 2D dispersion corresponding to a triangular lattice with nearest neighbor hopping parameters given by $t^{xy}_{CoO_2} = 0.1$ and $t^{xy}_{Li} = -0.3$ in arbitrary units with an on-site offset of $\delta = 0.9$ (in units of $\Gamma/6 = (t^{xy}_{CoO_2} + t^{xy}_{Li})$), *(Bottom left)* Surface charge per unit cell area as function of on-site parameter $\delta$; *(Bottom right)* Ratio of charge on Li to charge on $CoO_2$ surface layers from numerical TB and analytic equation Eq.(6) derived in the main text. In the limit of $\delta = 0$, this ratio approaches 1.

# Conclusions and outlook 9

In this thesis, we focused mainly on the topological aspects of the electronic structure of group-V 2D honeycomb structures like antimonene and arsenene. We not only systematically unraveled the series of topological transitions as the system evolves from completely flat to the equilibrium buckled form of free-standing Sb but also predict it has unique transport properties associated with these topological states, in particular, the phenomenon of goniopolarity. We found that even in its semiconducting buckled ground state, the system has non-trivial topology when viewed within the recent theories of obstructed atomic limit. These recent topology based concepts were reviewed in an introductory chapter. These show that Sb and As monolayers are not only topologically non-trivial but in fact, higher order and quadrupole systems with protected corner states. We further used tight-binding models to demonstrate the robustness of these topological states when subject to disorder. Then, we generalized the physics of Dirac cone merging by showing they necessarily lead to obstructed atomic limit insulators and applied the concepts of symmetry indicators, Zak phase and entanglement entropy to this system. Finally, in a completely different system of layered $LiCoO_2$, we discovered the occurrence of surface states hosting a 2D electron gas on the Li terminated surfaces and hole gas on the $CoO_2$ terminated surfaces and showed that this also finds its origin in topology. We did this in particular by mapping the system to a 3D extension of the quadripartite SSH4 model.

As mentioned in introduction, one of the unsolved problem regarding $LiCoO_2$ $NaCoO_2$ systems is that $NaCoO_2$ is a superconductor with a $T_c$ of about 4K[171], but only in the presence of $H_2O$. Its superconductivity in general is assumed to be unconventional *i.e.*, the Cooper pairs are not in a spin singlet state with $s$-wave symmetry, as with conventional superconductors. Adding to the mystery, experimental reports are often contradictory (see for instance Ref.[172–177]) and solid evidence for any particular pairing state remains lacking. This has thus made $NaCoO_2$ the record breaker for number of superconducting mechanisms proposed from spin-triplet state of $p-$ or $f-$wave pairing symmetry[172], simple singlet ordering [174], spin-triplet $f$-wave [178], the exotic $d_1 + id_2$ state caused by spin-charge separation[179], Resonating Valence Bond (RVB) [180] and various nesting related mechanisms [181], to name a few. One should note that all these are related to the strong correlations due to partially filled $d$ orbital of the system. Because of the humble layered nature of this system, people often considered modeling the Hamiltonians as correlations in 2D triangular $d-$shell.

As an outlook toward future work, the same physics of spin-polarized surface states also occurs in 3D layered $LiCoO_2$ and $NaCoO_2$, when the system is put under tensile strain along the direction normal to the layers. Beyond a certain critical strain, the Li breaks its symmetry of being equally associated with two layers and then behaves as a Li-terminated surface. We have already found that this transition occurs at a lower strain in $NaCoO_2$ than in $LiCoO_2$. Moreover, we see that the c-axis lattice value at which this criticality occurs is almost the same as the c-axis in superconducting hydrated $NaCoO_2$. Since such a strain can among other be produced by inserting $H_2O$ intercalated between the $CoO_2$ layers, this finding may shed new light on the superconductivity in NaCoO2.$H_2O$. The occurrence of this 2DEG of electrons in each layer has until now not been taken into account in attempts to explain the superconductivity. We plan to investigate whether this can resolve some of the findings on this system that have until now eluded explanation even when using strongly correlated DMFT theory.

# Publications 10

- "Topological band structure transitions and goniopolar transport in honeycomb antimonene as a function of buckling." Santosh Kumar Radha, and Walter RL Lambrecht, *Physical Review B 101.23 (2020): 235111.*

- "Buckled honeycomb group-$V$-$S_6$ symmetric $(d-2)$ higher order topological insulators." Santosh Kumar Radha, and Walter RL Lambrecht, arXiv preprint *arXiv:2003.12656 (2020).*

- "Topological quantum switch and controllable quasi 1D wires in antimonene." Santosh Kumar Radha, and Walter RL Lambrecht, arXiv preprint *arXiv:2005.06096 (2020).*

- "Spin-polarized two-dimensional electron/hole gases on $LiCoO_2$ layers." Radha, Santosh Kumar, and Walter RL Lambrecht, arXiv preprint *arXiv:2003.00061 (2020).*

- "Topological Obstructed Atomic Limit by annihilating Dirac fermions" Santosh Kumar Radha, *Manuscript in preparation*.

- "Distortion modes in halide perovskites: To twist or to stretch, a matter of tolerance and lone pairs." Santosh Kumar Radha, Churna Bhandari, and Walter RL Lambrecht, *Physical Review Materials 2, no. 6 (2018): 063605.*

- "Understanding the Crystallographic Phase Relations in Alkali-Trihalogeno-Germanates in Terms of Ferroelectric or Antiferroelectric Arrangements of the Tetrahedral GeX3 Units". Santosh Kumar Radha, and Walter RL Lambrecht, *Advanced Electronic Materials 6, no. 2 (2020): 1900887.*

- "Band Gaps and Stability of CsSiX3 Halides." Santosh Kumar Radha, and Walter RL Lambrecht, *physica status solidi (a) 216, no. 15 (2019): 1800962.*

- "Ordering in the mixed ZnGeN 2-GaN alloy system: Crystal structures and band structures of ZnGeGa 2 N 4 from first principles." Jayatunga, Benthara Hewage Dinushi, Sai Lyu, Santosh Kumar Radha, Kathleen Kash, and Walter RL Lambrecht , *Physical Review Materials 2, no. 11 (2018): 114602.*

- "Tuning Rashba spin–orbit coupling in gated multilayer InSe." Premasiri, Kasun, Santosh Kumar Radha, Sukrit Sucharitakul, U. Rajesh Kumar, Raman Sankar, Fang-Cheng Chou, Yit-Tsong Chen, and Xuan PA Gao. *Nano letters 18, no. 7 (2018): 4403-4408.*

- "Electrical Characterization and Charge Transport in Chemically Exfoliated 2D $LixCoO_2$ Nanoflakes" Kyle Crowley, Kevin Pachuta,

Santosh kumar Radha, Halyna Volkova, Alp Sehirlioglu, Walter Lambrecht, Marie-Helene Berger, Xuan Gao *Manuscript under review*.

- "Stochastic differential theory of cricket." Santosh Kumar Radha, arXiv preprint *arXiv:1908.07372 (2019). Manuscript under review*.

- "Classical-Gravity as curvature in Manifold." Santosh Kumar Radha, *Manuscript Submitted*.

# Bibliography

References in citation order.